\documentclass[rnote]{aa} 

\usepackage[normalem]{ulem}
\usepackage{natbib,twoopt}
\usepackage[breaklinks=true]{hyperref} 
\bibpunct{(}{)}{;}{a}{}{,} 
\makeatletter
\newcommandtwoopt{\citeads}[3][][]{\href{http://adsabs.harvard.edu/abs/#3}%
{\def\hyper@linkstart##1##2{}%
\let\hyper@linkend\@empty\citealp[#1][#2]{#3}}}
\newcommandtwoopt{\citepads}[3][][]{\href{http://adsabs.harvard.edu/abs/#3}%
{\def\hyper@linkstart##1##2{}%
\let\hyper@linkend\@empty\citep[#1][#2]{#3}}}
\newcommandtwoopt{\citetads}[3][][]{\href{http://adsabs.harvard.edu/abs/#3}%
{\def\hyper@linkstart##1##2{}%
\let\hyper@linkend\@empty\citet[#1][#2]{#3}}}
\newcommandtwoopt{\citeyearads}[3][][]%
{\href{http://adsabs.harvard.edu/abs/#3}
{\def\hyper@linkstart##1##2{}%
\let\hyper@linkend\@empty\citeyear[#1][#2]{#3}}}
\makeatother

\usepackage{graphicx}
\usepackage{txfonts}
\begin{document}

   \title{Atlas of monochromatic images of planetary nebulae}

   \author{W. A. Weidmann\inst{1,2}\fnmsep\thanks{Visiting Astronomer, Complejo 
Astron\'omico El Leoncito operated under agreement between the Consejo Nacional de Investigaciones Cient\'ificas y 
T\'ecnicas de la Rep\'ublica Argentina and the National Universities of La Plata, C\'ordoba and San Juan.}, 
E. O. Schmidt\inst{3}, R. R. Vena Valdarenas\inst{3}, 
J. A. Ahumada\inst{1},
M. G. Volpe\inst{1} 
and A. Mudrik\inst{4}
          }

   \institute{Observatorio Astron\'{o}mico, Universidad Nacional de C\'{o}rdoba,
Laprida 854, 5000 C\'{o}rdoba, Argentina     
            \and
             Consejo Nacional de Investigaciones Cient\'ificas y T\'ecnicas (CONICET)\\
             \email{walter@oac.unc.edu.ar}
            \and
Instituto de Astronom\'ia Te\'orica y Experimental (IATE), Universidad Nacional de C\'ordoba,
 CONICET, Observatorio Astronómico de Córdoba, Laprida 854, Córdoba, Argentina
             \and
             Facultad de Matem\'atica, Astronom\'ia y F\'isica, Universidad Nacional de C\'ordoba
             }


 
  \abstract
   {We present an atlas of more than one hundred original images of planetary nebulae (PNe). 
    These images were taken in a narrow-band filter centred on the nebular emission of the [\ion{N}{ii}] 
    during several observing campaigns using two moderate-aperture telescopes, at the Complejo Astron\'{o}mico El Leoncito
(CASLEO), and the Estaci\'{o}n
Astrof\'{\i}sica de Bosque Alegre (EABA), both in Argentina.
The data provided by this atlas represent one of the most extensive
  image surveys of PNe in [\ion{N}{ii}].
We compare the new images with those available in the literature, 
and briefly describe
 all cases in which our [\ion{N}{ii}] images reveal new and interesting structures.
  }
    

  \authorrunning{Weidmann et al.}\space\maketitle


\section{Introduction}\label{intro}

Currently, about 3000 Galactic planetary nebulae (PNe) are known, which is 
more than twice as many as a decade ago; unfortunately, most of them still remain poorly studied. 
The PNe are highly representative of the latest stages of stellar evolution 
for stars of between 0.8 and 8 M$_\odot$; 
however, there are still many details of their evolutionary scheme 
such as the mass-loss
processes, which are not yet fully understood.
In this context,
an extensive and complete atlas of narrow-band images 
centred on different nebular lines is essential 
because these images
trace the distribution of ionized mass and also  
provide information about the processes of mass ejection involved in PN formation.

Clearly, it is important for any study about extended astronomical objects,
and in particular for a study of PNe,
to have access to a sample of narrow-band images that is
as complete as possible. 
However,
this type of work is time-consuming, which makes it an ideal project for medium-sized 
telescopes like those of the Complejo Astron\'omico El Leoncito (CASLEO) and the
Estaci\'on Astrof\'isica de Bosque Alegre (EABA).

Despite the importance of these narrow-band surveys, 
there are not many works of this type. 
In addition, only a few studies contain observations in [\ion{N}{ii}] and 
in general cover only
a small number of objects.
The most relevant previous catalogues containing the [\ion{N}{ii}]  narrow band that have been published are by  
\citet[50 objects]{1987AJ.....94..671B},            
\citet[243 objects]{1996iacm.book.....M},            
\citet[19 objects]{1996ApJS..107..255T},            
\citet[22 objects]{1998A&AS..133..361H},             
\citet[101 objects]{1999A&AS..136..145G} and            
\citet[18 objects]{2007ApJS..169..289H}.

Here we present an atlas of images comprising 108 PNe in the [\ion{N}{ii}] narrow band. 
The goal is to increase the amount
of observational data of these	amazing and highly complex objects.
We hope that this work (together with the previous ones) will  
become a useful database for future research.


\section{Sample, observations, and data reduction}

We have observed a sample of 108 true PNe selected from the catalogues of
\citet{1992secg.book.....A} and \citet{2005MNRAS.362..689P}.
The selected nebulae did not have previous observations in [\ion{N}{ii}],
except for a small group that was chosen to compare the quality of our data with that of previous works.
Our selection criterion requires that the angular size of the objects in the mentioned
catalogues be between 16  and 200 arcsec.
In this way, it should be possible to infer its morphology and, 
in some cases, to detect low-ionization regions (LIS).

\begin{figure}[ht!]
 \centering
  \includegraphics[width=0.4\textwidth]{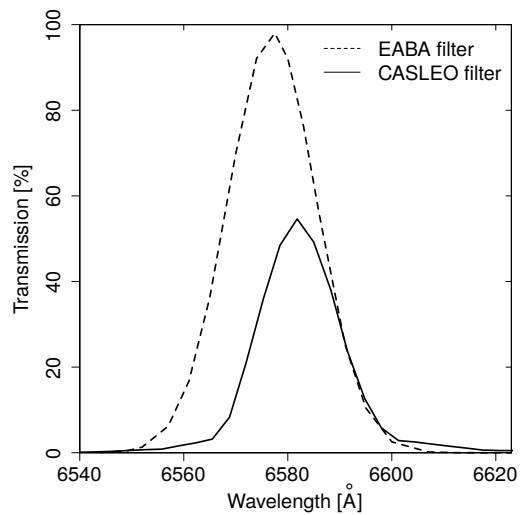}
   \caption{Effective transmission curves for the filters 
            (at 20 degrees Celsius).
}
    \label{Fig_fil}
\end{figure}

The observational data were obtained the 1.54 m telescope
at EABA (configured in the newtonian focus)
and the 2.15 m telescope at CASLEO. 
The imaging data for this survey were taken over the course of several
observing runs from 2013 to 2015, making a total of 7 nights at CASLEO and 30 at EABA.
The main features of the observational systems are described in Table~\ref{tab:star2}.
The transmission curves of the filters used are shown in Fig.~\ref{Fig_fil}.

In EABA, series of 20 consecutive six-minute exposures were obtained for most of the PNe.
On the other hand, 
the lower brightness objects were observed at CASLEO where, in
general, three exposures of ten minutes each were acquired.
Some objects, such as He~2-15, were purposely oversaturated to show the faintest regions of the nebula. 
Table~\ref{atlas} summarizes our sample of objects, all of which 
were observed at low zenith distance (z$<$40$^\circ$).

Image reduction was performed using standard procedures with IRAF\footnote{IRAF: the Image Reduction and 
Analysis Facility is distributed by
the National Optical Astronomy Observatories, which is operated by the
Association of Universities for Research in Astronomy, Inc. (AURA)
under cooperative agreement with the National Science Foundation (NSF).}. 
Thus, the data were first trimmed, bias-subtracted using
a series of ten averaged bias frames and dark-subtracted
(only in frames of EABA).
The data were then flatfielded
using a series of twilight flat fields.
After the reduction was complete, 
the individual frames were spatially aligned to a common system using the IRAF task \texttt{imalign}. 
Typically, four stars were used to obtain this transformation.
After that, all images were combined, cosmic rays removed, and 
the signal-to-noise  (S/N) improved
(IRAF task \texttt{imcombine}).
Finally the images were rotated so that north was
up and east to the left
(IRAF tasks \texttt{imtranspose} and \texttt{rotate}).

\begin{figure}[ht!]
 \centering
  \includegraphics[width=0.4\textwidth]{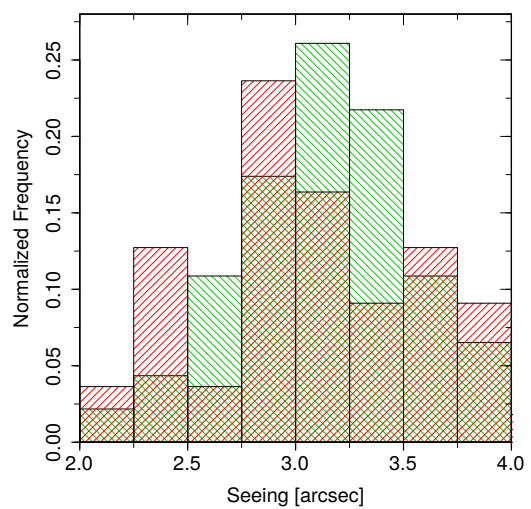}
   \caption{FWHM of stars in CASLEO images, measured
            in the final combined frame (green) and individual reduced images (red).
}
    \label{Fig_0}
\end{figure}

The sky conditions during each night varied greatly.
The combined effects of occasional poor telescope guiding and
variable dome and sky seeing resulted in the stellar FWHM
 ranging from 2$\arcsec$ to 5$\arcsec$, as illustrated in
Fig.~\ref{Fig_0} for CASLEO.

Moreover, since the final images suffered a 
slight degradation in quality due to the spatial
alining process, this effect is shown in Fig.~\ref{Fig_0}.
The combined images with a seeing greater than 4$\arcsec$ were rejected.
All the objects in the atlas have a $S/N>4$. 
More details about the images that were obtained at EABA are reported in
\cite{2015BAAA...57..197V}.

\begin{table}
\caption[]{Summary of the main features of the observational systems.
}
\label{tab:star2}
\centering
\setlength{\tabcolsep}{0.5mm}
\begin{tabular}{l|c|c}
\hline\hline\noalign{\smallskip}
                             & EABA                          & CASLEO                     \\
\noalign{\smallskip}\hline\noalign{\smallskip}
telescope size [m]           &  1.54                         &  2.15                      \\
f-ratio                      &  4.9                          &  8.6                       \\
CCD  brand                   &  Apogee                       & Roper                      \\
detector  size   [px]        &  3072 $\times$ 2048           &    2048 $\times$ 2048   \\
binning [px]                 &  3 $\times$ 3                 &  5 $\times$ 5              \\
pixel scale [arcsec/px]      &  0.74                         &  0.75                      \\
field of view [arcmin]       &  12 $\times$ 8                &  5.1 $\times$ 5.1          \\
central $\lambda$ [\AA]\tablefootmark{a}    &  6578                  &  6582                      \\
bandpass filter [\AA]\tablefootmark{a}      &  20                    &  18                         \\
filter transmission  $T_{\lambda6585}$   & 0.92              &  0.53                   \\
zero-point $k_0$ [mag]       & $-3.38\pm0.10$                &  $-1.77\pm0.03$            \\               
ext.\ coeff.\ $k_1$ [mag/airmass]   &   $-0.08\pm0.04$       &  $-0.11\pm0.02$           \\
lim.\ magnitude [mag~arcmin$^{-2}$] &   12.1                  & 12.6            \\
  \hline
\end{tabular}
\tablefoot{The default gain for each instrument was used in all runs.  
\tablefoottext{a}{see Fig.~\ref{Fig_fil}.}}
\end{table}

\subsection{Photometric calibration and detection limit}

To assess the detection limit at each observatory, 
two additional nights were devoted to the observation of two planetary
nebulae and of a number of spectrophotometric
standard stars to first derive the extinction coefficients for the filters and
 the sensibility of the instrumental systems.
The  observed nebulae are \emph{i)} the faint PN~G242.5$-$05.9 (on the night of 15--16
October 2015 from CASLEO), and \emph{ii)} the brighter PN~G002.7$-$52.4 (IC~5148, on the night
of 17--18 October 2015 from EABA). During each night, several 
secondary and tertiary spectrophotometric standard stars from
\citet{hamuy94} were also observed at airmasses 
in the ranges 1.01--1.88 (CASLEO) and 1.01--1.41 (EABA). Exposures
for the stars reach from a few to 180 seconds. Several images of 600  seconds
 of PN~G242.5$-$05.9 at  airmasses of $\sim$1.01,
and of 360 seconds at $\sim$1.15 for PN~G002.7$-$52.4 were
obtained and combined to improve the S/N relation.

Aperture photometry was performed on nebulae and stars with 
 the package \textsc{daophot} in IRAF. 
For standard stars, apertures with radii
 of four times the FWHM of the stellar images were adopted without
 recourse to aperture correction. For PN~G242.5$-$05.9 an aperture of 
 30 arcsec of radius centred on  
 $\alpha = 07^\textup{h}26^\textup{m}04\fs9$,
 $\delta = -28\degr58'23\farcs6$ (J2000.0)
 was used, and for PN~G002.7$-$52.4 an 
 aperture of 79 arcsec of radius centred on 
 $\alpha = 21^\textup{h}59^\textup{m}35\fs1$,
 $\delta = -39\degr23'07\farcs1$ (J2000.0)
 was chosen. 
In the first case,
 four stars that appear inside the nebula and a fifth one
 very near its apparent limits were successfully subtracted
 before performing the photometry.  
 No star was subtracted from IC~5148.

The transformation equations are of the form:
\begin{equation}
 M(\lambda6585) = m(\lambda6585) + k_0 + k_1\times X,
\end{equation}
where $M(\lambda6585)$ is the calibrated monochromatic magnitude
 at $\lambda6585$,
$m(\lambda6585)$ is the instrumental magnitude
 with an arbitrary zero~point of 21.0, $k_0$ and
$k_1$ are the zero-point and extinction coefficients,
 and $X$ is the airmass. It is worth mentioning that
 the system sensitivity, including the effects of telescope
 optics and detector response, and the filter 
 transmission at $\lambda6585$ 
 ($T_{\lambda6585}$,  listed in Table~\ref{tab:dosnebs}),
 are included in $k_0$.
The values of 
 $M(\lambda6585)$
 for the observed spectrophotometric
standards were obtained by interpolating in
 Tables~6 and~7 of \citet{hamuy94}
and are listed in Table~\ref{tab:flest}.
The derived coefficients
and their errors, as given by \textsc{daophot}, are given in
Table~\ref{tab:star2}. 
 The monochromatic magnitudes of the nebulae are shown in Table~\ref{tab:dosnebs};
 these are related to the flux density $f_\nu$ at $\lambda6585$ 
 by 
\begin{equation}
M(\lambda6585) = -2.5 \times \log_{10} (f_\nu) - 48.590,
\end{equation}
\citep{hamuy94}, where  $f_\nu$ is expressed 
in erg~cm$^{-2}$~s$^{-1}$~Hz$^{-1}$. 
This flux density has to be
multiplied by the factor $c/(6585)^2$ ($c$ is the speed of light)
 to give it as $f_\lambda$, that is, in the 
units erg~cm$^{-2}$~s$^{-1}$~\AA$^{-1}$.
 After propagating the errors adequately,
the flux densities of the nebulae
and their logarithms are those listed in  Table~\ref{tab:dosnebs}.

\begin{table}
\caption[]{Adopted magnitudes at  $\lambda6585$  of spectrophotometric standards.}
\label{tab:flest}
\centering
\setlength{\tabcolsep}{0.5mm}
\begin{tabular}{l c}
\hline\hline\noalign{\smallskip}
Star & $M(\lambda6585)$ \\
\noalign{\smallskip}\hline\noalign{\smallskip}
EG 21 & 11.936 \\
HR 718 & 4.611 \\
HR 1544 & 4.634 \\
HR 9087 & 5.414 \\
LTT 377 & 11.106 \\
LTT 1020 & 11.320 \\
LTT 2415 &  12.125 \\
  \hline
\end{tabular}
 \end{table}

\begin{table}
\caption[]{Calibrated magnitudes and fluxes at $\lambda6585$ of two planetary nebulae.}
\label{tab:dosnebs}
\centering
\setlength{\tabcolsep}{0.5mm}
\begin{tabular}{l|c|c}
\hline\hline\noalign{\smallskip}
 & PN~G002.7$-$52.4 & PN~G242.5$-$05.9   \\
 & (EABA) & (CASLEO) \\
\noalign{\smallskip}\hline\noalign{\smallskip}
 $M(\lambda6585)$ & $8.78\pm0.04$ & $14.20\pm0.12$ \\
 $f_\lambda$\tablefootmark{a} & $(7.8\pm0.3)\times10^{-13}$ 
  &$(5.2\pm0.6)\times10^{-15}$   \\
 $\log f_\lambda$ & $-12.11\pm0.04$ & $-14.3\pm0.1$ \\
 surface brightness\tablefootmark{b} &  $10.36\pm0.05 $ &  $12.56\pm 0.14$\\
 surface brightness\tablefootmark{c} & $(5.0\pm0.6)\times10^{-17}$ & $(6.6\pm1.0 )\times10^{-18}$  \\
  \hline
\end{tabular}
\tablefoot{\tablefoottext{a}{erg~cm$^{-2}$\!~s$^{-1}$\!~\AA$^{-1}$.}
 \tablefoottext{b}{mag~arcmin$^{-2}$.}
\tablefoottext{c}{erg~cm$^{-2}$\!~s$^{-1}$\!~\AA$^{-1}$arcsec$^{-2}$.}
}
 \end{table}

The surface brightnesses were calculated 
by adopting for circular shapes and areas of 0.217~arcmin$^2$
 (PN~G242.5$-$05.9) and 4.313~arcmin$^2$ (PN~G002.7$-$52.4)
for the nebulae.
For the first, a surface brightness of 12.56~mag~arcmin$^{-2}$ at a
 S/N~$\sim$~4 for an exposure time of 600~seconds is estimated, while
 for IC~5148 the value is 10.36~mag~arcmin$^{-2}$ with S/N~$\sim$~20 and
 360~seconds. These brightnesses, expressed in magnitudes 
 per squared arcmin and also in
  erg~cm$^2$~s$^{-1}$~\AA$^{-1}$~arcsec$^{-2}$, are listed in Table~\ref{tab:dosnebs}. 
 While the surface brightness of the faint nebula observed at CASLEO can
be adopted as the limiting magnitude  for the images
taken there, IC~5148 is rather
bright, therefore we divided its flux  by 5 to obtain a
S/N relation of $\sim4$ like that measured in a combination 
 of 360-second exposures
 of the faint  nebula Hf~39 (Figure~\ref{Fig_11}).
 The magnitude calculated with this new flux, 12.1~mag~arcmin$^{-2}$,
can now be considered as the limiting magnitude 
 in the images obtained at EABA for this survey.

 It must be pointed out, finally, that the transformation derived in this
 section is in general not applicable to our collection of images,
 since observing on photometric nights was not a requirement for
 the original purposes of this work.


\section{Considerations about the survey}

\subsection{Narrow-band filters}

When interference filters are used in a converging beam, the primary effect is 
a shift of the transmittance peak towards shorter wavelengths
\citep{1969ApOpt...8..227E}.

Another important phenomenon that affects our observations is the bandpass shift 
due to variations of the ambient temperature \citep{1992ASPC...23..195P}.
It moves towards the red or the blue, depending on whether
the temperature increases or decreases, respectively,
following a linear temperature dependence.
For a narrow-band filter, the shift in mean wavelength due to temperature
is about 0.18\AA\ per degree Celsius.
An additional problem is posed by the age of filters: 
after 6-10 years they usually deteriorate, mainly losing transmittance.

On the other hand, we have to take into account the radial velocity of the objects.
The radial velocity of a PN could cause the [\ion{N}{ii}] emission to falls 
outside of the effective bandpass of the filter.
Unfortunately, many objects included in our atlas do not have radial velocity determinations,
in particular the objects extracted from the MASH catalogue \citep{2005MNRAS.362..689P}.

The filter peak transmittance was determined at a temperature of 20$^\circ$C, 
which means that
those objects that were observed from EABA in winter
(where the temperature may drop to 5$^\circ$)
and with positive radial velocity
might be contaminated by  H$\alpha$ emission.
These objects are labelled in Table~\ref{atlas}, where 
we used the radial velocity data from \citet{1998A&AS..132...13D}
as a reference.

Finally, it is necessary to consider another possible nebular emission in this wavelength range.
Fortunately, there are only very few nebular lines 
(\ion{C}{ii} at 6578.05 and 6582.88\AA)
that could contaminate our data.
Nevertheless, contamination by continuous nebular emission
will always be present.


\subsection{Comparison with other observations}

Our work has objects in common with the catalogues of
\cite{1987AJ.....94..671B} and \cite{1998A&AS..133..361H}.
These objects are labelled in Table~\ref{atlas}.
With the former, the only nebula in common is NGC~2610,
which we observed from EABA. 
In this case the Balick images are deeper.

Our atlas includes ten PNe that were observed by \cite{1998A&AS..133..361H}. 
The comparison in this case is more difficult, since 
Hua used three different telescopes and we used two.
However, the images of those PNe that were observed from CASLEO 
are of a quality and depth comparable to those obtained by \cite{1998A&AS..133..361H}
(e.g. He~2-70 and K~1-3).


\section{Notes on individual objects}\label{nio}

PN G002.7$-$52.4: ring-like PN with
 an inner highly elongated elliptical ring with several irregularities in the brightness
and an external round fainter nebulosity.
More images can be found in
\cite{2010MNSSA..69..229S},
\cite{1991PASP..103..275M}, and
\cite{1987ApJS...64..529C}.

PN G003.1$+$02.9:  \cite{1993A&A...279..521S} classified this PN as irregular.
We disagree with this classification because
in our image we can see an elliptical structure with prominent jets.
More images can be found in
\cite{2009A&A...505..249M},
\cite{2000RMxAC...9..201L},
\cite{1998ApJS..117..341Z},
\cite{1997ApJ...487..304H},
\cite{1997ApJ...485..697L}, and
\cite{1996A&A...313..913C}.

PN G013.8$-$02.8: bipolar PN.
More images can be found in \cite{1987A&AS...69..527S}.

PN G014.8$-$25.6:
faint nebulosity with no obvious symmetry.

PN G019.4$-$13.6: middle elliptical according to the morphological scheme of
\cite{1989IAUS..131...83B}.
More images can be found in \cite{2003RMxAA..39..149B}.

PN G043.5$-$13.4: 
PN with no clear symmetry, roughly elliptical with a nearly rectangular shape.
More images can be found in
\cite{1992A&AS...96...23S}.

PN G118.8$-$74.7:  a bright and well-studied object.
\cite{2003MNRAS.344..501P} classified this PN as round.
Our image shows that it has an elliptical shape with
an inner filamentary structure and intense patches to the SE.
We suggest that the object needs to be reclassified.
Images of this object are found in
\cite{2013ApJ...771..114P},
\cite{2009AJ....138..691C},
\cite{2007ApJ...670..442H},
\cite{2007JBAA..117..204M},
\cite{2003ApJ...594..874S}, and
\cite{2003RMxAC..15...84S}.

PN G209.1$-$08.2: 
in the MASH images the object is very faint and does not reveal any morphology.
In our image it shows a clearly round morphology, with non-uniform brightness.

PN G215.6$+$11.1:
curious morphology, nebula with a nearly rectangular shape.
Possible bipolar PN.

PN G217.2$+$00.9: probable interaction with the ISM.

PN G224.3$-$03.4: probable interaction with the ISM.

PN G225.4$+$00.4:
intriguing morphology, it displays a non-concentric multishell, similar to the 
 mysterious rings of supernova 1987A.

PN G226.7$+$05.6: 
all authors agree that the object shows a bipolar morphology.
\cite{1993A&A...276..463S} classified it as bipolar with an outer structure.
More images can be found in
\cite{2000ApJ...544..889H},
\cite{1998A&A...332..721P},
\cite{1995A&A...293..871C},
\cite{1992A&A...264L...1S},
\cite{1992A&AS...96...23S}, and
\cite{1993A&A...278..247C}.

PN G233.0$-$10.1: ring-like PN.
More images can be found in
\cite{1987A&AS...69..527S}.
 
PN G236.0$-$10.6: object poorly studied, morphology
type one-side \citep[see][]{2000Ap&SS.271..245A}
with radial filaments. More images can be found in
\cite{1987A&AS...69..519H}.

PN G236.7$-$01.6: our image shows two areas of lower brightness 
that are not visible in the H$\alpha$ image of the MASH catalogue.

PN G239.6$-$12.0: round with internal structure and intense patches.
More images can be found in \cite{1984A&AS...56..325F}.

PN G239.6$+$13.9:
\cite{1989IAUS..131...83B} classified this PN as elliptical (early-middle subtype).
Additional images can be found in \cite{1992A&AS...96...23S}.

PN G243.8$-$37.1:
\cite{2012A&A...545A.146B} performed a morphological analysis of this PN.

PN G247.5$-$04.7: 
there is an HII region towards the SW, and it is unclear whether it is part of the PN or
is the optical counterpart of the molecular cloud BRAN~63.

PN G249.3$-$05.4:
ring-like PN with a very faint CSPN. It shows
several irregularities in the brightness of the ring.

PN G249.8$+$07.1:
this seems to be two overlapping PN. 
It is difficult to explain the homogeneous and extensive overdensity of material towards the NW.

PN G250.4$-$01.3: classical bipolar PN with an arc towards the NW. It is
also visible in the [\ion{N}{ii}] images of \cite{2000PASP..112..542K}.

PN G257.5$+$00.6: poorly studied
planetary nebula with intriguing morphology. The PN displays a ring-like structure
of low surface brightness towards the NE and a emission region towards the SSE that is difficult to explain.
It probably is a bipolar PN. This object requires a more detailed study.

PN G258.0$-$15.7: \cite{2002RMxAC..13..119G} show an [\ion{N}{ii}] image, but it is not as deep as ours.
In both cases the huge jet-like features are clearly visible.
Additional images can be found in
\cite{1996A&A...313..913C} and
\cite{1999ApJ...523..721C}.

PN G259.1$+$00.9: 
it is difficult to perform a morphological classification of this Type~I PN.
A complete analysis of this object was made by \cite{2014A&A...562A..89J}. 
Our [\ion{N}{ii}] image does not show any
differences with the [\ion{N}{ii}]+H$\alpha$ published in that article.
A H$\alpha$ image is presented by \cite{1999A&AS..136..145G}.

PN G260.7$-$03.3: the H$\alpha$ image of MASH catalogue displays a more extended HII region than our image.
According to its morphology, it would not be a PN. 
It could be a filament of the supernova remnant (SNR) Puppis~A
\citep{1978A&A....62..283G}.

PN G261.6$+$03.0: 
\cite{1995A&A...293..871C} classify this Type~I PN as a possible/probable bipolar.
Our [\ion{N}{ii}] image shows a well-defined bipolar morphology with a pronounced waist,
thus we confirm this classification.
It is not evident from our image if the two saturated regions define the torus of the waist.
If this were the case, the waist would not be
perpendicular to the bipolar lobes. 
To answer this question, it is necessary to perform a detailed spectroscopic study.

PN G261.9$+$08.5: this Type~I PN has been classified by
\cite{1983IAUS..103..233P} as a filamentary bipolar.
More images can be found in
\cite{2012ApJ...751..116V},
\cite{2003RMxAA..39..149B} and
\cite{1984ApJ...287..341D}.

PN G262.6$-$04.6: 
it presents a roughly round shape with two intense oval patches.
Additional images can be found in \cite{1980MNRAS.193..521L} and
\cite{1992A&AS...96...23S}.

PN G263.3$-$08.8:
this object does not have published narrow-band images.
Our image shows a round morphology with a dark band crossing the core, with a PA of 135$^\circ$.
Additional images
 can be found in \cite{1984A&AS...56..325F}.

PN G264.1$-$08.1: well-defined elliptical nebula with a high
surface brightness and an evident central star.

PN G265.1$-$04.2: ring-like PN with some knots on the edge.
It is unclear if the off-centre bright star is indeed its ionizing source.
Additional images can be found in \cite{1984A&AS...56..325F}.

PN G265.4$+$04.2:
round PN with two diametrically opposed condensations. Probable bipolar PN.

PN G268.6$+$05.0: 
object with no obvious symmetry.
It displays an oval ring-like internal structure
with opposite faint ansae.
It might be a bipolar PN.

PN G268.9$-$00.4:  
this fascinating object has the appearance of a steering wheel.
There is a nebulosity in the NW that is not visible in the MASH H$\alpha$ image.
It is unclear if it belongs to the nebula.
The central bright star could be the ionizing source.
This object deserves a more detailed study.

PN G274.6$+$03.5: round PN with a bright bar along its diameter.
Additional images can be found in \cite{1999A&AS..136..145G}.

PN G276.1$-$03.3: object that probably has a double envelope.

PN G277.1$-$03.8:
\cite{1989IAUS..131..179L} have classified this PN as an evolved bipolar, but
\cite{1987A&AS...70..201L} have classed it as peculiar.
It has an overall bow-shock morphology, which was also noted
by \cite{1999A&A...347..169R}.
Perhaps the gaseous envelope has interacted with the ISM.
A detailed analysis of this object was presented by \citep{1991A&A...241..526L}.
More narrow-band images can be found in 
\cite{2014MNRAS.440.2036D}, 
\cite{1998A&A...332..721P},
\cite{1995A&A...293..871C}, and
\citep{1991A&A...241..526L}.

PN G277.7$-$03.5: 
a round morphology with a complex ionization structure was revealed in our image.
This object deserves a more detailed study.

PN G278.5$-$04.5: this object has been very poorly studied.
Our image shows an elliptical envelope
with an internal structure of point-symmetric type.
In addition, we observe two pairs of beams in the E-W 
direction, similar to those of the Egg nebulae.

PN G279.1$-$03.1: roughly elliptical with two knots diametrically opposed.

PN G279.6$-$03.1:  
bipolar \citep{1995A&A...293..871C}, intense knots are superimposed at each end of the minor axis.
In addition, it has a pair of arcs that make it an S-shaped nebula \citep{2007AJ....133..987L}.
A detailed study of this object has been made by \cite{1993A&A...273..247C}.
More images can be found in
\cite{2007ApJ...665..341L}, 
\cite{1998A&A...332..721P} and 
\cite{1992A&AS...96...23S}.

PN G280.0$+$02.9: round or point-like nebula.
We do not detect any outer structure.

PN G285.5$-$03.3: 
strange morphology for a PN. Perhaps it is indeed the galaxy
LEDA~2792457 and/or LEDA~2792455.

PN G286.3$-$00.7: \cite{2007AJ....134..846S} showed a [\ion{N}{ii}]+H$\alpha$ narrow-band image of this PN.
It is a clear ring PN.

PN G286.3$+$02.8: ring-like PN with a bright condensation to the SW.

PN G288.4$-$01.8: bipolar, it seems that its waist is a torus.

PN G288.4$+$00.3: it probably has two symmetrical jets at PA 100$^\circ$.

PN G288.9$-$00.8: possible WR star of population I \citep{1994A&A...281..833S}.
Additional images can be found in 
	\cite{1987A&A...182..229S} and
	\cite{2014MNRAS.440.1080F}.

PN G290.1$-$00.4: clear bipolar morphology with an bright waist.
Moreover, in our image two ansae perpendicular to the minor axis are evident,
which are not clear in the image published by  \cite{1998A&AS..133..361H}.

PN G290.5$+$07.9: the central star is visible.
Well-studied object \citep[see][]{1996A&A...307..225P, 1993A&A...267..194L},
deeper images can be found in \cite{2012Sci...338..773B}.

PN G292.6$+$01.2: very bright PN with a complex internal structure. 
Bipolar appearance with a faint external shell.
More images can be found in \cite{1999A&AS..136..145G}.

PN G293.6$+$01.2: possible bipolar, with a very brilliant waist.

PN G293.6$+$10.9: well-defined irregular disc with a central hole.
A double-shell structure can be seen.
Aditional images can be found in \cite{1999A&A...347..169R}.

PN G294.6$+$04.7:  very complex object that has been studied in detail.
It shows jets and knots \citep{1999ApJ...523..721C},
an extended halo \citep{2003MNRAS.340..417C,2004A&A...417..637C}, and a
bow-shock feature \citep{2013A&A...557A.121G}.
Additional narrow-band images can be found in
\cite{1992A&AS...96...23S} and \cite{1991ApJ...377..210R}.

PN G294.9$-$00.6:
round morphology, an exterior elliptical structure may be present (PA=0$^\circ$).
The central region is probably obscured by dust.

PN G296.6$-$20.0: elliptical morphology that resembles the Owl nebula.

PN G298.2$-$01.7: \cite{1995A&A...293..871C} classified it as a probable or possible bipolar PN. 
Our image allows us to confirm this classification.
More images can be found in \cite{1999A&AS..136..145G}.

PN G299.0$+$03.5: roughly elliptical with a bright waist. The central star is evident.

PN G299.0$+$18.4: ring-like with a diffuse double-shell; the central star is visible.
Additional narrow-band images can be found in \cite{1992A&AS...96...23S}.

PN G302.6$-$00.9: object poorly studied even though it displays a curious morphology.
The PN exhibits an elliptical ring-like  main structure.
It seems to have a pair of symmetric jets in PA=135$^\circ$.

PN G304.2$+$05.9: a ring-like object with a strange arc towards the NW.
This is a very poorly studied object, it requires a more detailed analysis.

PN G307.2$-$03.4:
PN with a puzzling morphology and a very hot [WO1] central star.
\cite{1983A&A...117...33P} reported five sets of ansae, but
\cite{1984MNRAS.206...71R} described this situation as a ring of condensations.
\cite{1986ApJ...301..772Z} classified this PN as irregular (with condensations and filaments), but
\cite{1995A&A...293..871C} classified it as a probable or possible bipolar PN and
\cite{1990A&A...232..184P} as a late butterfly.
Finally, \cite{2012RMxAA..48..165S} performed a detailed study of this object and classified it
as a quadrupolar nebula with multiple sets of symmetrical condensations.
Moreover, a spherical halo is evident in the [\ion{O}{iii}] image presented by \cite{1998A&AS..133..361H}. 
In our image we can distinguish two mass-loss events.
The first is an irregular structure of intense knots, and the other is
an internal and discontinuous bar.
More images can be found in 
\cite{1999A&AS..136..145G},
\cite{1997A&A...318..571P},
\cite{1994A&AS..107..481P}, and
\cite{1987A&A...174..243G}.

PN G309.0$-$04.2: we agree with the morphological description of
  \cite{1993A&A...279..521S}, who classified it as an elliptical with internal structure.
\cite{1991PASP..103..275M} reported a nebulosity to the NW that we do not see.
More images can be found in \cite{1989ApJS...70..213K} and \cite{1992A&AS...96...23S}.

PN G309.2$+$01.3: this PN is interacting with the ISM \citep{2012A&A...541A..98A}.

PN G310.7$-$02.9: narrow-band images are provided in
\cite{1999A&AS..136..145G}  and \cite{1991PASP..103..275M}.

PN G310.8$-$05.9: round PN.
More narrow-band images are shown in \cite{1980MNRAS.193..521L}.

PN G311.7$-$00.9: bipolar, the waist looks like a torus. 
The central star is evident.

PN G315.0$-$00.3: the H$\alpha$ image from \cite{1999A&AS..136..145G}  shows a condensation 
towards the NW that is not visible in our image.
Other images are presented by 
\cite{1991PASP..103..275M},
\cite{1989A&A...223..277M}, and
\cite{1987RMxAA..14..520M}.

PN G317.1$-$05.7:  
the MASH H$\alpha$ image shows an external envelope that is not visible in ours.
On the other hand, the H$\alpha$ image \citep{1991PASP..103..275M} does not 
reveal the internal structure that appears in our [\ion{N}{ii}] image.
In the [\ion{O}{iii}] image \citep{1999A&AS..136..145G} none of these structures are visible.
Other narrow-band images can be found in \cite{1987RMxAA..14..520M}.

PN G317.8$+$03.3: poorly studied PN that presents
 a ring-like inner structure with several irregularities in the brightness.
A diffuse nebulosity is superimposed at each end of the minor axis.

PN G318.3$-$02.0: all authors agree with the classsification of this PN as bipolar.
More images can be found in
\cite{1999A&AS..136..145G}, 
\cite{1995A&A...293..871C}, 
\cite{1991PASP..103..275M},
\cite{1987A&AS...70..201L}, and
\cite{1987RMxAA..14..520M}.

PN G318.3$-$02.5: well-defined elliptical nebula with two intense patches.
More images can be found in
\cite{1999A&AS..136..145G},
\cite{1991PASP..103..275M}, and
\cite{1987RMxAA..14..520M}.

PN G318.4$-$03.0: ring-like PN.

PN G319.6$+$15.7: 
 Type~I PN with a [WR] core.
 This PN, according to its appearance, is a typical bipolar, but
\cite{1986ApJ...301..772Z} described this object as a
disc of approximately uniform brightness, elliptical shape, and ansae.     
\cite{1998A&AS..133..361H} reported some internal structures.
Several images of this PN have been taken in multiple works, for example
\cite{2002AJ....123.3329O},
\cite{1998ApJS..117..341Z},
\cite{1995A&A...293..871C}, and 
\cite{1992A&AS...96...23S}.

PN G321.8$+$01.9:
ring-like type with a clear ansae to the W. Moreover, two symmetric condensations towards N and S
are visible. We classify this object as a PE according to the
classification scheme of \cite{1997A&A...318..256G}.
On the other hand, this PN has been classified as elliptical by \cite{2003MNRAS.344..501P}.
Additional narrow-band images can be found in \cite{1999A&AS..136..145G}.

PN G322.4$-$02.6: 
in the [\ion{N}{ii}]$+$H$\alpha$ image published by \cite{1998A&A...329..683M},
the knots at the edge of the object towards the NW are not visible, 
but they are clear in our [\ion{N}{ii}] image and may be LIS.
H$\alpha$ and [\ion{N}{ii}] images shown by \cite{1992A&AS...96...23S} are overexposed.
They do not show 
the complex internal structure that is evident in our [\ion{N}{ii}] image.

PN G327.8$-$07.2: the
H$\alpha$ image shows an elliptical ring of a faint nebulosity \citep{1991PASP..103..275M}.
Our [\ion{N}{ii}] image is similar to the [\ion{O}{iii}] image of \cite{1991PASP..103..275M}, 
showing a disc-like PN.
In addition, a faint nebulosity to the N and S of the inner region
is visible in both images.

PN G328.9$-$02.4: round PN,
the diffuse nebulosity to the S that we see in our image is not visible in the [\ion{O}{iii}] images shown by
\cite{1999A&AS..136..145G}.
It may be an ultracompact HII~region \citep{1997MNRAS.291..261W}.
There is a young stellar object candidate near to this object,
reinforcing the hypothesis of an HII~region.

PN G329.3$-$02.8: the
H$\beta$ image \citep{1999A&AS..136..145G} does not show the two arcs 
(N and S) visible in our [\ion{N}{ii}] image.
This structure  could be a LIS.
The central star is not at the geometric centre,
probably because of an interaction with the ISM.

PN G331.5$-$03.9: elliptical shape with internal structures.
Additional images can be found in \cite{1999A&AS..136..145G}.

PN G332.0$-$03.3: round shape with several irregularities in the brightness.
Additional images can be found in \cite{1999A&AS..136..145G}.

PN G334.3$-$09.3: \cite{1997A&A...318..256G} classified this object as a bipolar PN, subclass BE, and
we agree with this classification.
On the other hand, \cite{1993A&A...279..521S} classified it as elliptical with outer structure.
The H$\alpha$ image published by \cite{1992A&AS...96...23S} shows a round halo.

PN G340.8$+$10.8: 
diffuse nebulosity, elliptical shape.
Two symmetric arcs are easily discernible superimposed at the edge of the nebula (PA=70$^\circ$).
This feature is not evident in the H$\alpha$ image published by 
\cite{1999A&AS..136..145G}.

PN G341.8$+$05.4:
our image is different from the H$\alpha$ image published by 
\cite{1999A&AS..136..145G}, perhaps because of the seeing effects.
It is difficult to perform a morphological classification, although it displays an 
elliptical shape \citep{1997A&A...318..256G}.
We can distinguish in our image an internal structure.
More images can be found in 
\cite{2000MNRAS.312..585L} and \cite{1986A&A...155..397P}.

PN G342.7$+$00.7: it is probably a bipolar PN.

PN G342.9$-$02.0: the
H$\alpha$ image \citep{1999A&AS..136..145G} displays a double-shell nebulosity.
We can see 
in our [\ion{N}{ii}] image that there is an arc towards the S, still far away from the nebula.

PN G342.9$-$04.9: this object shows a 
morphology similar to PM$~$1-242 \citep{2009AJ....137.4140M}.
Additional images can be found in \cite{1999A&AS..136..145G}.

PN G345.4$+$00.1: ring-like internal structure, with an external faint shell, 
intense condensations are superimposed towards the SE, and 
we can distinguish a knot towards the NE in our image.
\cite{1993A&A...279..521S} classified it as elliptical with an internal structure,
\cite{1997A&A...318..256G} classified it as a probable elliptical.
Additional images can be found in \cite{1992A&AS...96...23S}.

PN G346.9$+$12.4:
taking into account the peculiar morphology of this bright nebula, it is surprising
that it has been very poorly studied.
It consists of a bright  ring-like internal structure with
intense knots that are superimposed at each end of the minor axis.
In addition, this object has an outer shell structure that is not
perpendicular to the former.
\cite{1998A&AS..133..361H} suggested that it is a bipolar PN, but we
are unable to confirm this assertion.

PN G352.9$+$11.4: round PN with a [WCL] core. 
The morphology of this objects together with
its brilliant CSPN make it similar
to a nebula around a WC star of population~I.
Additional images can be found in
\cite{1992A&AS...96...23S}.

PN G353.7$-$12.8:
this object could be the galaxy IRAS~18232$-$4031
according to its spiral morphology.


\section{Results}

\begin{figure*}[!ht]
  \centering
\includegraphics[width=0.42\textwidth]{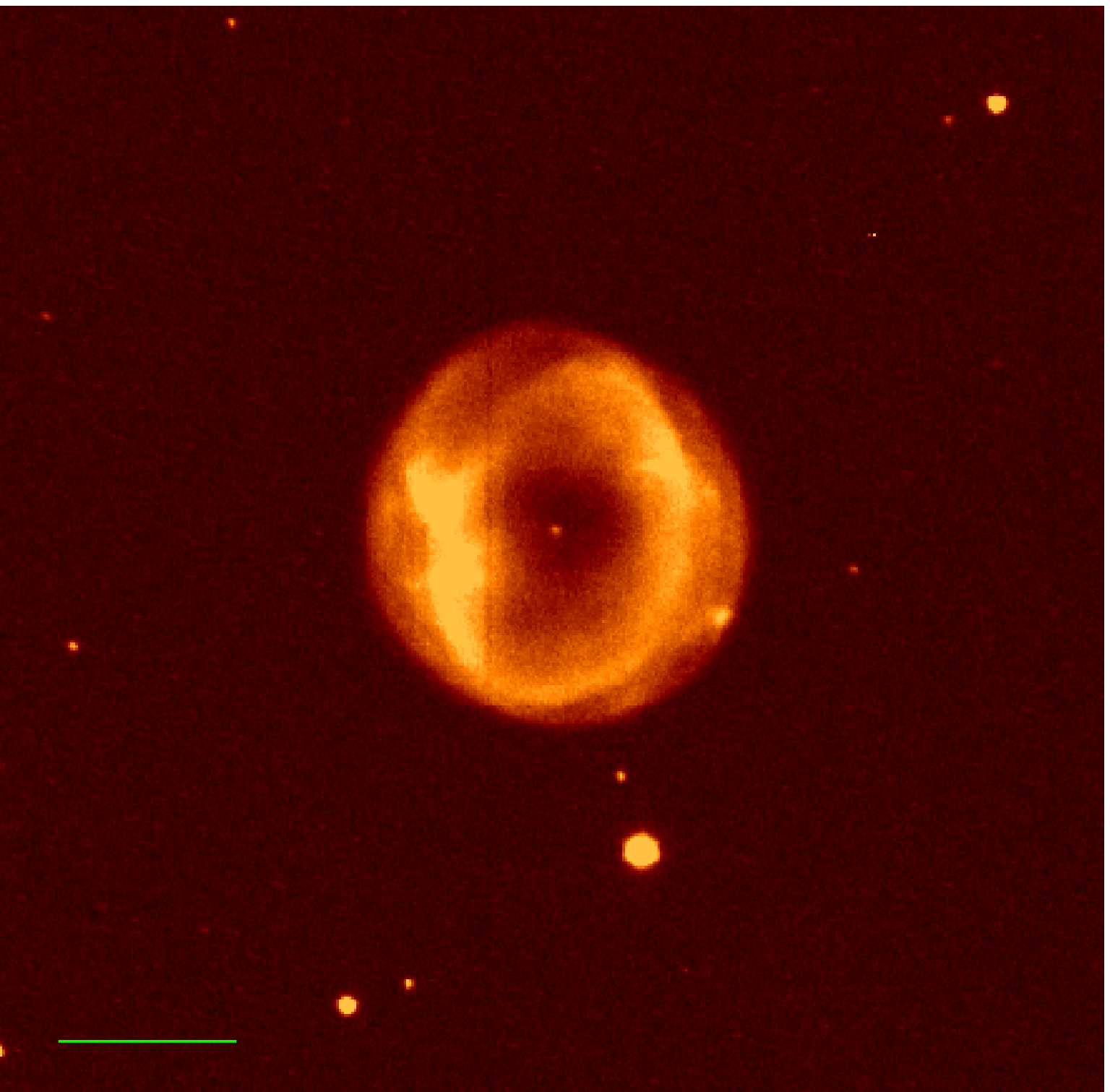}       \hspace{0.8cm}
\includegraphics[width=0.42\textwidth]{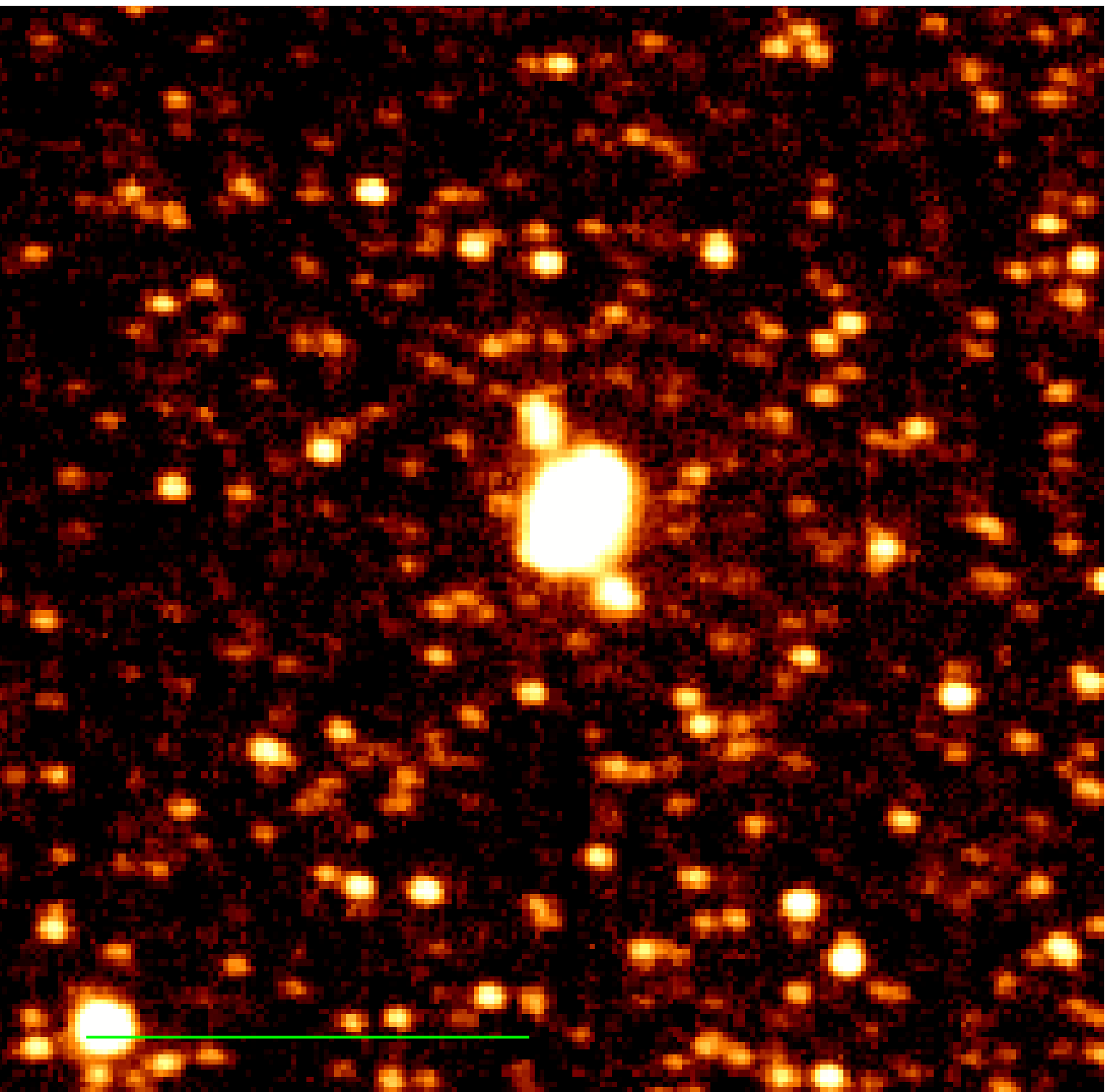}      \vspace{0.25cm}\\
\includegraphics[width=0.42\textwidth]{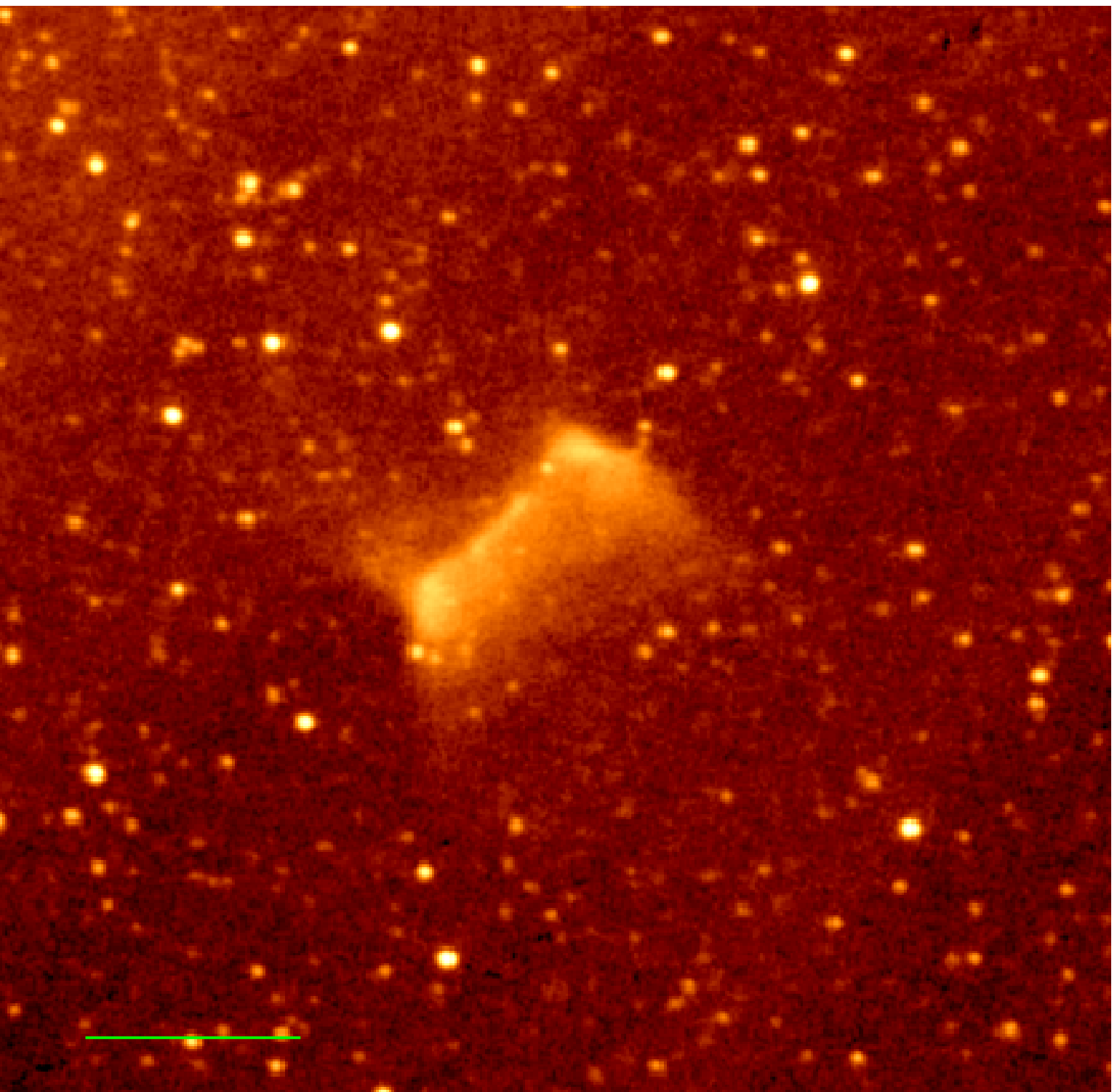}             \hspace{0.8cm}
\includegraphics[width=0.42\textwidth]{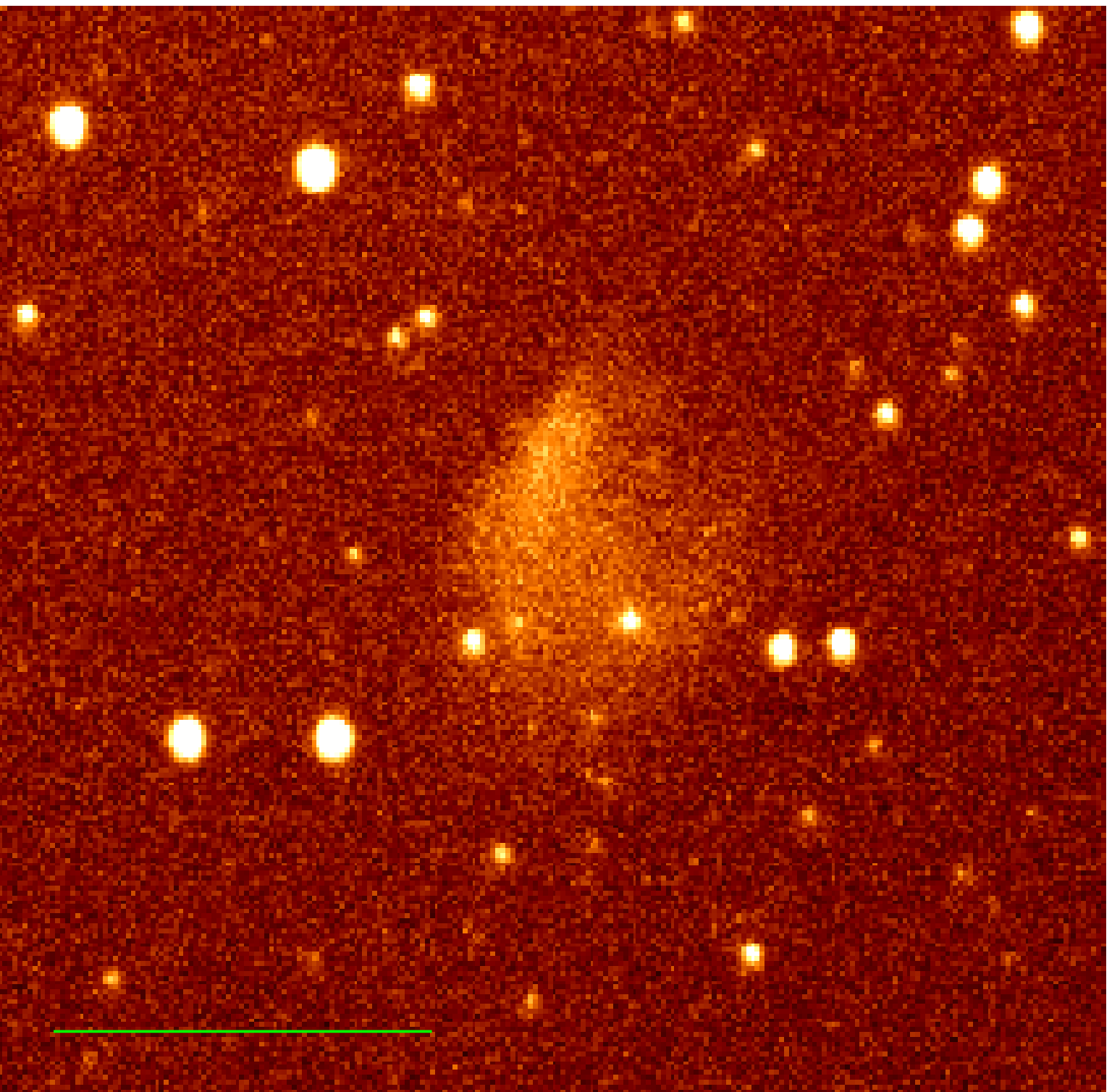}             \vspace{0.25cm}\\
\includegraphics[width=0.42\textwidth]{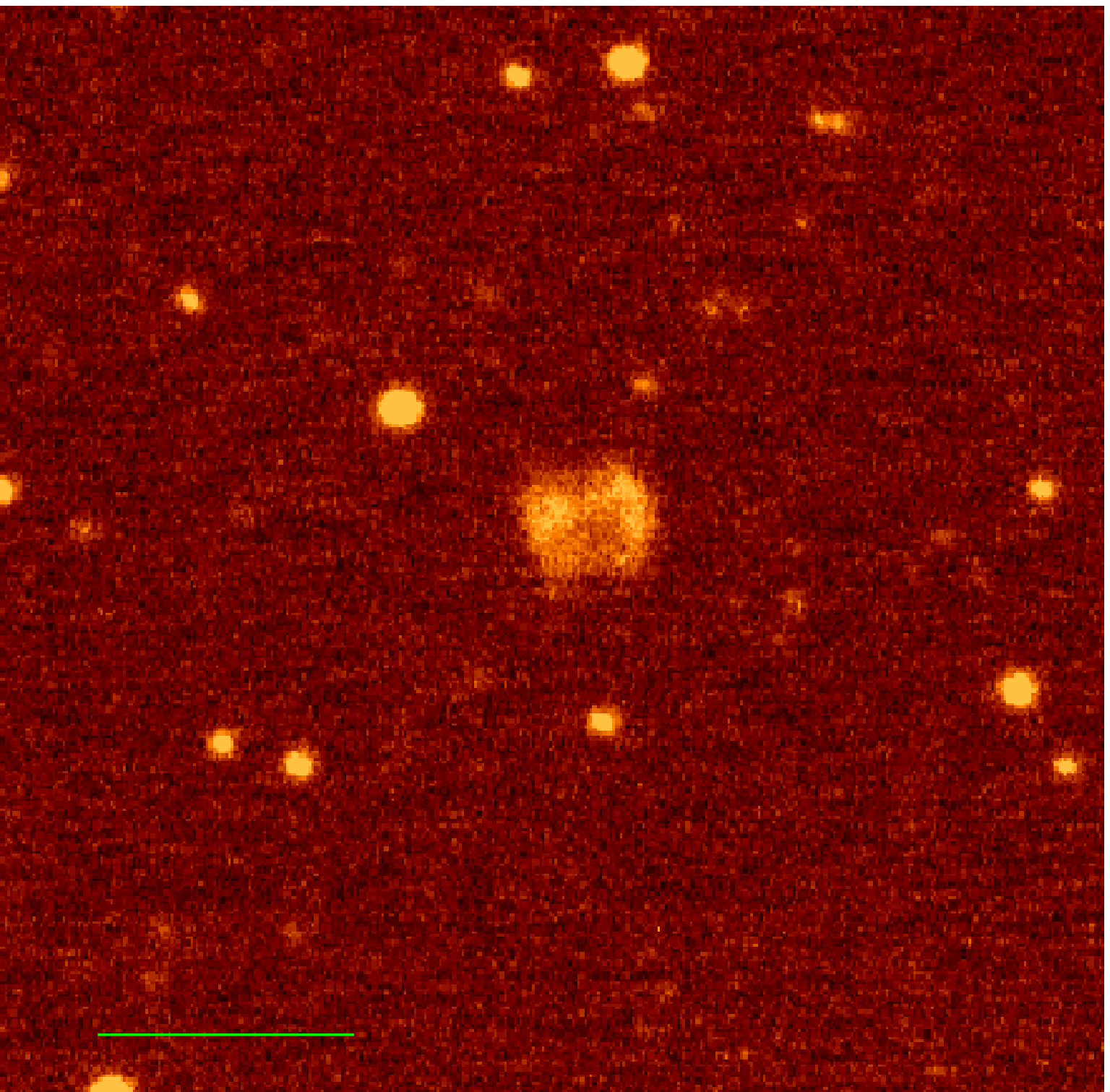}     \hspace{0.8cm}
\includegraphics[width=0.42\textwidth]{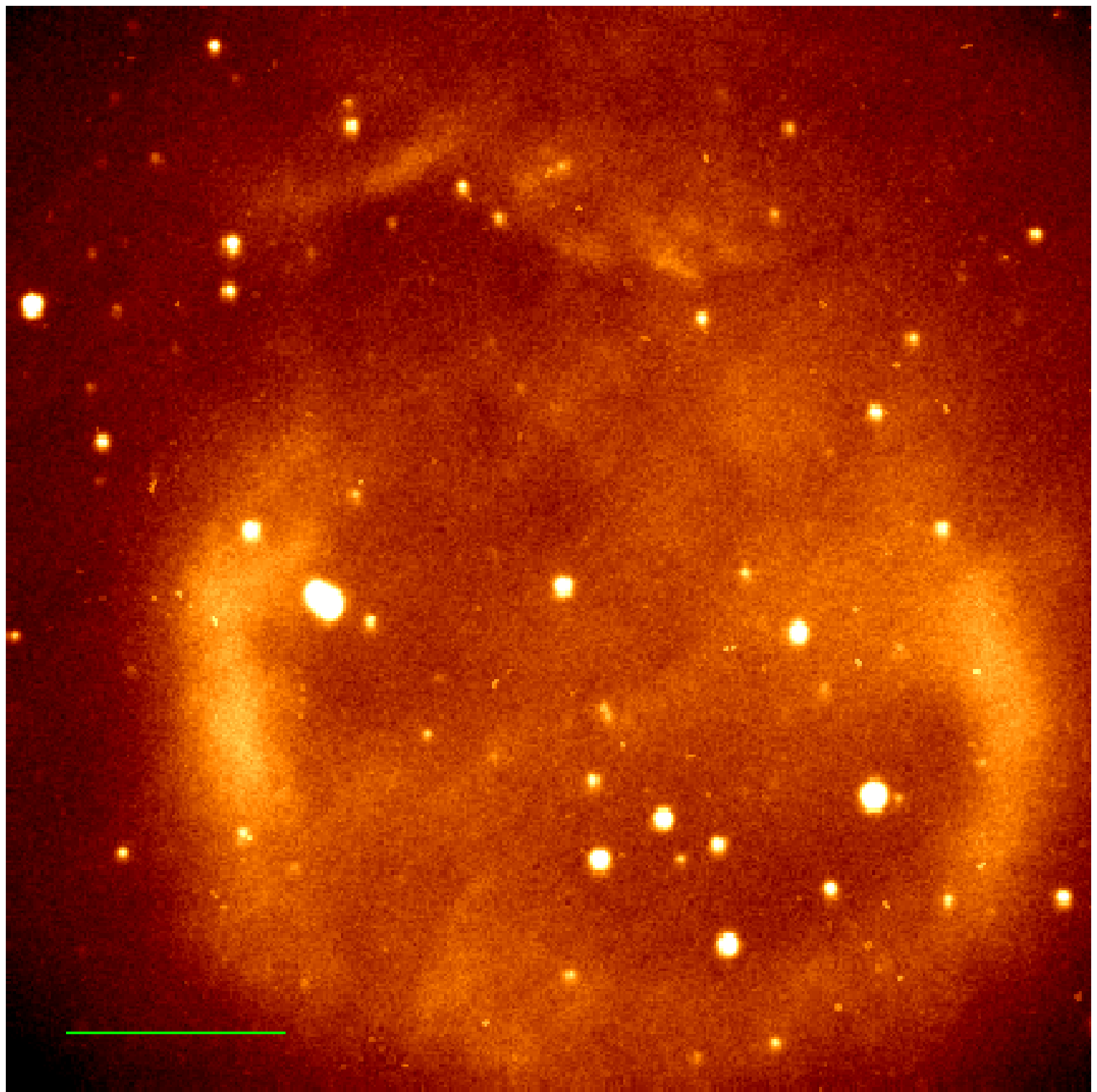}
\caption{[\ion{N}{ii}] images of planetary nebulae. North is up and east to the left.
The scale is indicated by a bar representing 60 arcseconds.    
From left to right, 
top IC~5148$-$50 and Hb~4 (L), middle SaWe~3 and HDW~12, bottom DeHt~3 and A~66.}
  \label{Fig_1}
\end{figure*}

\begin{figure*}[!ht]
  \centering
\includegraphics[width=0.42\textwidth]{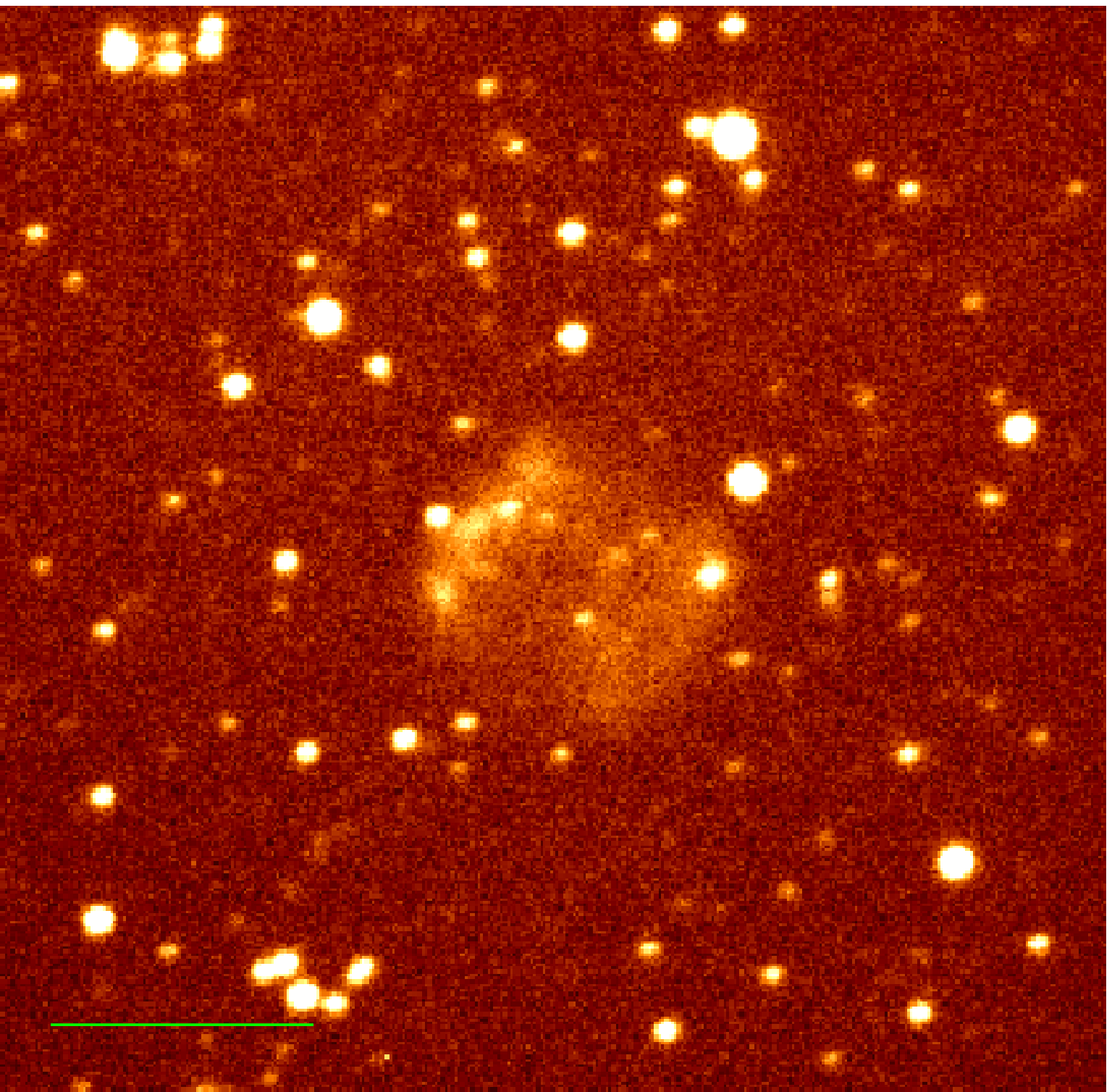}	             \hspace{0.8cm}
\includegraphics[width=0.42\textwidth]{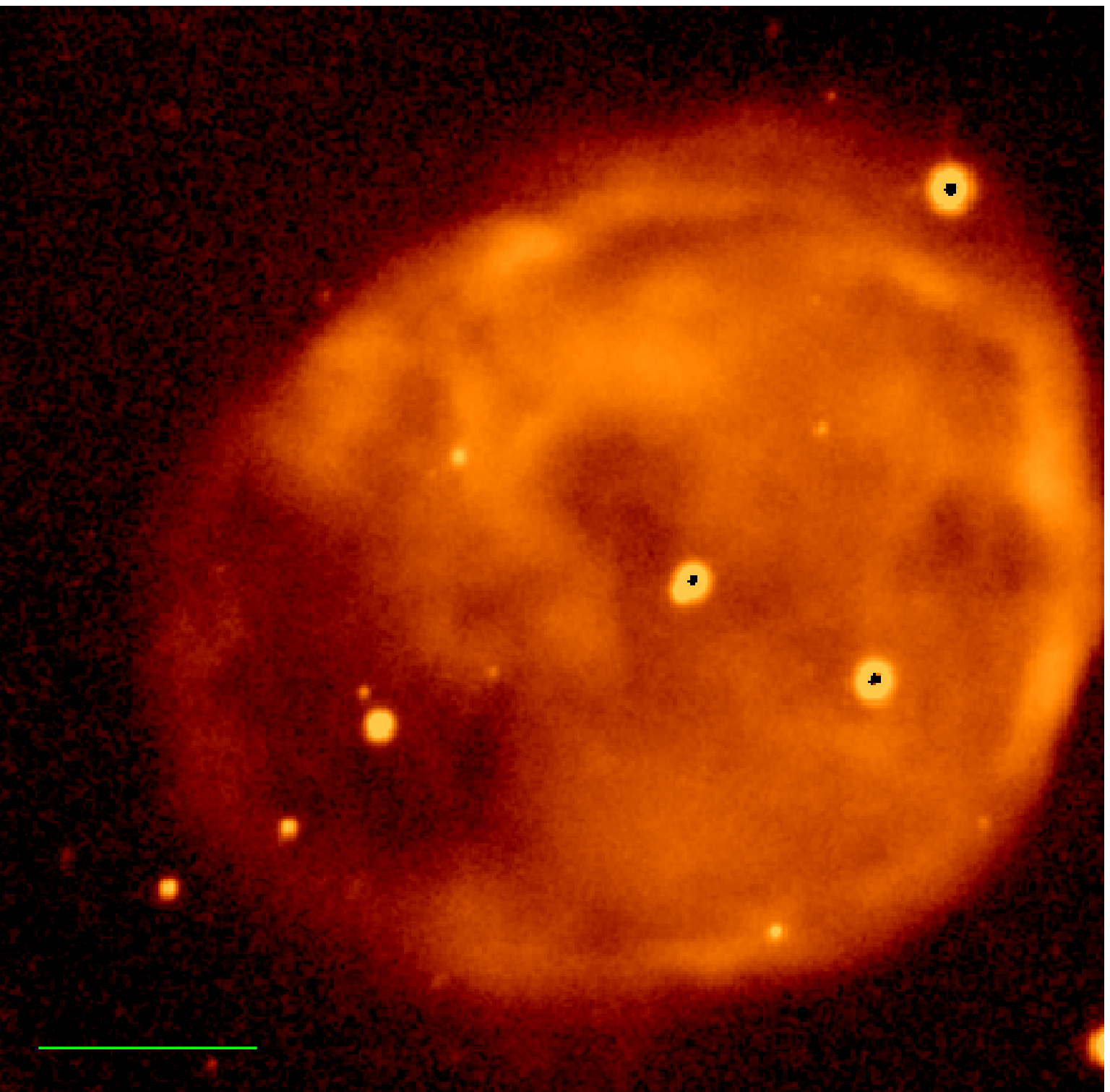}	     \vspace{0.25cm}\\
\includegraphics[width=0.42\textwidth]{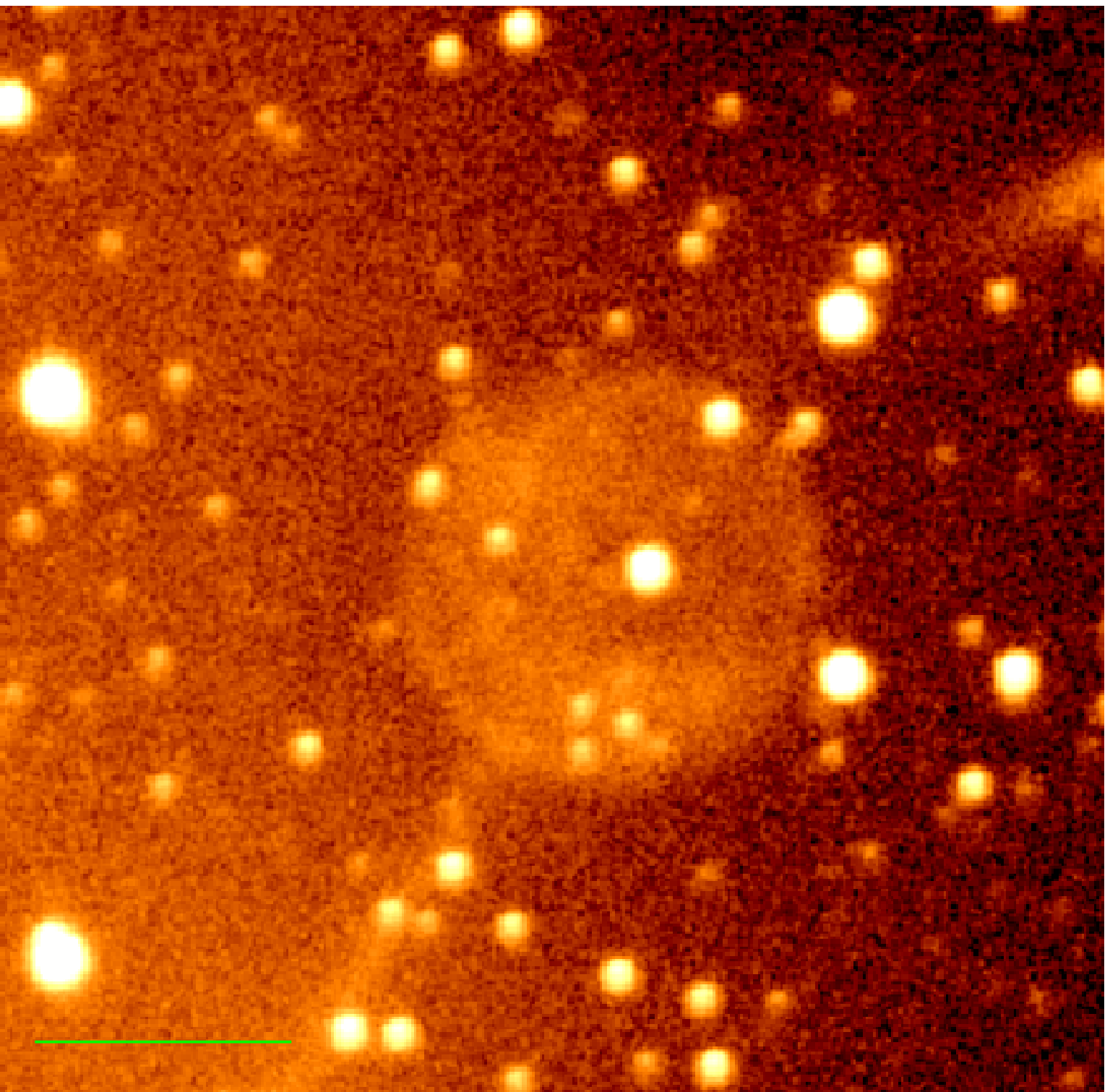}	     \hspace{0.8cm}
\includegraphics[width=0.42\textwidth]{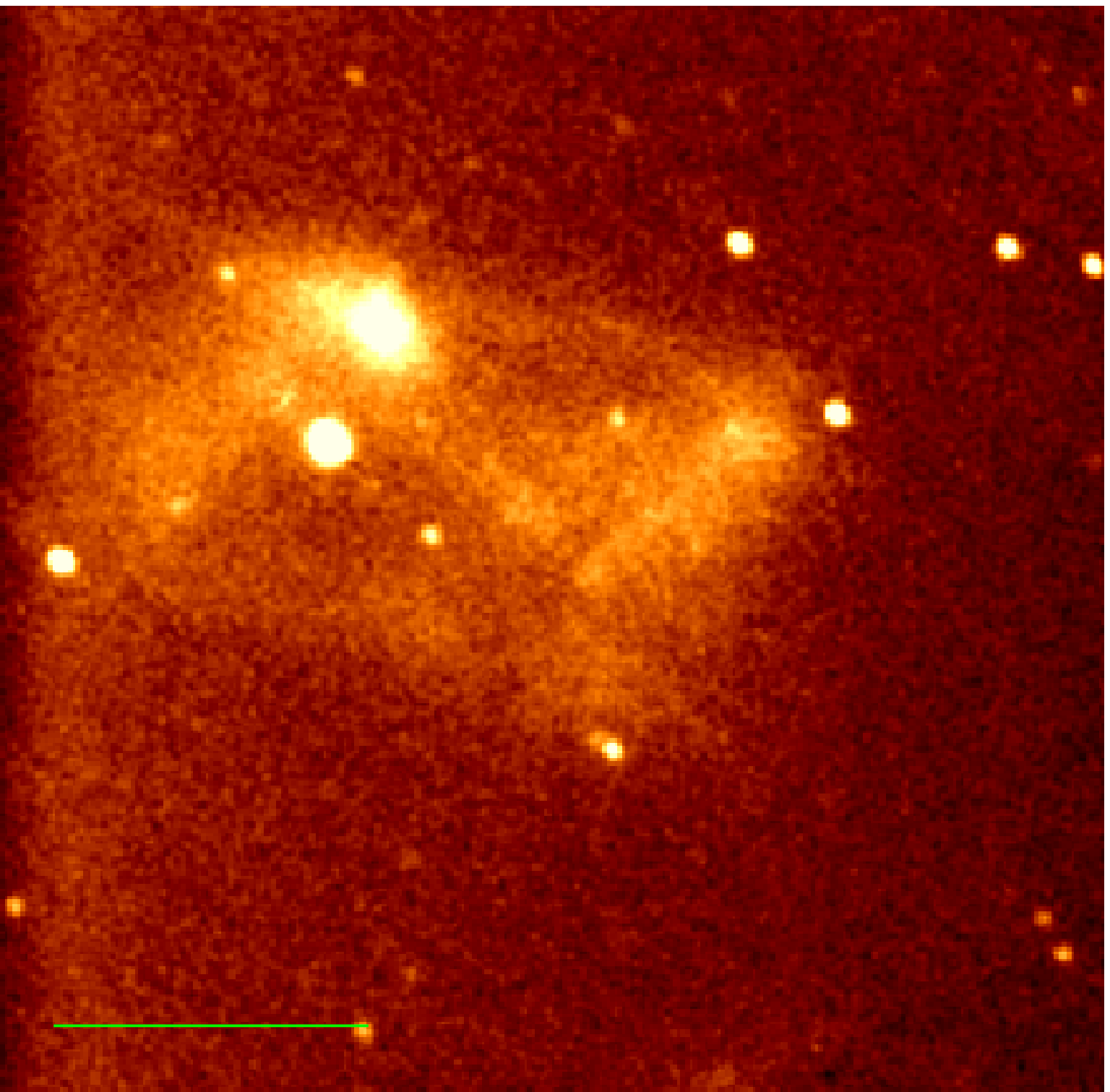}	     \vspace{0.25cm}\\
\includegraphics[width=0.42\textwidth]{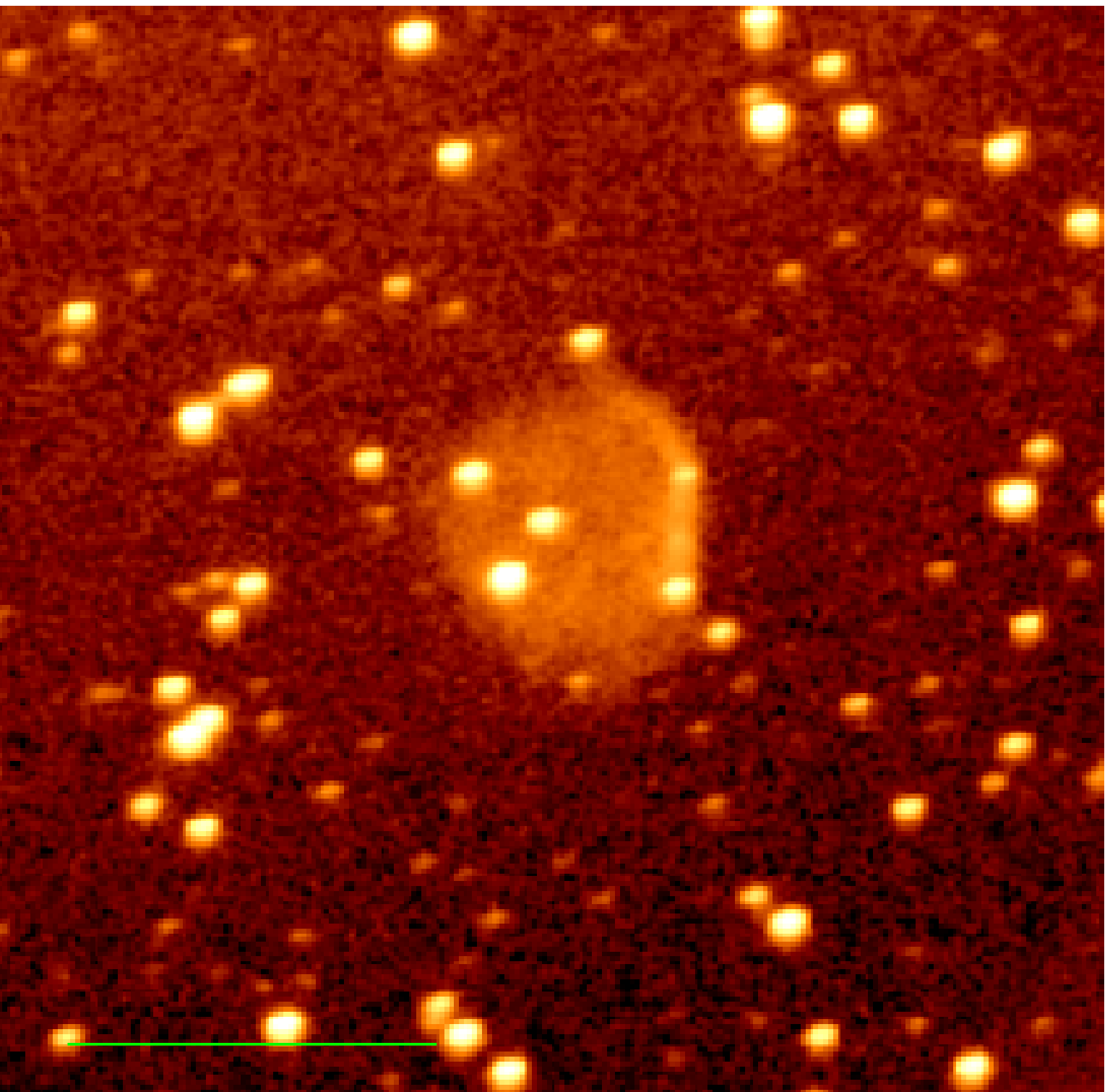}     \hspace{0.8cm}
\includegraphics[width=0.42\textwidth]{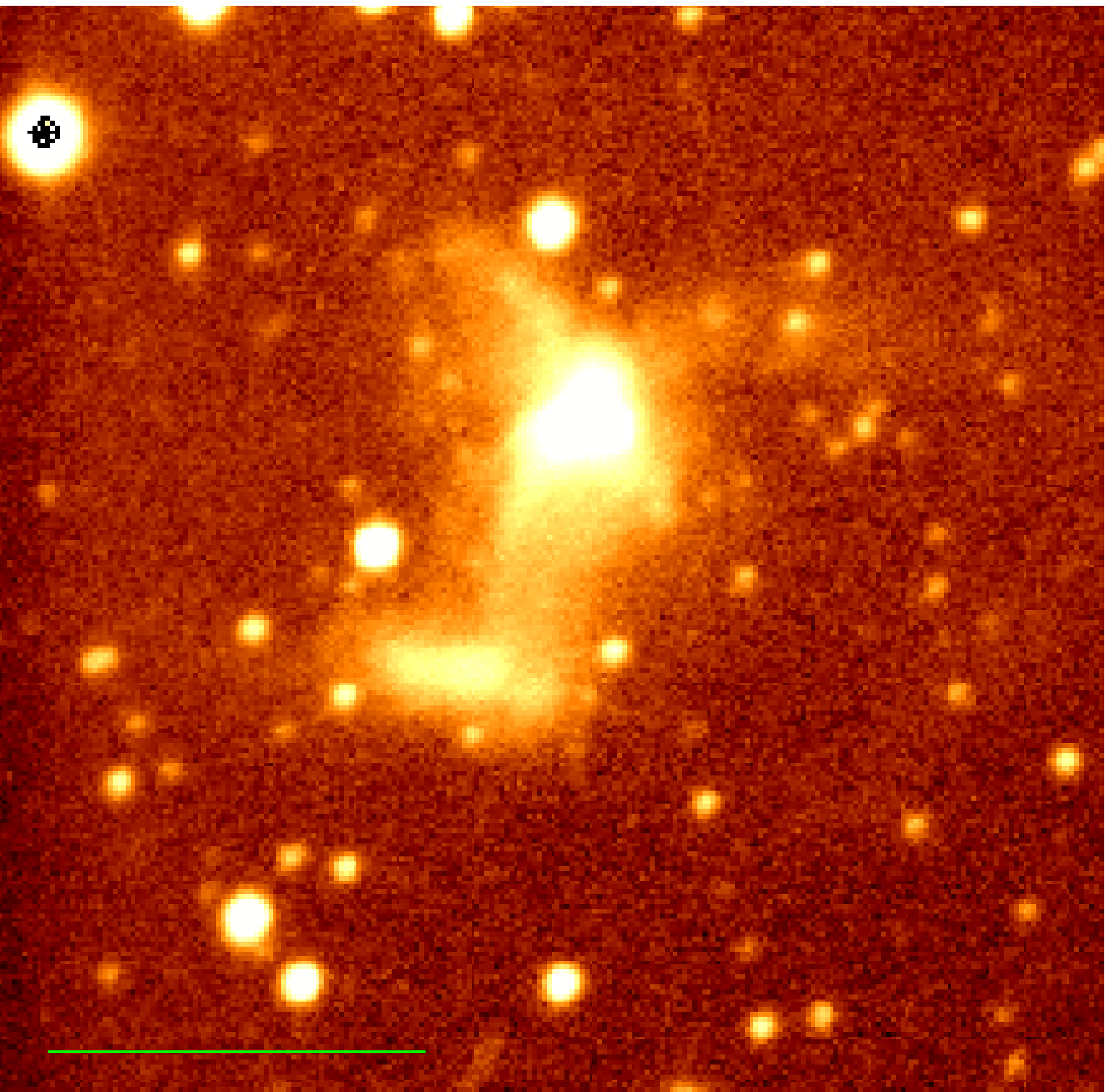}
\caption{Same as Fig.~\ref{Fig_1},
top A~67 and NGC~246 (L), middle  PHR0615$-$0025 (L) and K~1$-$11 , bottom  PHR0702$-$0324 (L) and PHR0705$-$0924 (L).}
  \label{Fig_2}
\end{figure*}

\begin{figure*}[!ht]
  \centering
\includegraphics[width=0.42\textwidth]{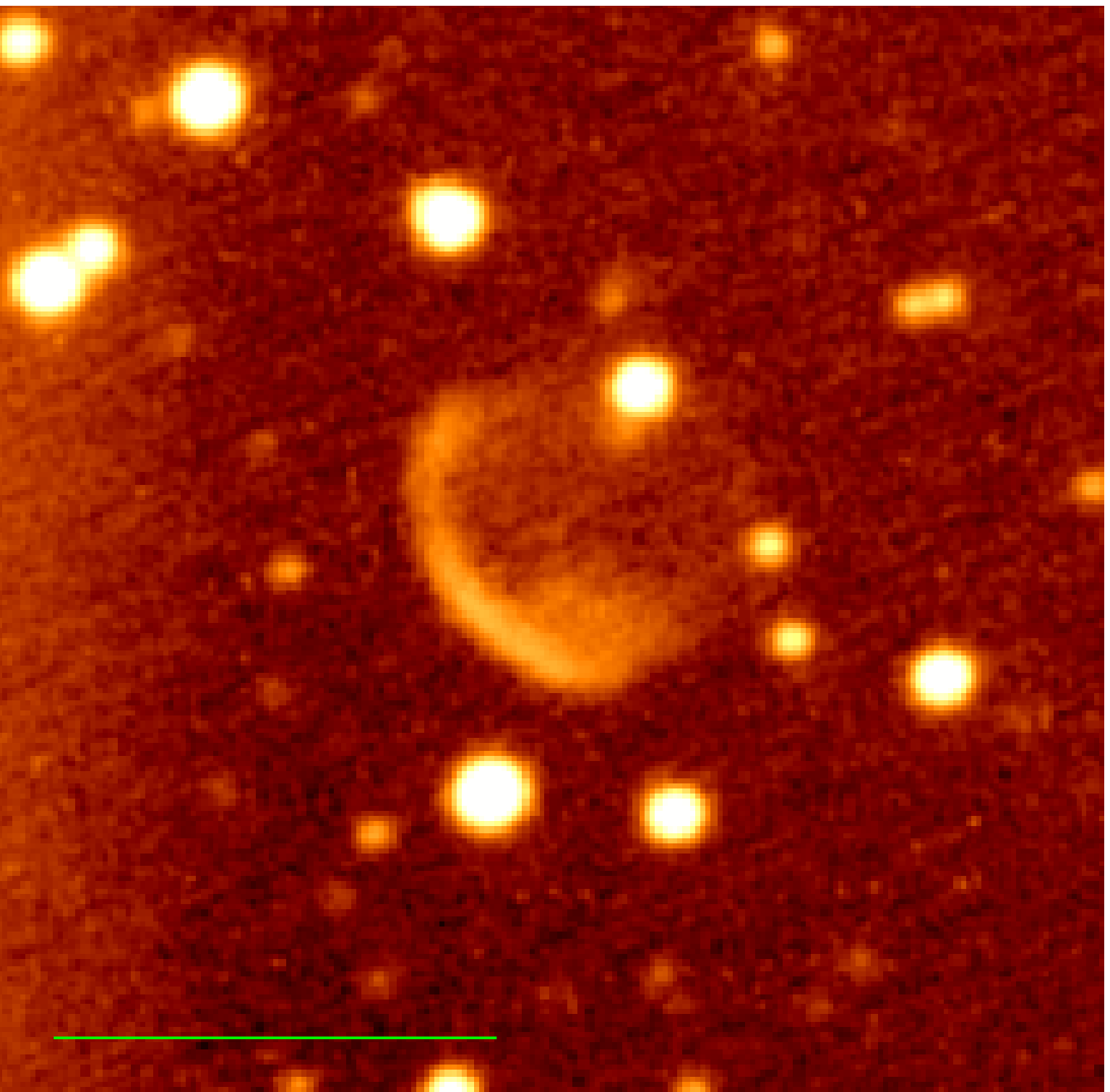}	\hspace{0.8cm}
\includegraphics[width=0.42\textwidth]{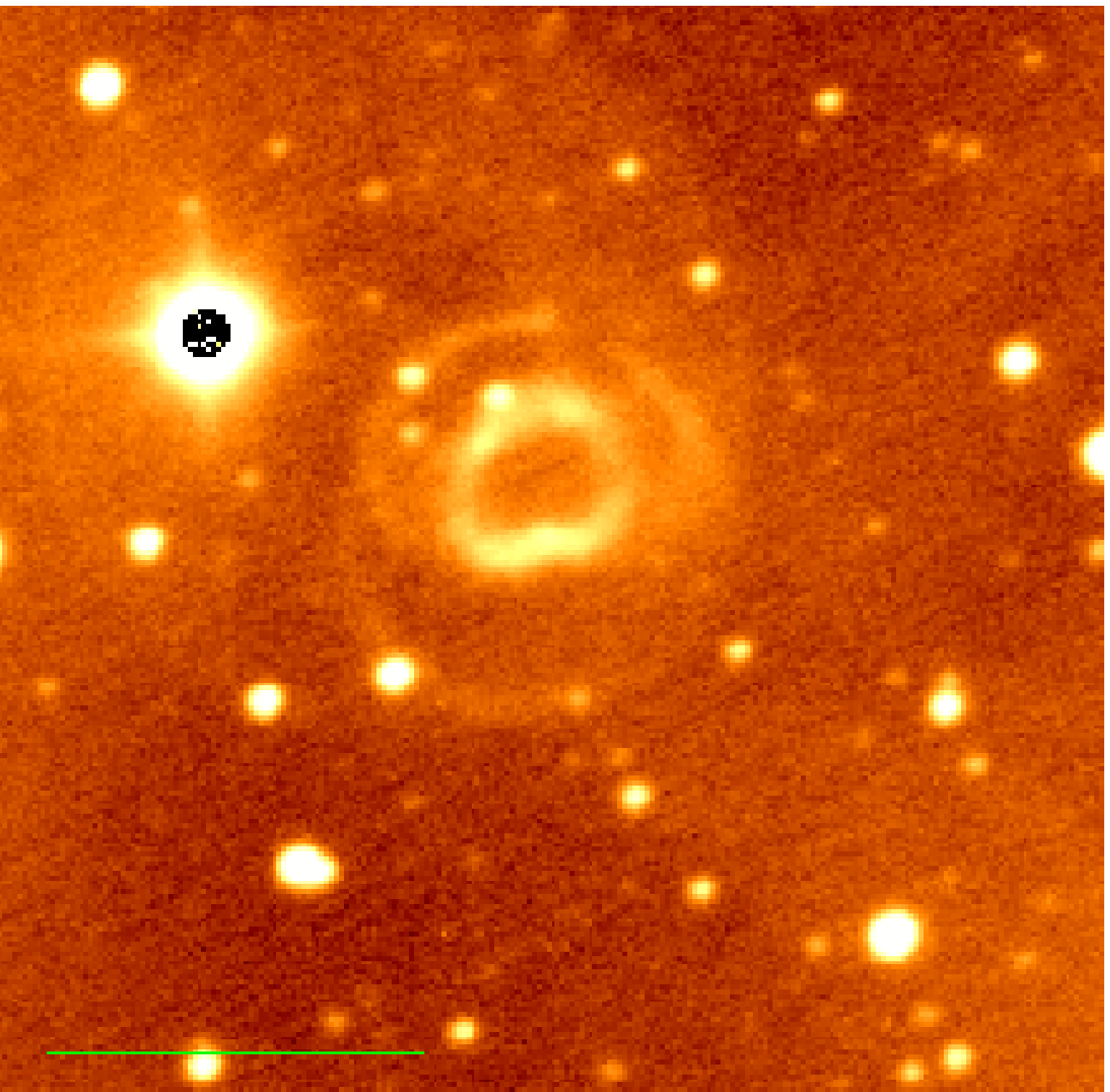}		\vspace{0.25cm}\\
\includegraphics[width=0.42\textwidth]{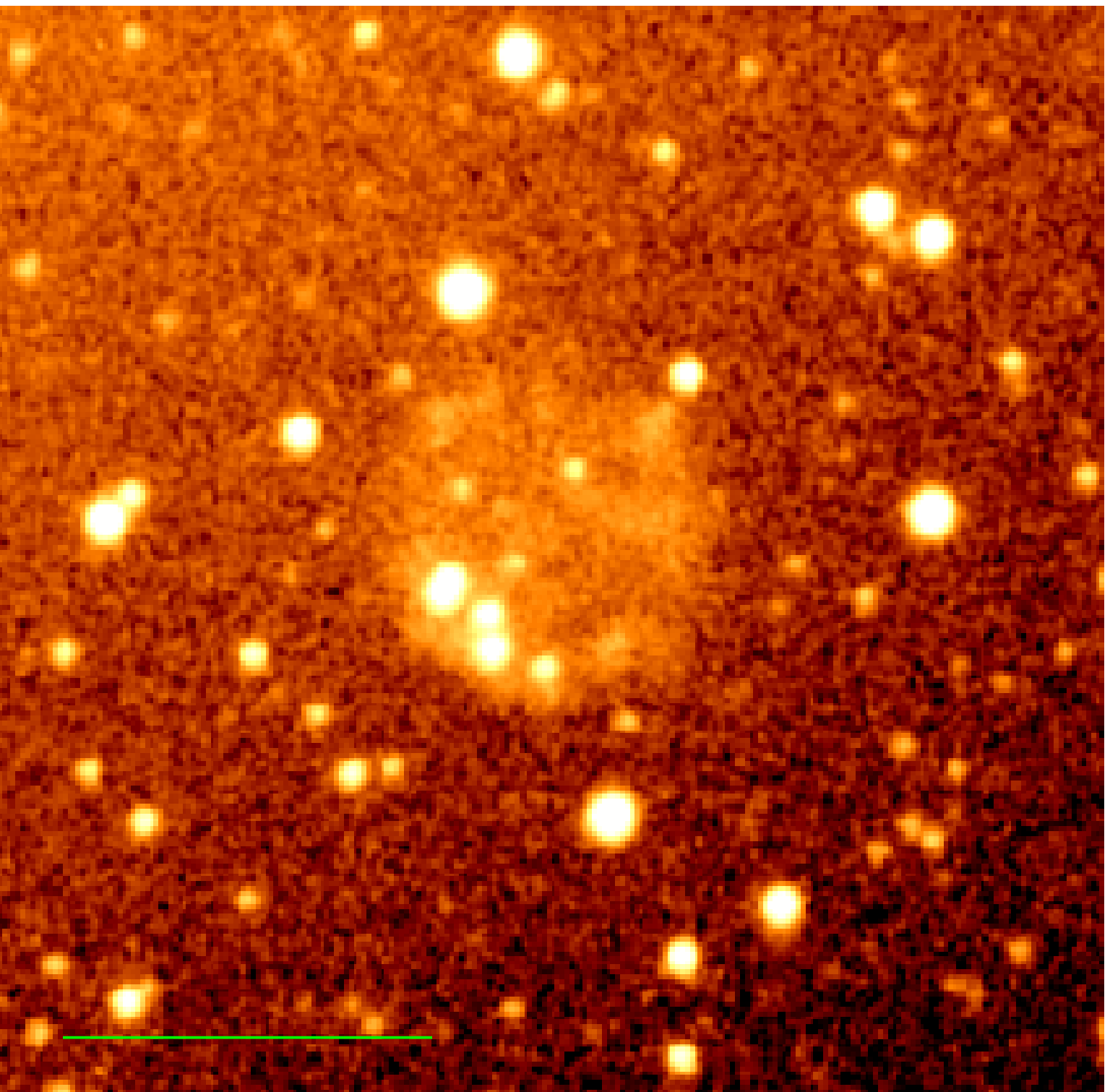}		\hspace{0.8cm}
\includegraphics[width=0.42\textwidth]{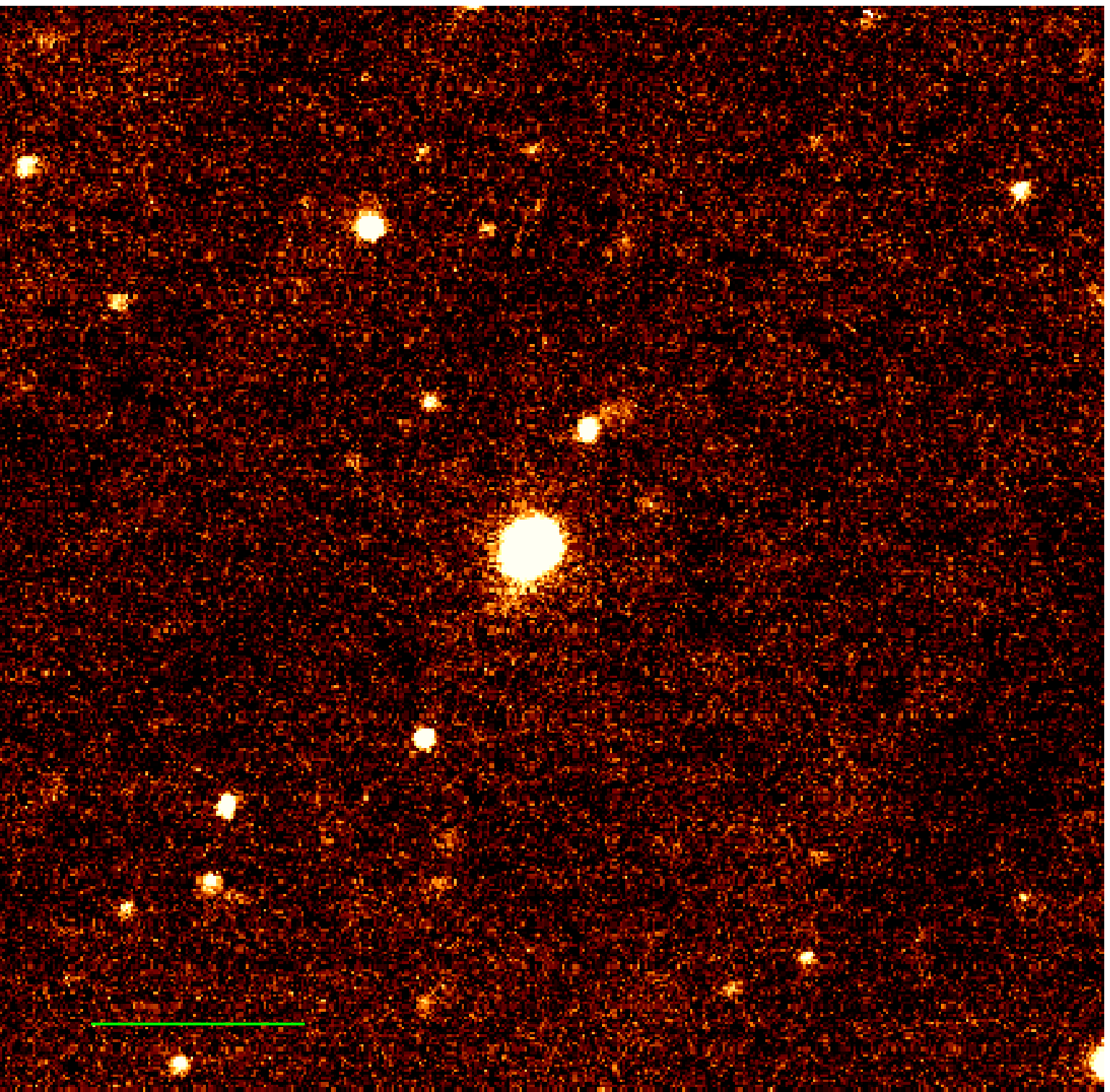}	\vspace{0.25cm}\\
\includegraphics[width=0.42\textwidth]{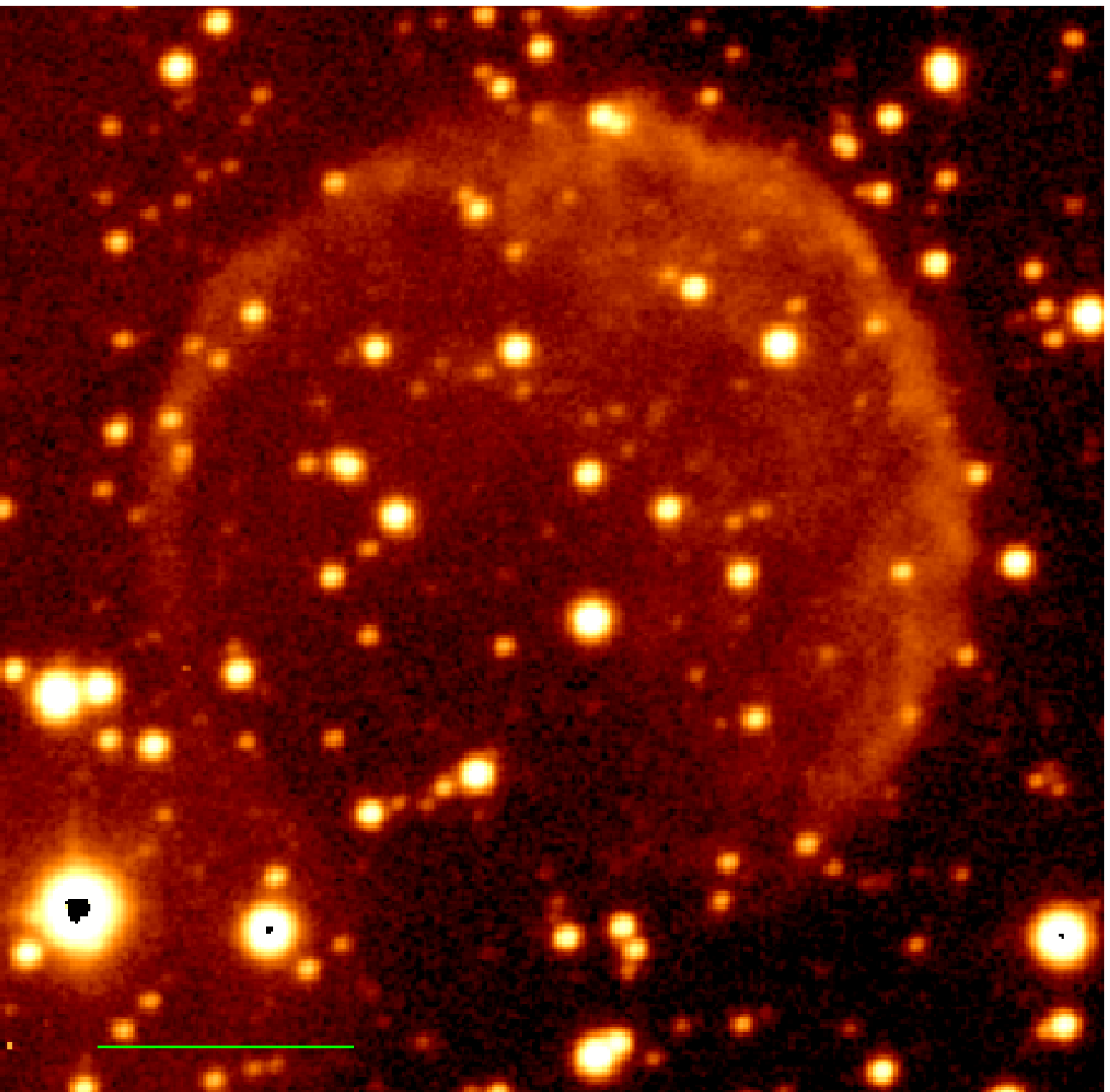}		\hspace{0.8cm}
\includegraphics[width=0.42\textwidth]{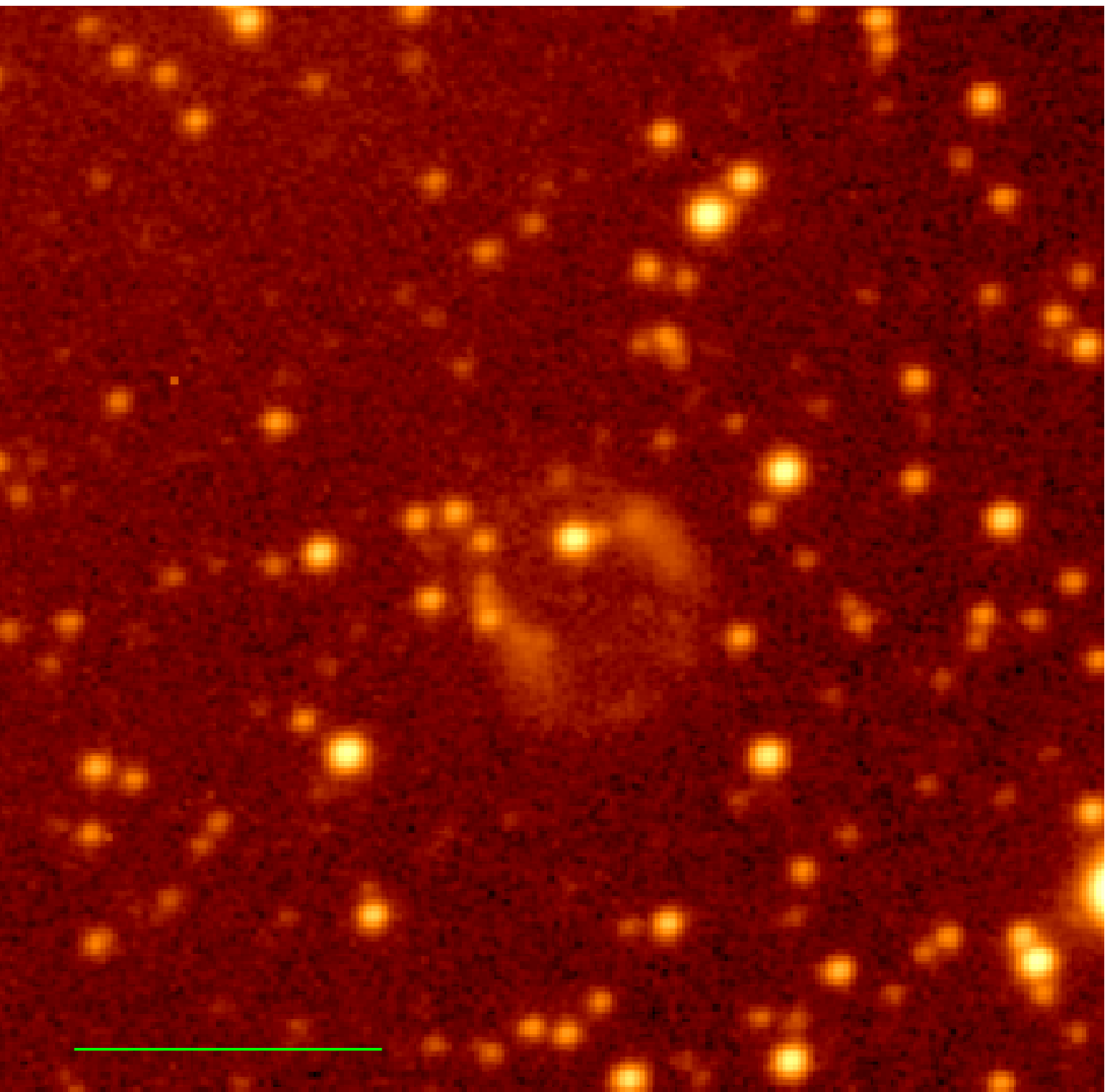}
\caption{ Same as Fig.~\ref{Fig_1},
top PHR0700$-$1143 (L) and PHR0716$-$1053, middle PHR0711$-$1238 (L) and M~1$-$6, bottom  PHR0719$-$1222 (L) and PHR0727$-$1259 (L).}
  \label{Fig_3}
\end{figure*}

\begin{figure*}[!ht]
  \centering
\includegraphics[width=0.42\textwidth]{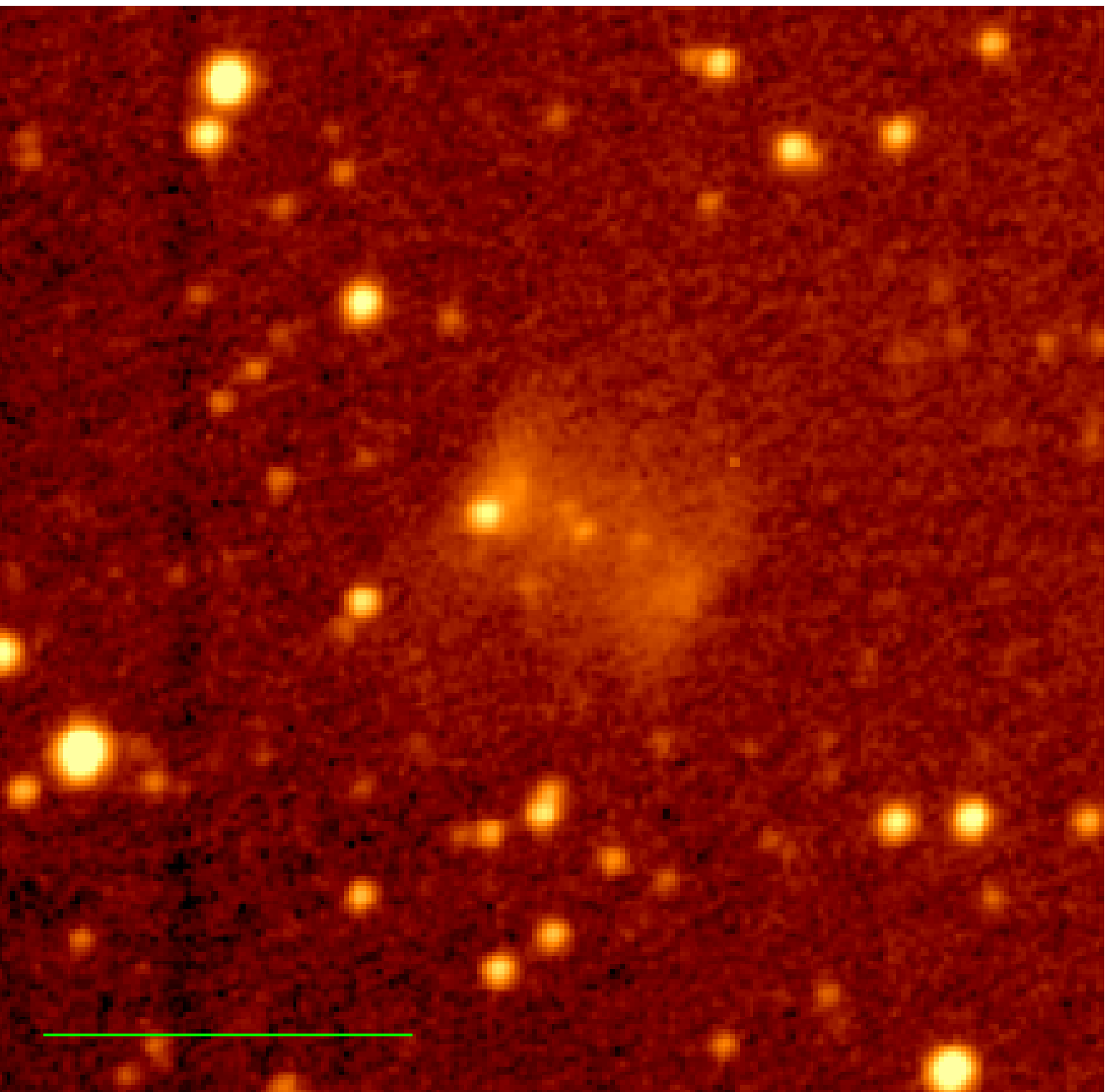}		\hspace{0.8cm}
\includegraphics[width=0.42\textwidth]{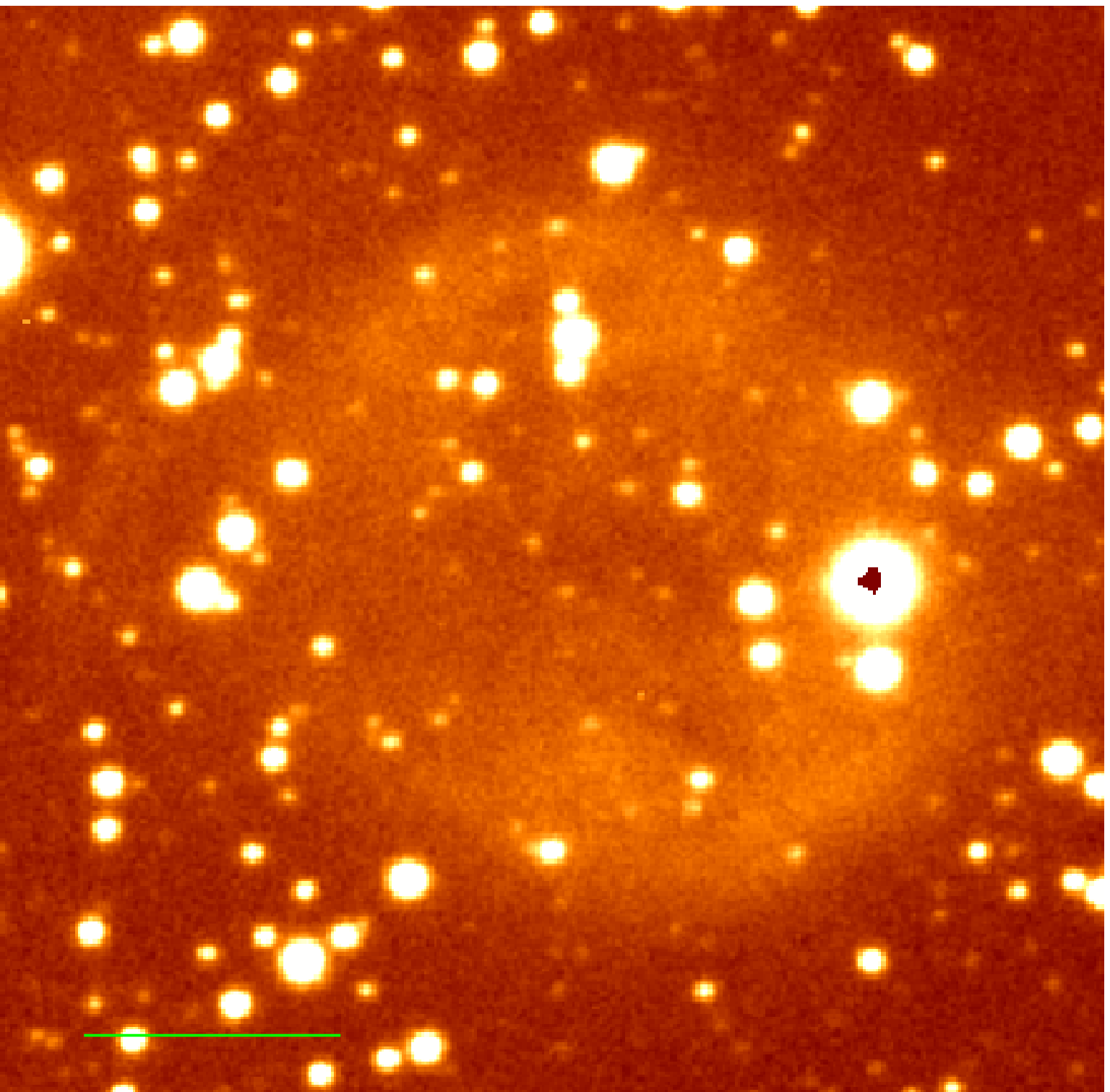}		\vspace{0.25cm}\\
\includegraphics[width=0.42\textwidth]{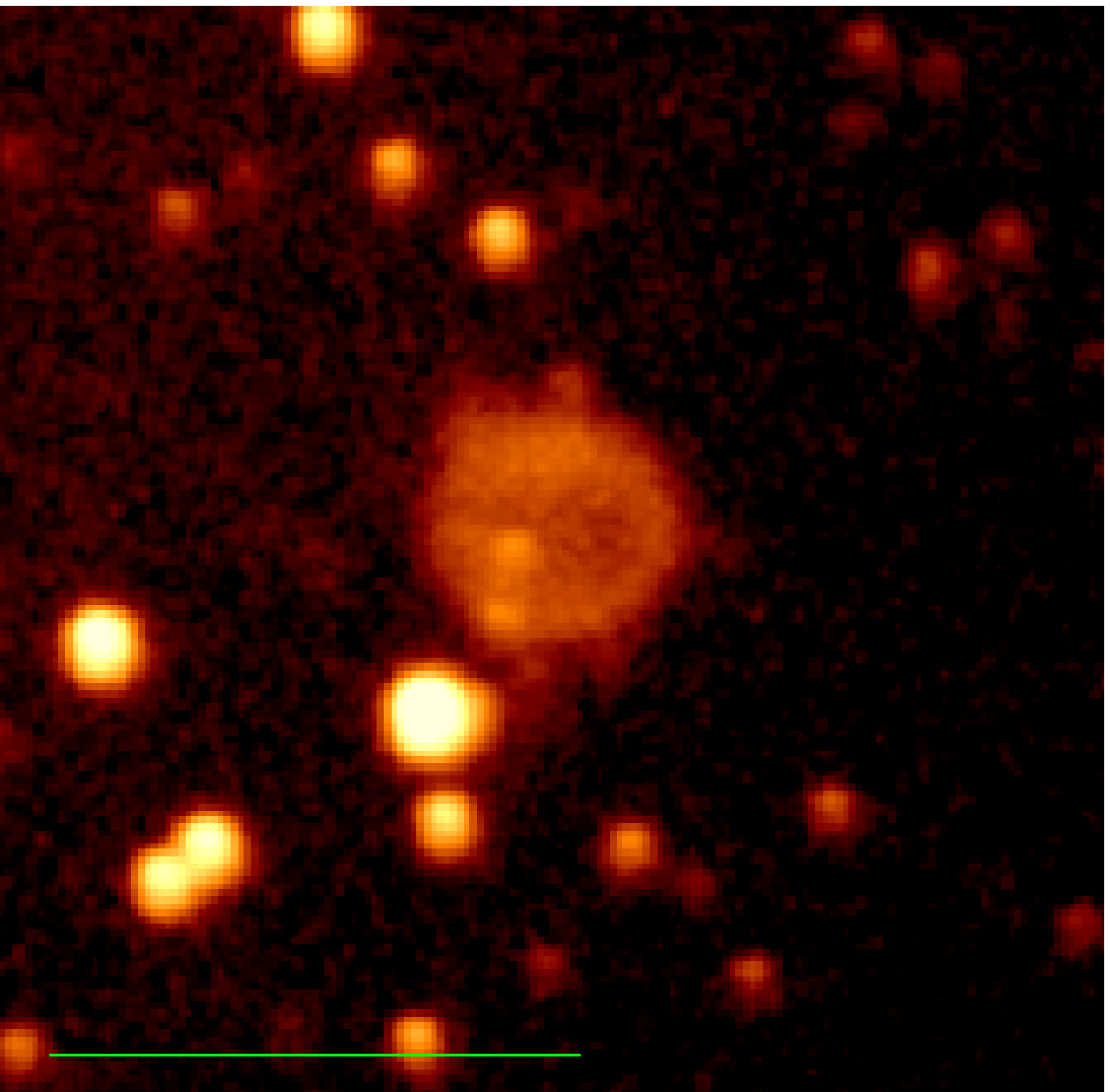}		\hspace{0.8cm}
\includegraphics[width=0.42\textwidth]{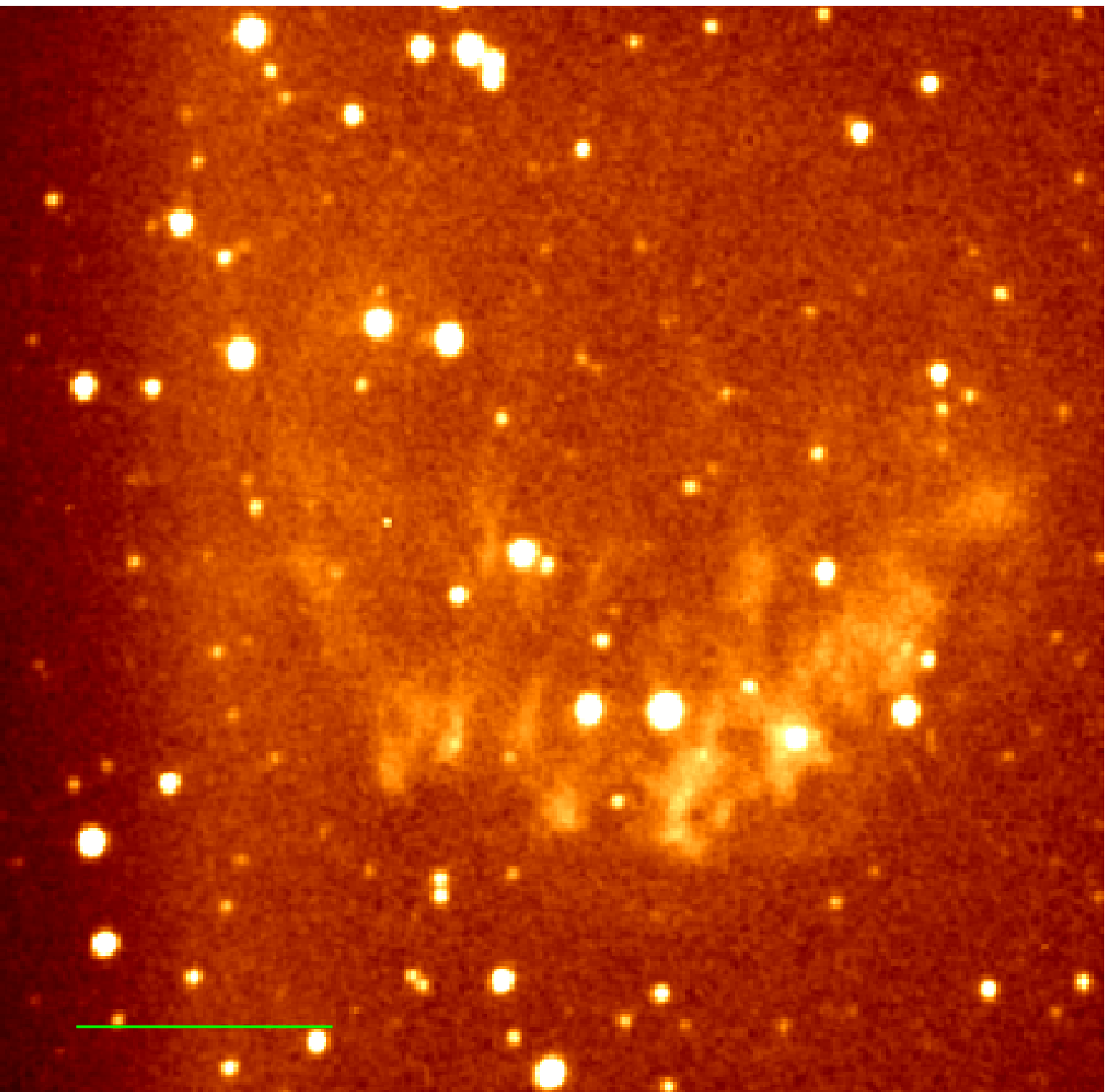}		\vspace{0.25cm}\\
\includegraphics[width=0.42\textwidth]{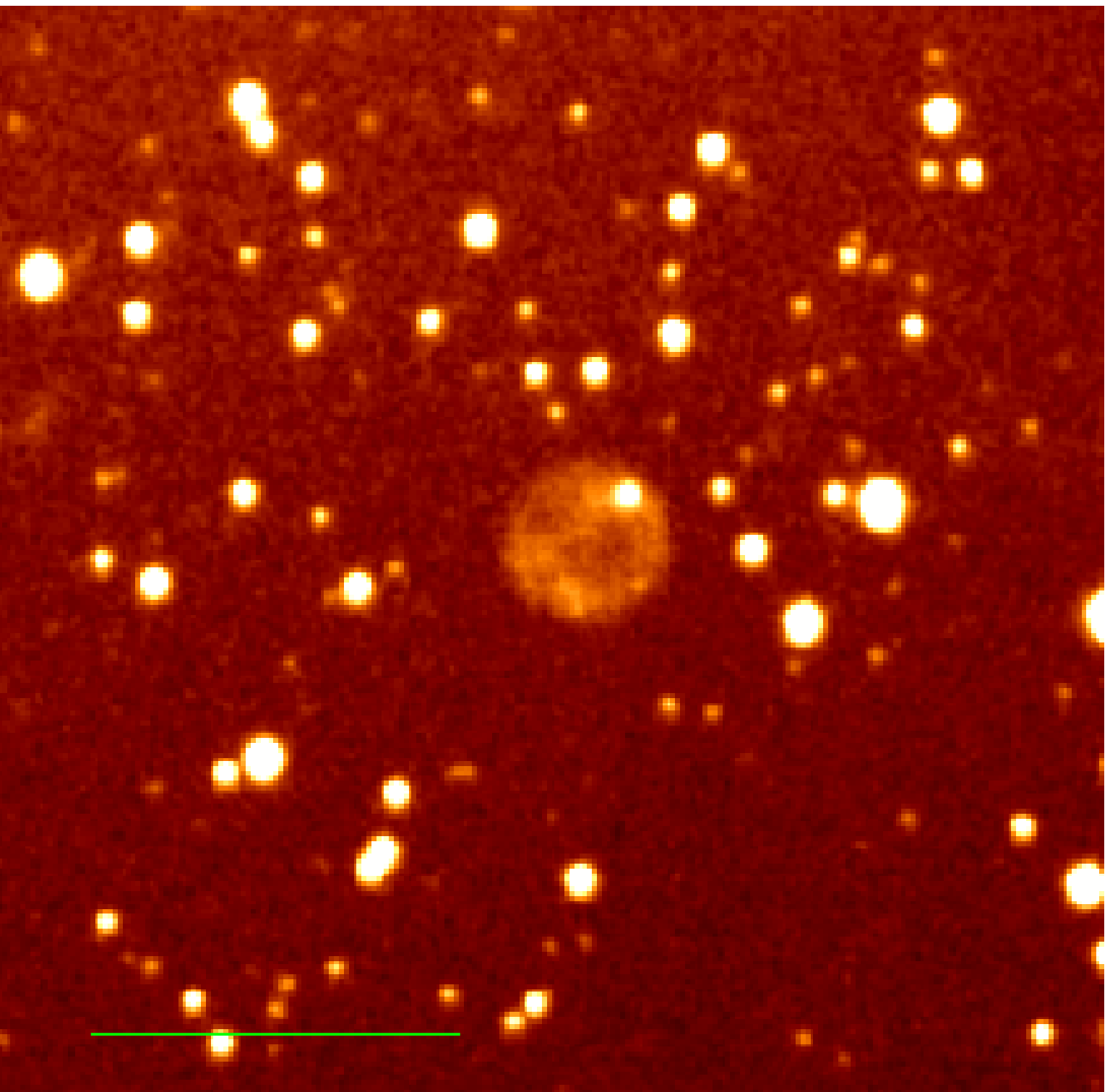}		\hspace{0.8cm}
\includegraphics[width=0.42\textwidth]{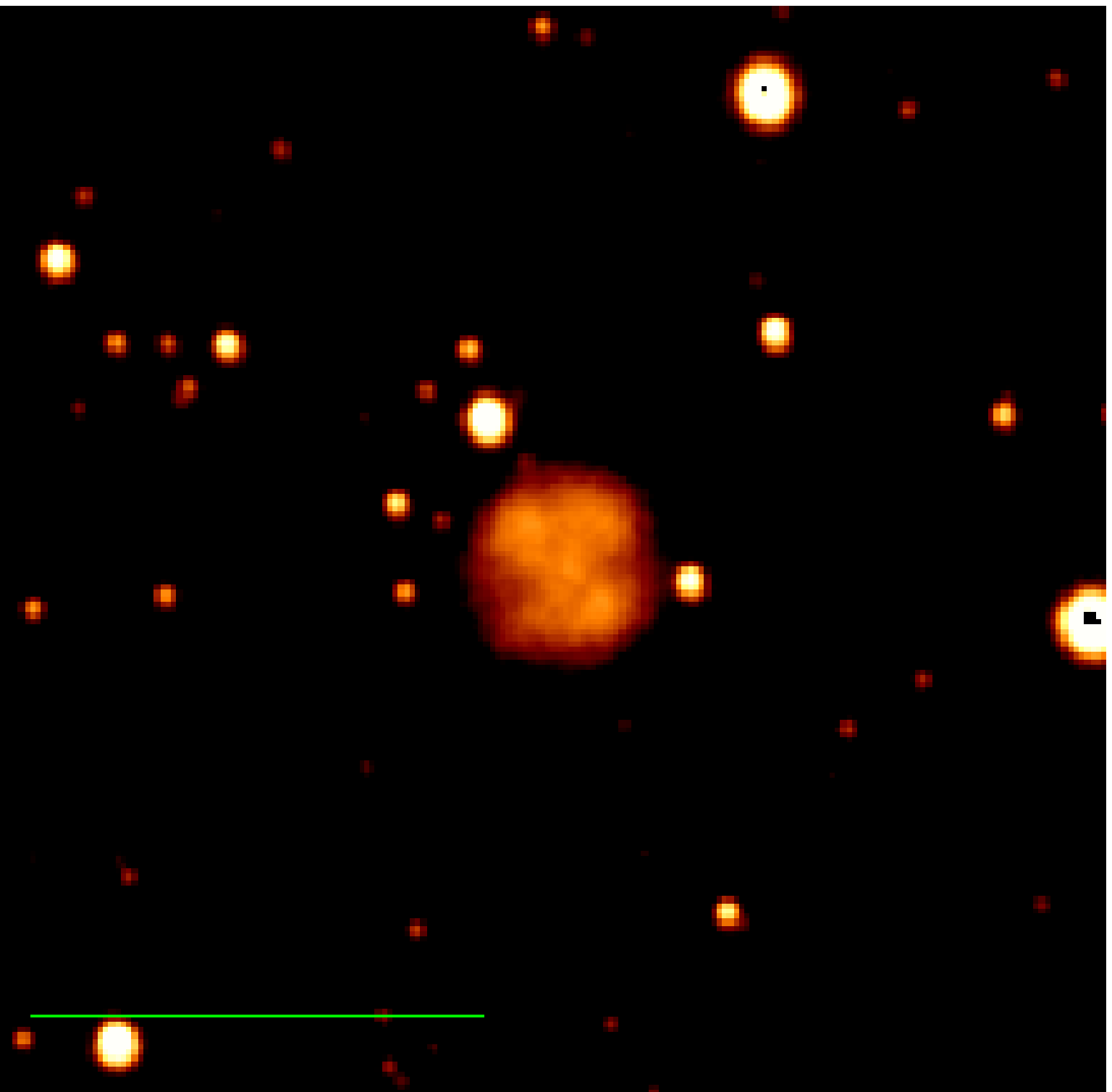}
\caption{Same as Fig.~\ref{Fig_1},
top PHR0727$-$1707 (L) and PHR0724$-$1757, middle SaWe~1 and HaWe~9, bottom PHR0730$-$2151 and ESO~427$-$19 (L).}
  \label{Fig_4}
\end{figure*}

\begin{figure*}[!ht]
  \centering
\includegraphics[width=0.42\textwidth]{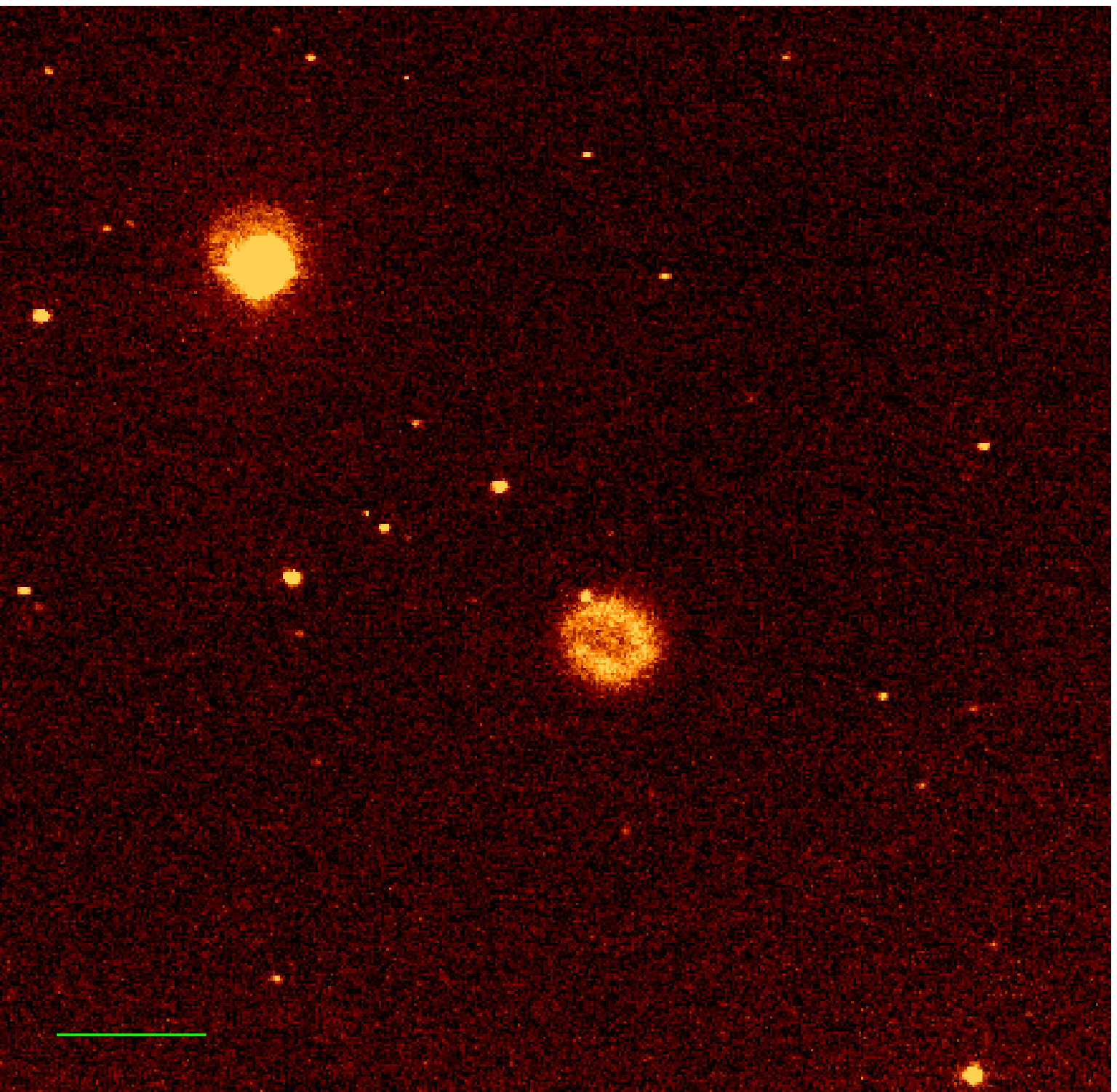}		\hspace{0.8cm}
\includegraphics[width=0.42\textwidth]{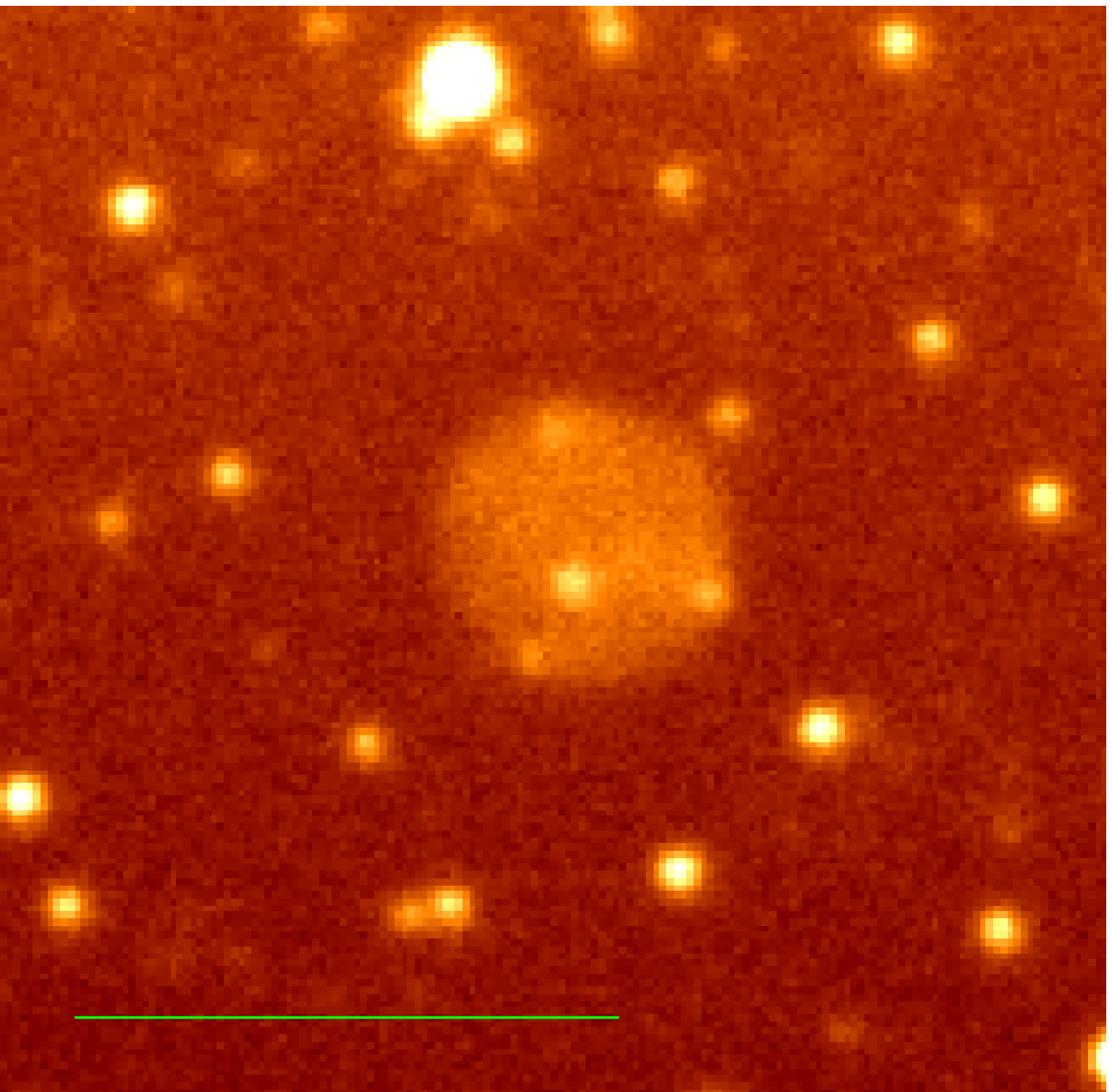}			\vspace{0.25cm}\\
\includegraphics[width=0.42\textwidth]{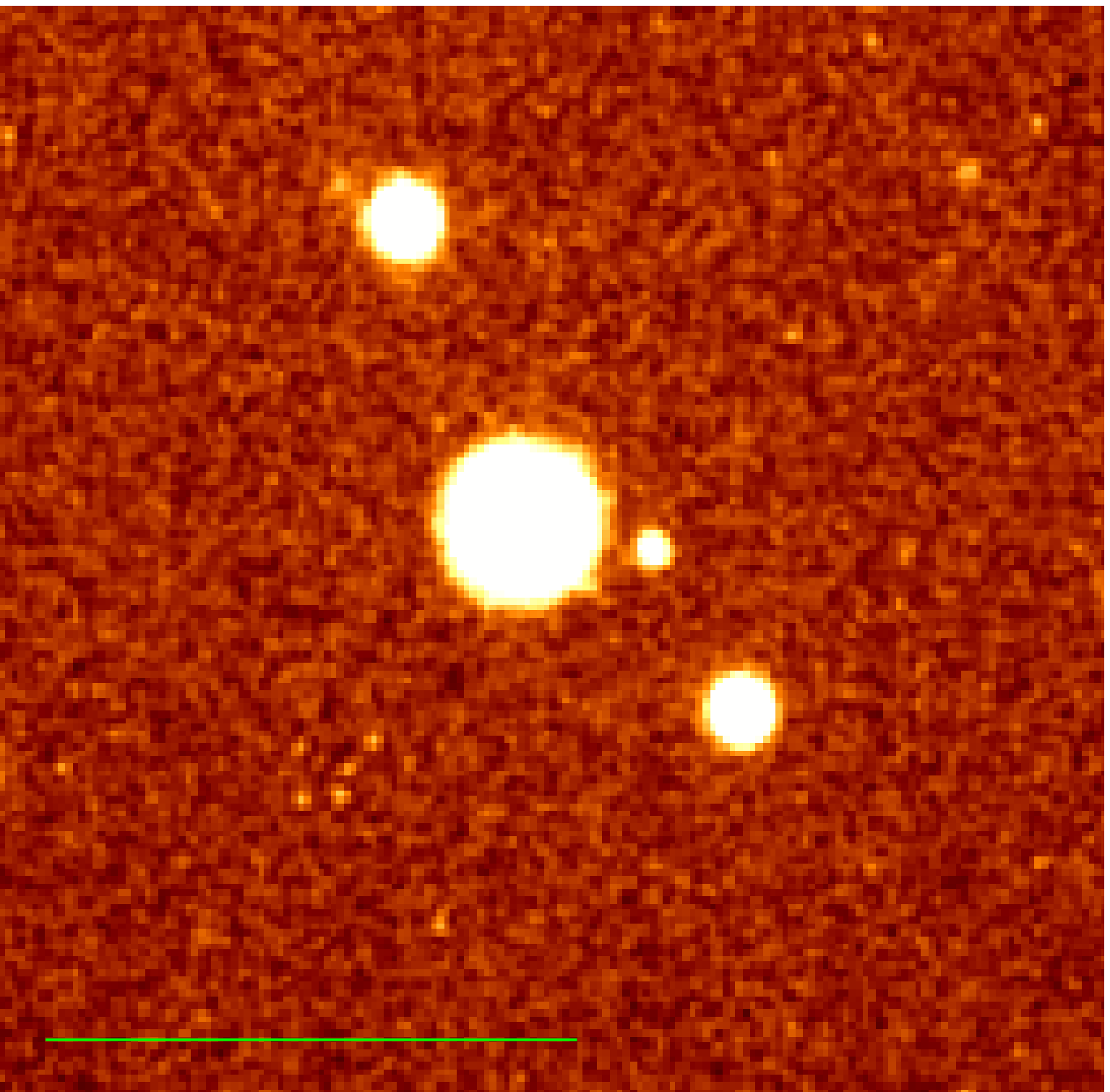}		\hspace{0.8cm}
\includegraphics[width=0.42\textwidth]{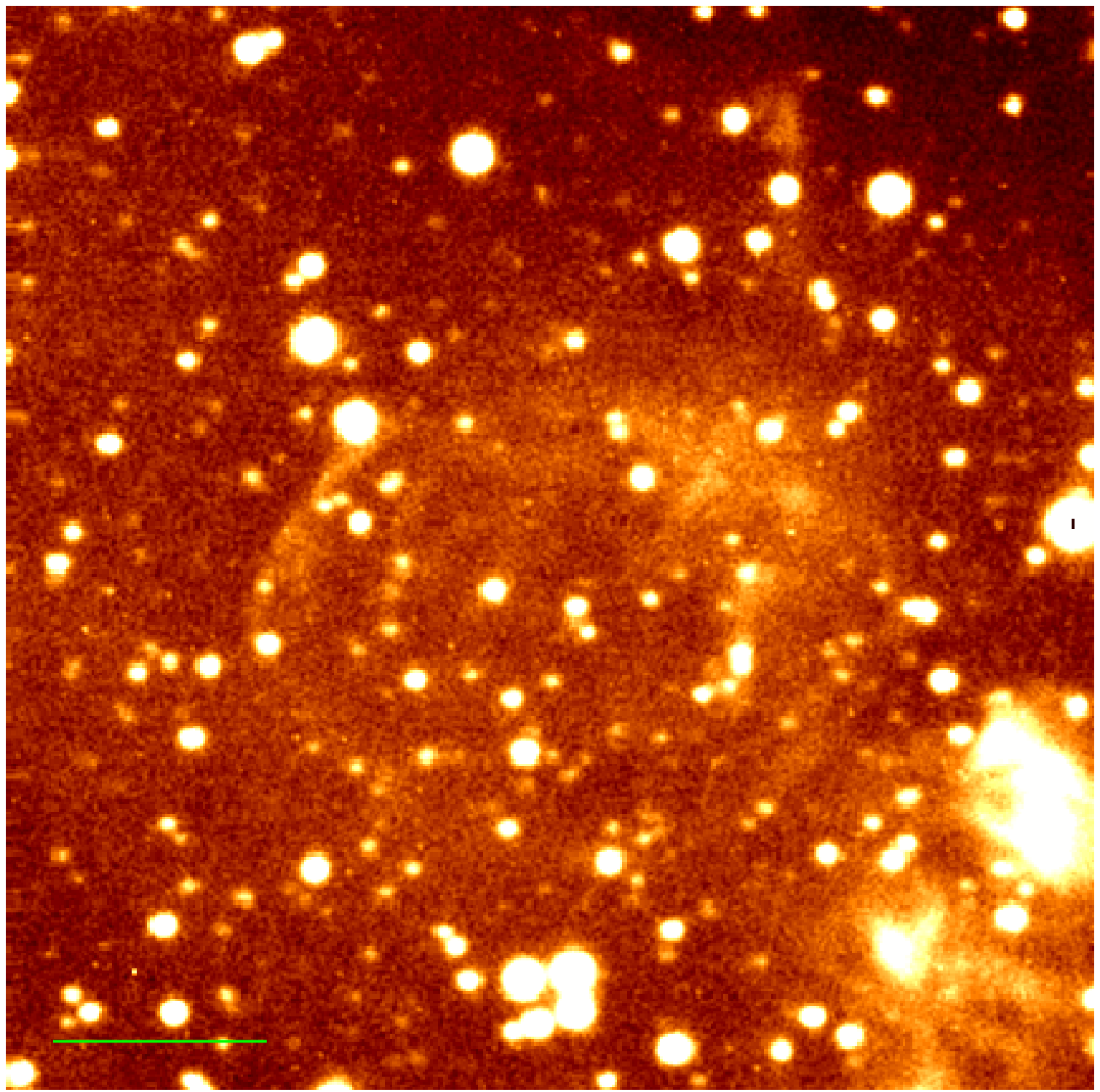}		\vspace{0.25cm}\\
\includegraphics[width=0.42\textwidth]{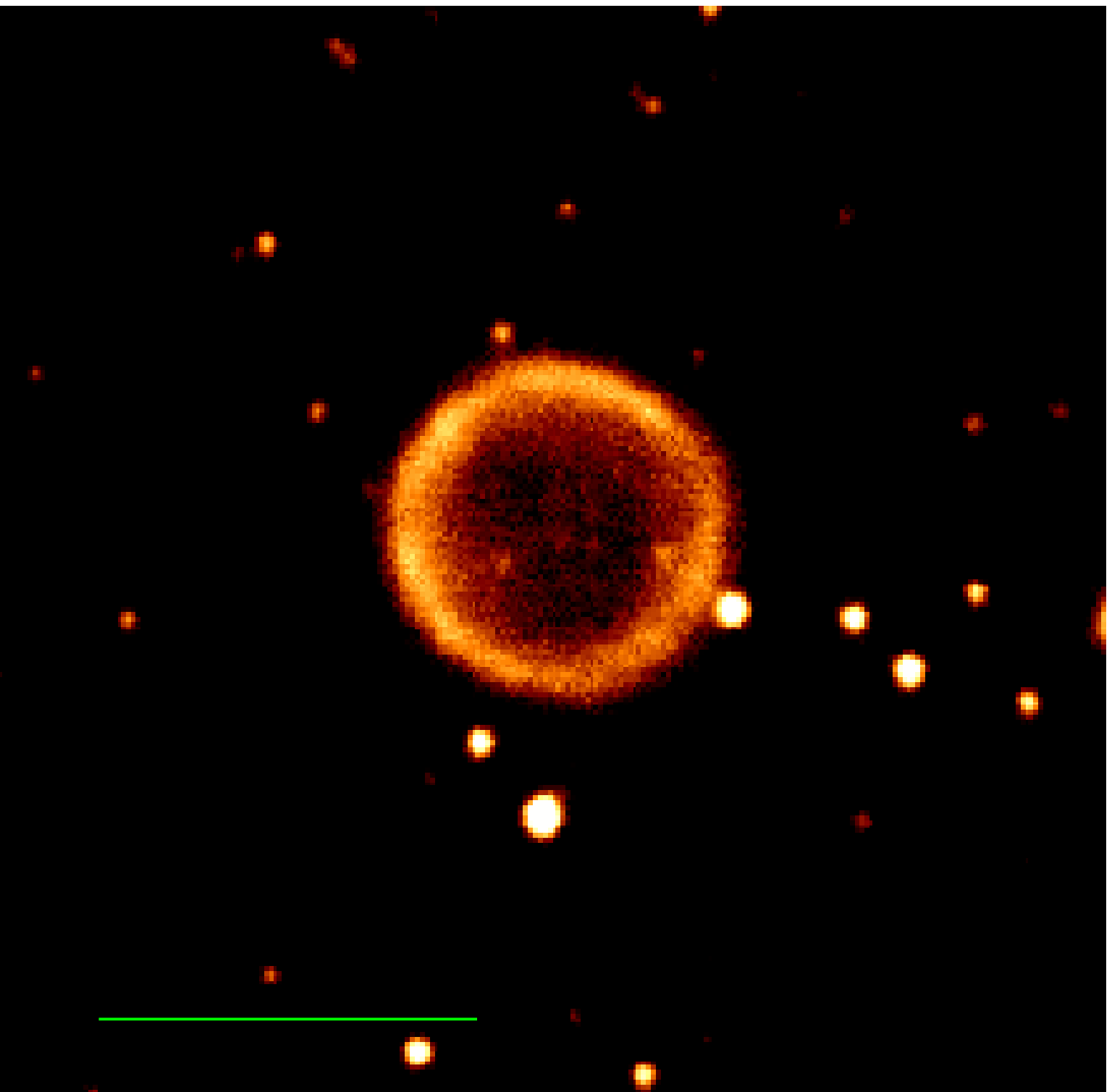}		\hspace{0.8cm}
\includegraphics[width=0.42\textwidth]{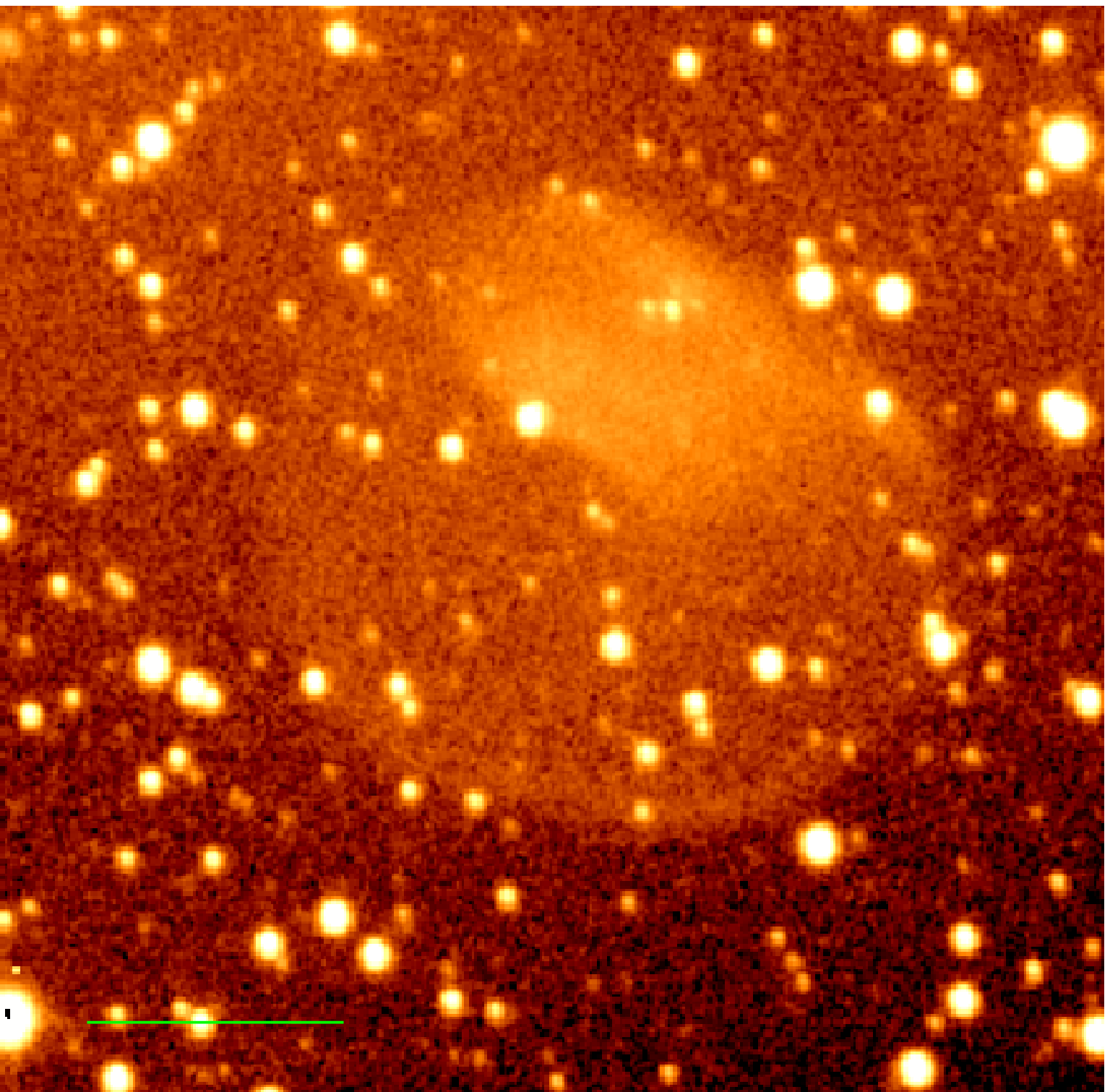}
\caption{Same as Fig.~\ref{Fig_1},
top NGC~2610 and PHR0726$-$2858, middle PRTM~1 and PHR0742$-$3247, bottom A~23 and PHR0834$-$2819 (L).}
  \label{Fig_5}
\end{figure*}

\begin{figure*}[!ht]
  \centering
\includegraphics[width=0.42\textwidth]{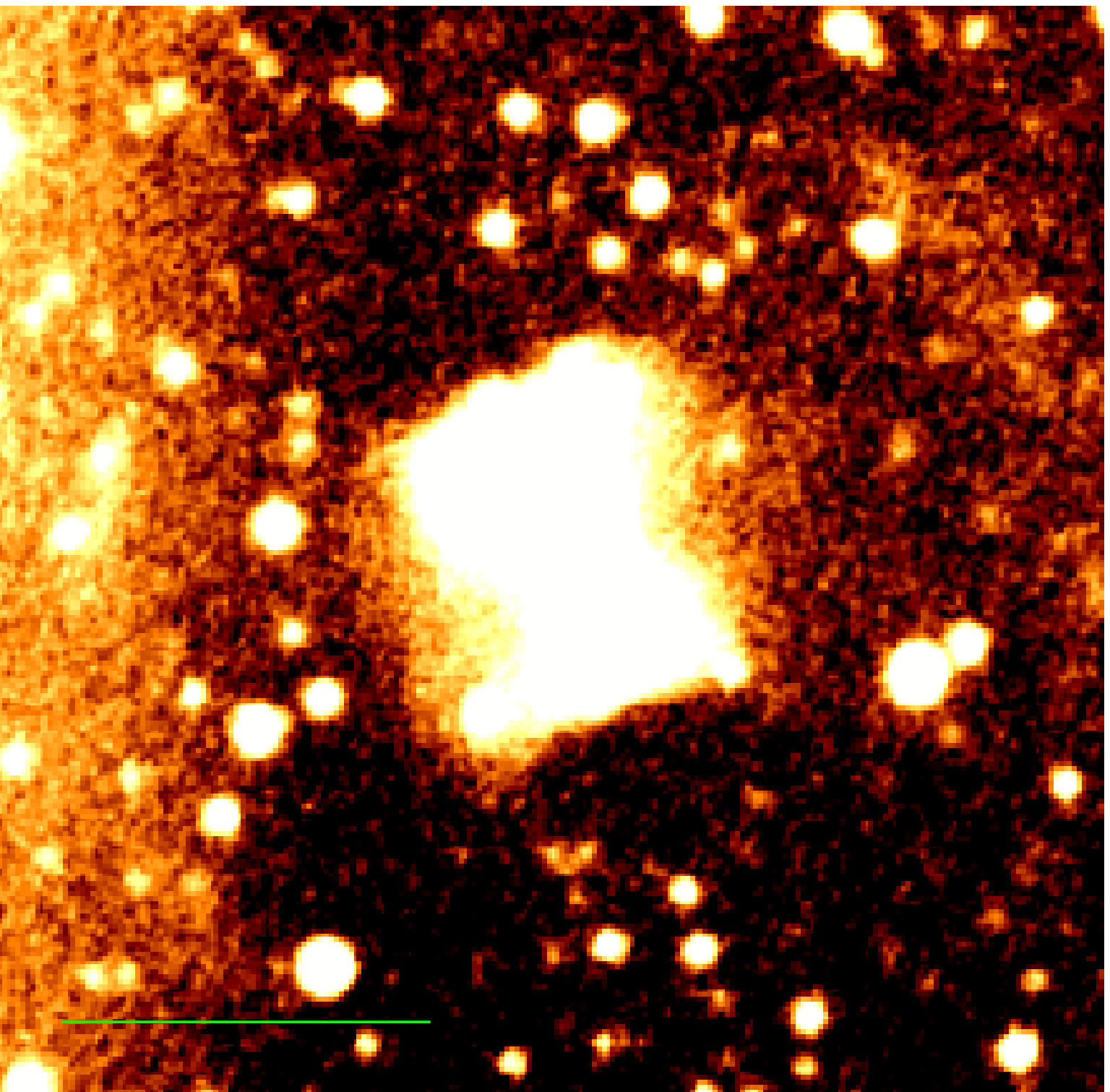}		\hspace{0.8cm}
\includegraphics[width=0.42\textwidth]{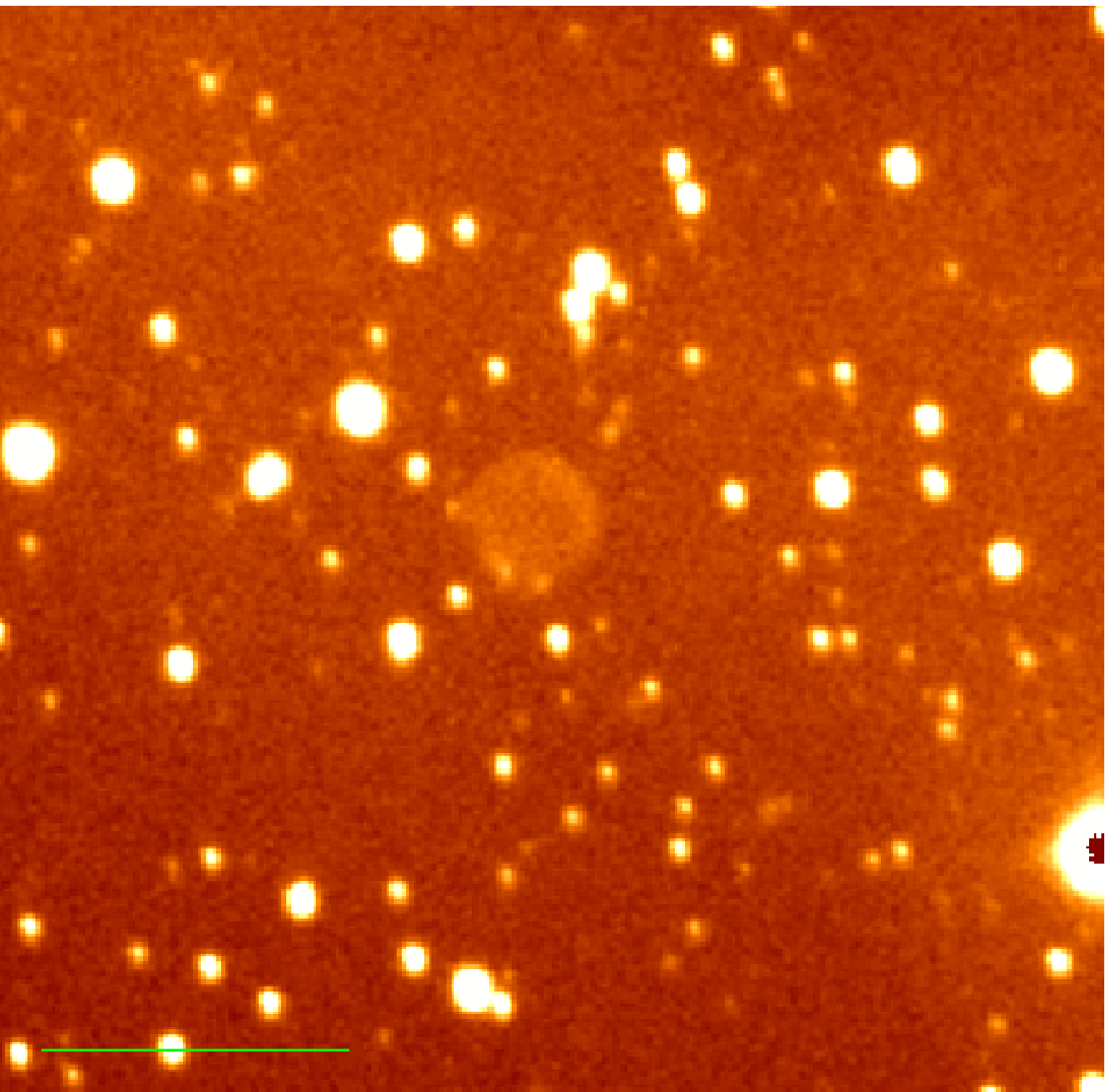}		\vspace{0.25cm}\\
\includegraphics[width=0.42\textwidth]{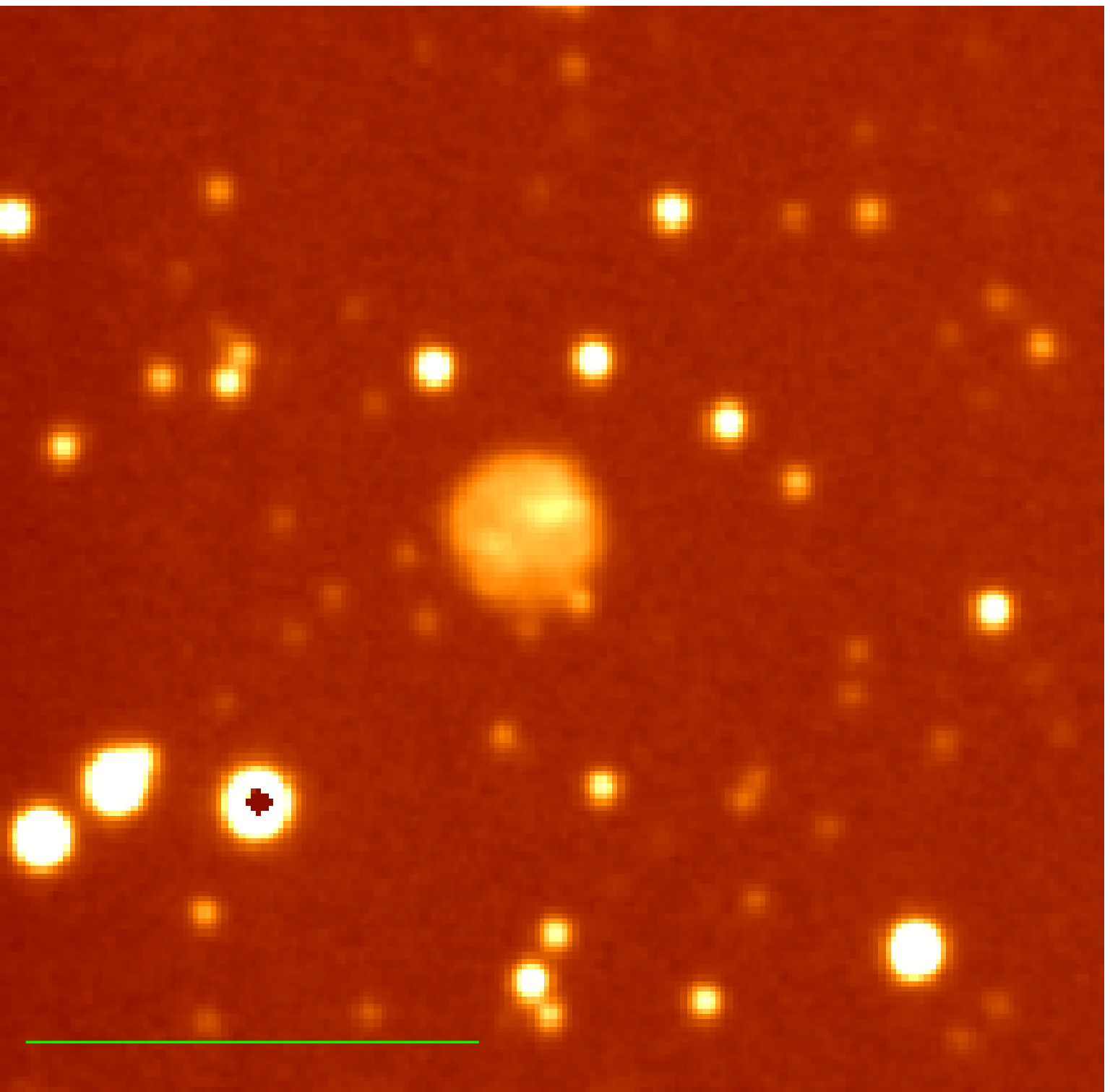}		\hspace{0.8cm}
\includegraphics[width=0.42\textwidth]{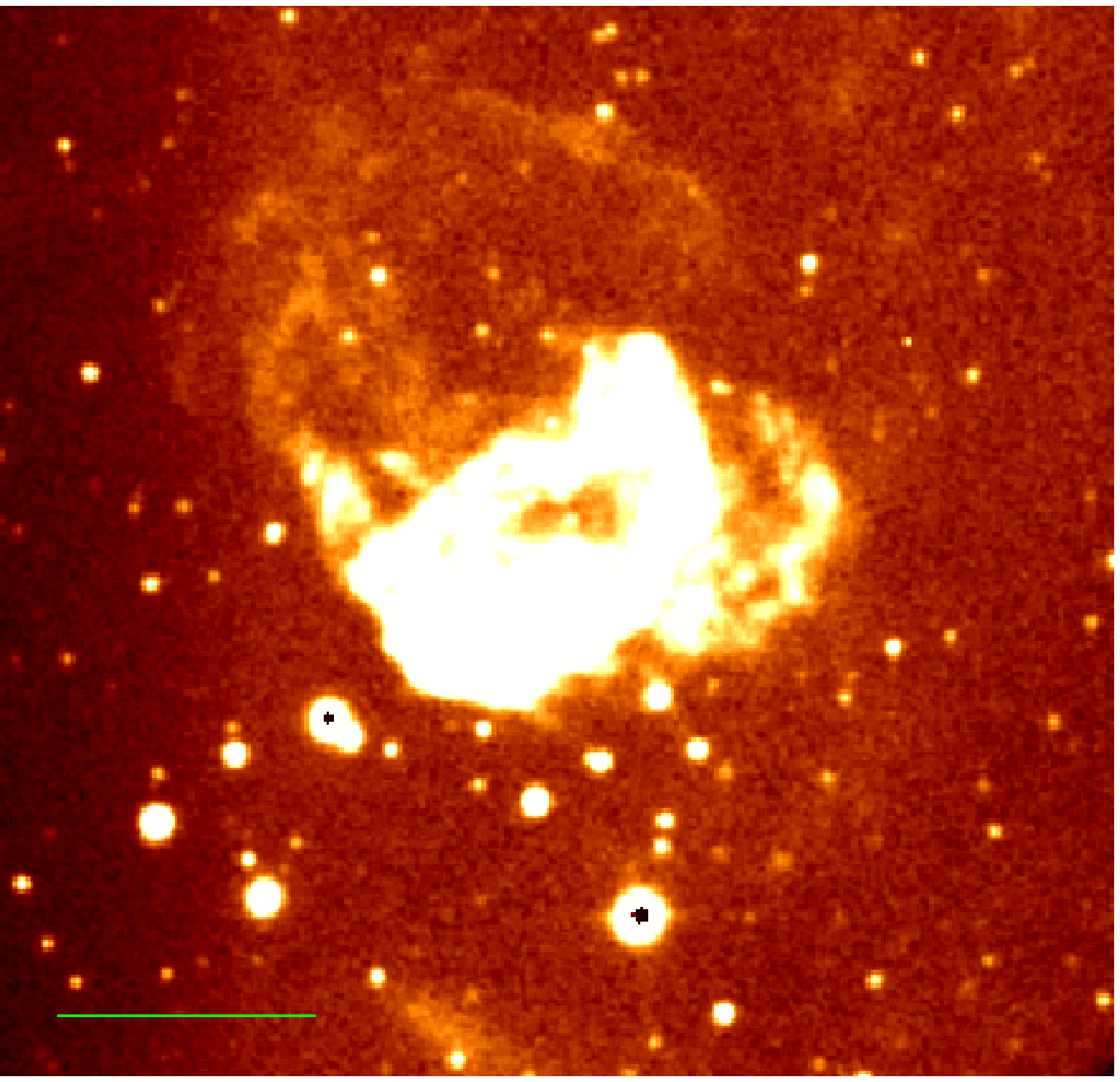}		\vspace{0.25cm}\\
\includegraphics[width=0.42\textwidth]{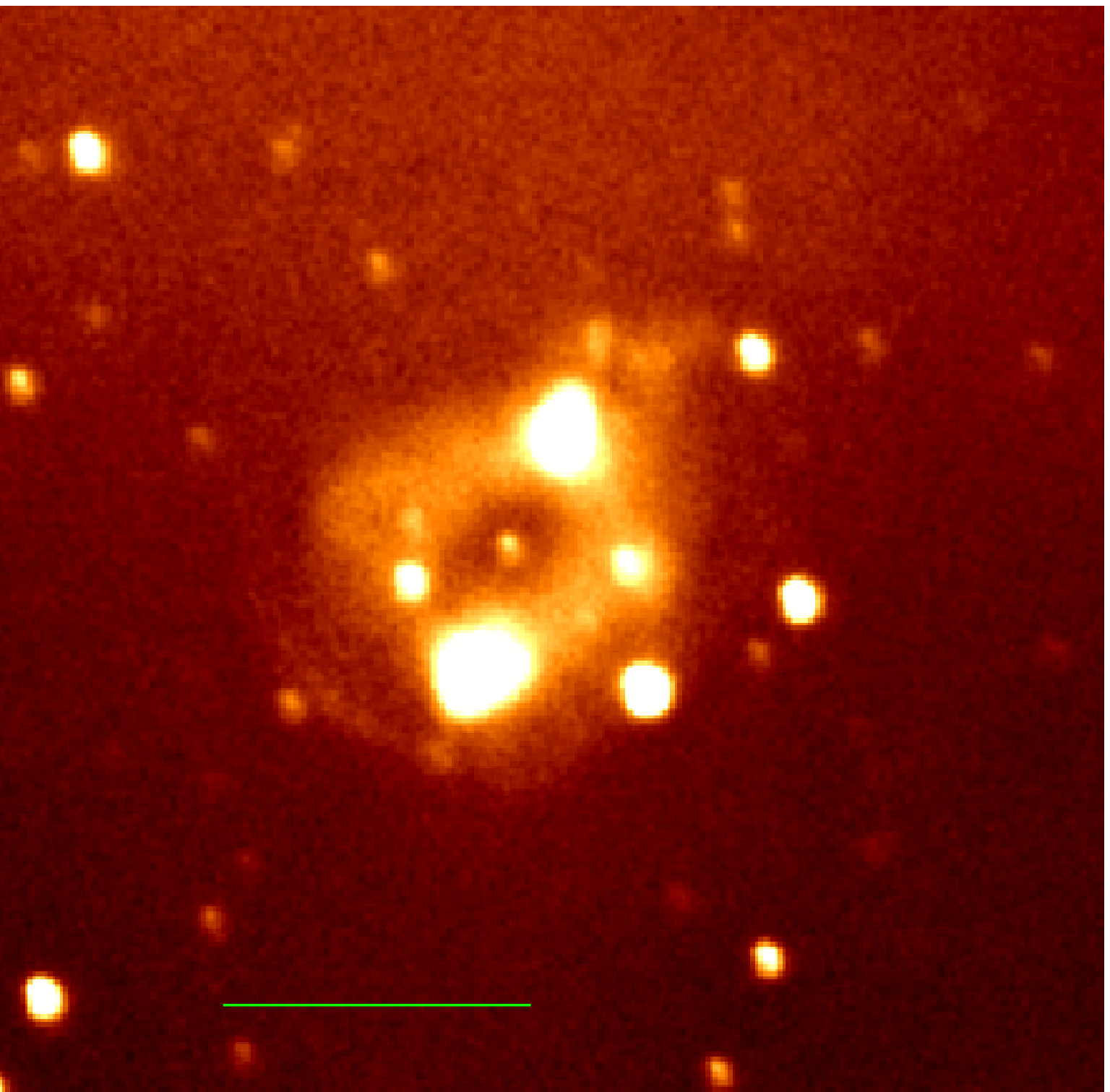}		\hspace{0.8cm}
\includegraphics[width=0.42\textwidth]{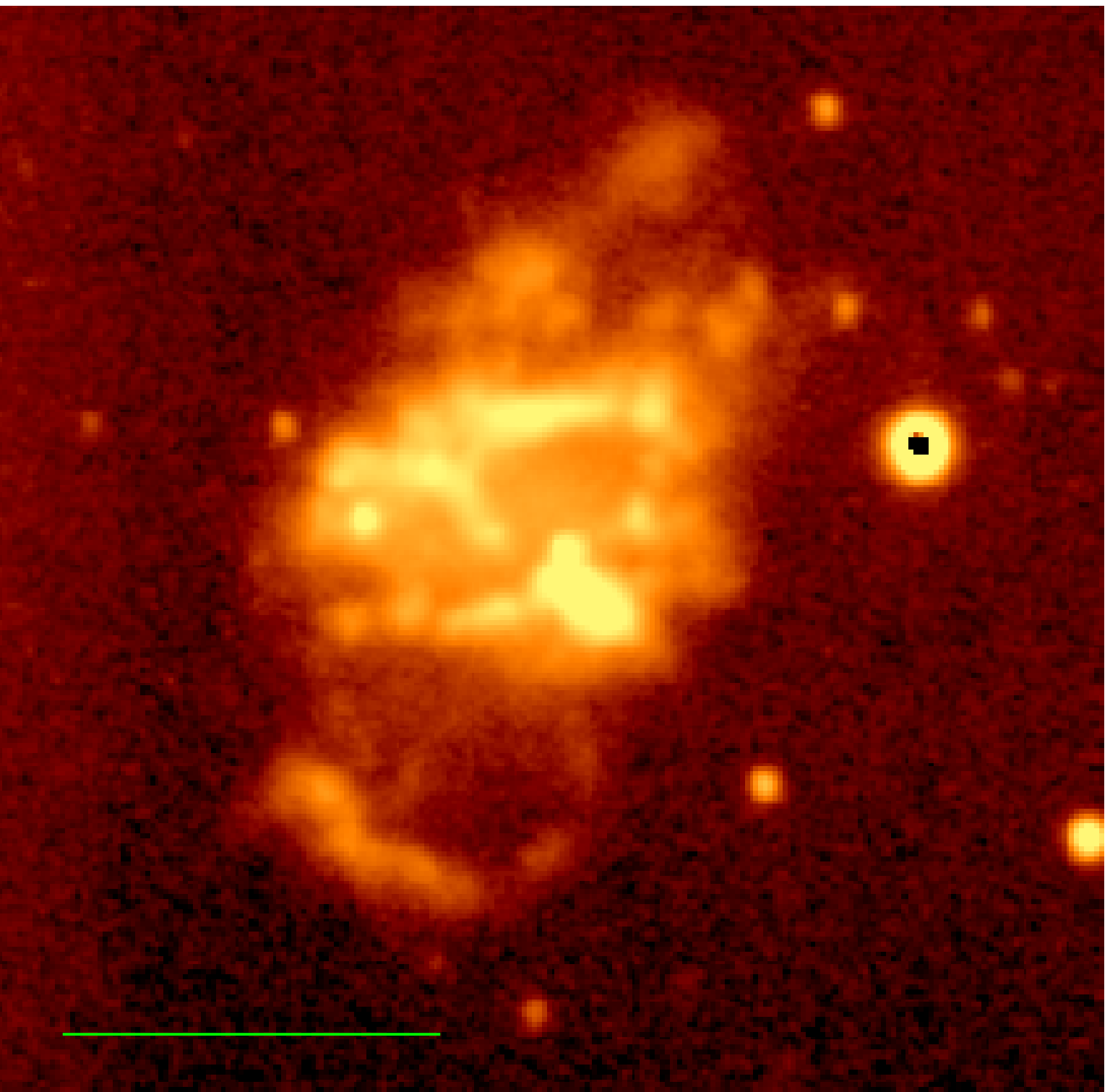}
\caption{Same as Fig.~\ref{Fig_1},
top PHR0803$-$3331 (L) and PHR0736$-$3901, middle PHR0820$-$3516 and VBRC~1, bottom  Wray~17$-$1 and He~2$-$11 (L).}
  \label{Fig_6}
\end{figure*}

\begin{figure*}[!ht]
  \centering
\includegraphics[width=0.42\textwidth]{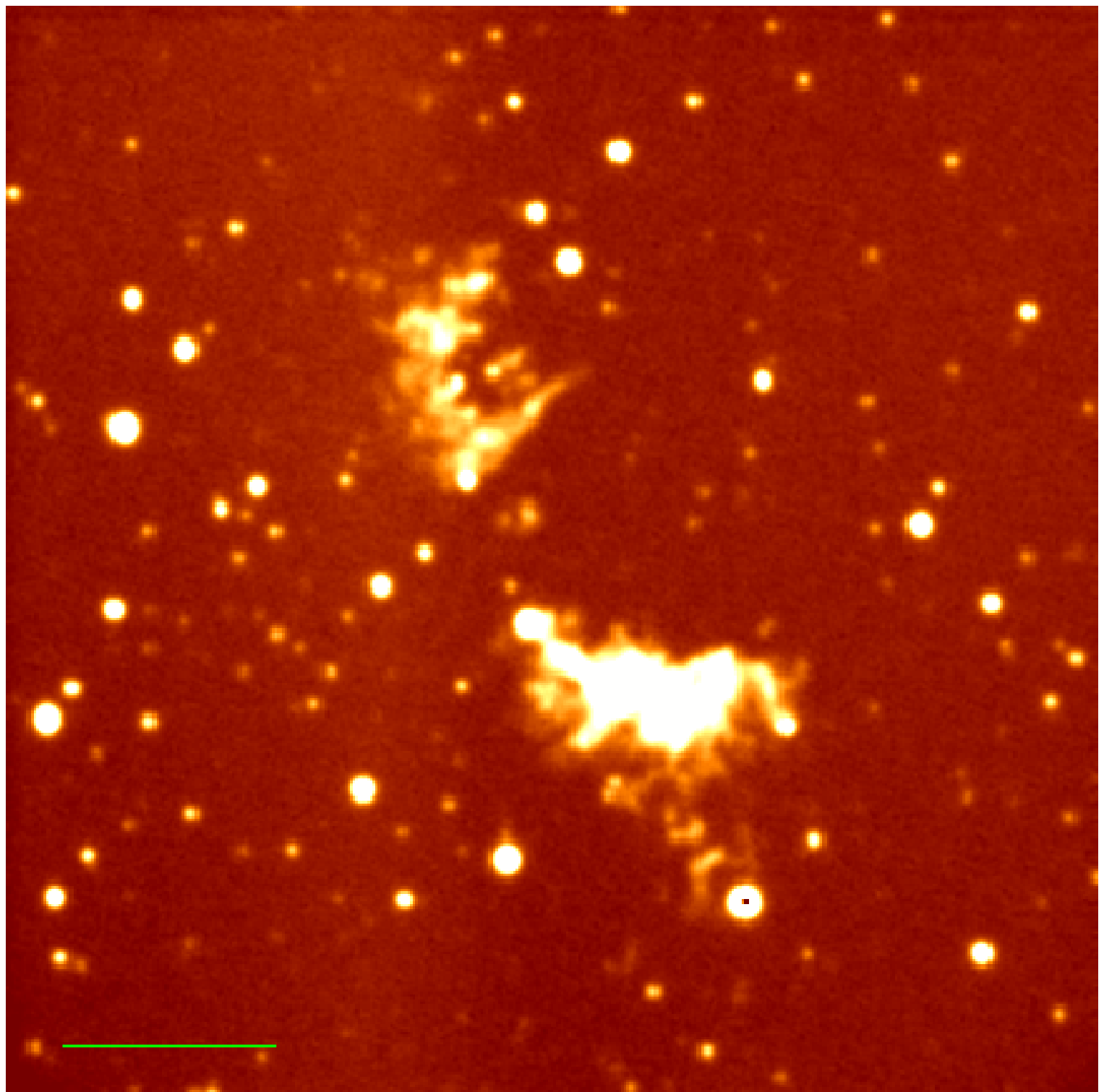}		\hspace{0.8cm}
\includegraphics[width=0.42\textwidth]{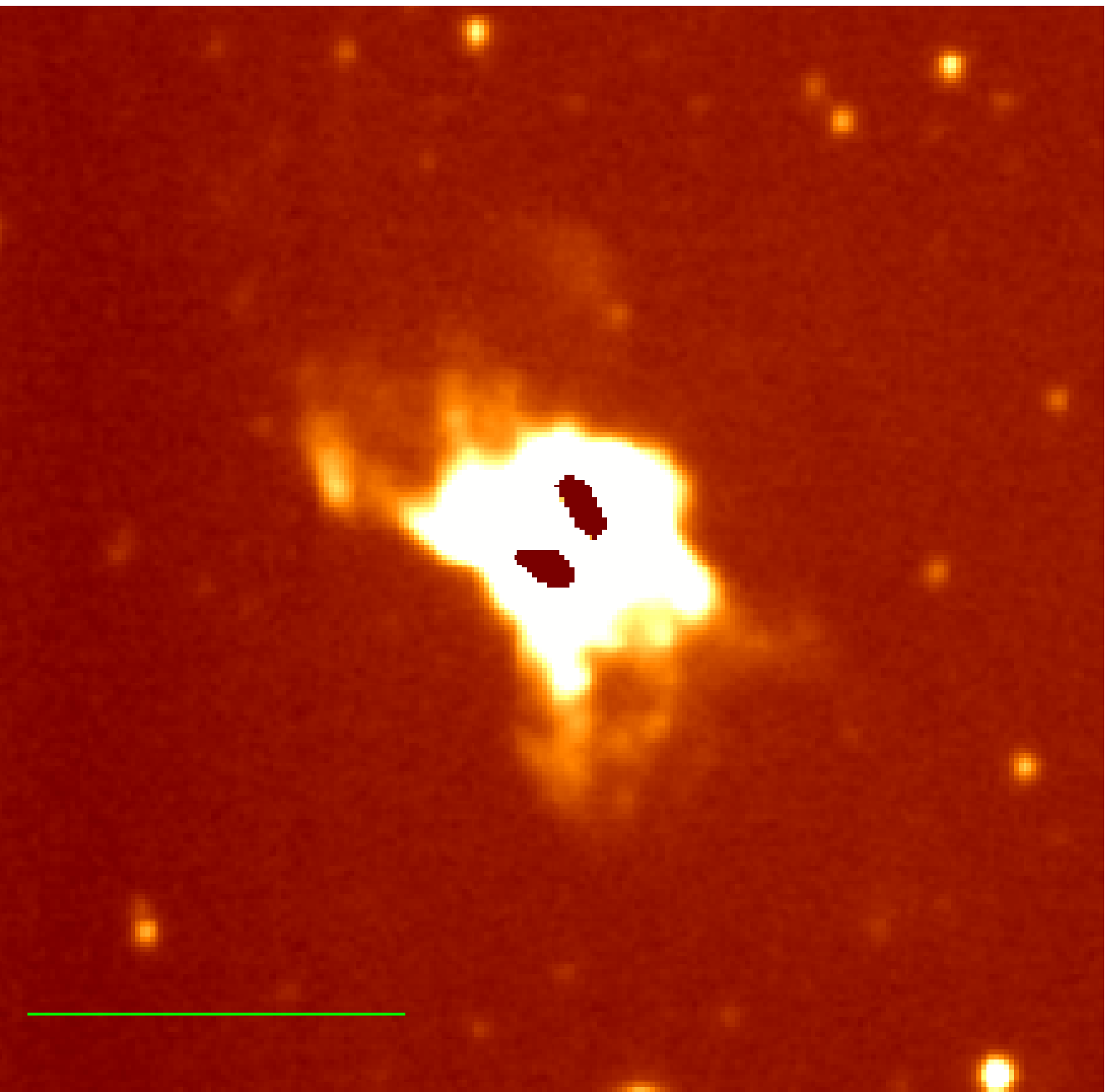}		\vspace{0.25cm}\\
\includegraphics[width=0.42\textwidth]{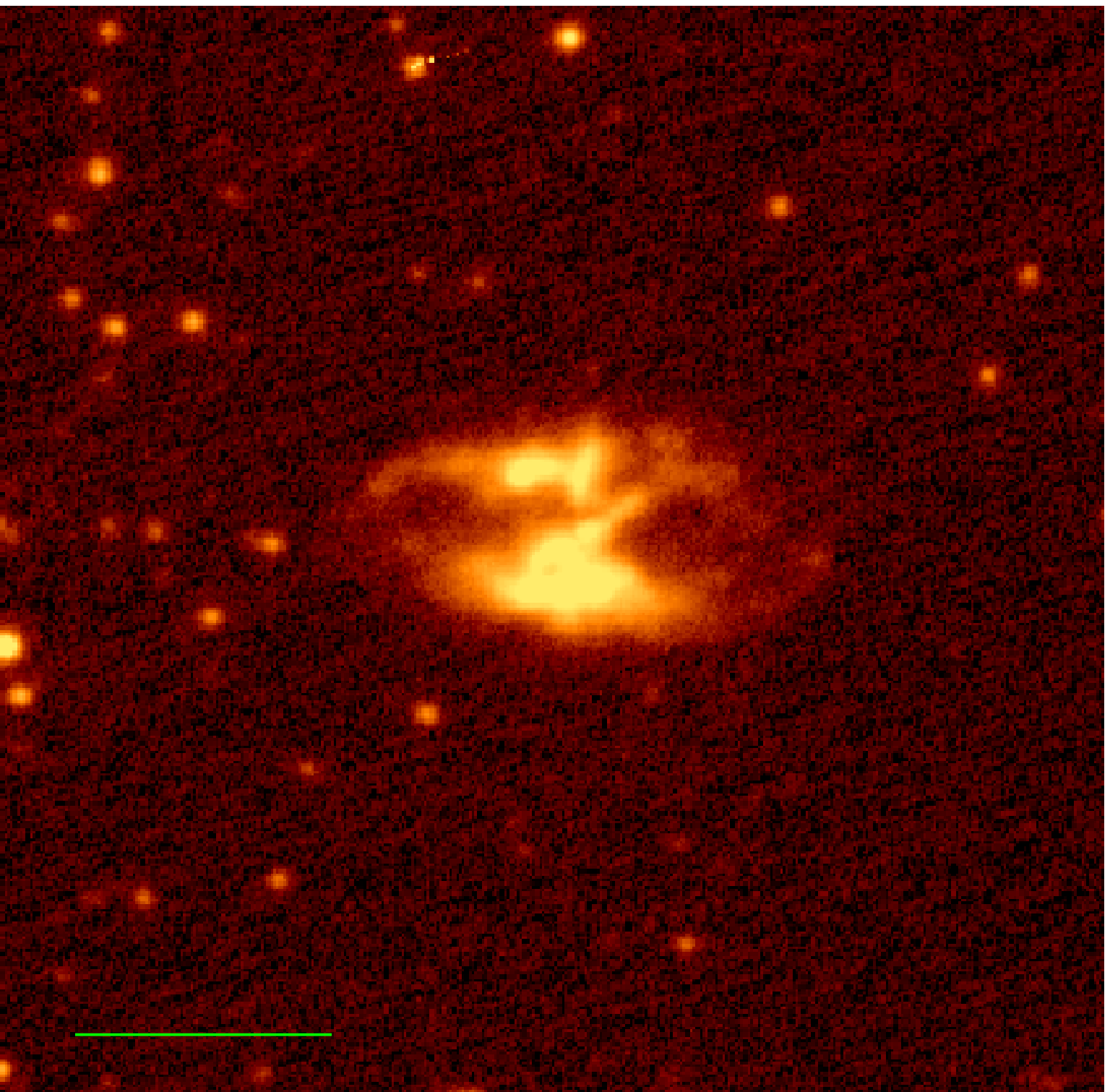}	\hspace{0.8cm}
\includegraphics[width=0.42\textwidth]{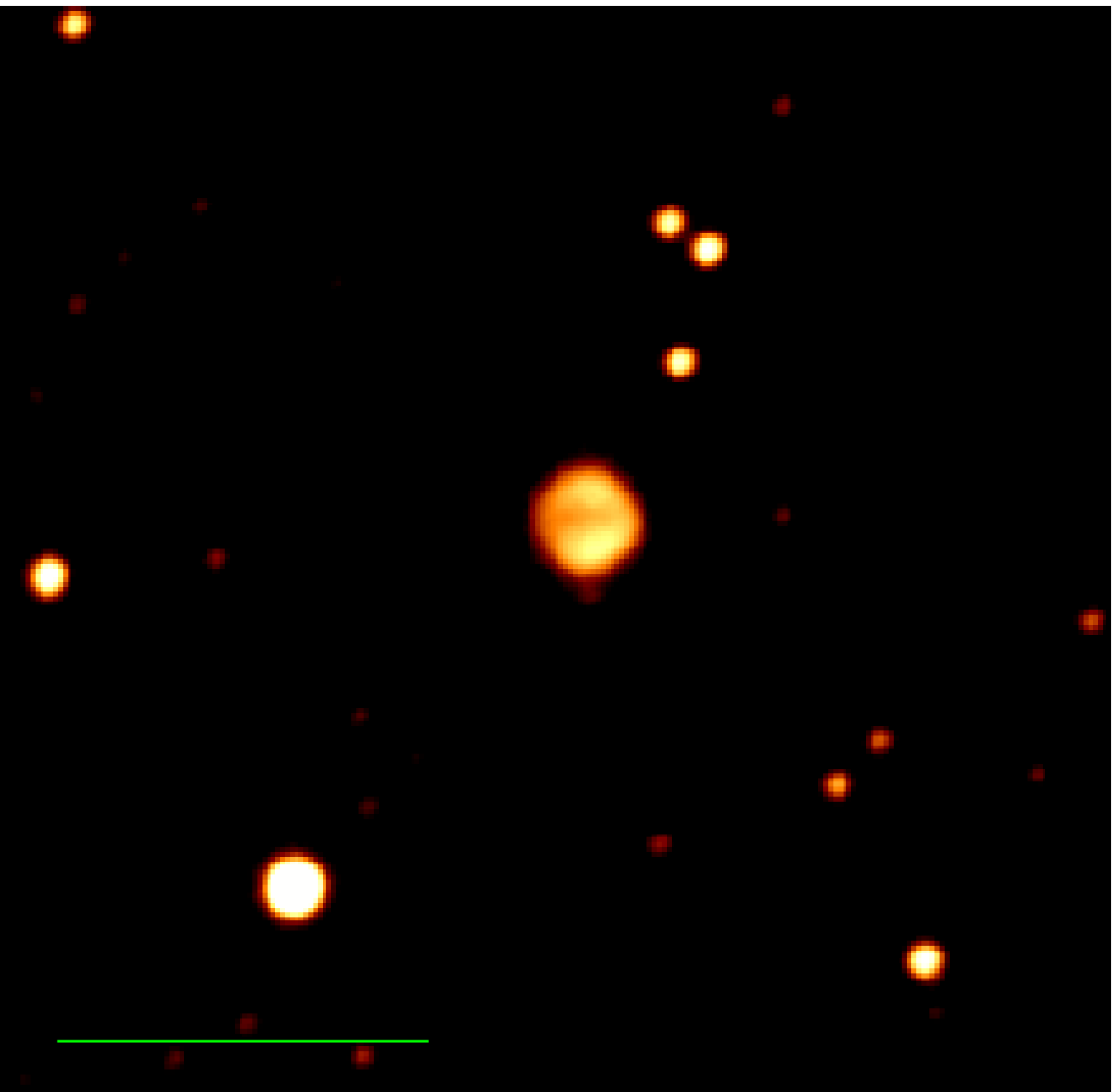}		\vspace{0.25cm}\\
\includegraphics[width=0.42\textwidth]{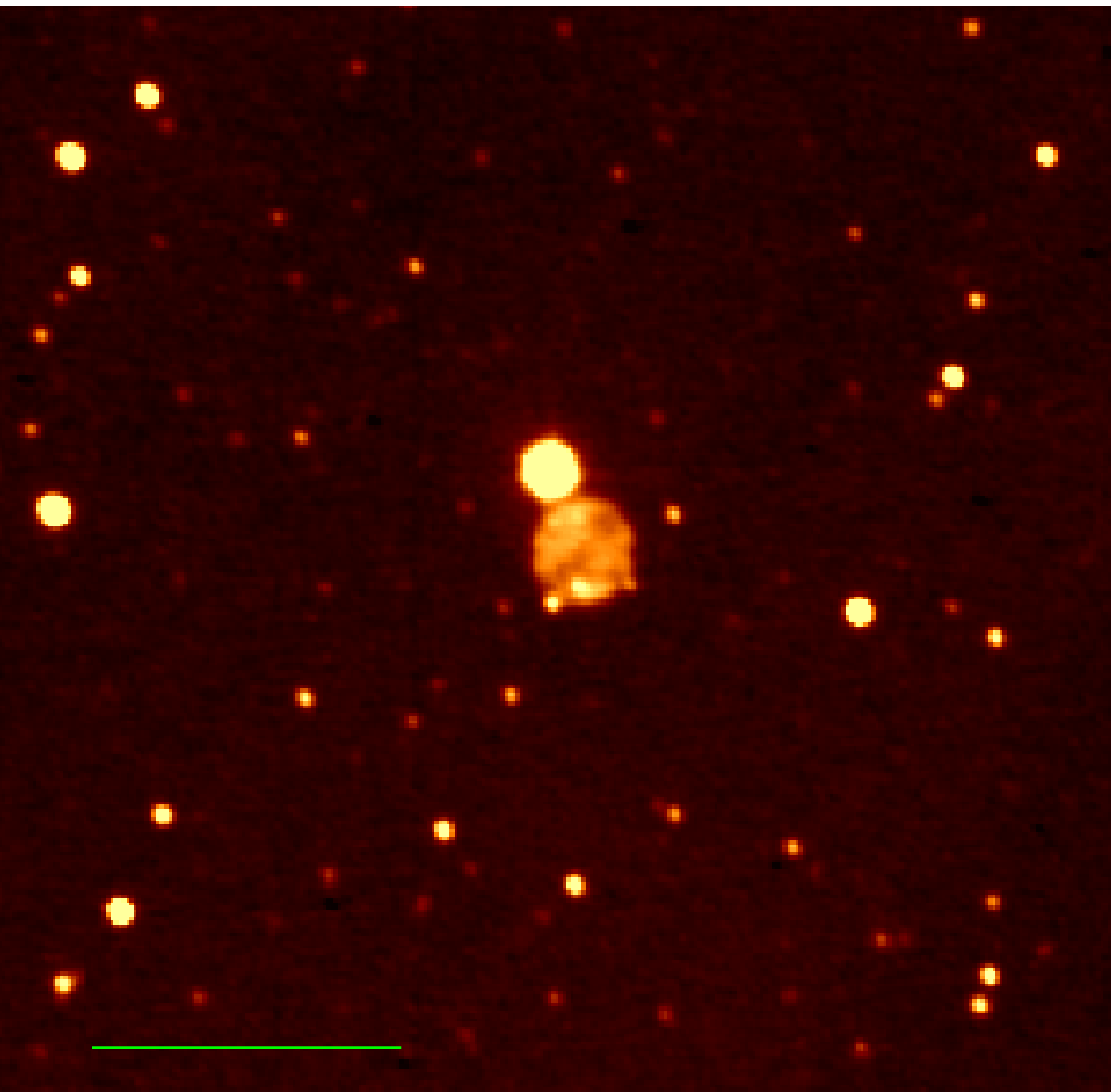}		\hspace{0.8cm}
\includegraphics[width=0.42\textwidth]{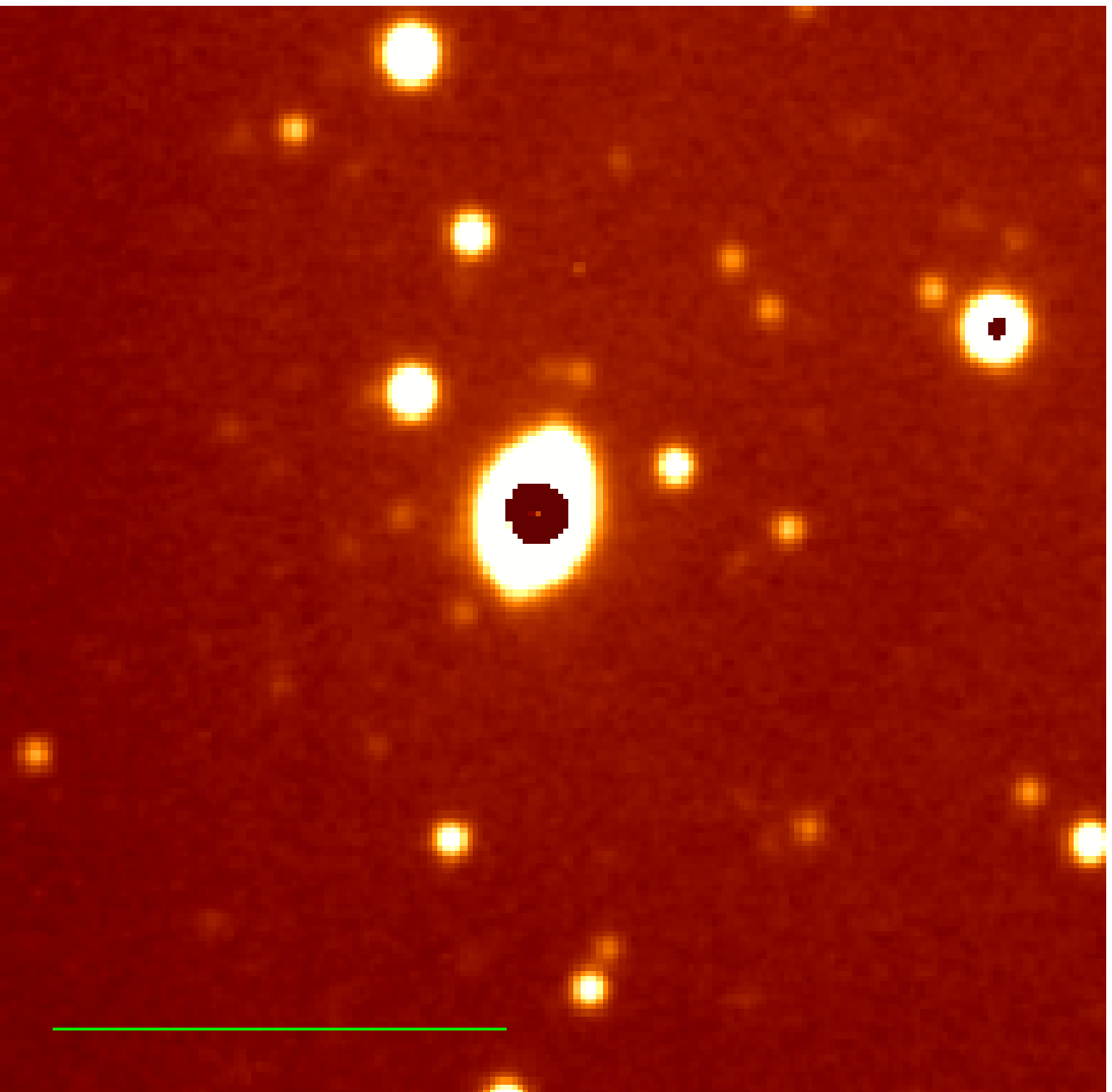}
\caption{Same as Fig.~\ref{Fig_1},
top Wray~16$-$20 and He~2$-$15, middle NGC~2818 (L) and Wray~17$-$18 (L), bottom ESO~209$-$15 and He~2$-$7.}
  \label{Fig_7}
\end{figure*}

\begin{figure*}[!ht]
  \centering
\includegraphics[width=0.42\textwidth]{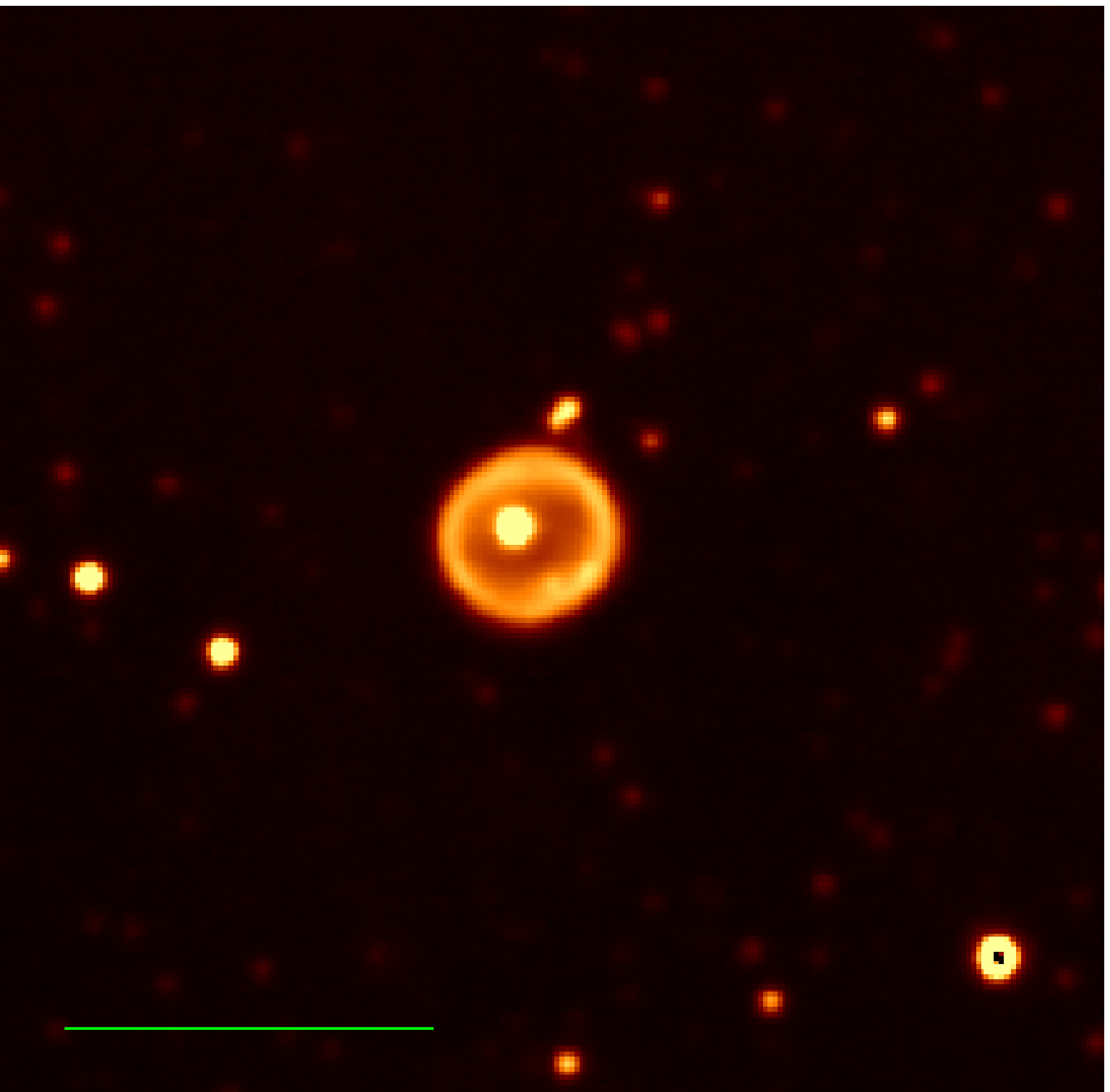}		\hspace{0.8cm}
\includegraphics[width=0.42\textwidth]{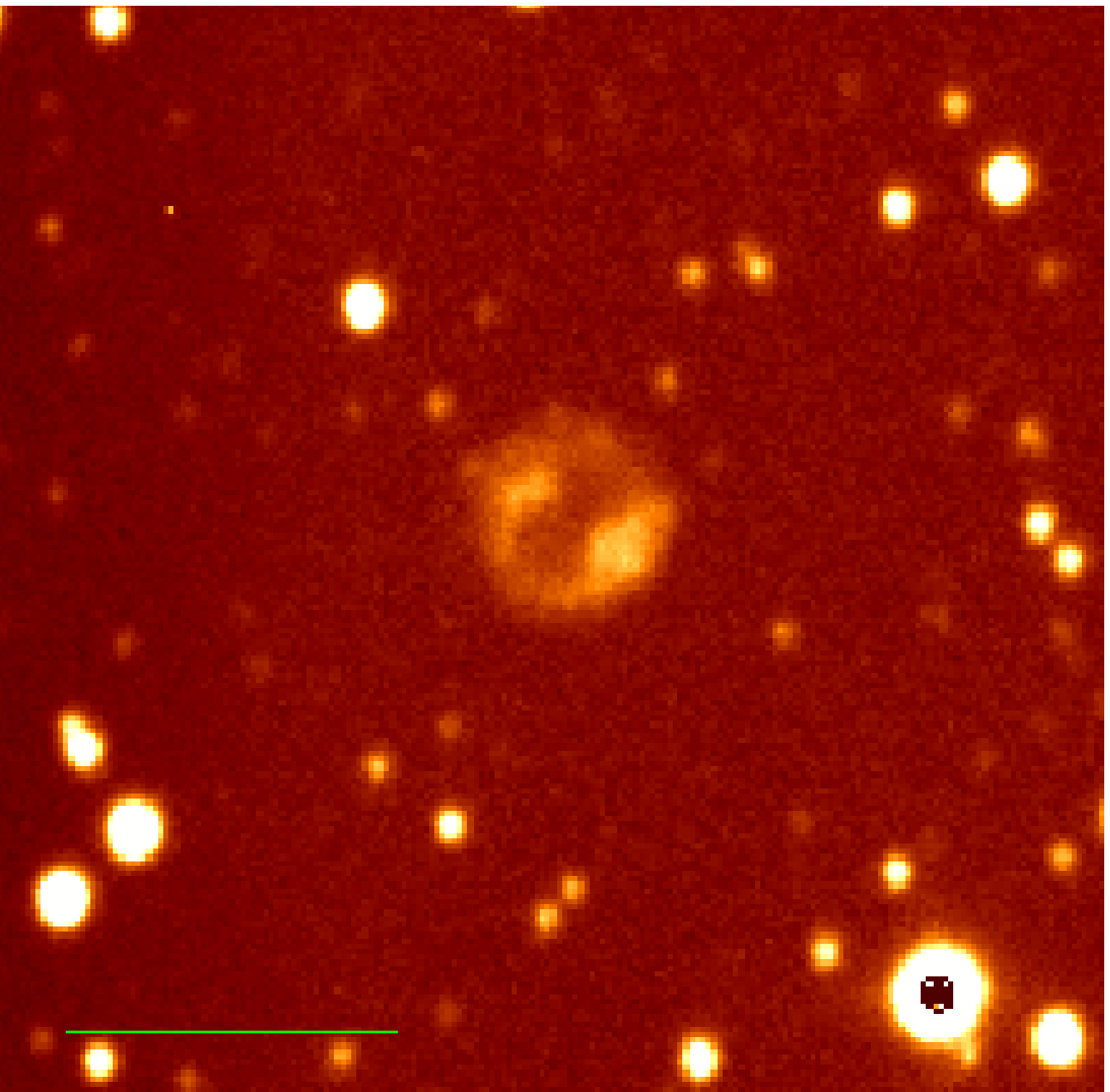}	\vspace{0.25cm}\\
\includegraphics[width=0.42\textwidth]{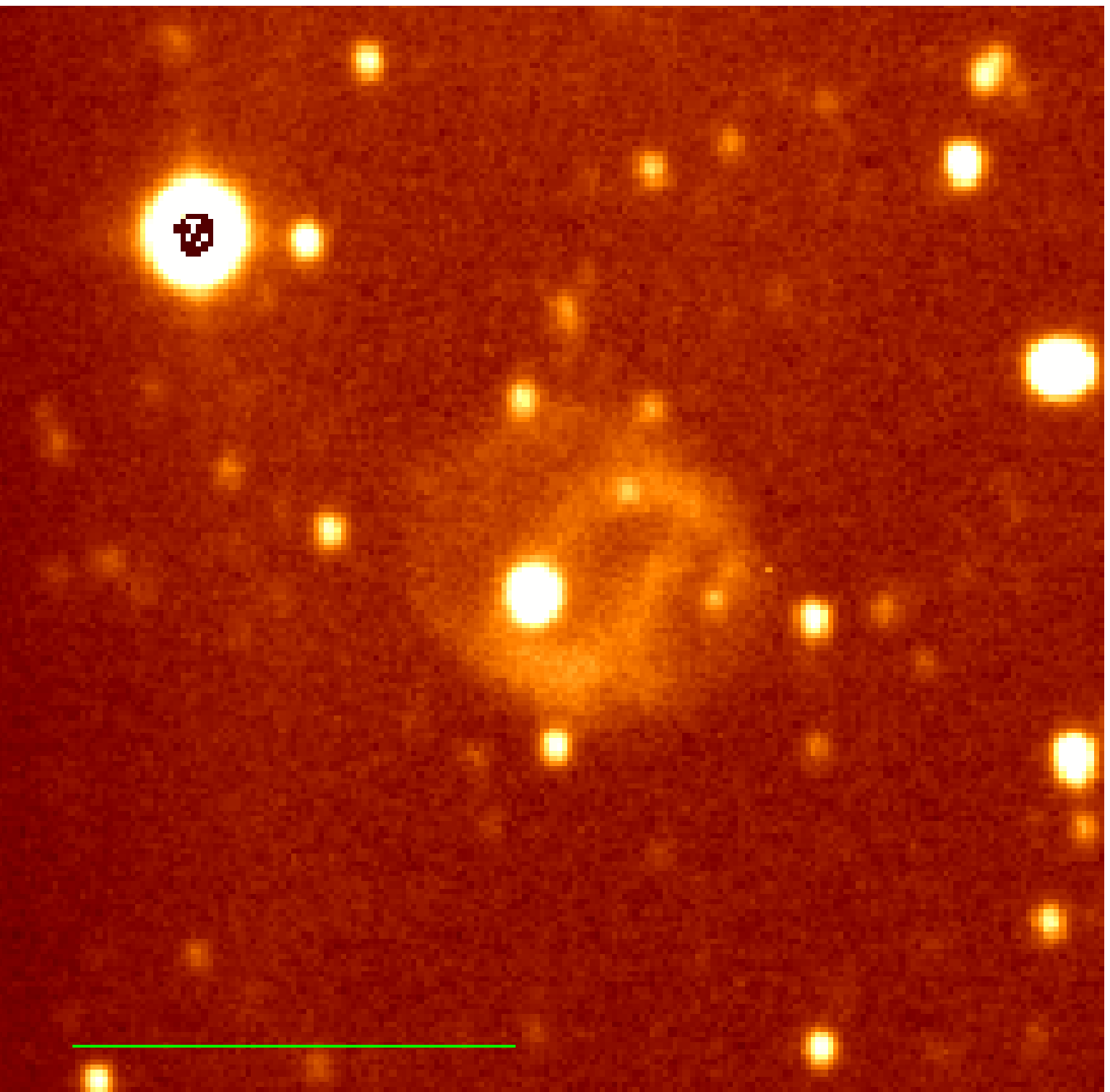}	\hspace{0.8cm}
\includegraphics[width=0.42\textwidth]{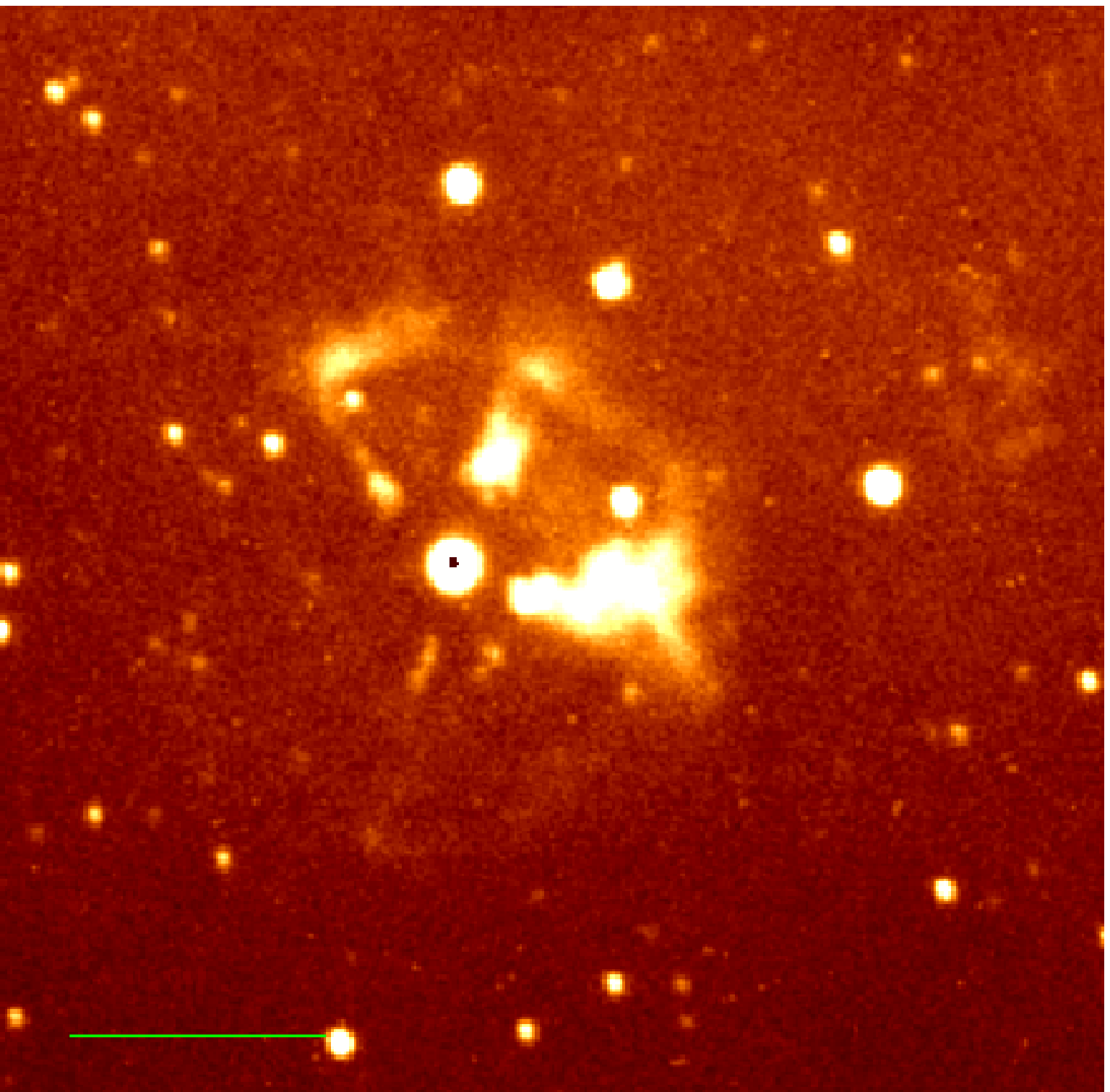}		\vspace{0.25cm}\\
\includegraphics[width=0.42\textwidth]{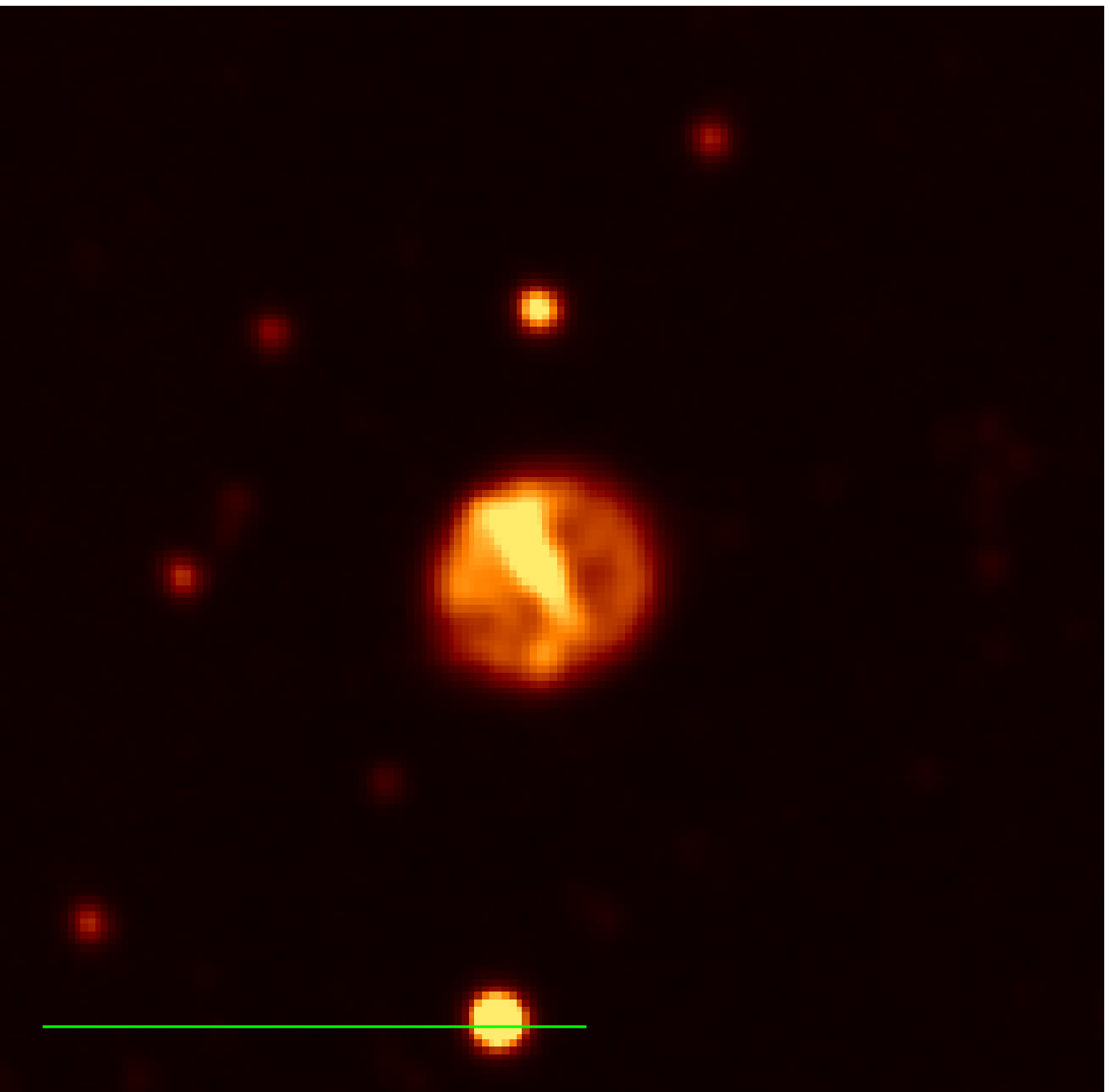}		\hspace{0.8cm}
\includegraphics[width=0.42\textwidth]{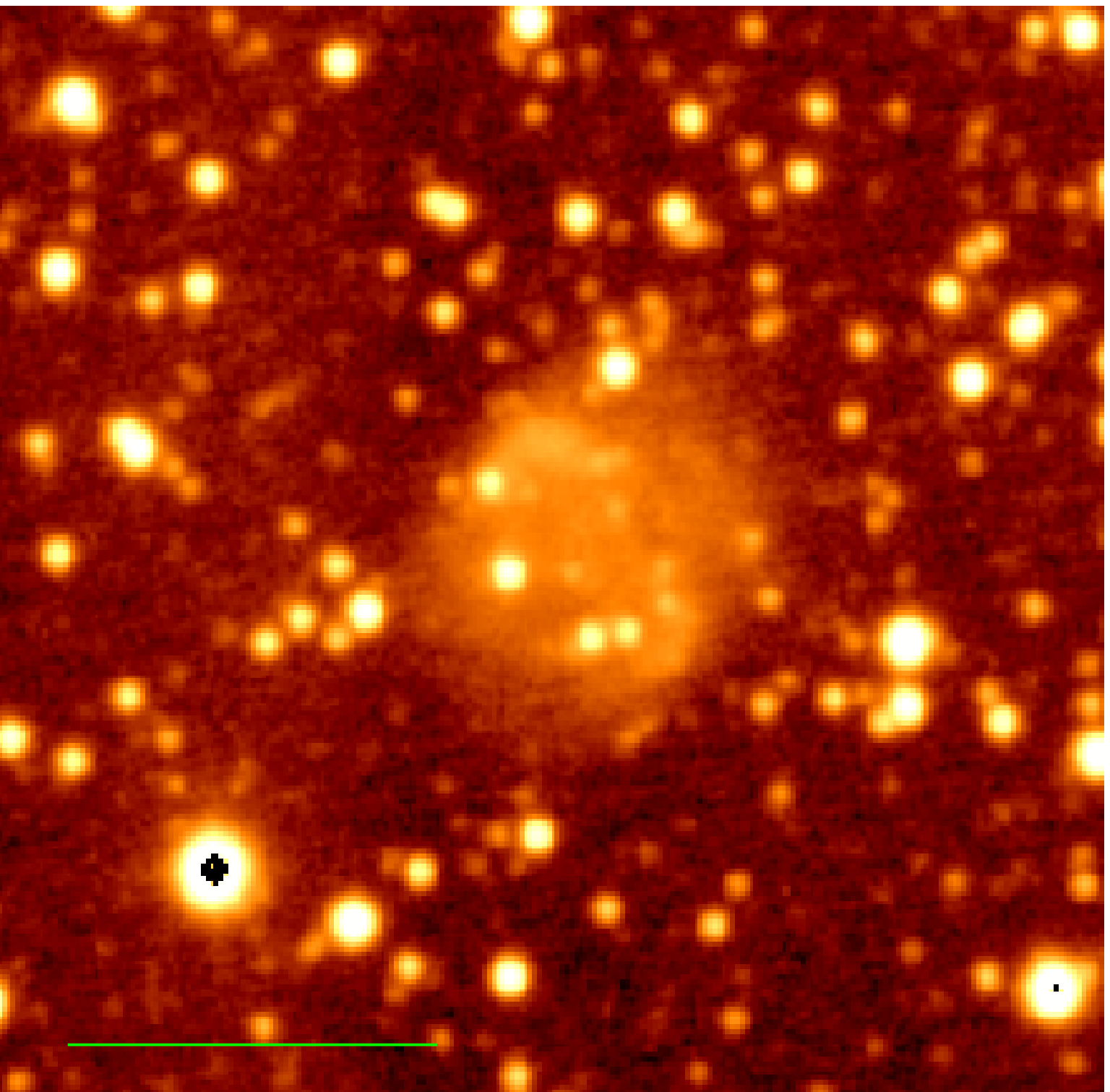}
\caption{Same as Fig.~\ref{Fig_1},
top ESO~259$-$10 and PHR0911$-$4205, middle PHR0927$-$4347 and  PHR0905$-$4753, bottom He~2$-$37 and PHR0924$-$5506.}
  \label{Fig_8}
\end{figure*}

\begin{figure*}[!ht]
  \centering
\includegraphics[width=0.42\textwidth]{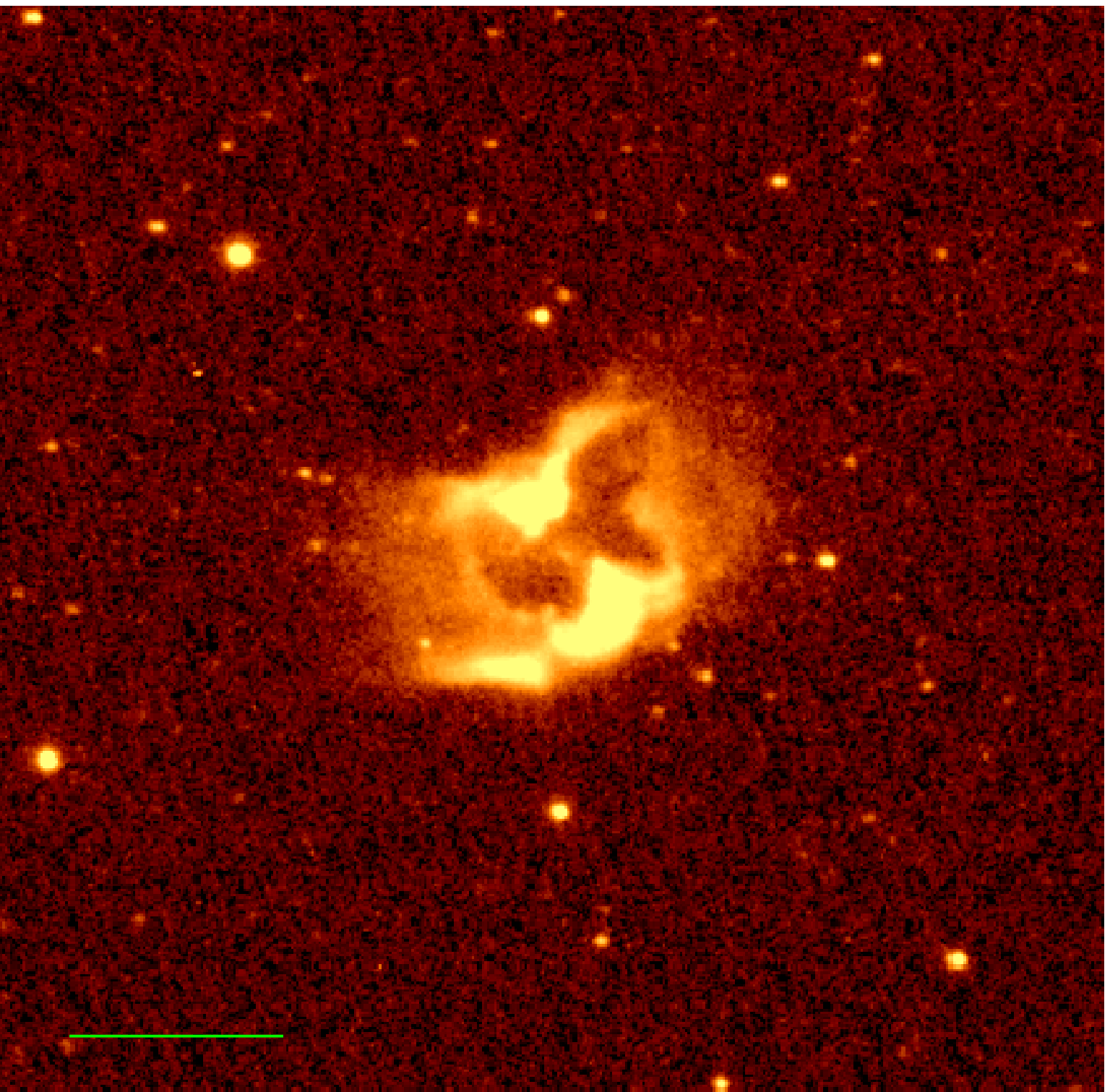}		\hspace{0.8cm}
\includegraphics[width=0.42\textwidth]{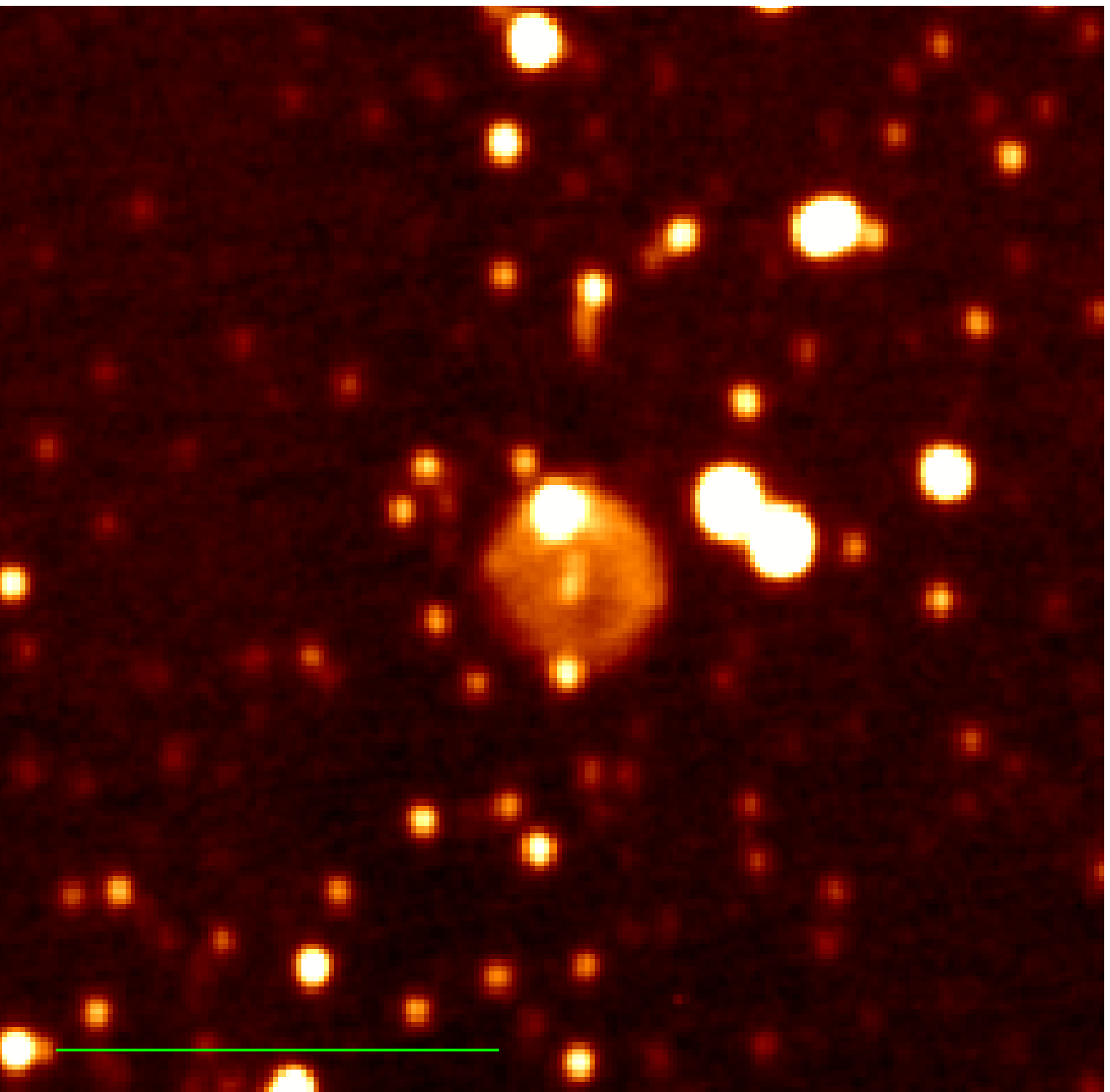}		\vspace{0.25cm}\\
\includegraphics[width=0.42\textwidth]{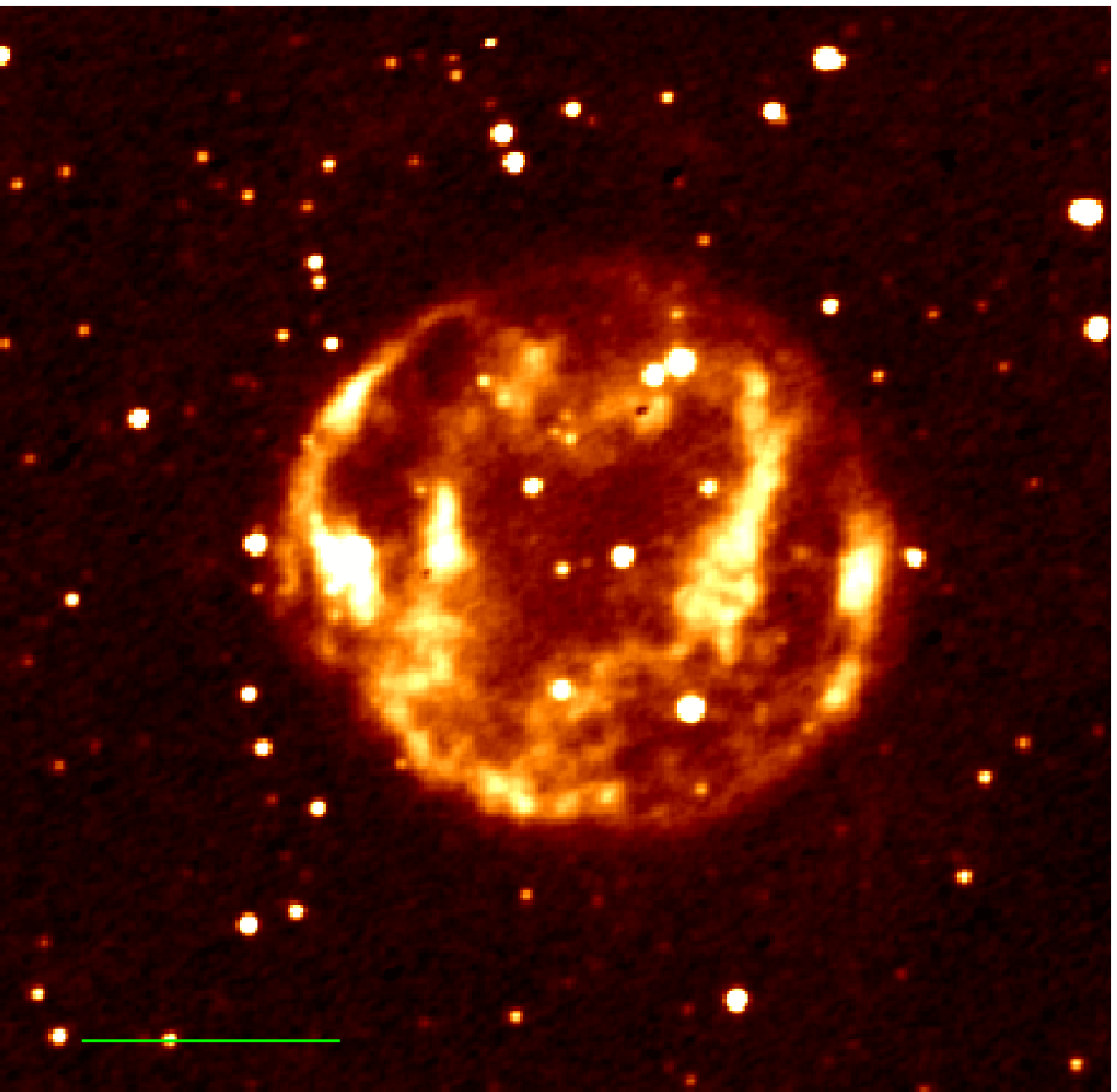}		\hspace{0.8cm}
\includegraphics[width=0.42\textwidth]{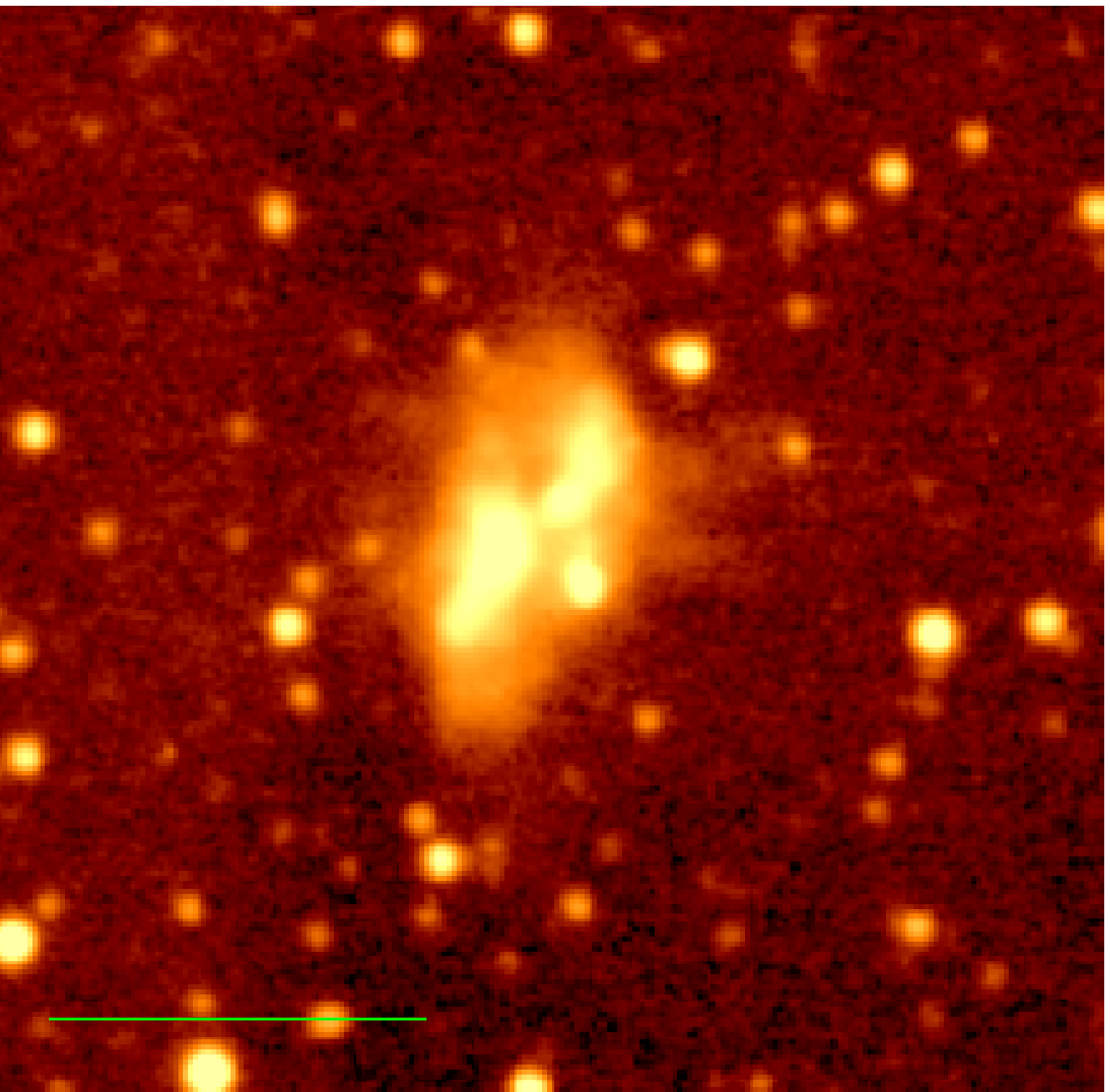}		\vspace{0.25cm}\\
\includegraphics[width=0.42\textwidth]{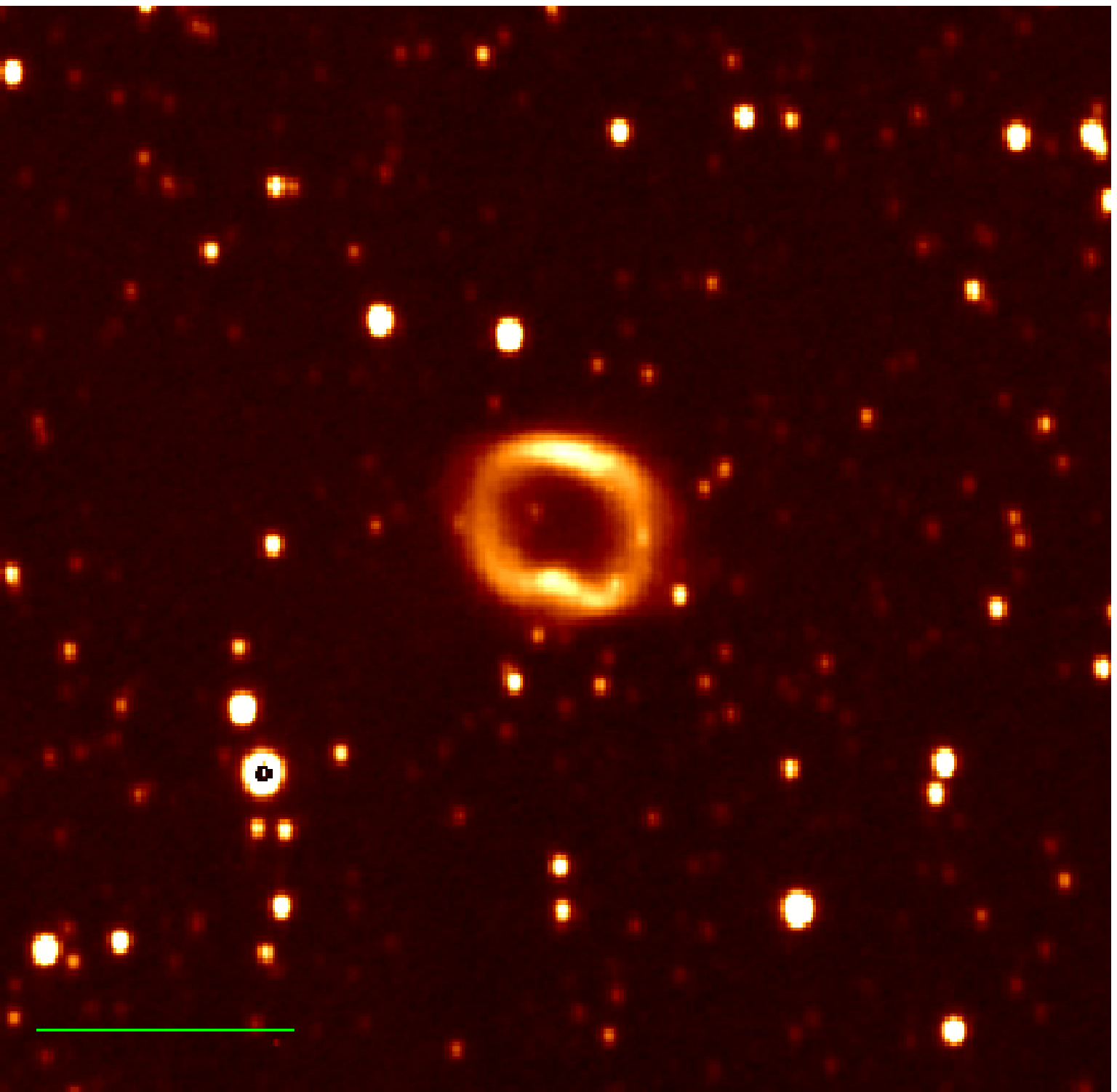}		\hspace{0.8cm}
\includegraphics[width=0.42\textwidth]{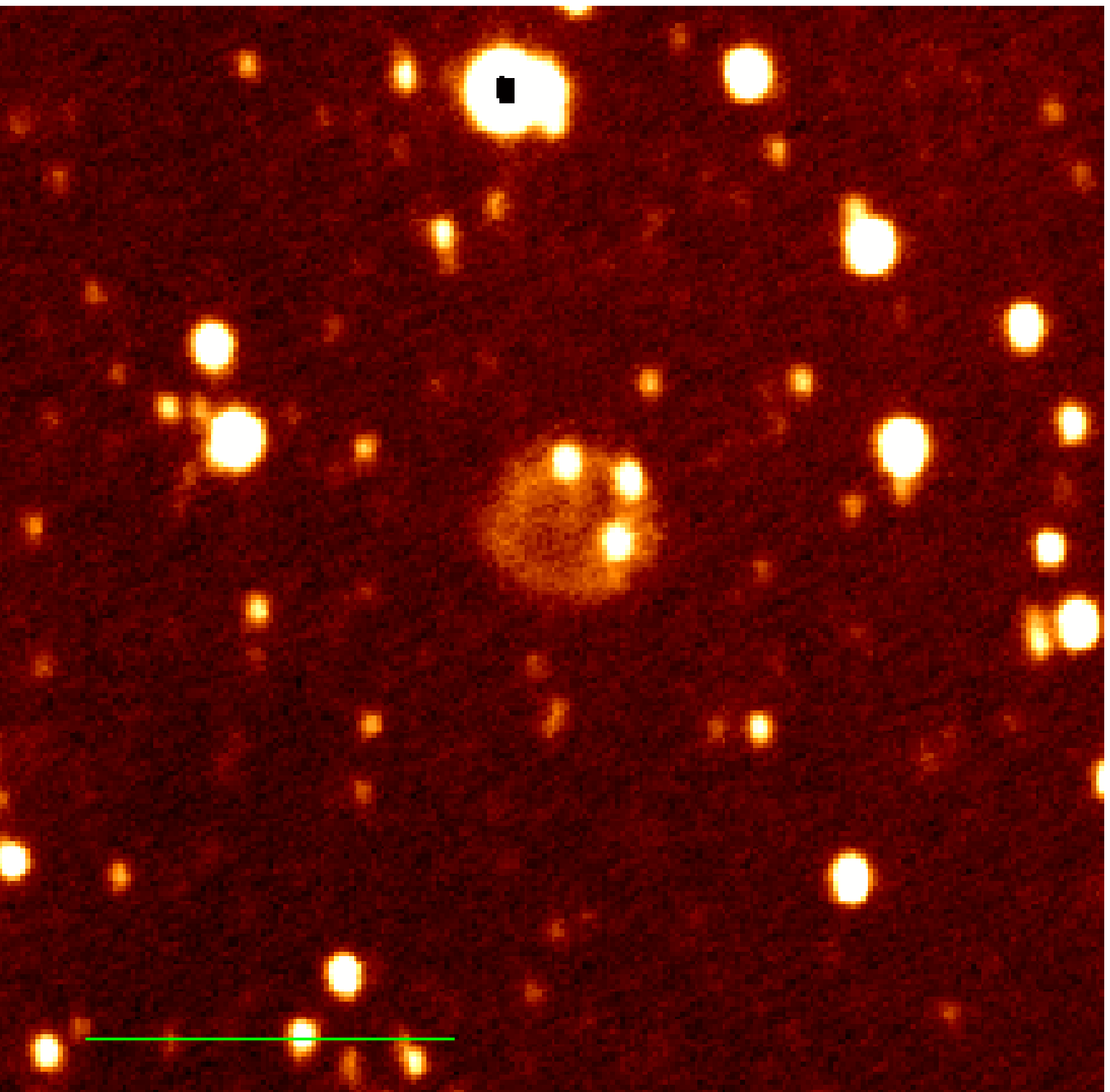}
\caption{Same as Fig.~\ref{Fig_1},
top NGC~2899 (L) and PHR0958$-$5039, middle Wray~17$-$31 and He~2$-$32 (L), bottom PHR0940$-$5658 and PHR1010$-$5146.}
   \label{Fig_9}
\end{figure*}

\begin{figure*}[!ht]
  \centering
\includegraphics[width=0.42\textwidth]{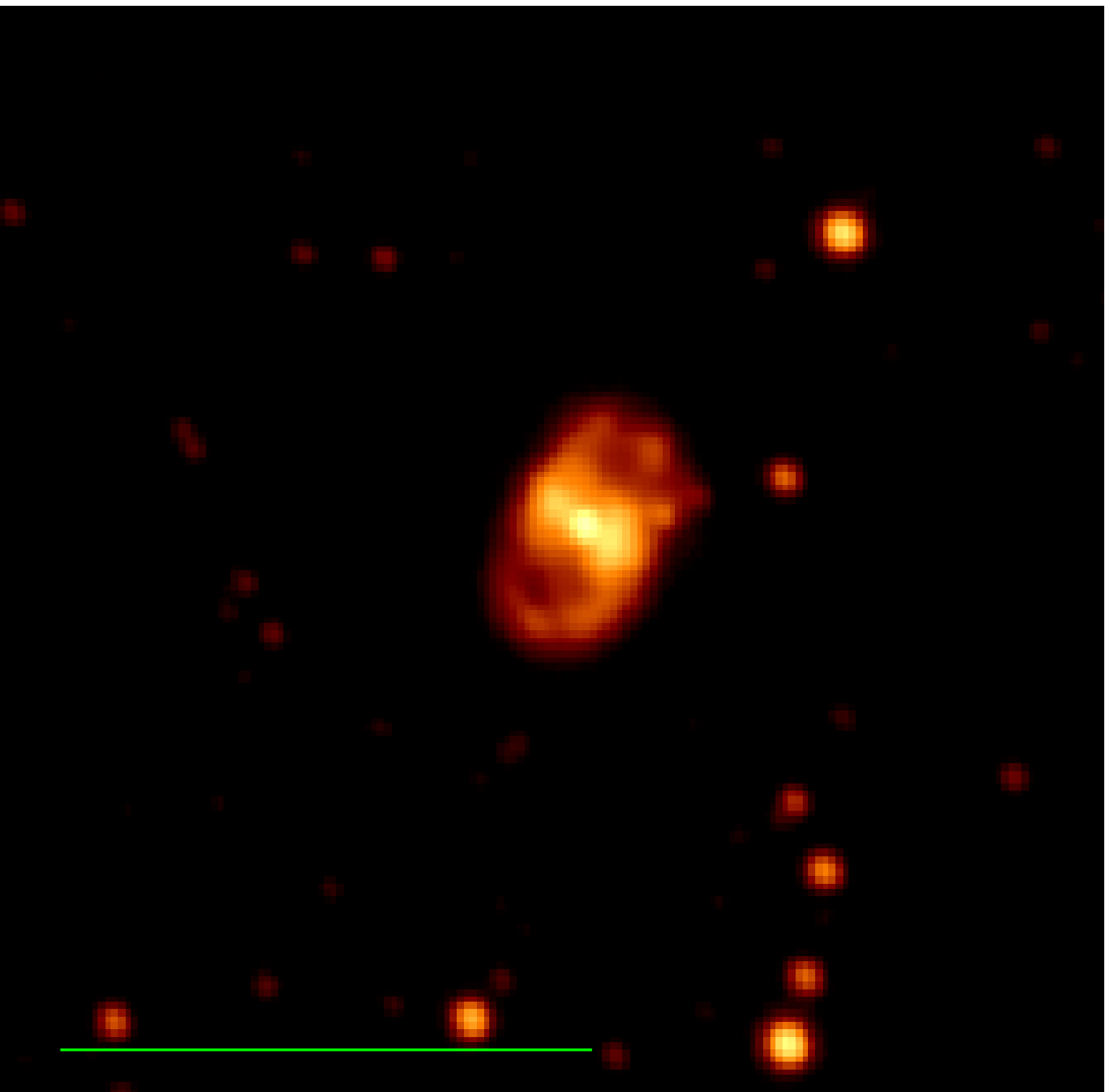}		\hspace{0.8cm}
\includegraphics[width=0.42\textwidth]{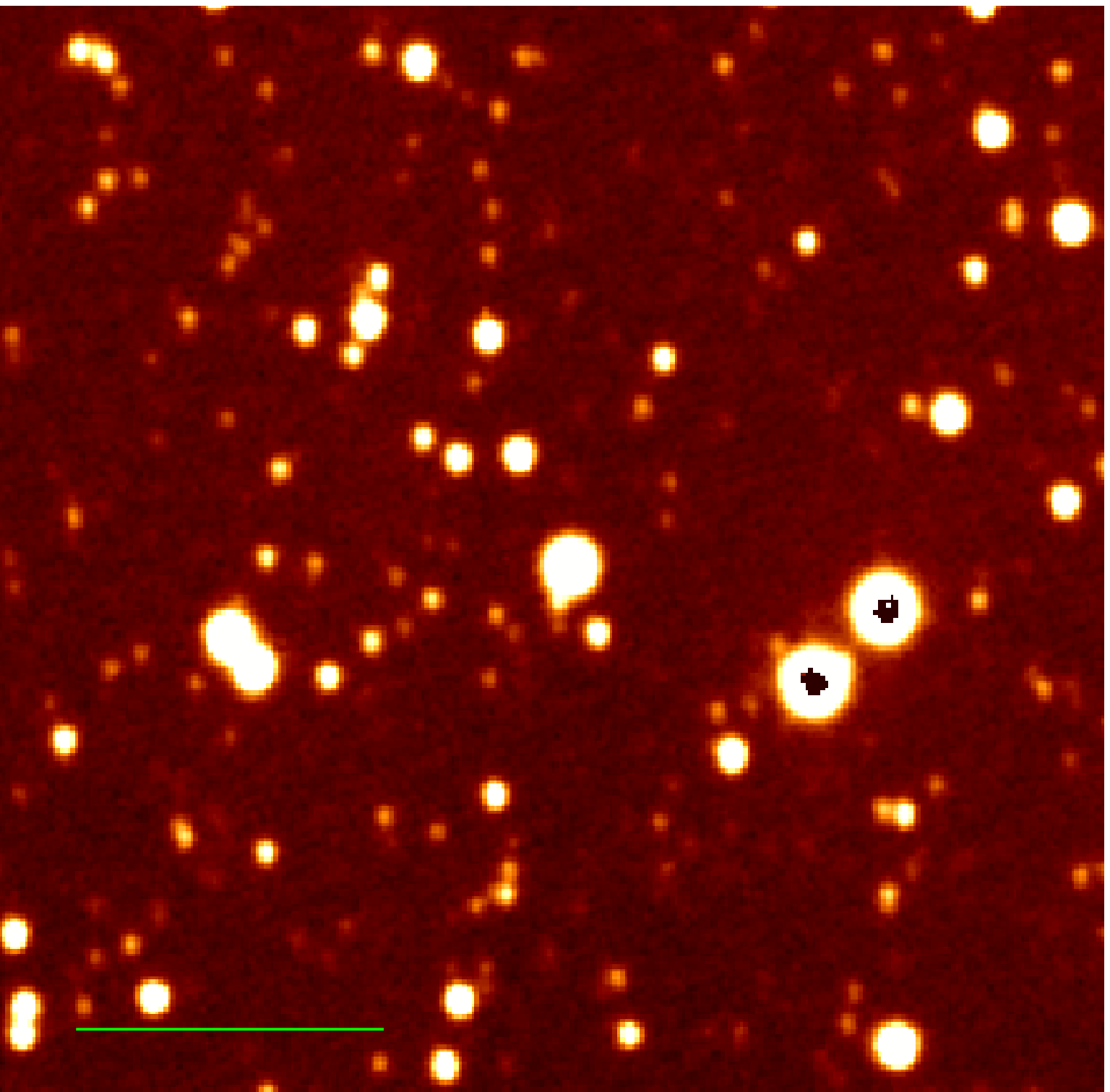}	\vspace{0.25cm}\\
\includegraphics[width=0.42\textwidth]{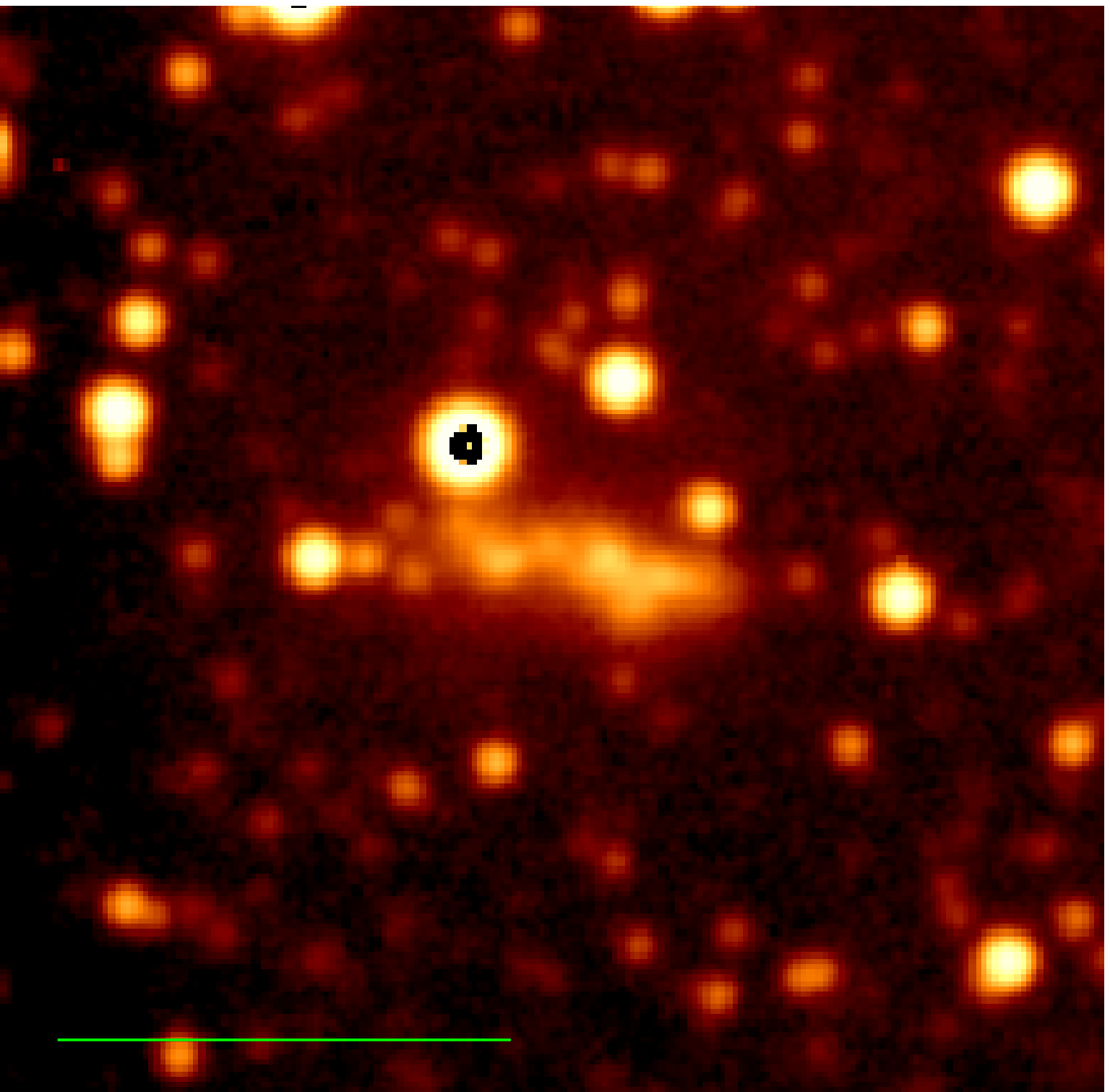}     	\hspace{0.8cm}
\includegraphics[width=0.42\textwidth]{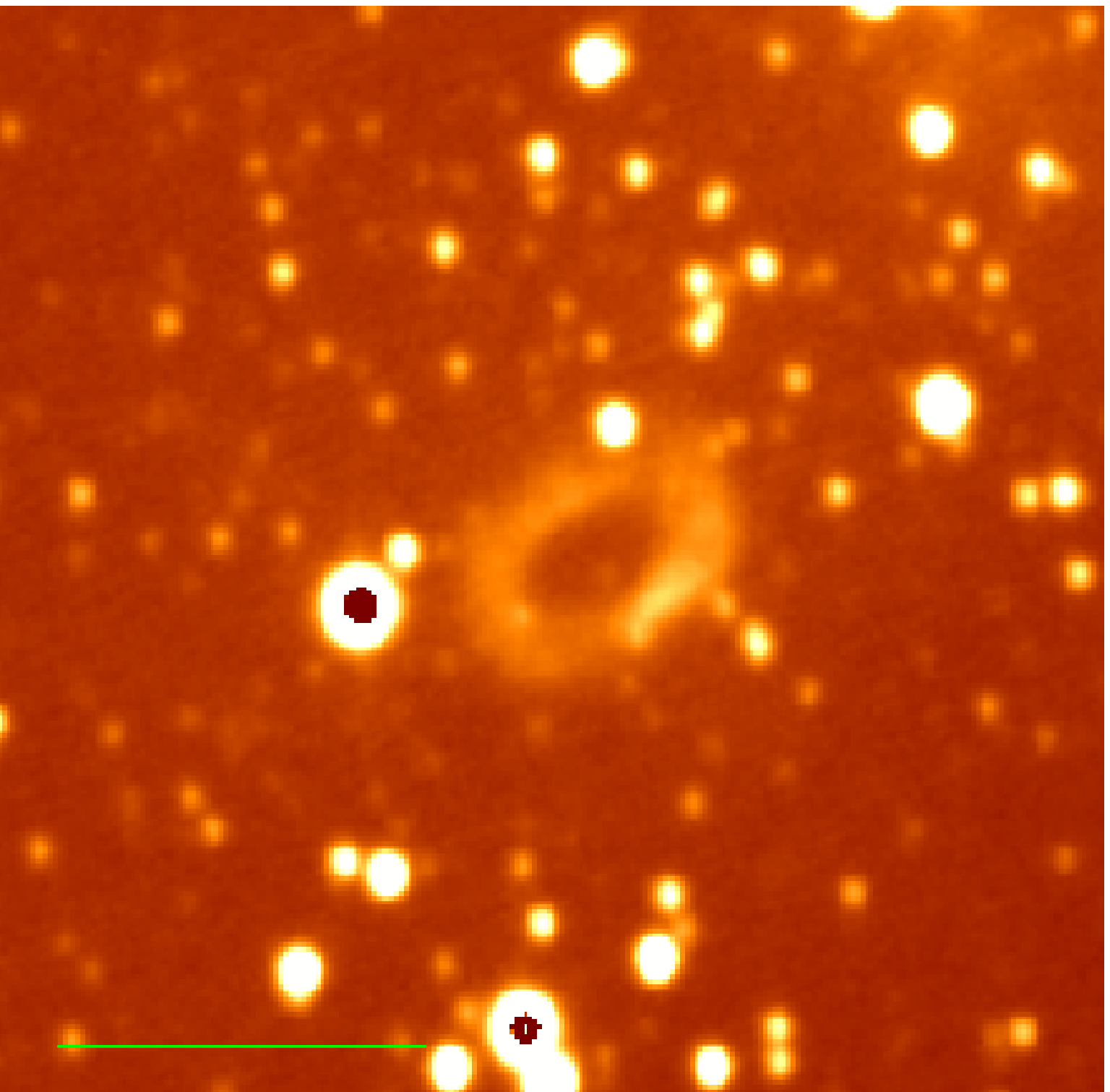}		\vspace{0.25cm}\\
\includegraphics[width=0.42\textwidth]{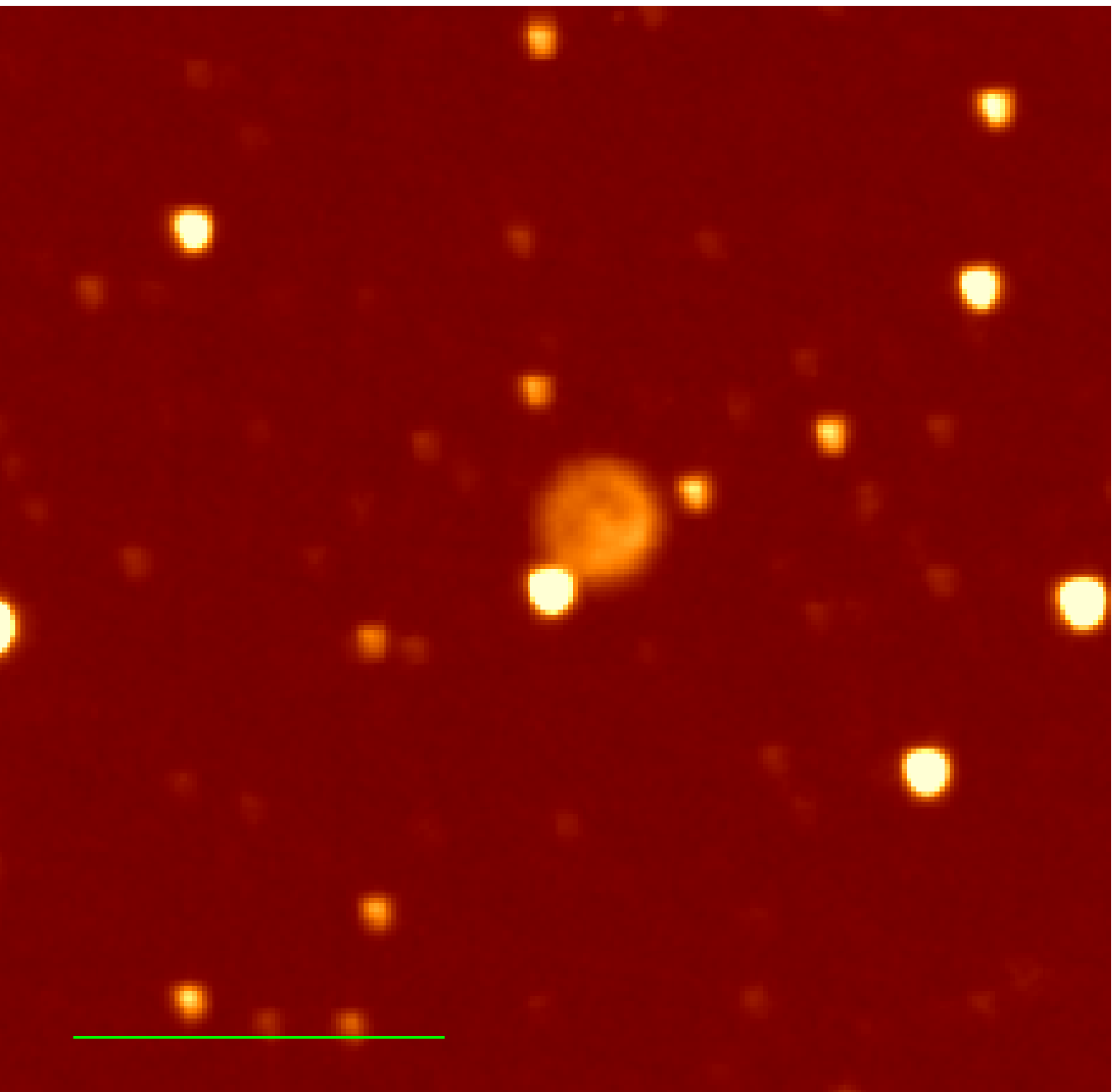}		\hspace{0.8cm}
\includegraphics[width=0.42\textwidth]{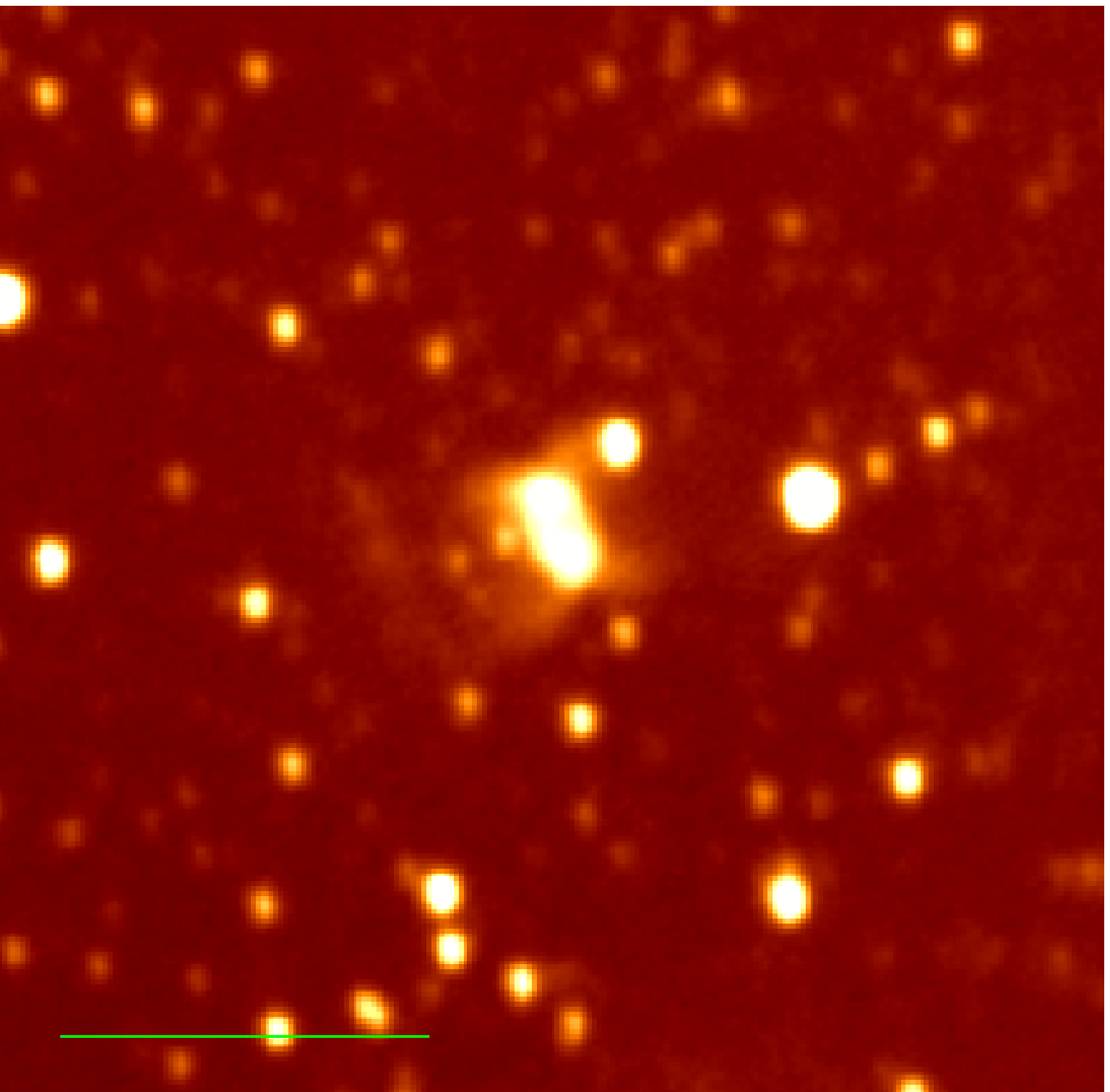}
\caption{Same as Fig.~\ref{Fig_1},
top He~2$-$36 (L) and Ste~2$-$1, middle PHR1019$-$6059 (L) and PHR1036$-$5909, bottom He~2$-$55 and PHR1046$-$6109.}
   \label{Fig_10}
\end{figure*}

\begin{figure*}[!ht]
  \centering
\includegraphics[width=0.42\textwidth]{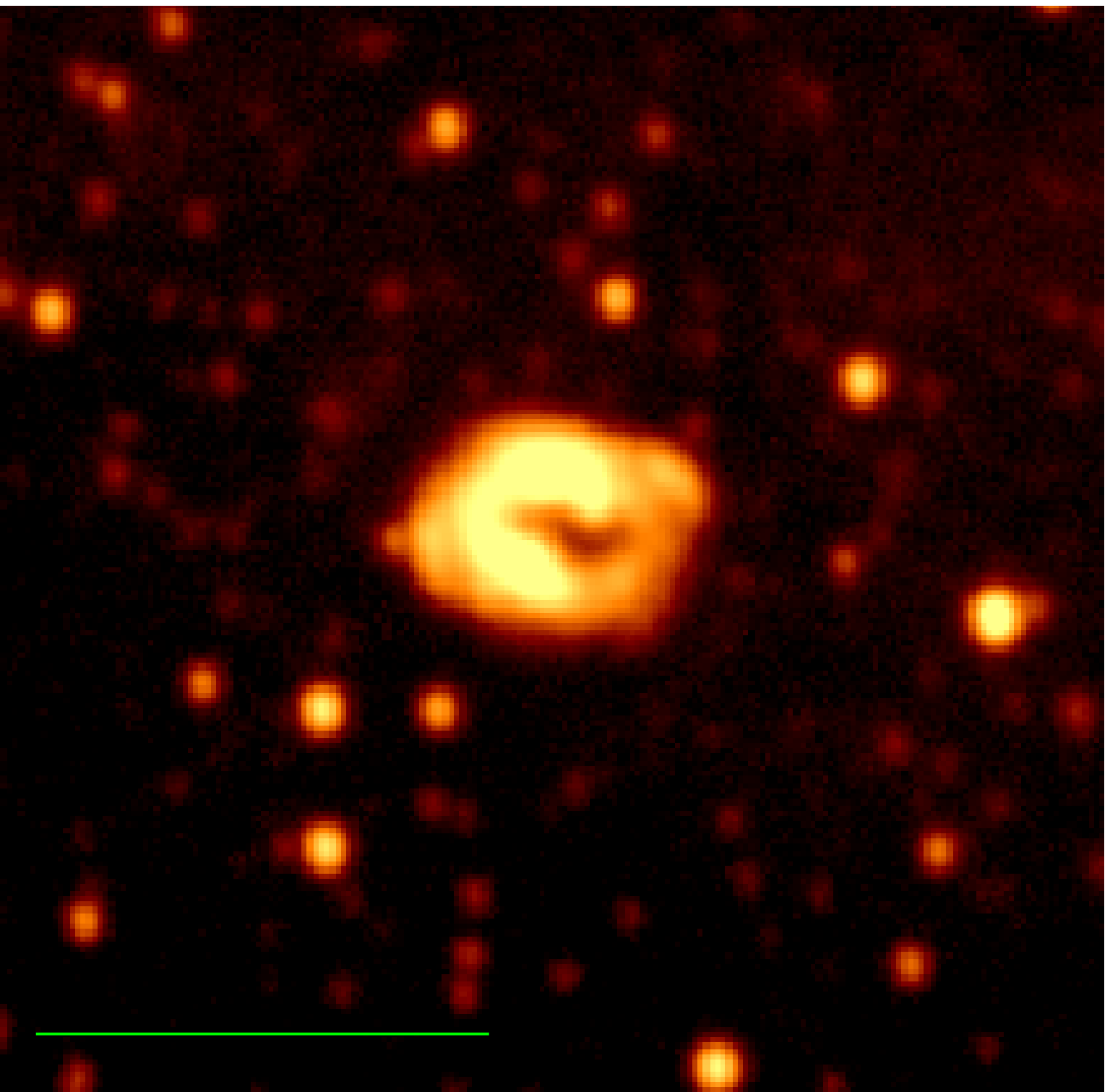}	\hspace{0.8cm}
\includegraphics[width=0.42\textwidth]{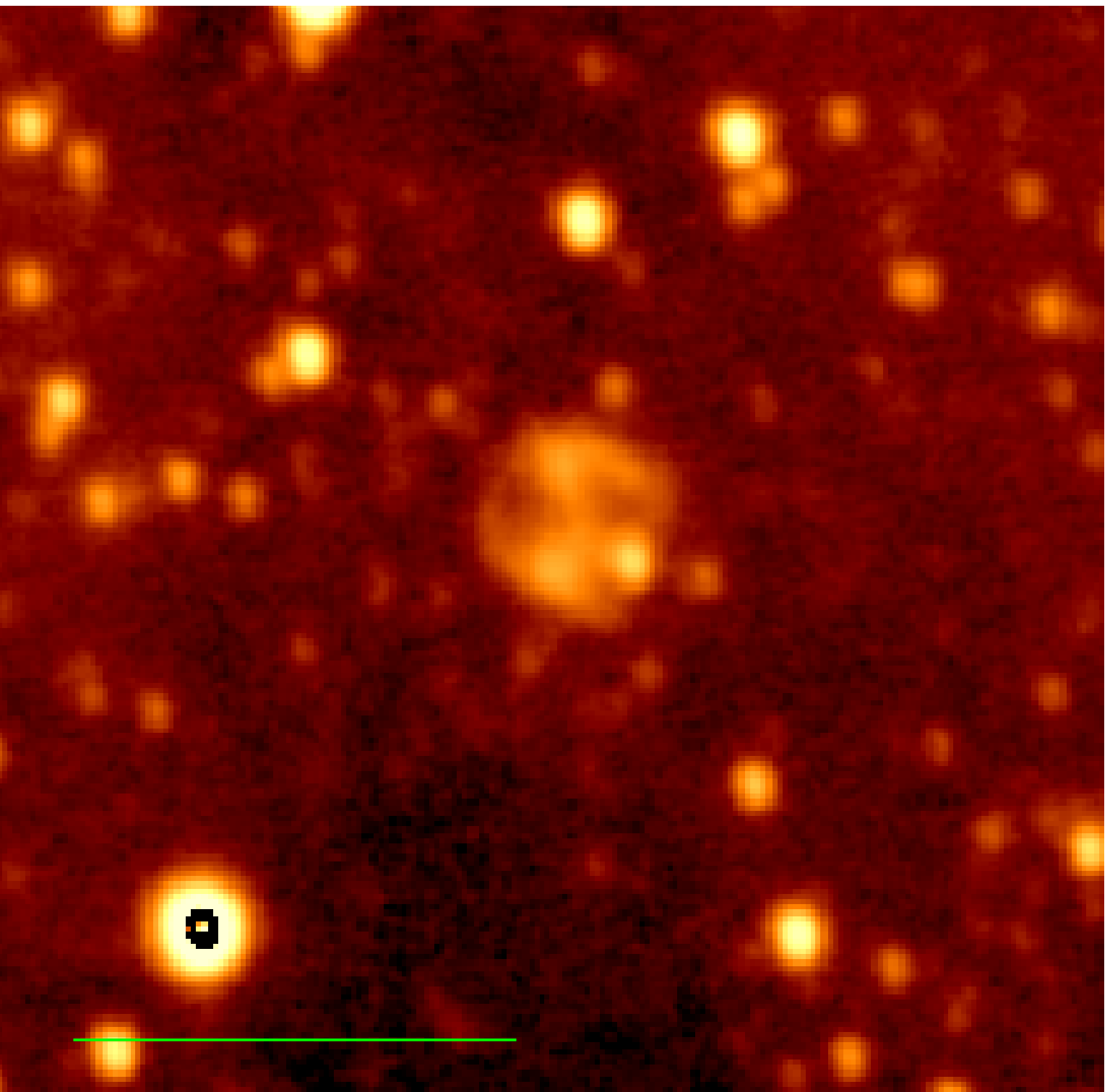}	\vspace{0.25cm}\\
\includegraphics[width=0.42\textwidth]{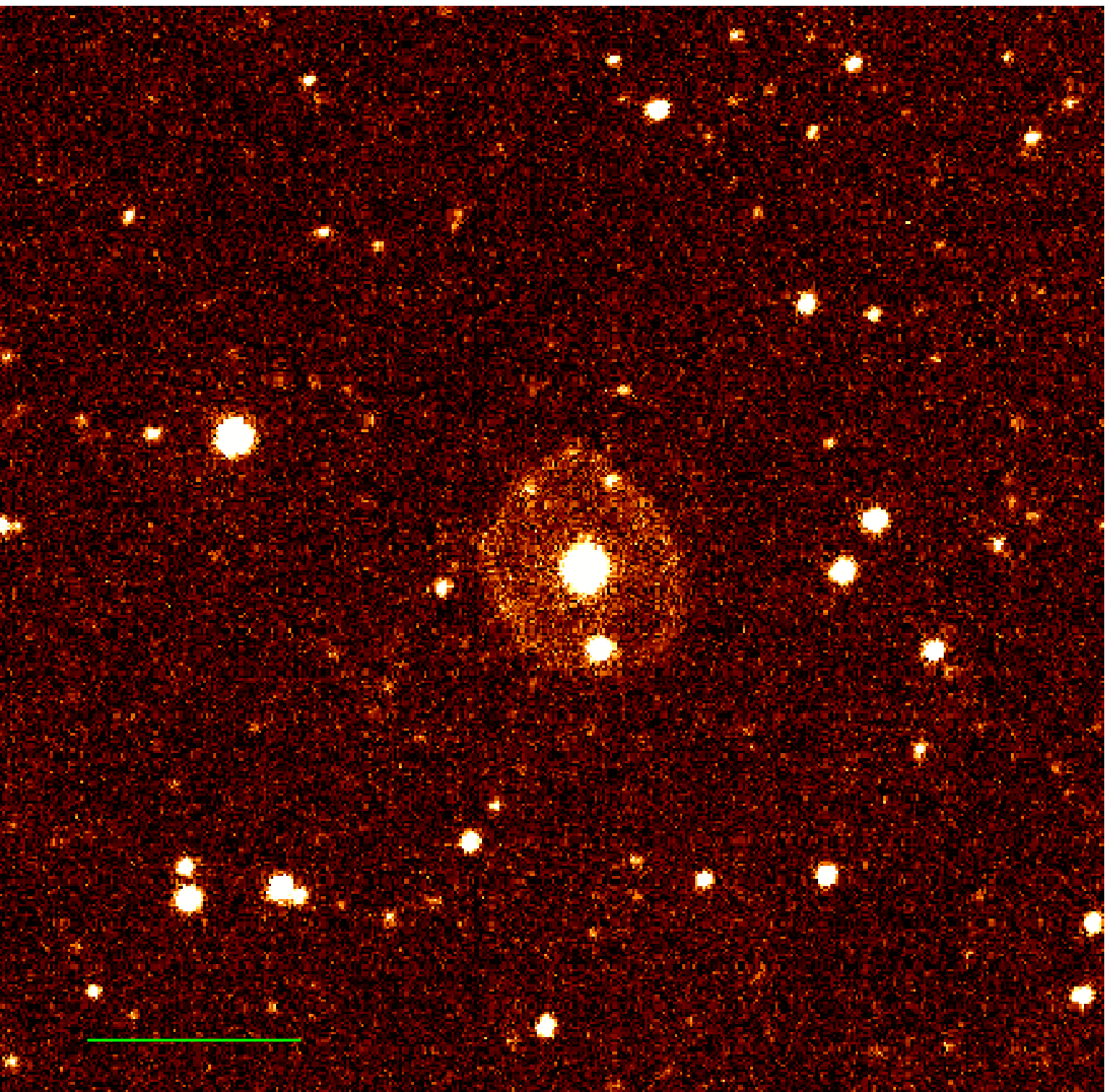}	\hspace{0.8cm}
\includegraphics[width=0.42\textwidth]{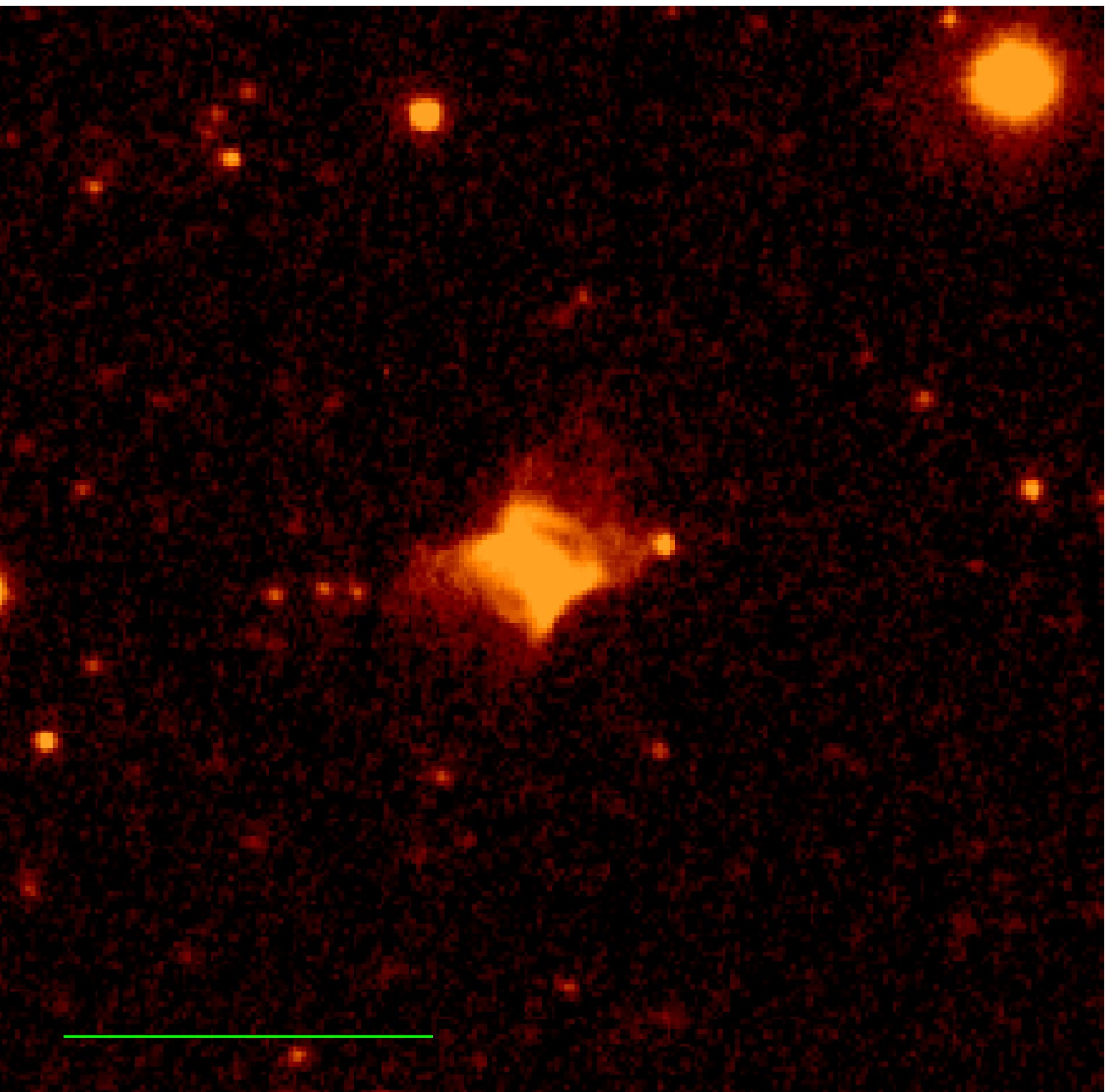}	\vspace{0.25cm}\\
\includegraphics[width=0.42\textwidth]{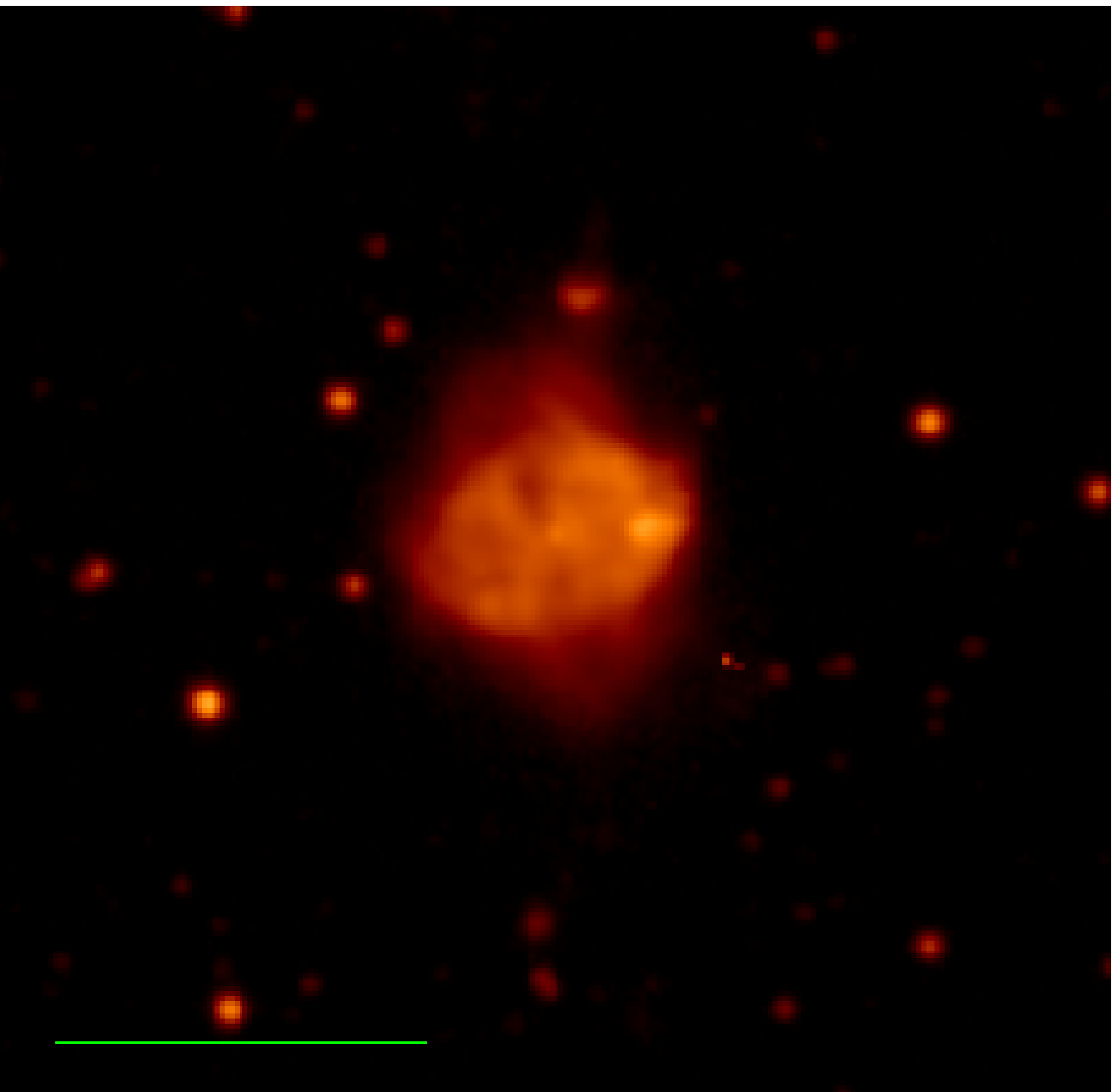}	\hspace{0.8cm}
\includegraphics[width=0.42\textwidth]{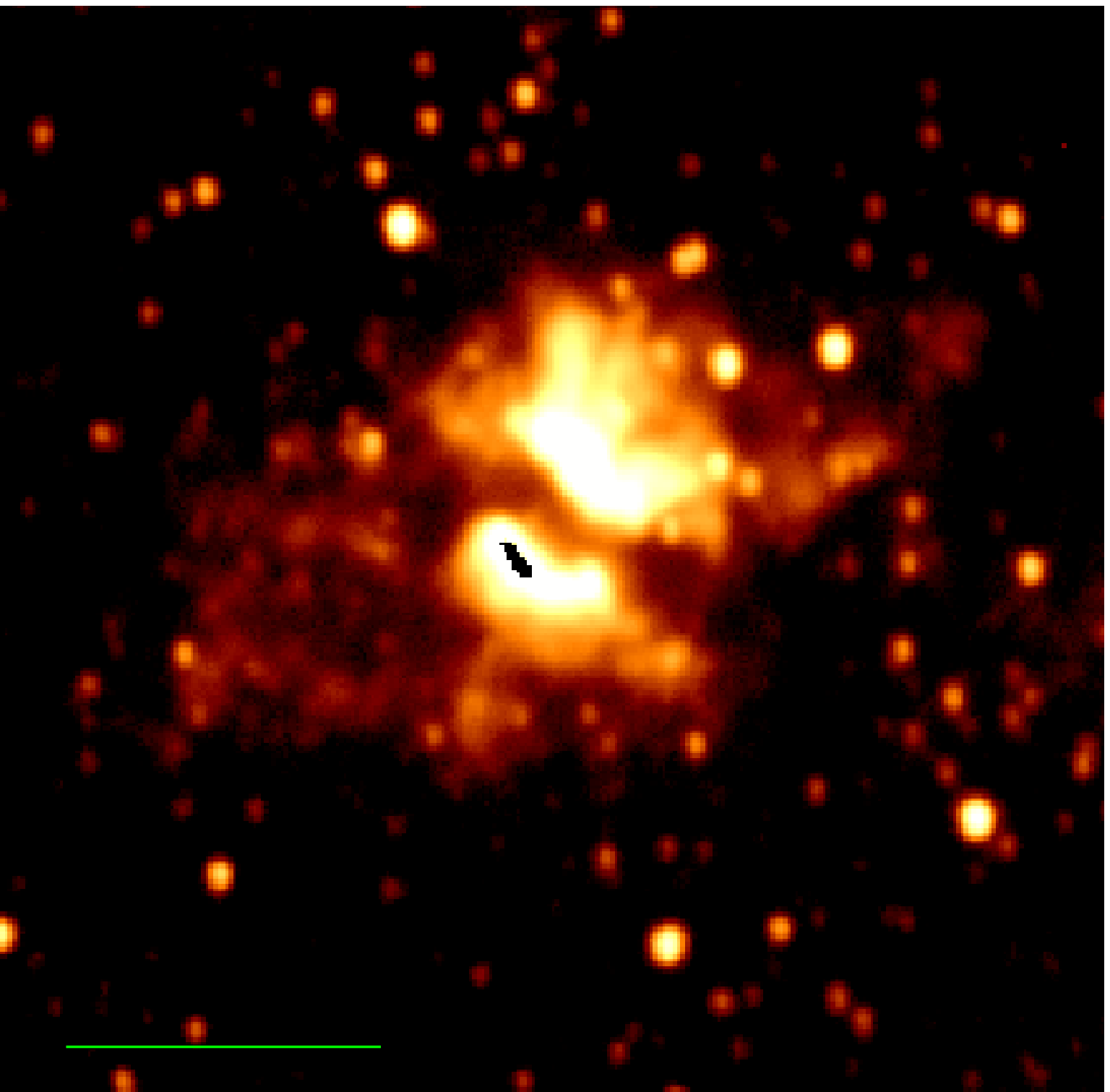}
\caption{Same as Fig.~\ref{Fig_1},
top Hf~38 (L) and PHR1058$-$5853 (L), middle Hf~39 and Hf~48 (L), bottom Fg~1 (L) and NGC~3699 (L).}
  \label{Fig_11}
\end{figure*}

\begin{figure*}[!ht]
  \centering
\includegraphics[width=0.42\textwidth]{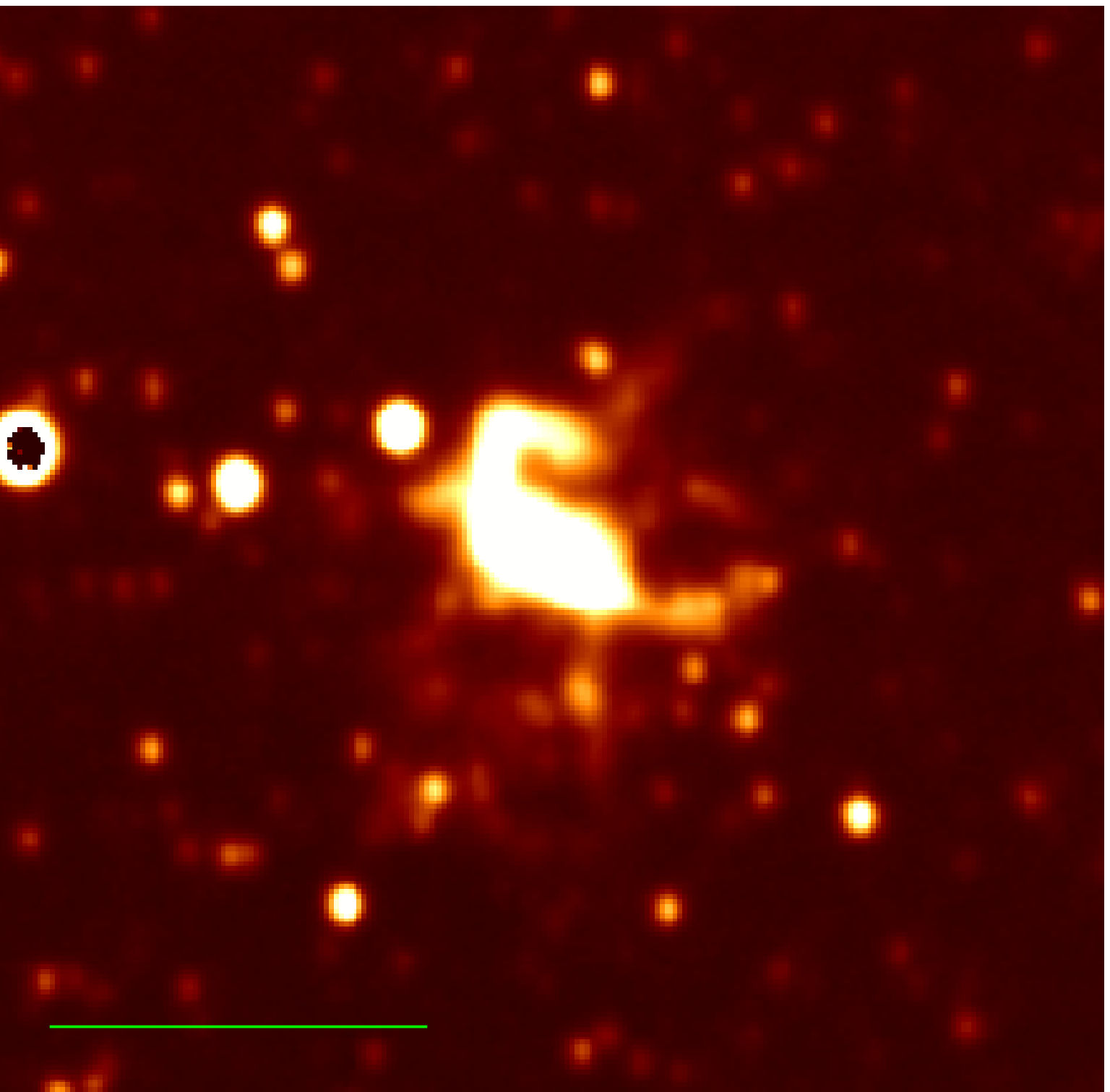}	\hspace{0.8cm}
\includegraphics[width=0.42\textwidth]{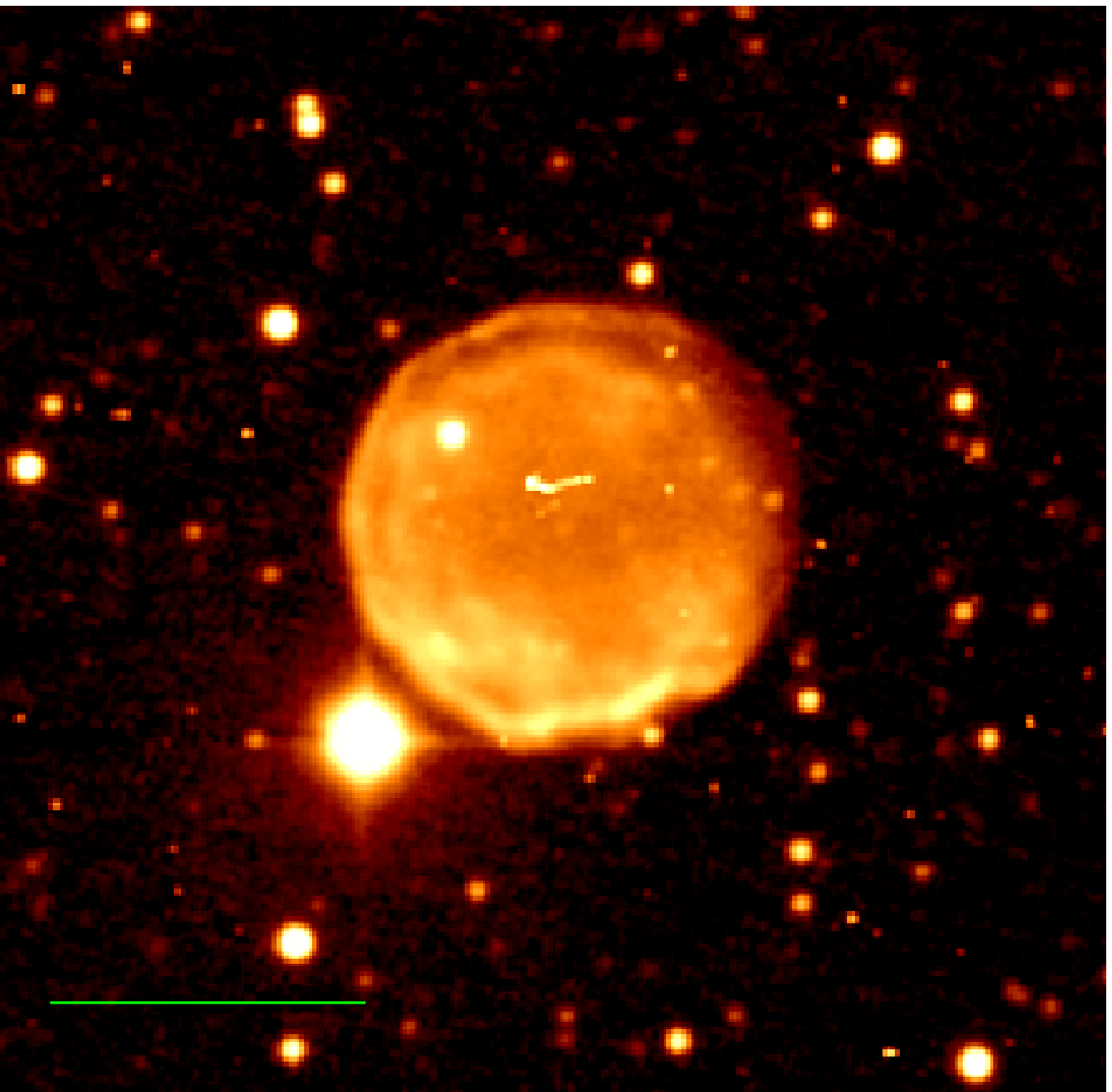}	\vspace{0.25cm}\\
\includegraphics[width=0.42\textwidth]{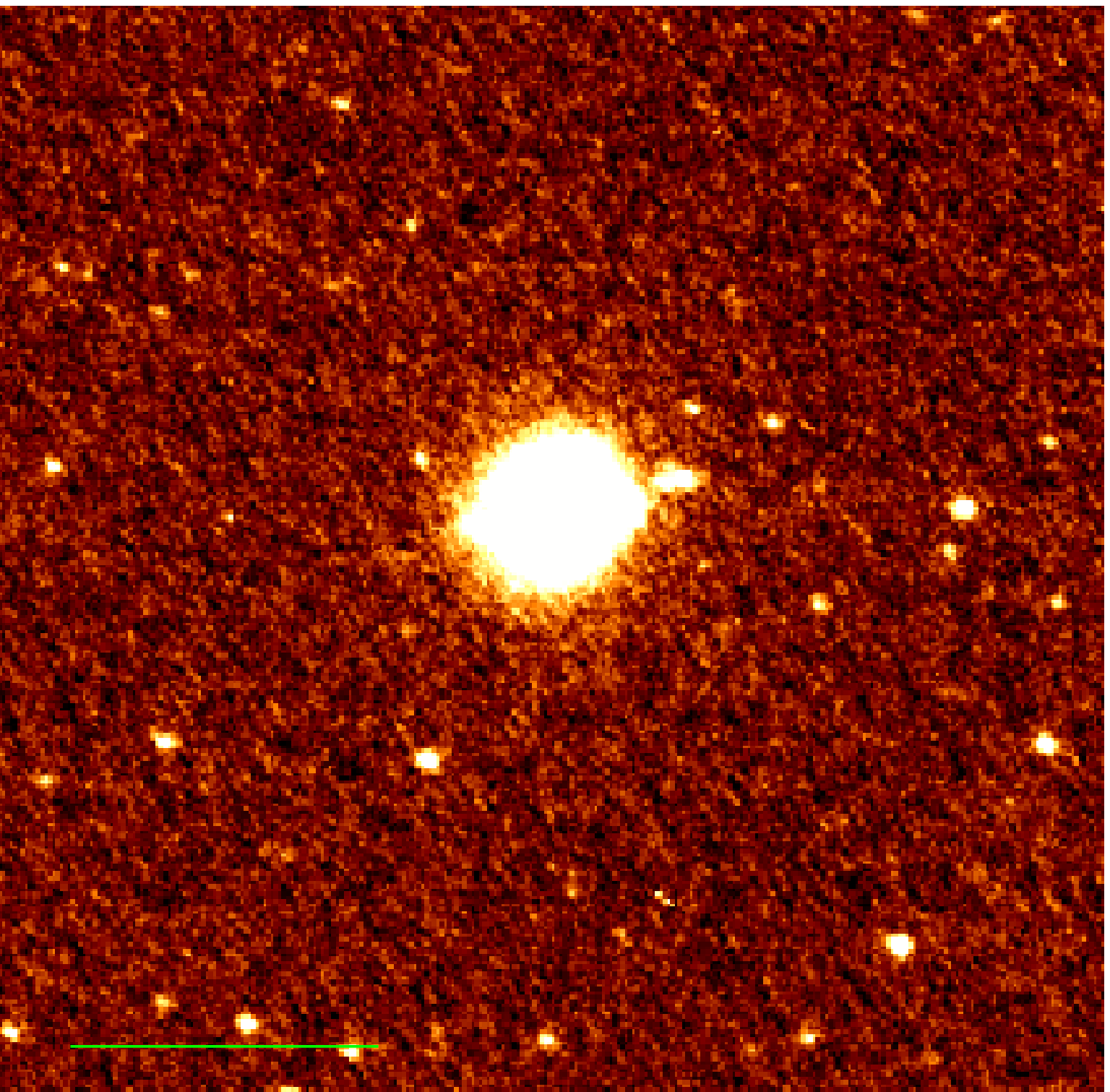}	\hspace{0.8cm}
\includegraphics[width=0.42\textwidth]{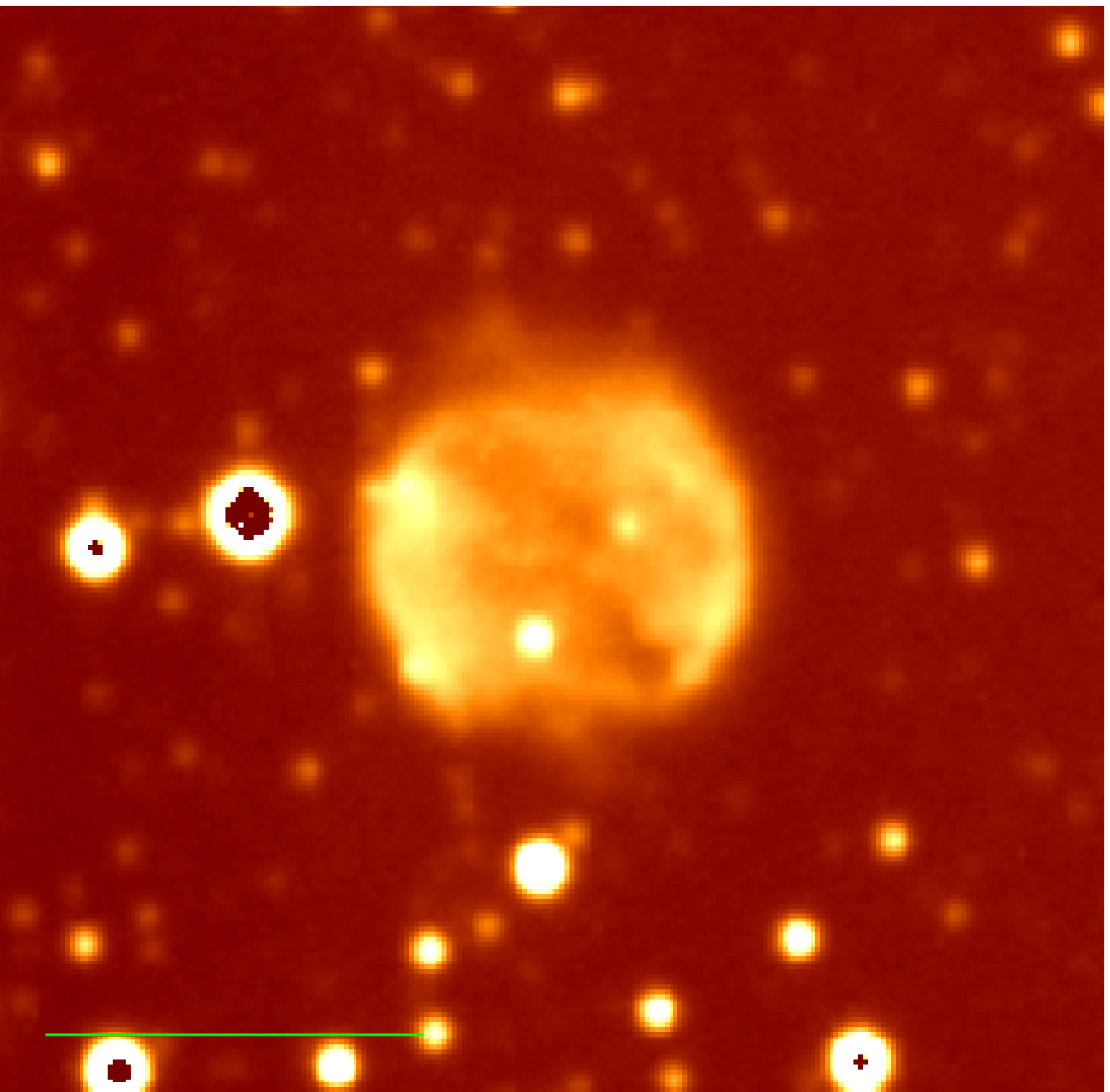}	\vspace{0.25cm}\\
\includegraphics[width=0.42\textwidth]{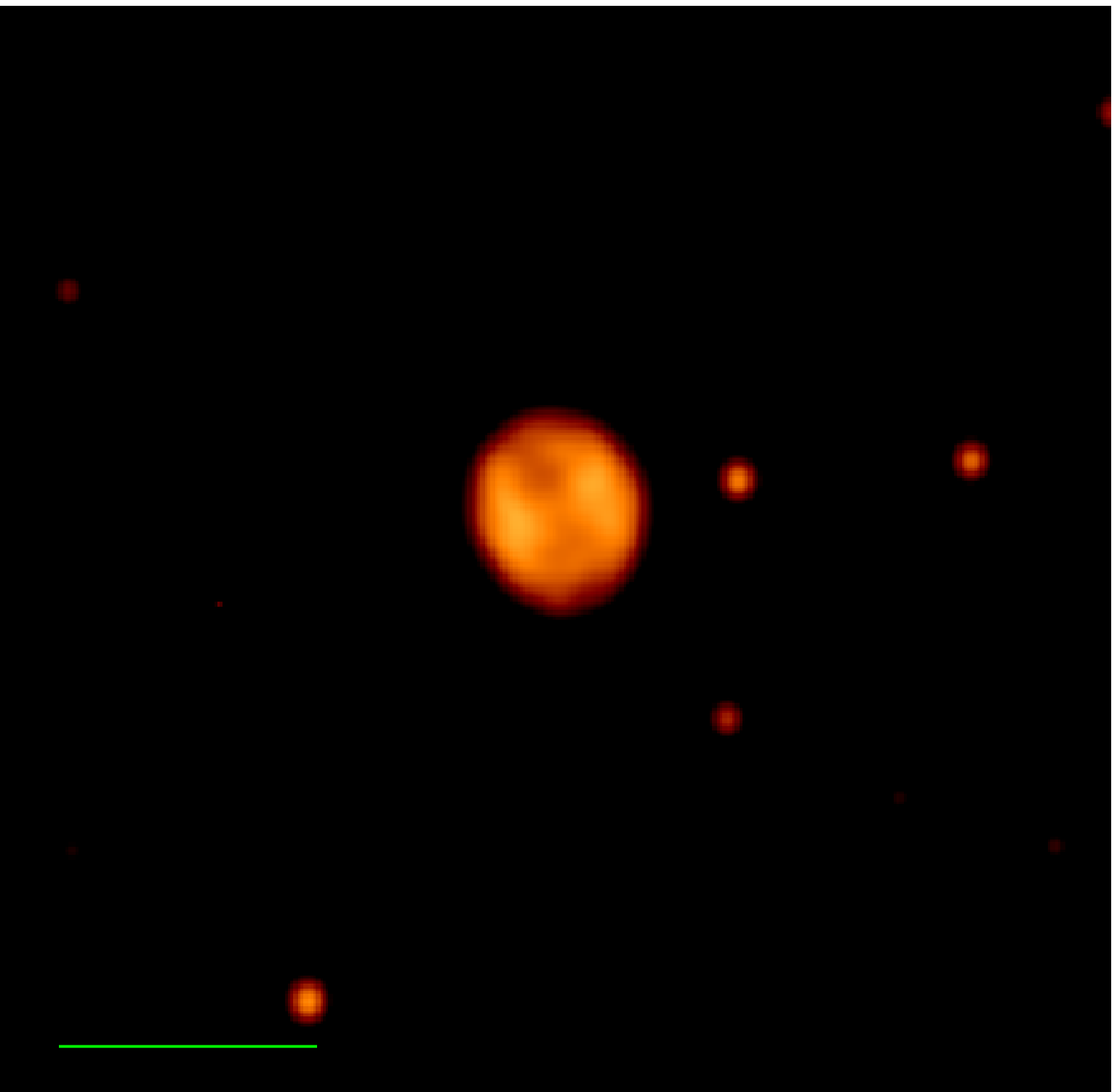}	\hspace{0.8cm}
\includegraphics[width=0.42\textwidth]{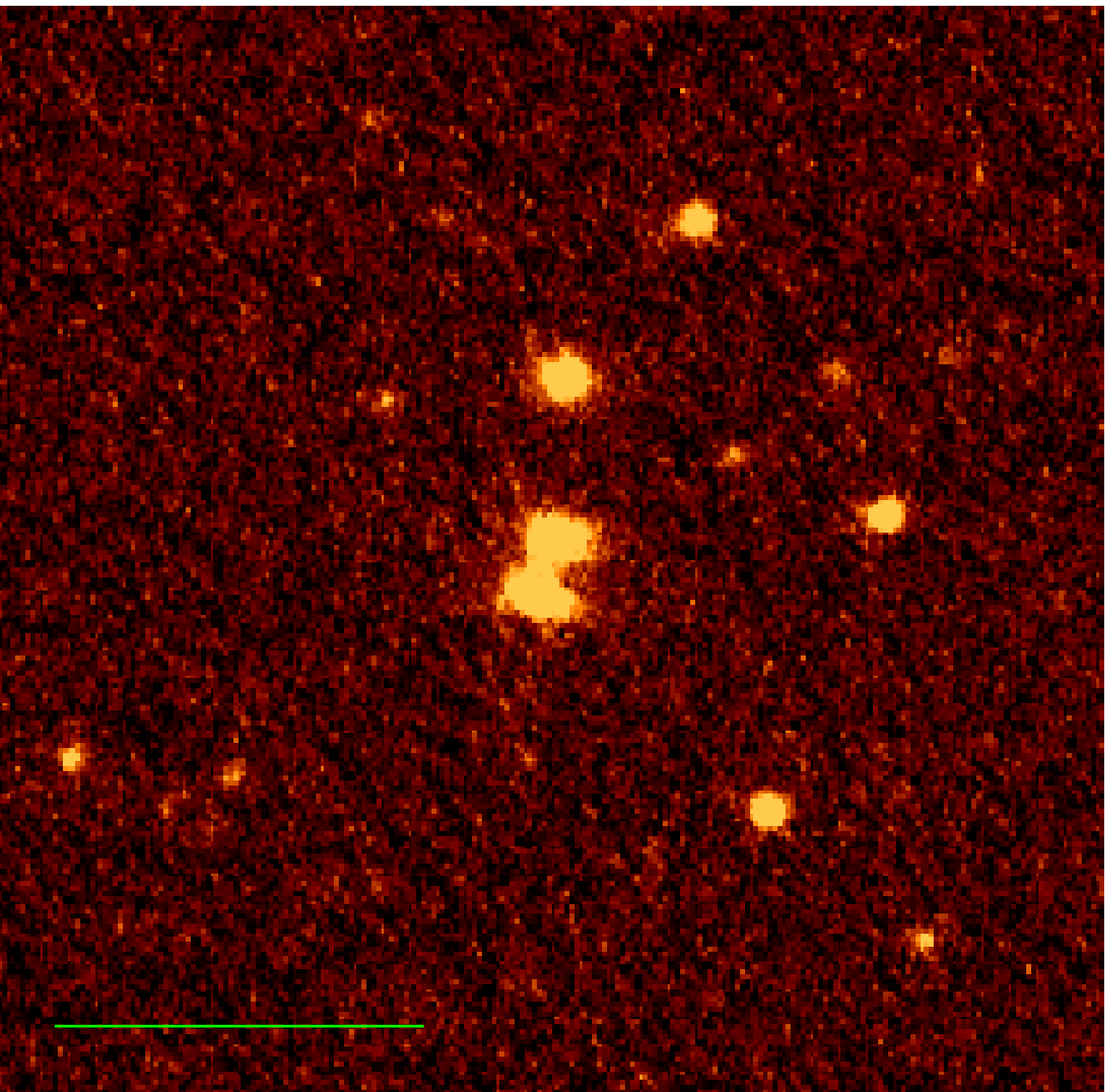}
\caption{Same as Fig.~\ref{Fig_1},
top He~2$-$70 and BlDz~1, middle NGC~3918 and He~2$-$72, bottom NGC~3195 (L) and He~2-76.}
  \label{Fig_12}
\end{figure*}

\begin{figure*}[!ht]
  \centering
\includegraphics[width=0.42\textwidth]{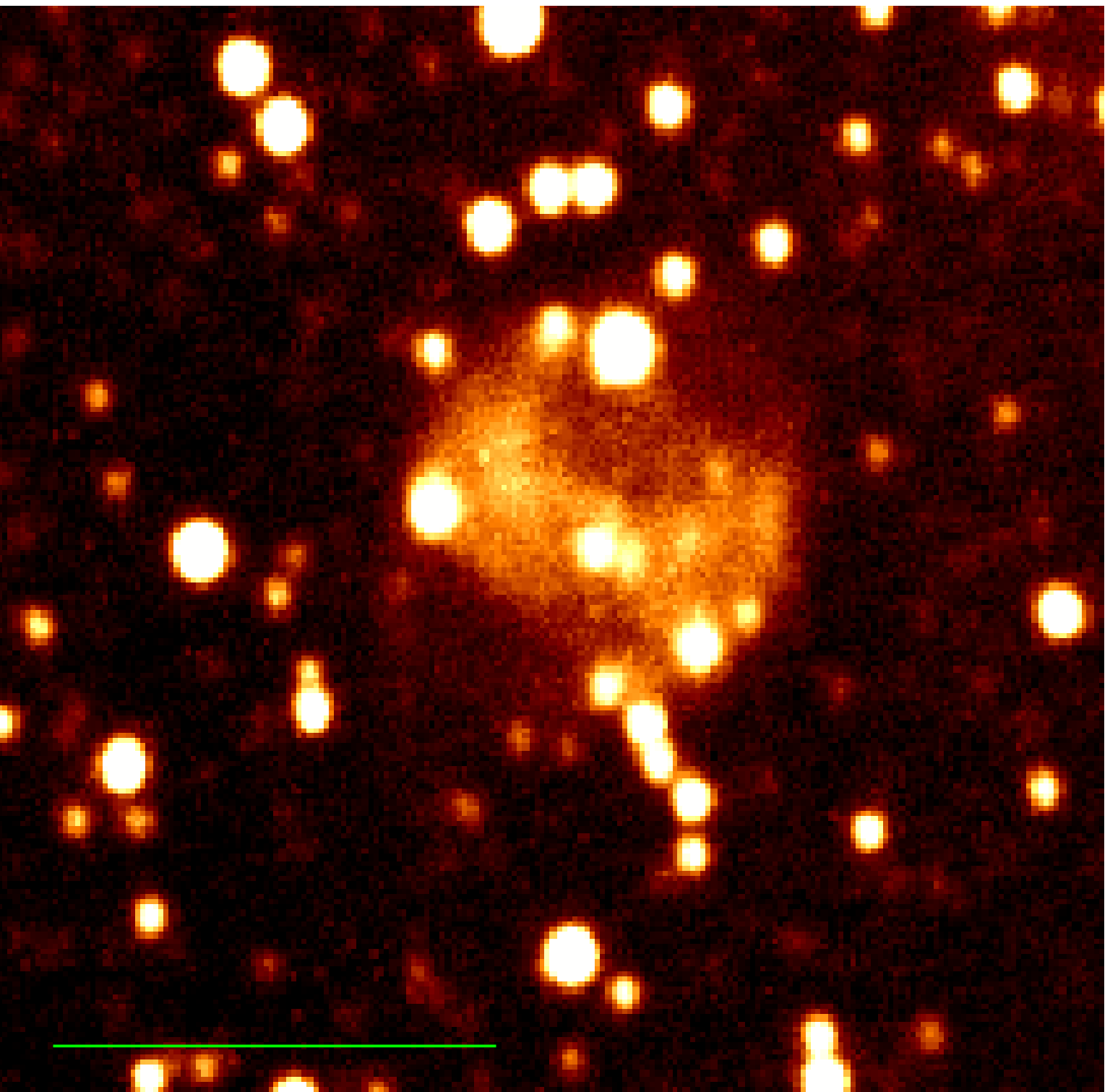}	\hspace{0.8cm}
\includegraphics[width=0.42\textwidth]{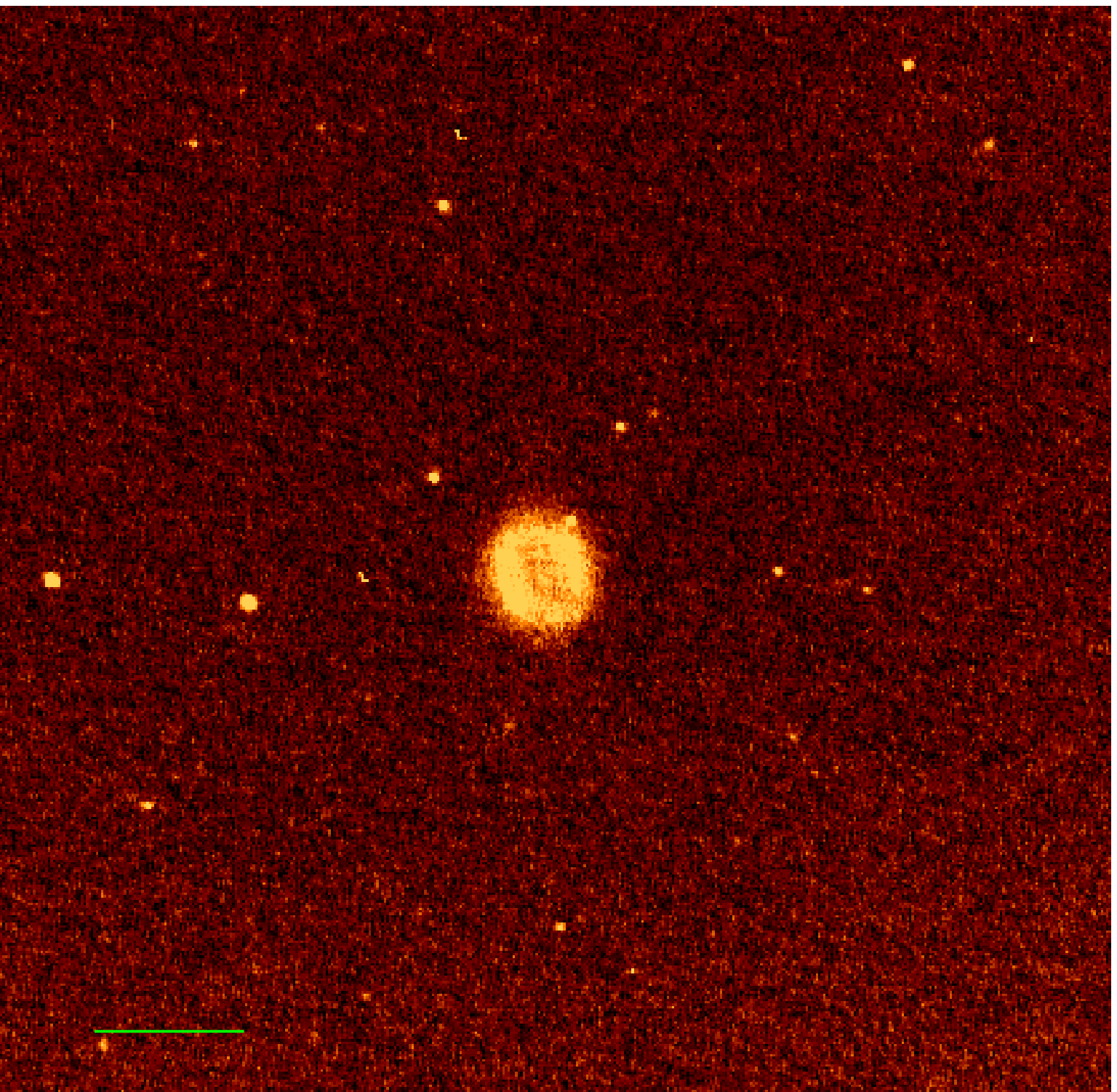}	\vspace{0.25cm}\\
\includegraphics[width=0.42\textwidth]{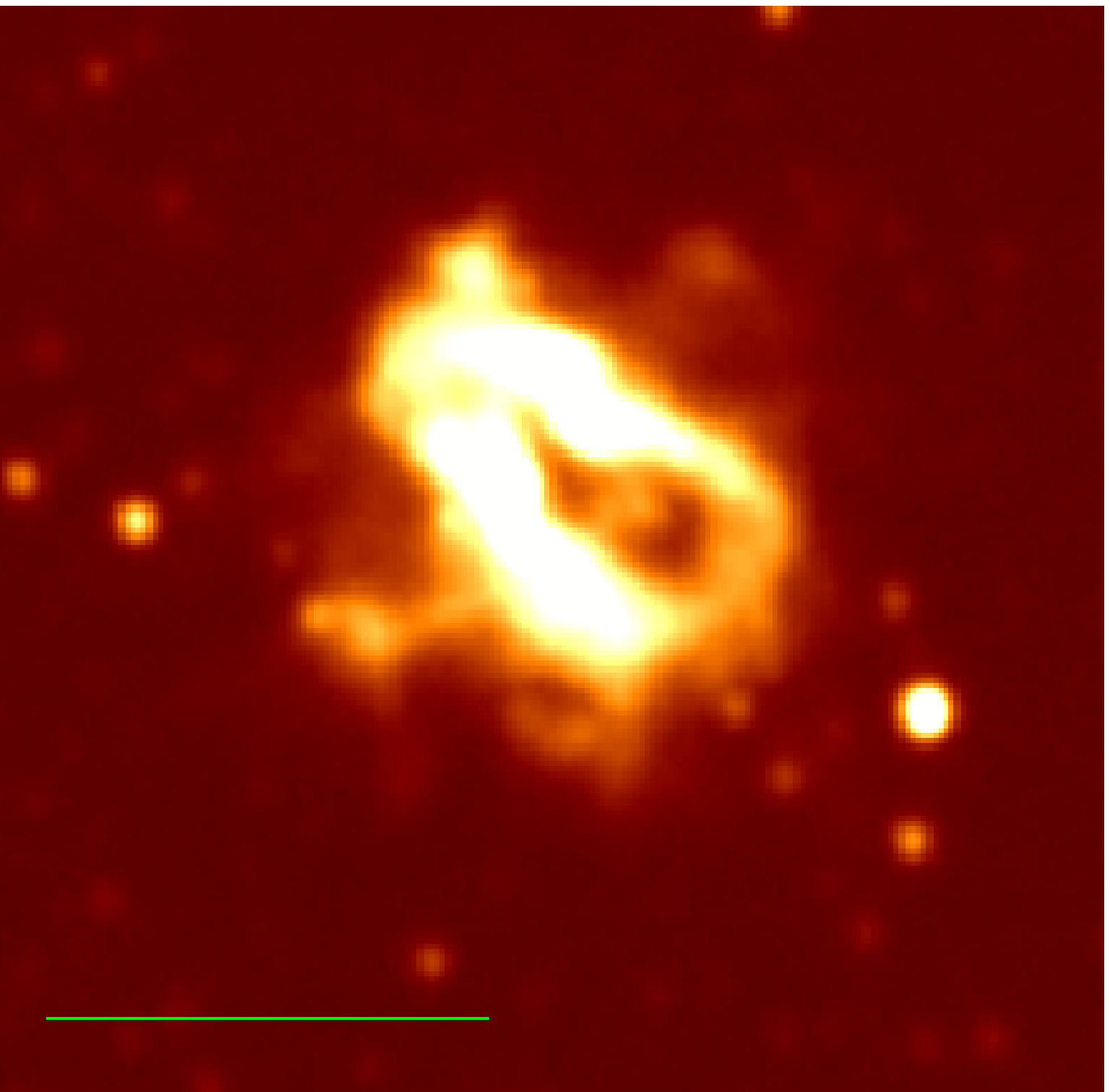}	\hspace{0.8cm}
\includegraphics[width=0.42\textwidth]{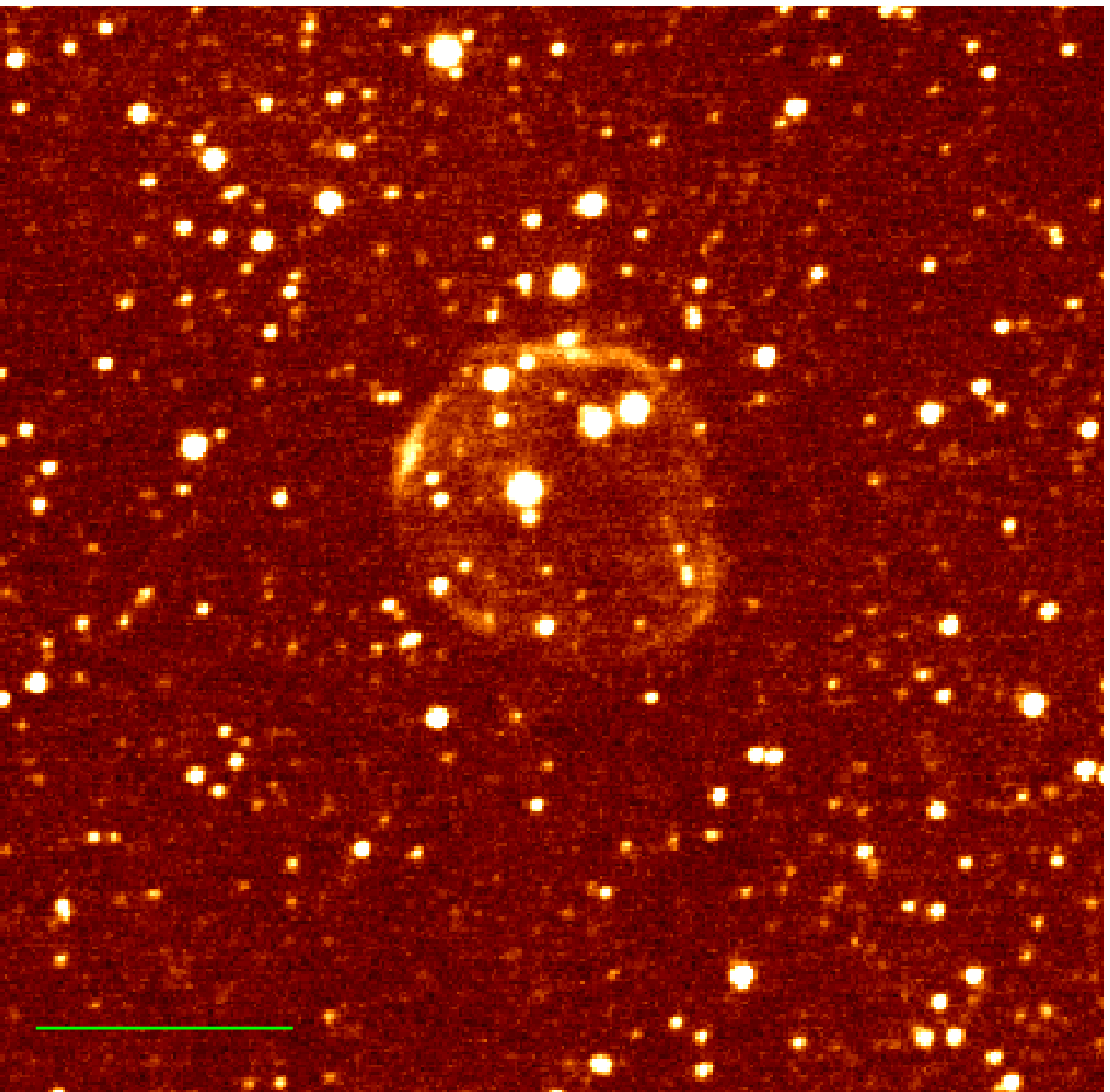}	\vspace{0.25cm}\\
\includegraphics[width=0.42\textwidth]{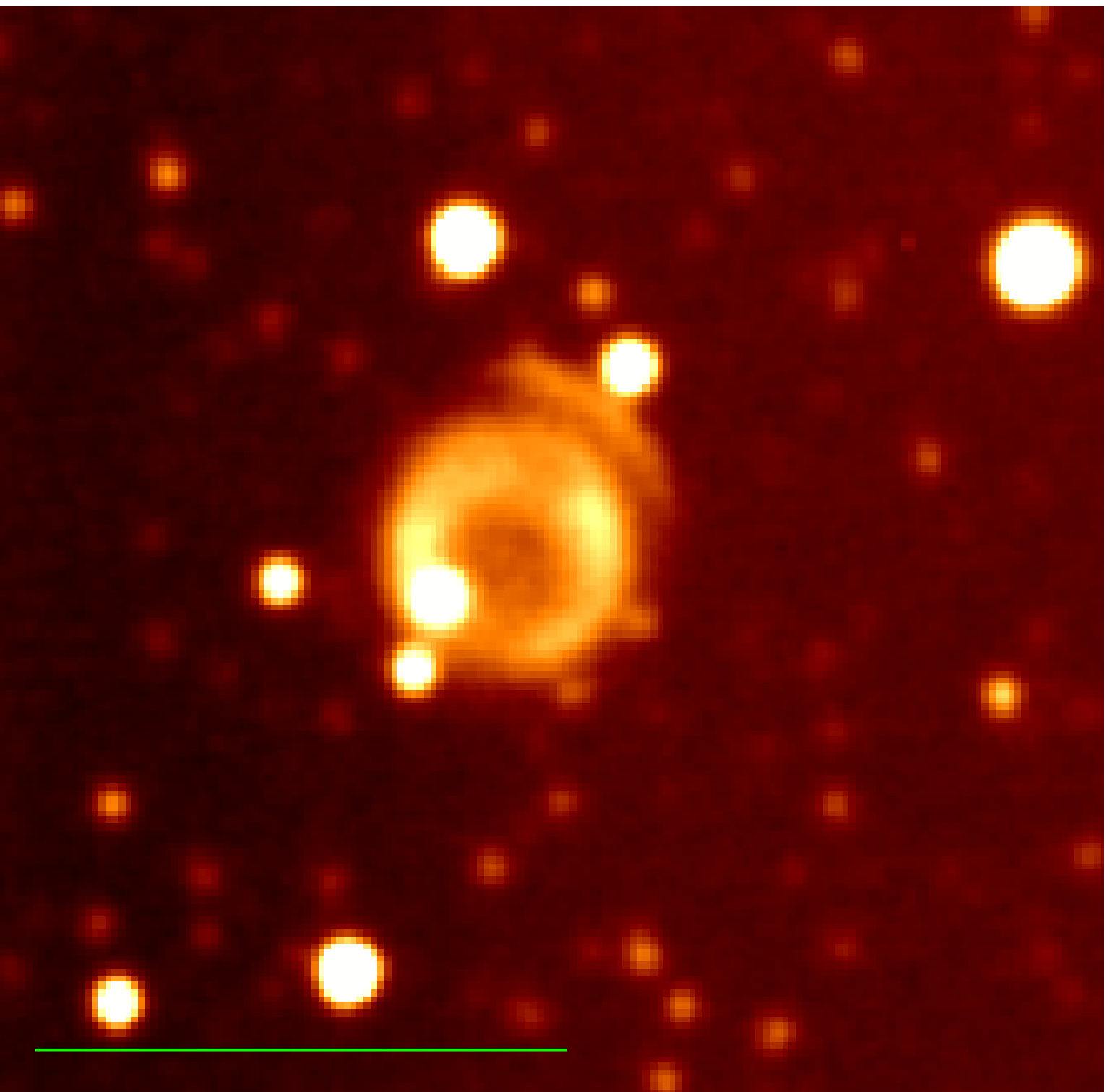}	\hspace{0.8cm}
\includegraphics[width=0.42\textwidth]{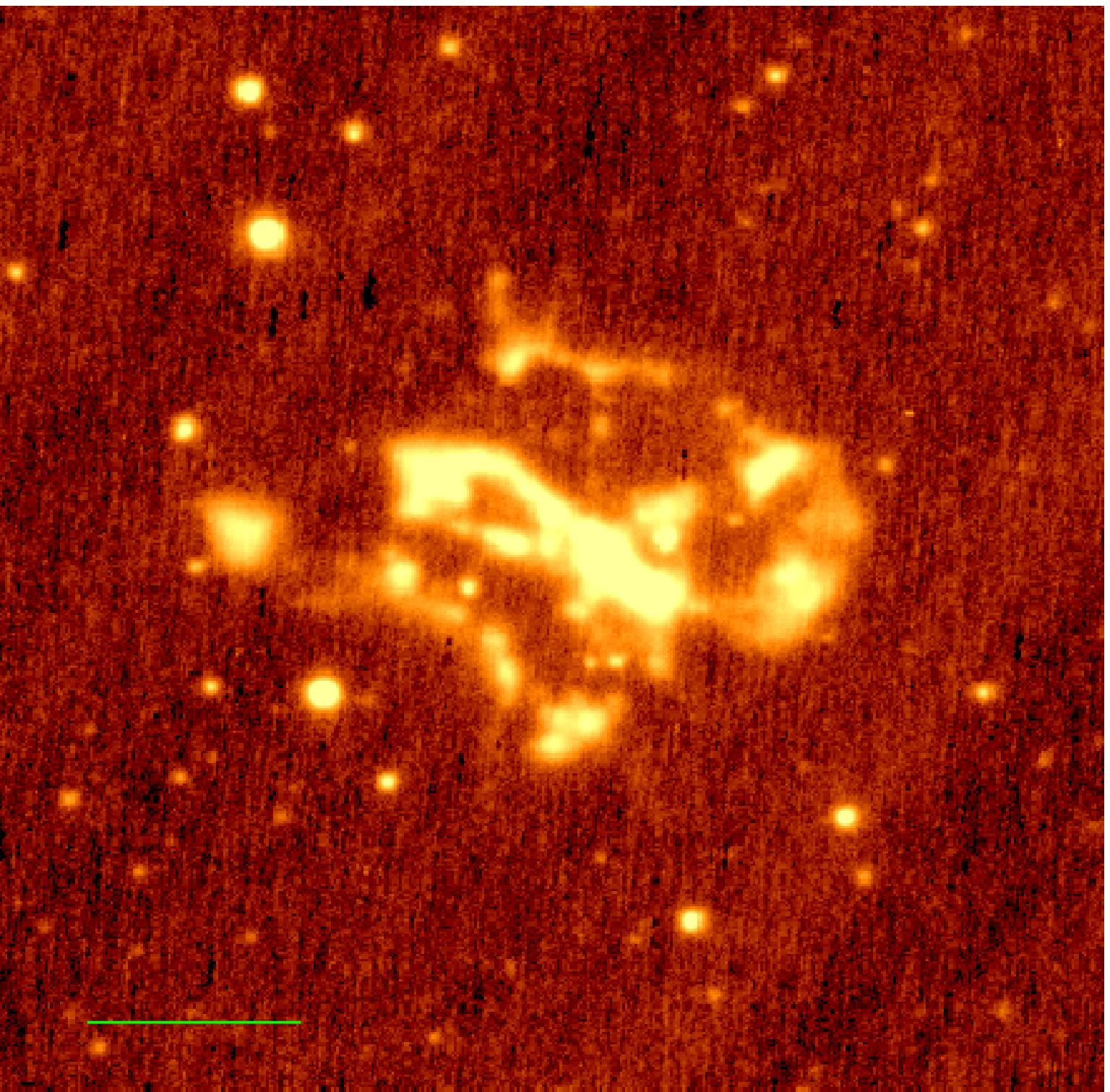}		
\caption{Same as Fig.~\ref{Fig_1},
top PHR1221$-$5907 and K~1$-$23, middle Wray~16$-$121, and PHR1250$-$6346, bottom Wray~16$-$122 and NGC~5189 (L).}
  \label{Fig_13}
\end{figure*}

\begin{figure*}[!ht]
  \centering
\includegraphics[width=0.42\textwidth]{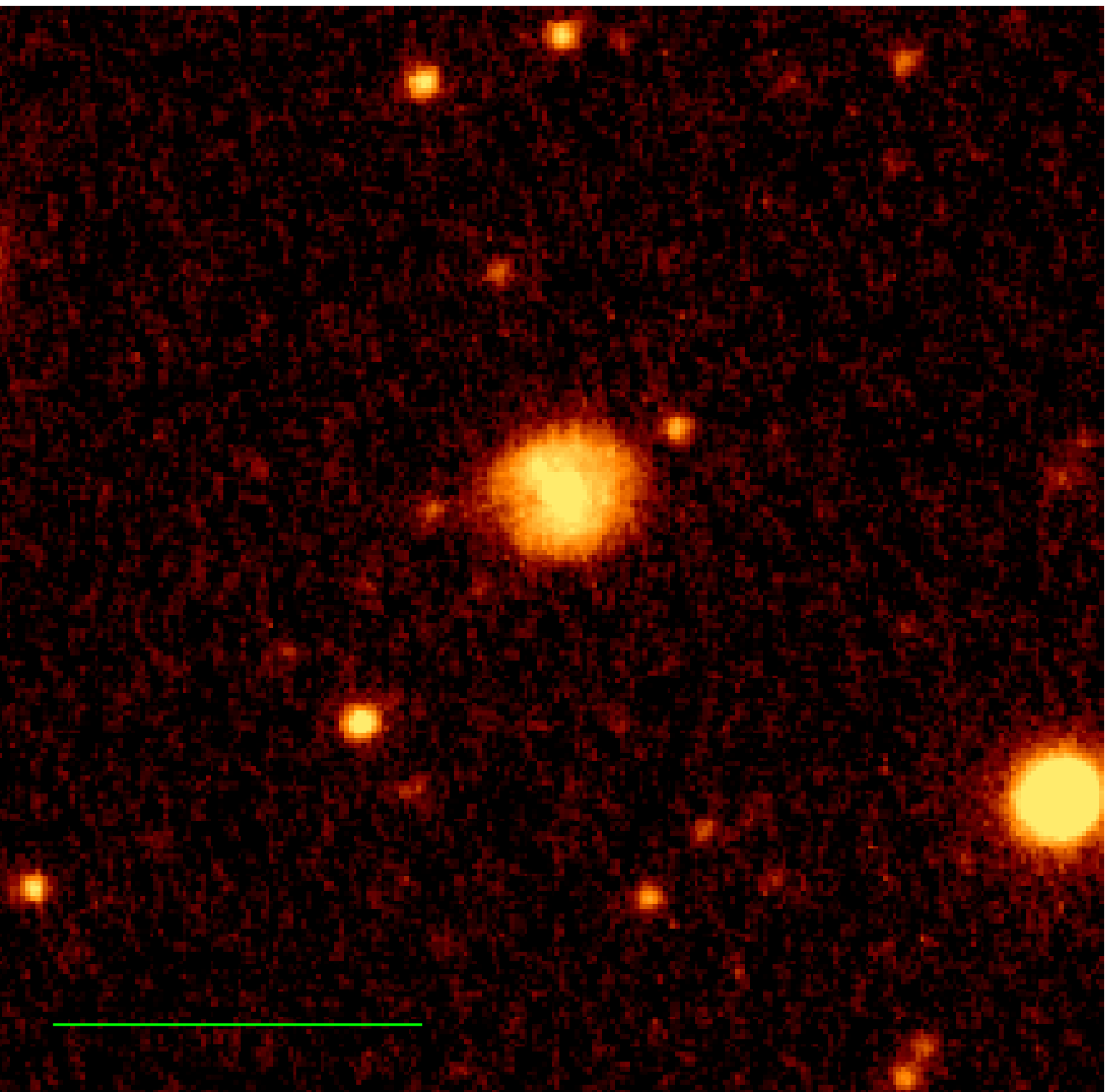}	\hspace{0.8cm}
\includegraphics[width=0.42\textwidth]{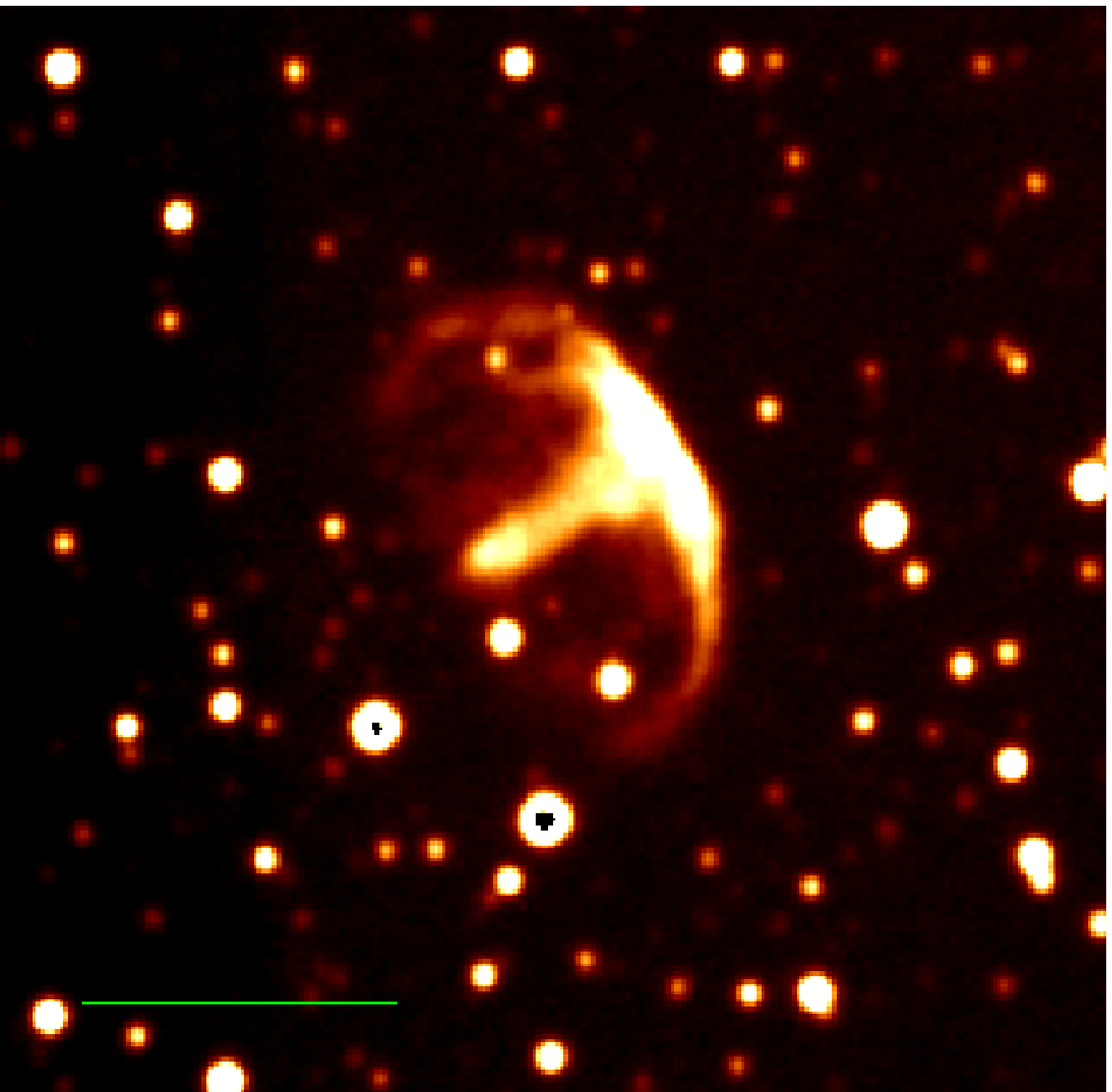}	\vspace{0.25cm}\\
\includegraphics[width=0.42\textwidth]{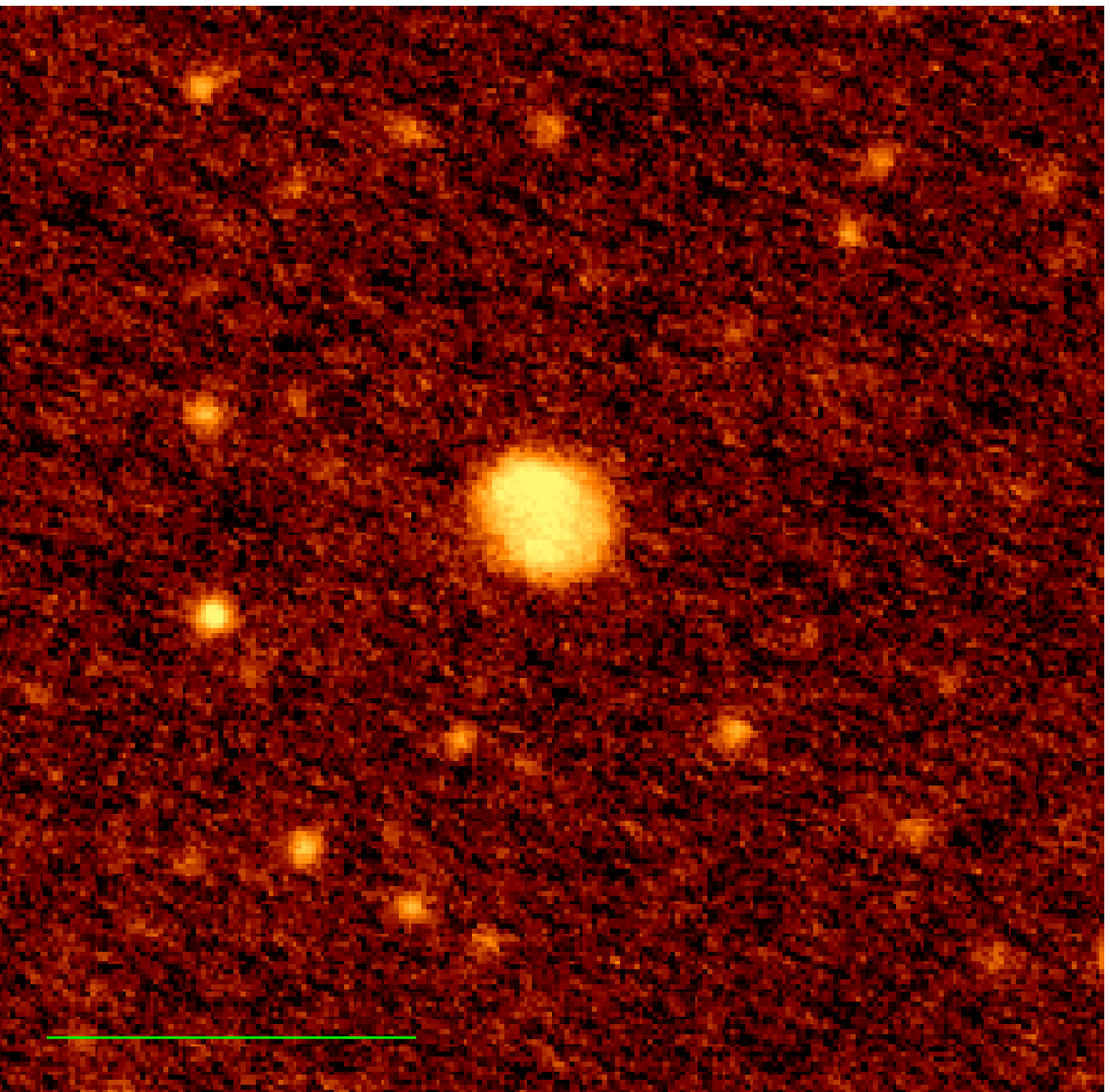}	\hspace{0.8cm}
\includegraphics[width=0.42\textwidth]{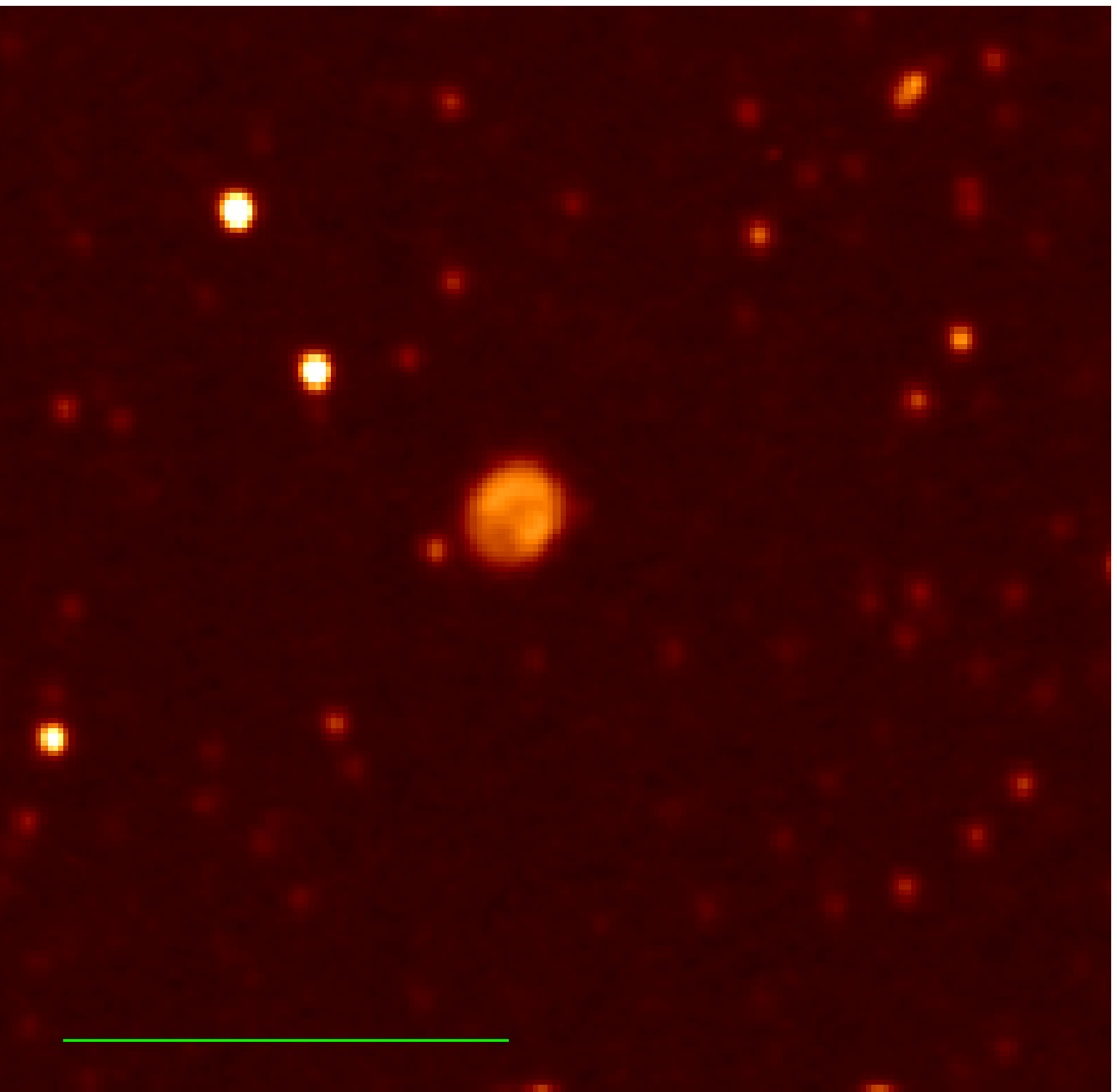}	\vspace{0.25cm}\\
\includegraphics[width=0.42\textwidth]{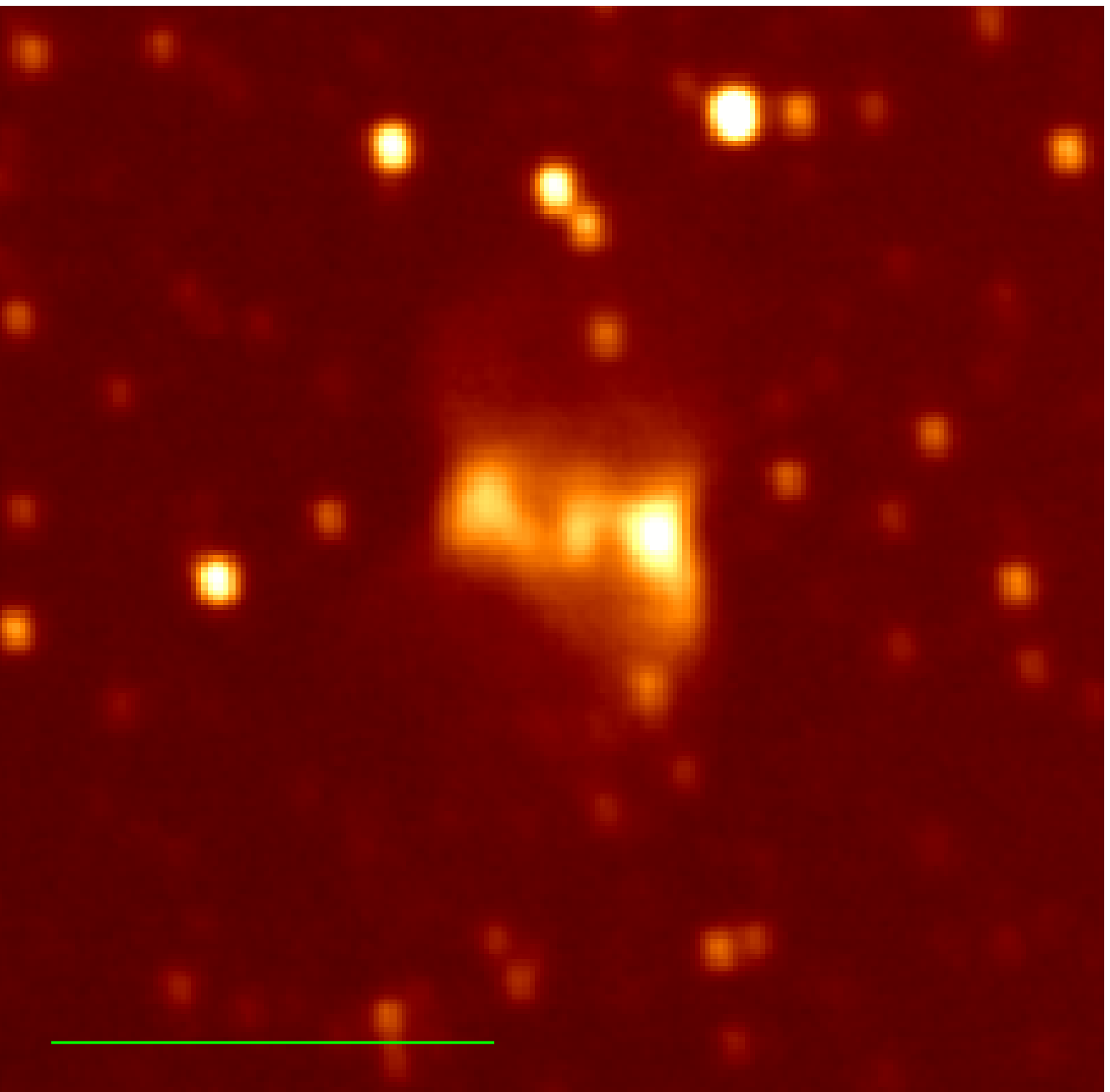}	\hspace{0.8cm}
\includegraphics[width=0.42\textwidth]{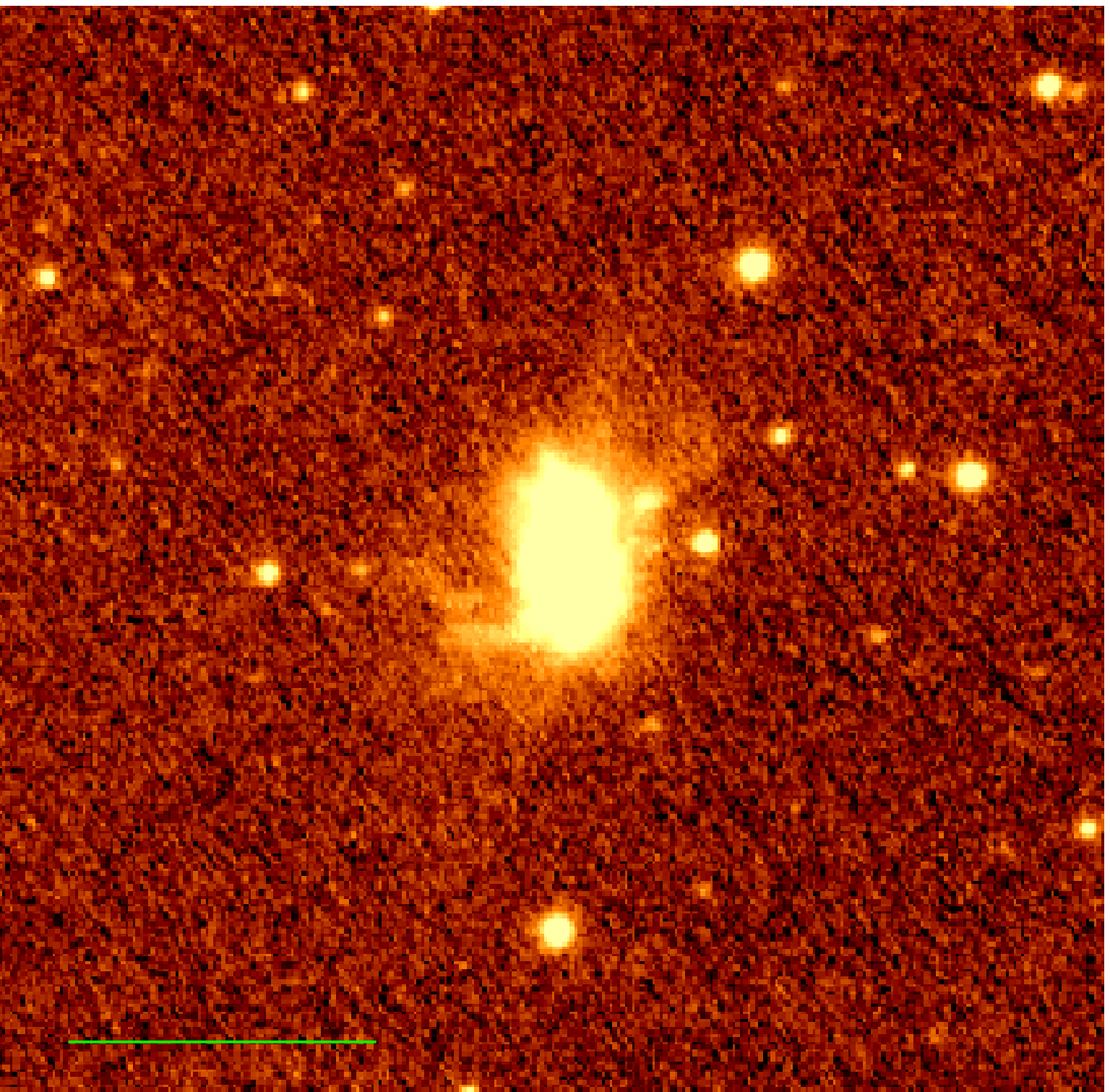}	
\caption{Same as Fig.~\ref{Fig_1},
top He~2$-$99 (L) and  VBRC~5, middle He~2$-$103 (L) and LoTr~7, bottom PHR1408$-$6229, and He~2$-$111 (L).}
  \label{Fig_14}
\end{figure*}

\begin{figure*}[!ht]
  \centering
\includegraphics[width=0.42\textwidth]{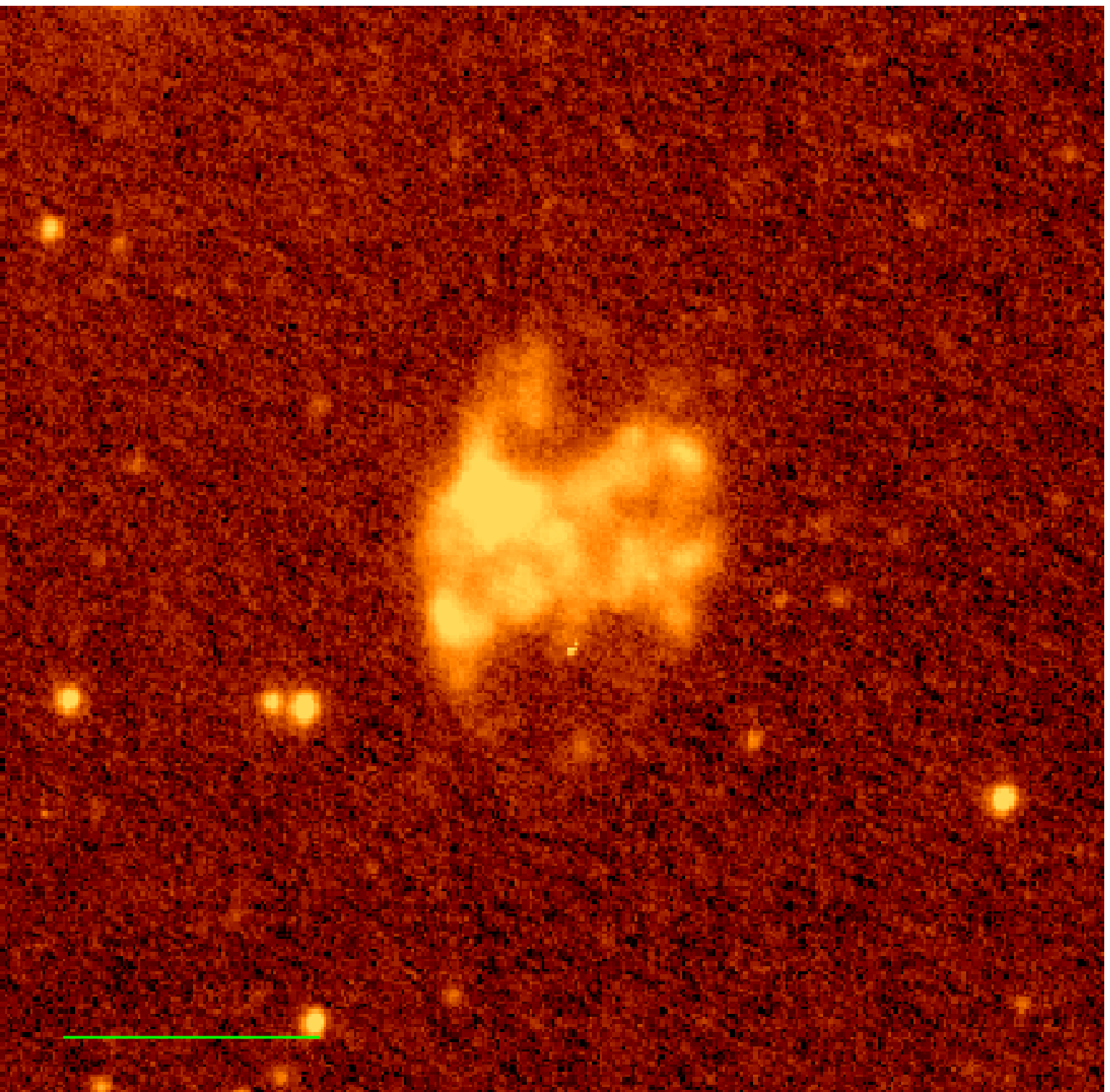}	\hspace{0.8cm}
\includegraphics[width=0.42\textwidth]{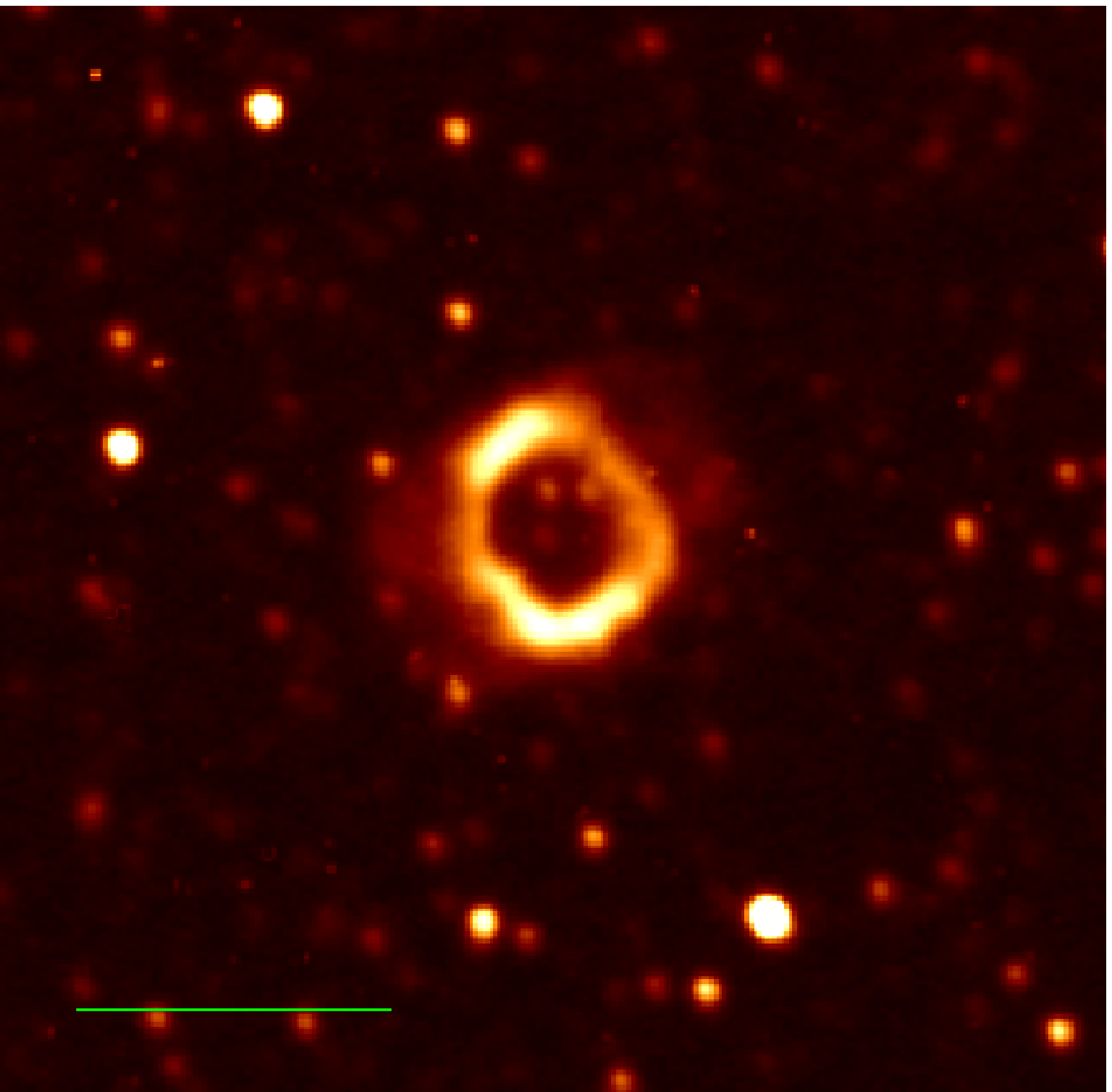}	\vspace{0.25cm}\\
\includegraphics[width=0.42\textwidth]{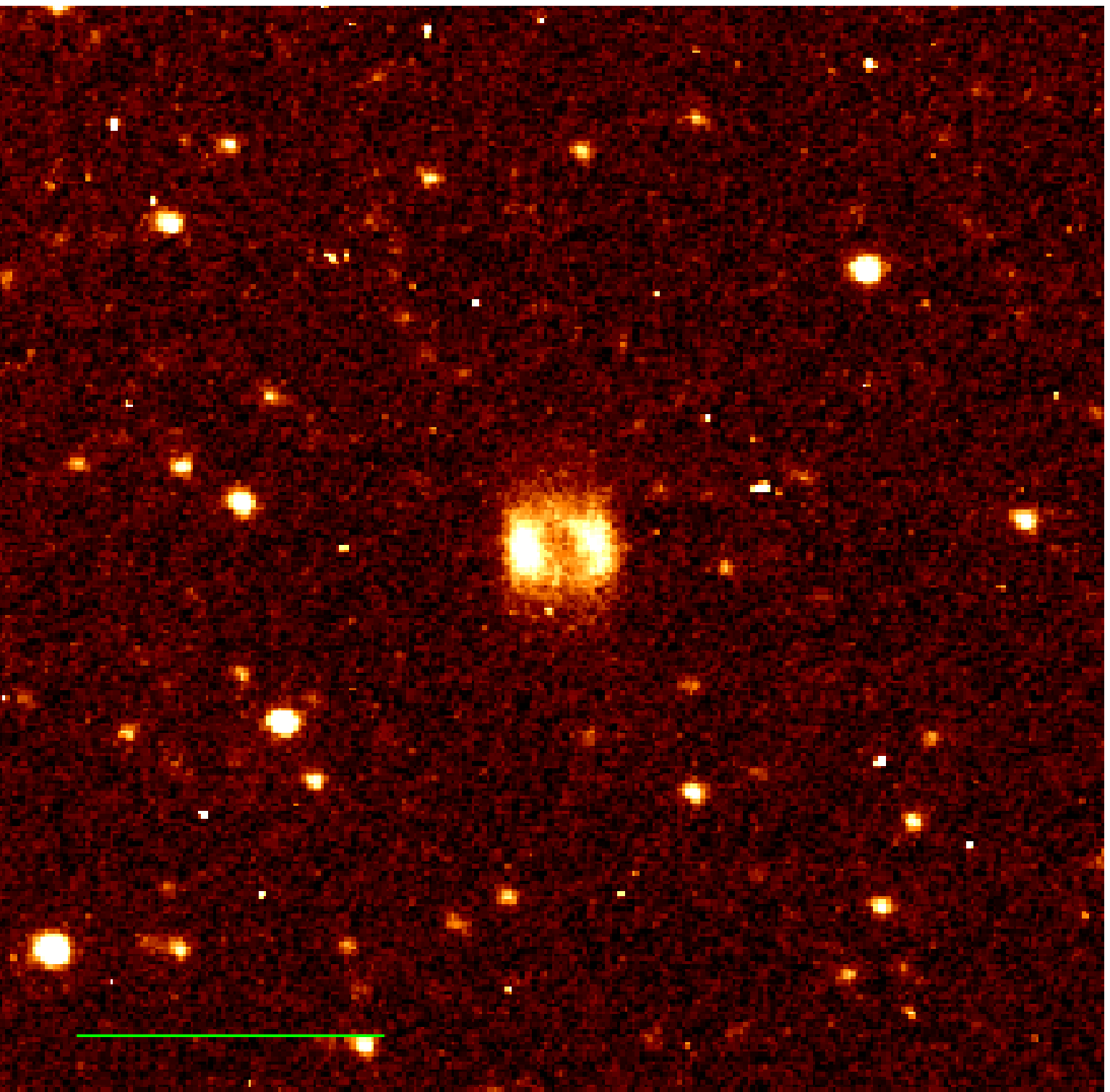}	\hspace{0.8cm}	
\includegraphics[width=0.42\textwidth]{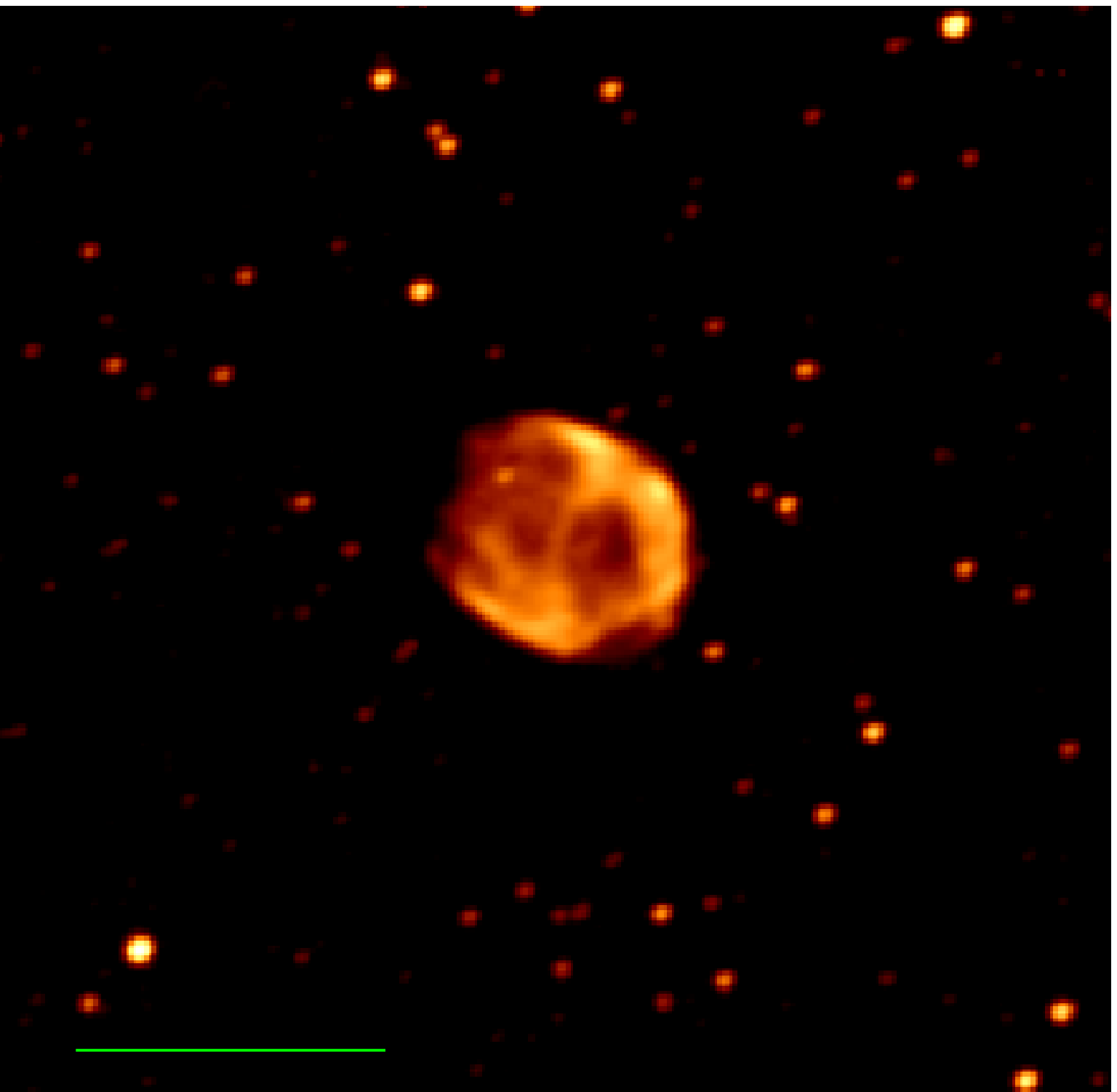}	\vspace{0.25cm}\\	
\includegraphics[width=0.42\textwidth]{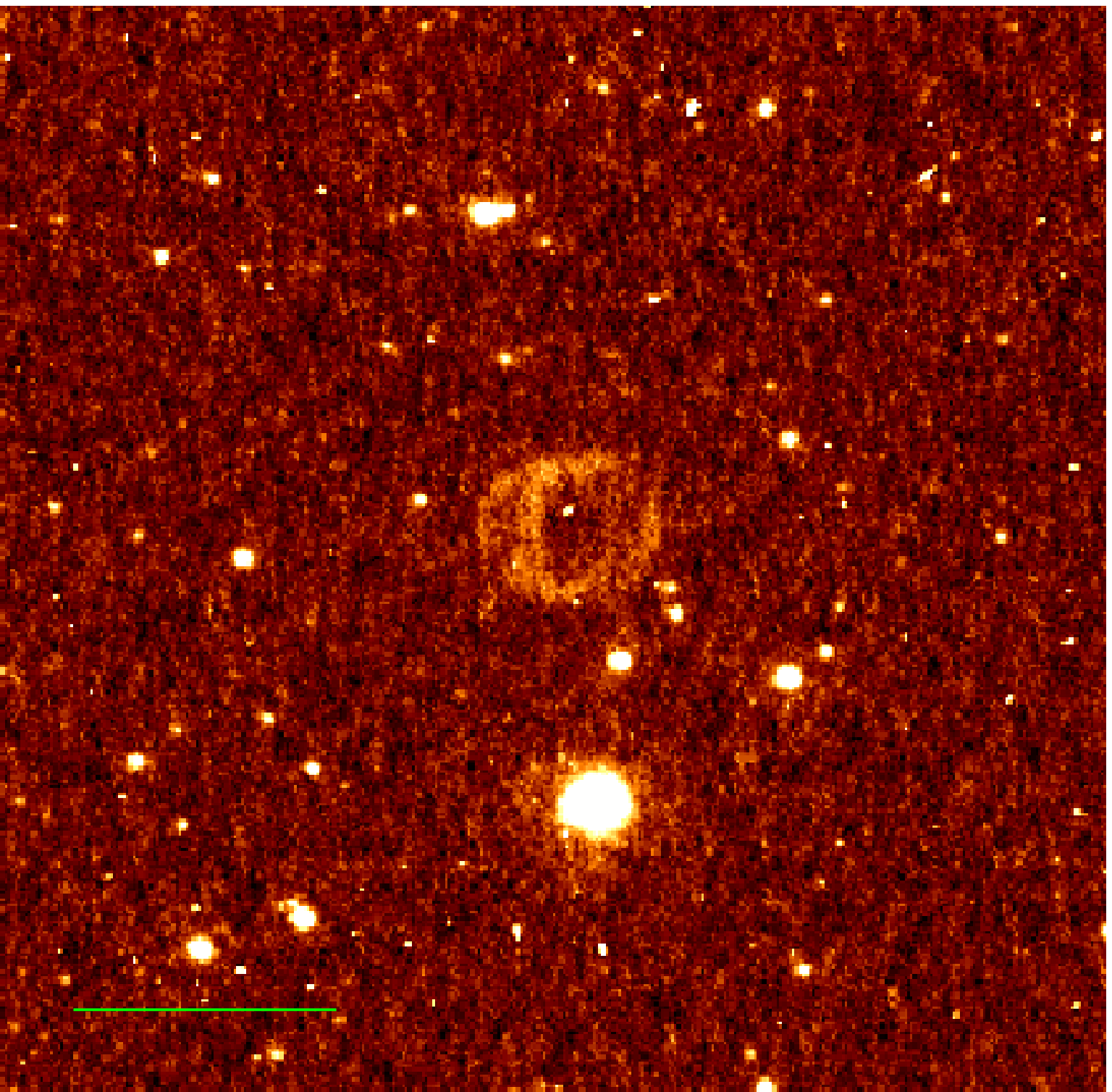}	\hspace{0.8cm}	
\includegraphics[width=0.42\textwidth]{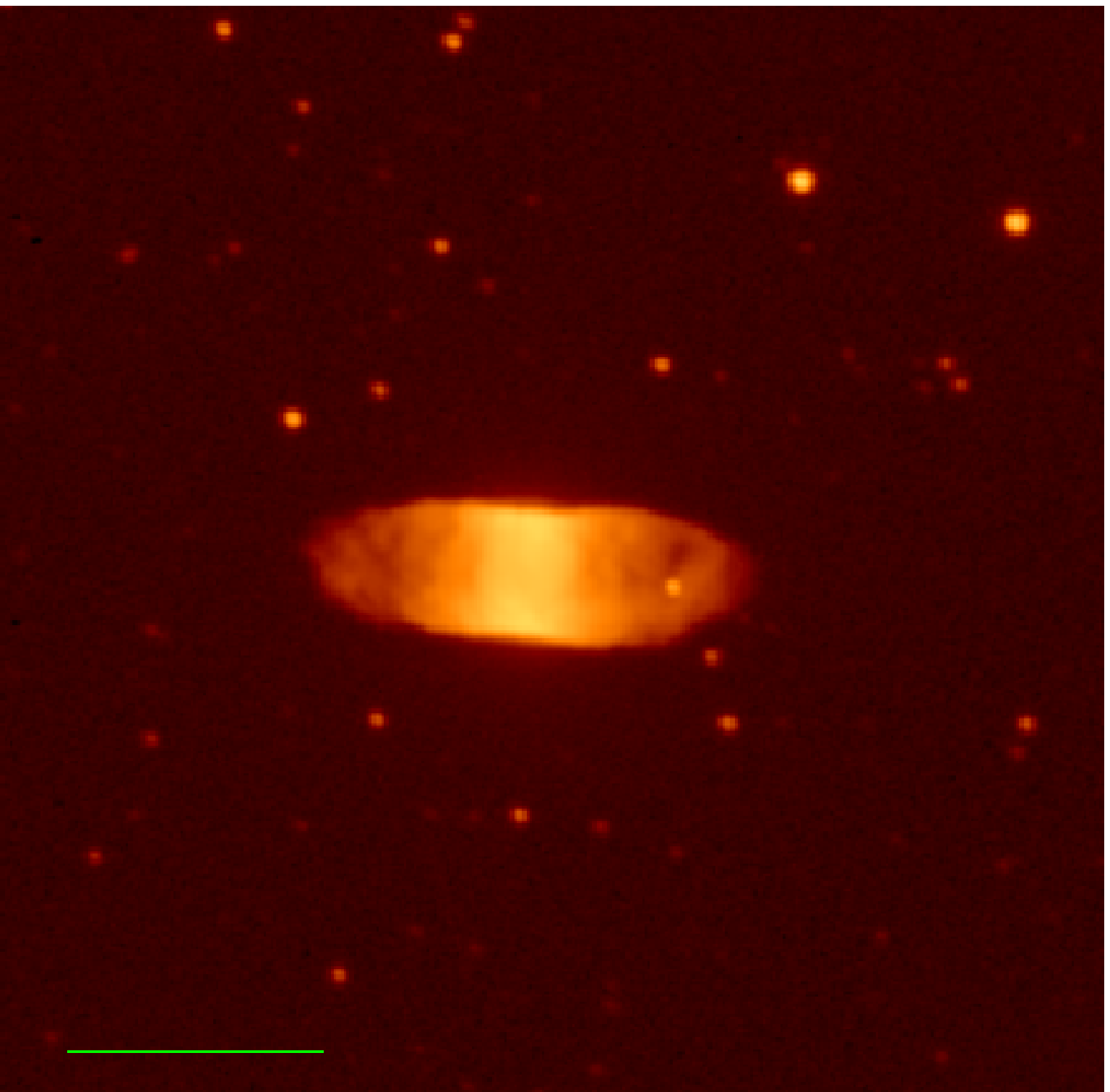}
\caption{Same as Fig.~\ref{Fig_1},
top He~2$-$119 (L) and VBRC~6, middle He~2$-$114 and He~2$-$116 (L), bottom ESO~135$-$04 and IC~4406 (L).}
  \label{Fig_15}
\end{figure*}

\begin{figure*}[!ht]
  \centering
\includegraphics[width=0.42\textwidth]{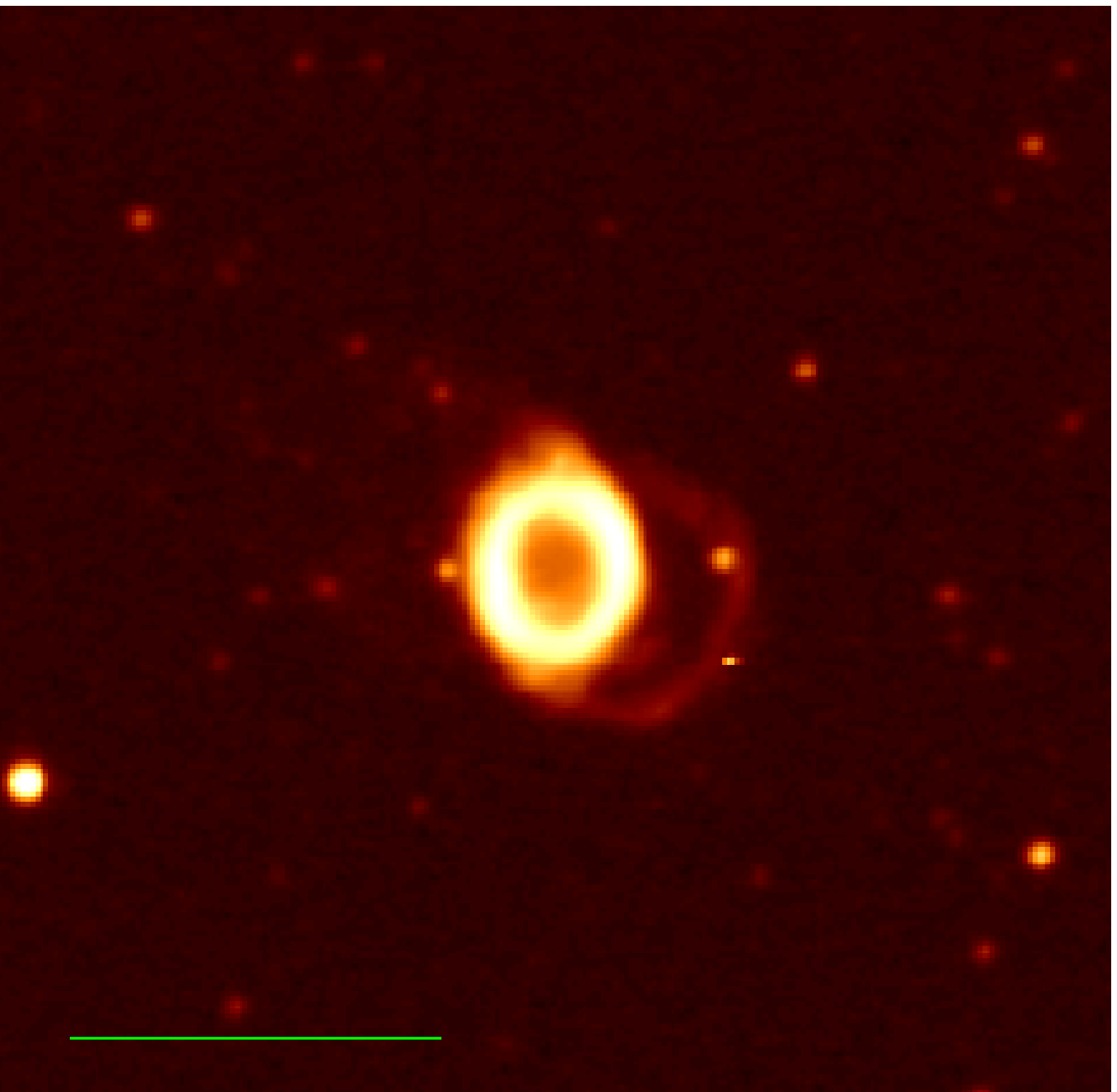}		\hspace{0.8cm}
\includegraphics[width=0.42\textwidth]{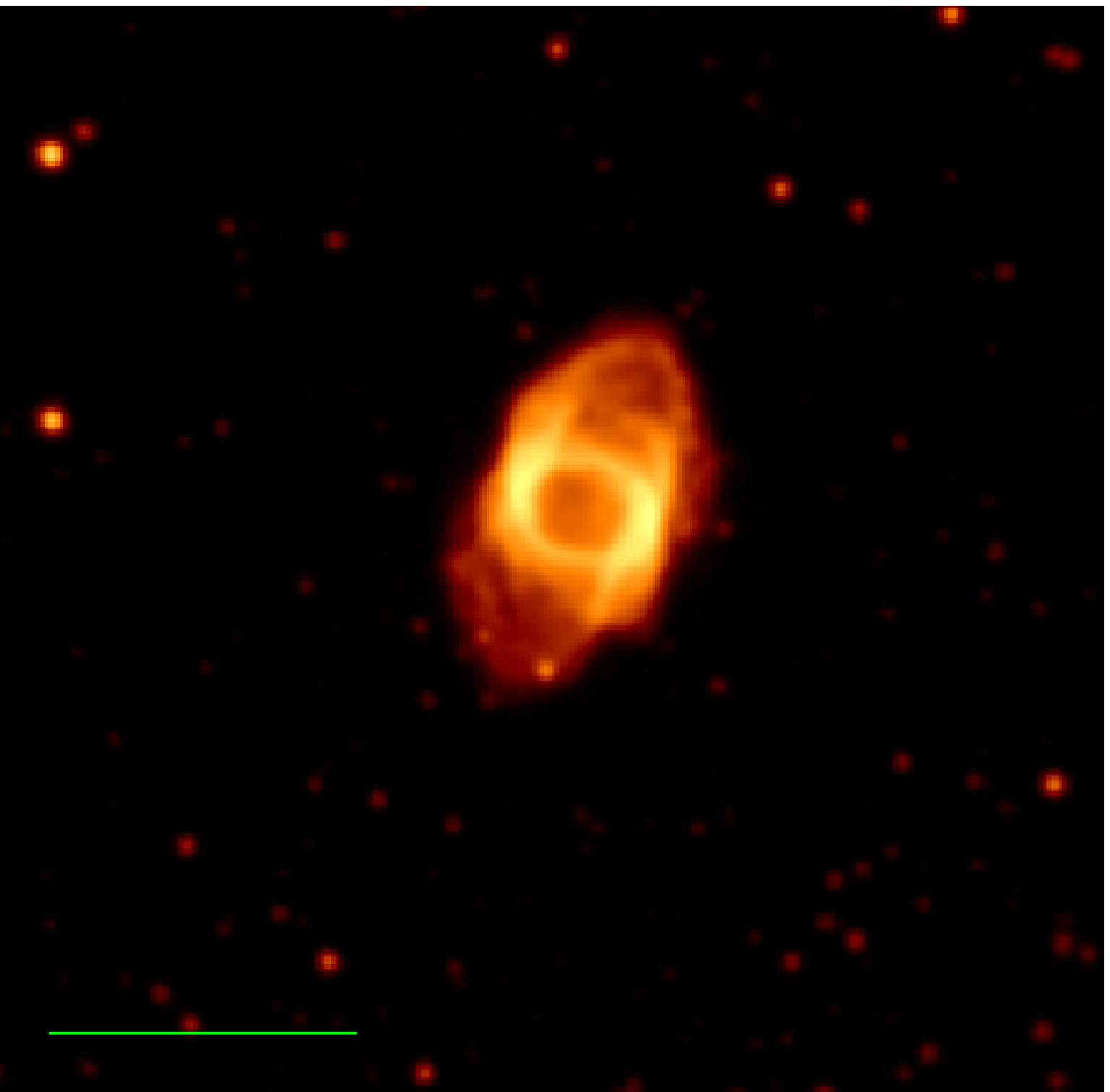}	\vspace{0.25cm}\\
\includegraphics[width=0.42\textwidth]{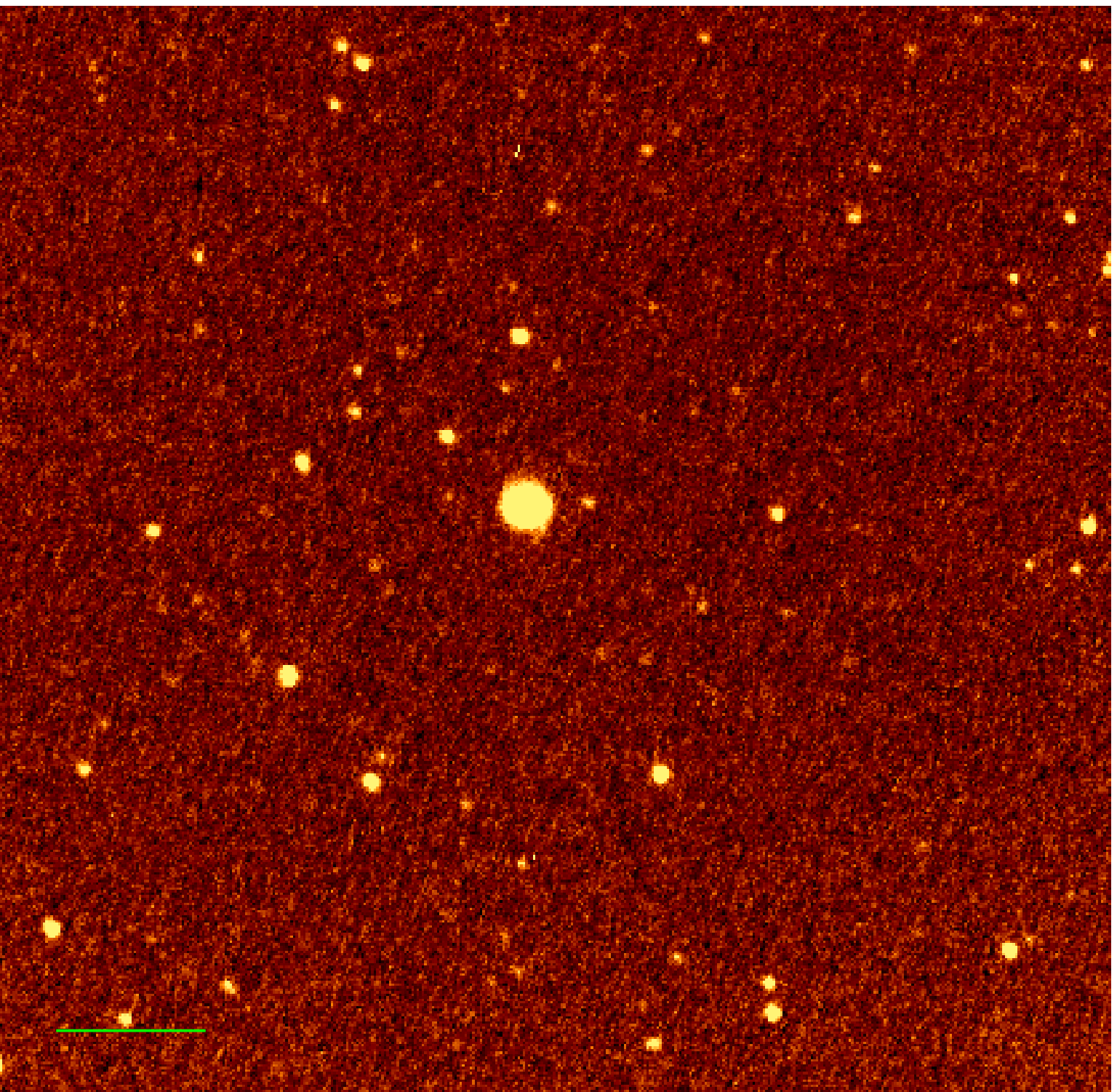}		\hspace{0.8cm}
\includegraphics[width=0.42\textwidth]{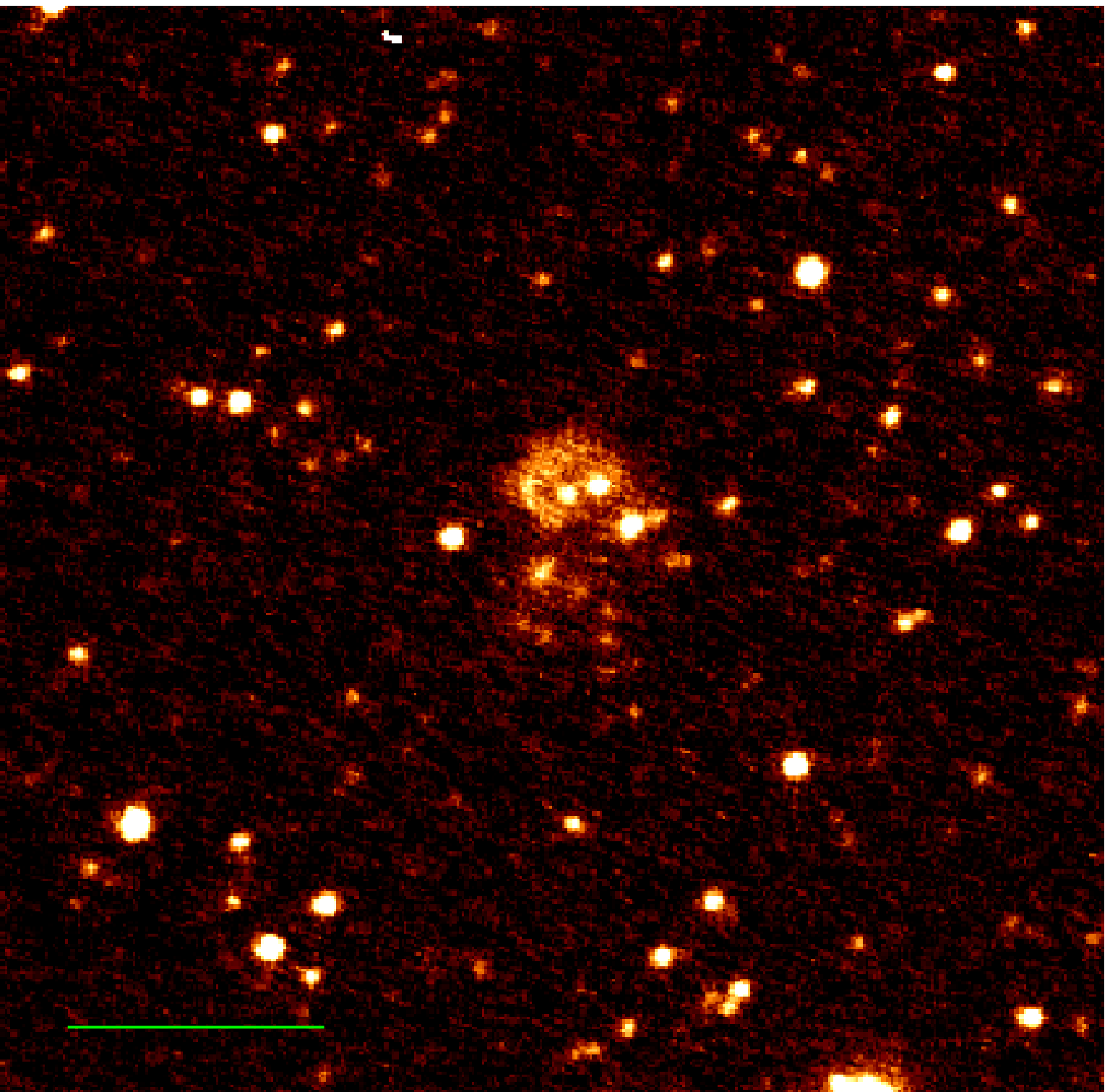}		\vspace{0.25cm}\\
\includegraphics[width=0.42\textwidth]{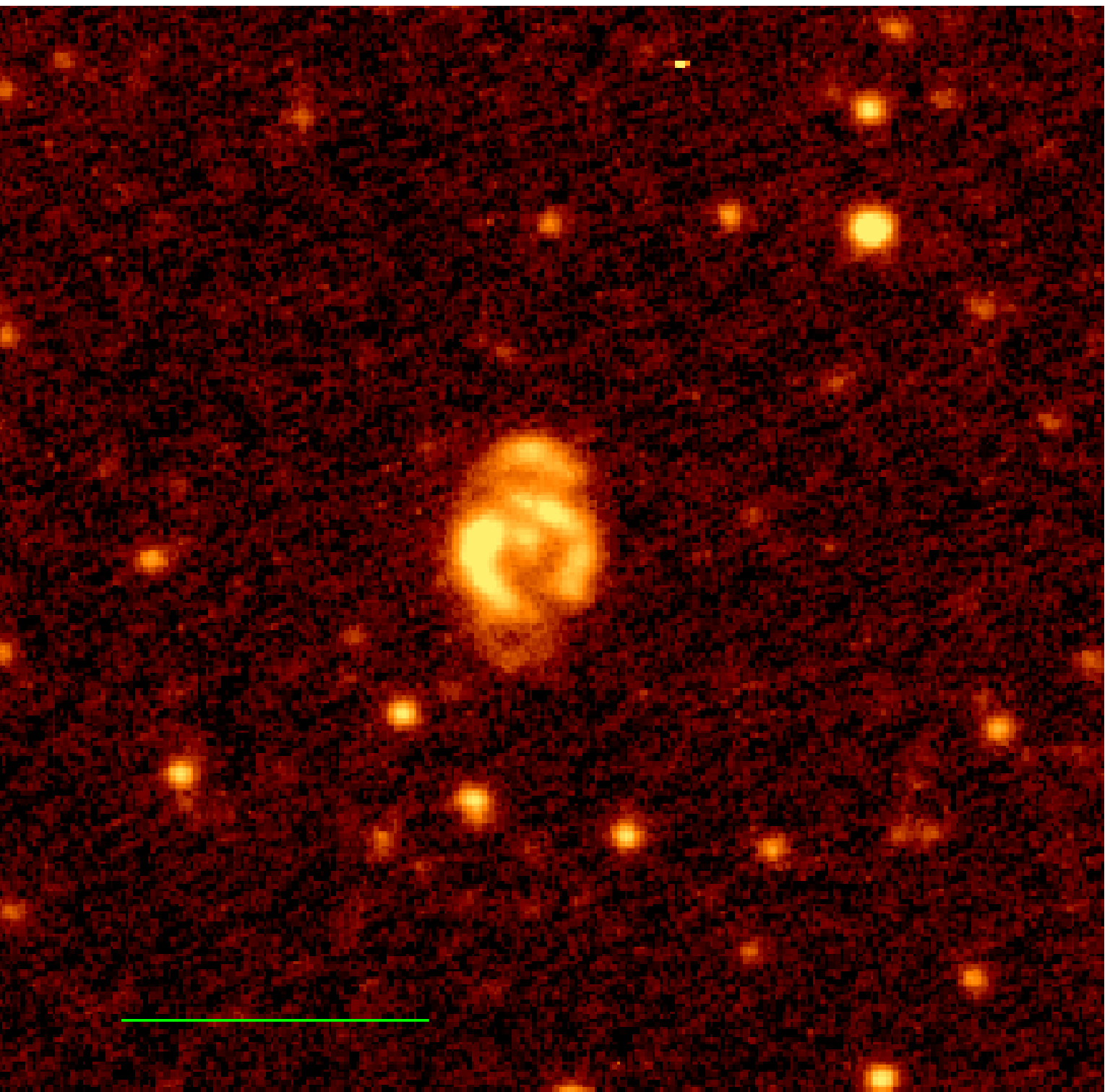}		\hspace{0.8cm}
\includegraphics[width=0.42\textwidth]{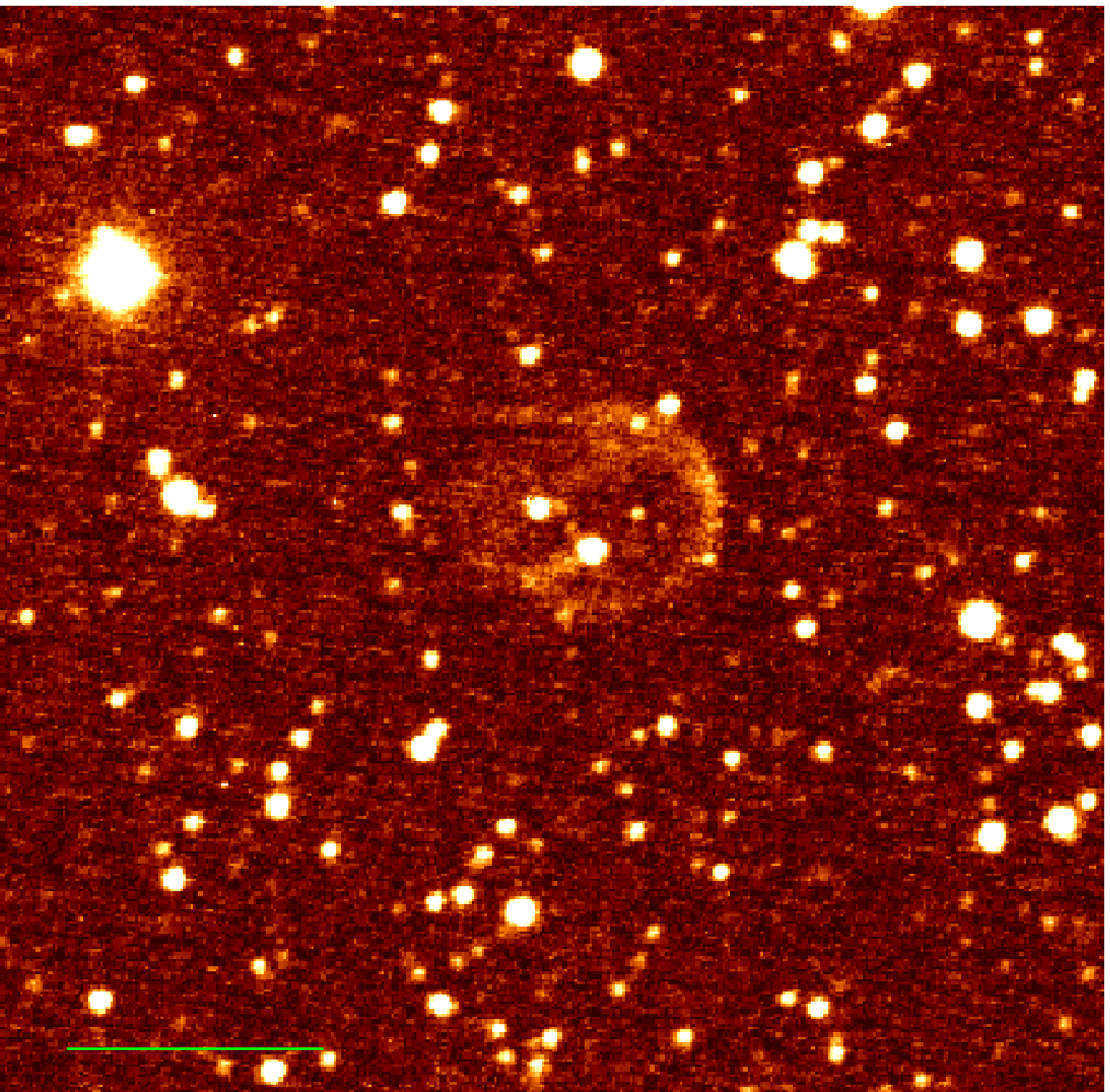}
\caption{Same as Fig.~\ref{Fig_1},
top He~2$-$120 (L) and Mz~1 (L), middle He~2$-$163 and He~2$-$146, bottom Mz~2 (L) and PHR1557$-$5128.}
  \label{Fig_16}
\end{figure*}

\begin{figure*}[!ht]
  \centering
\includegraphics[width=0.42\textwidth]{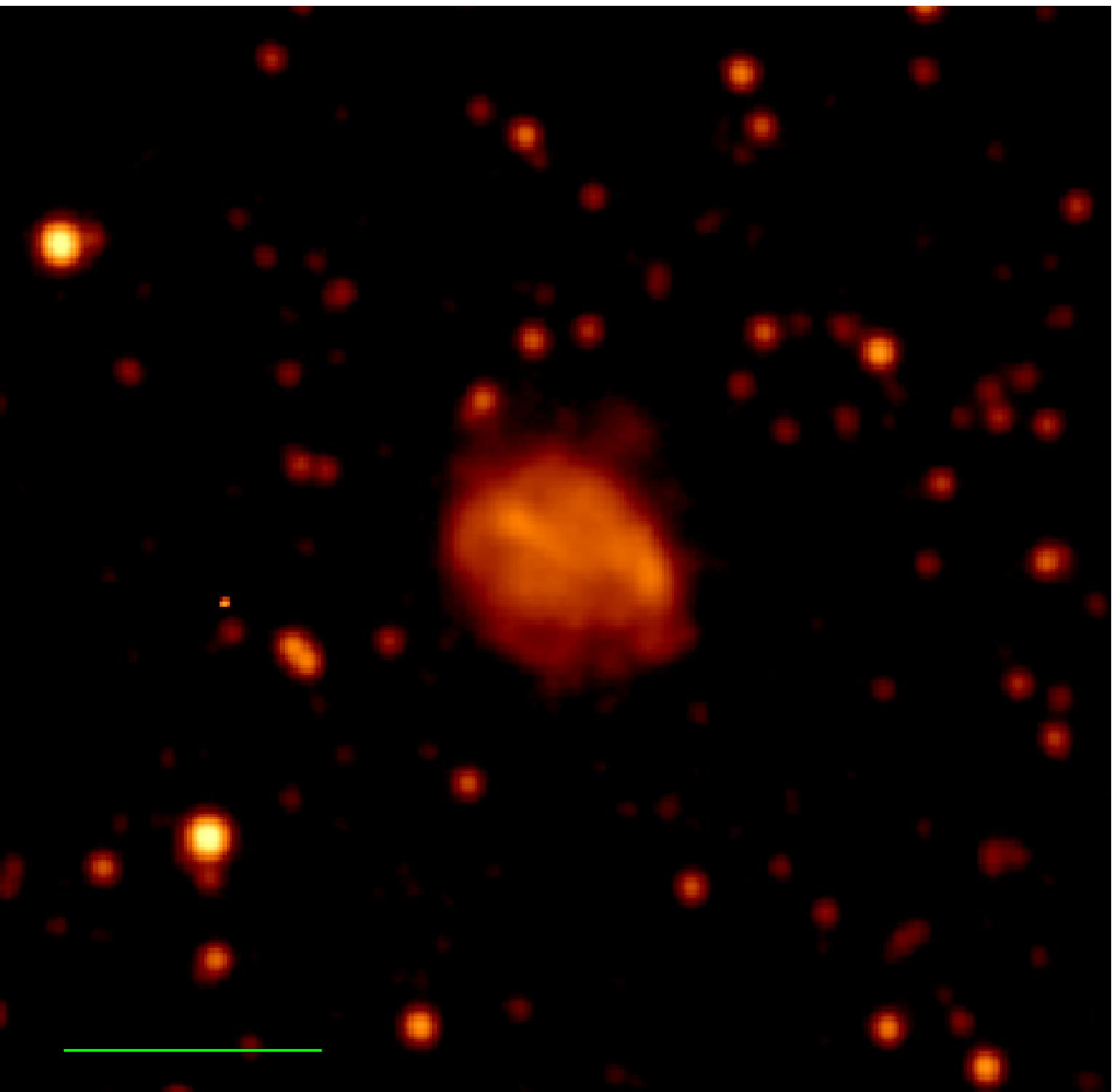}	\hspace{0.8cm}
\includegraphics[width=0.42\textwidth]{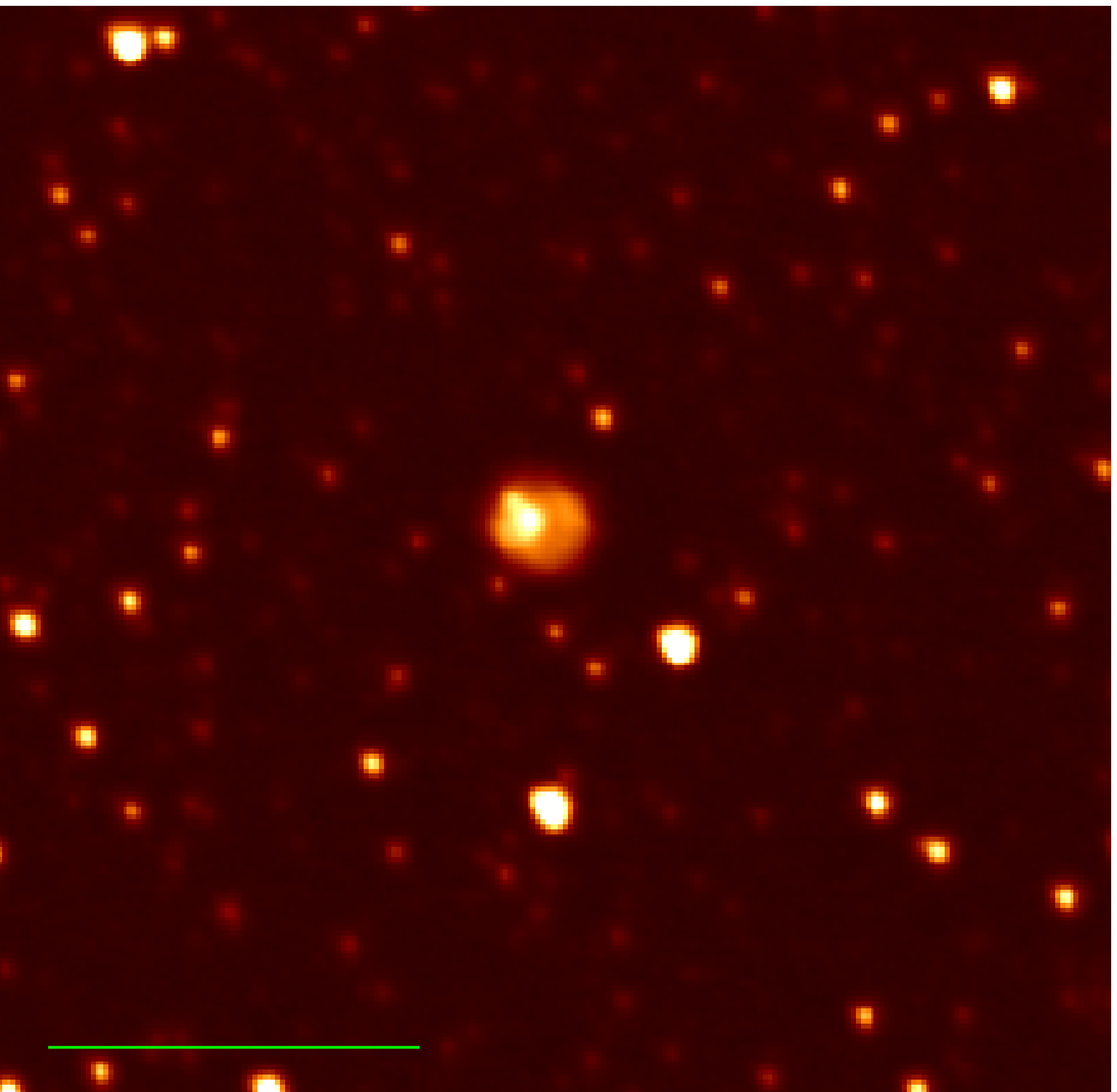}	\vspace{0.25cm}\\
\includegraphics[width=0.42\textwidth]{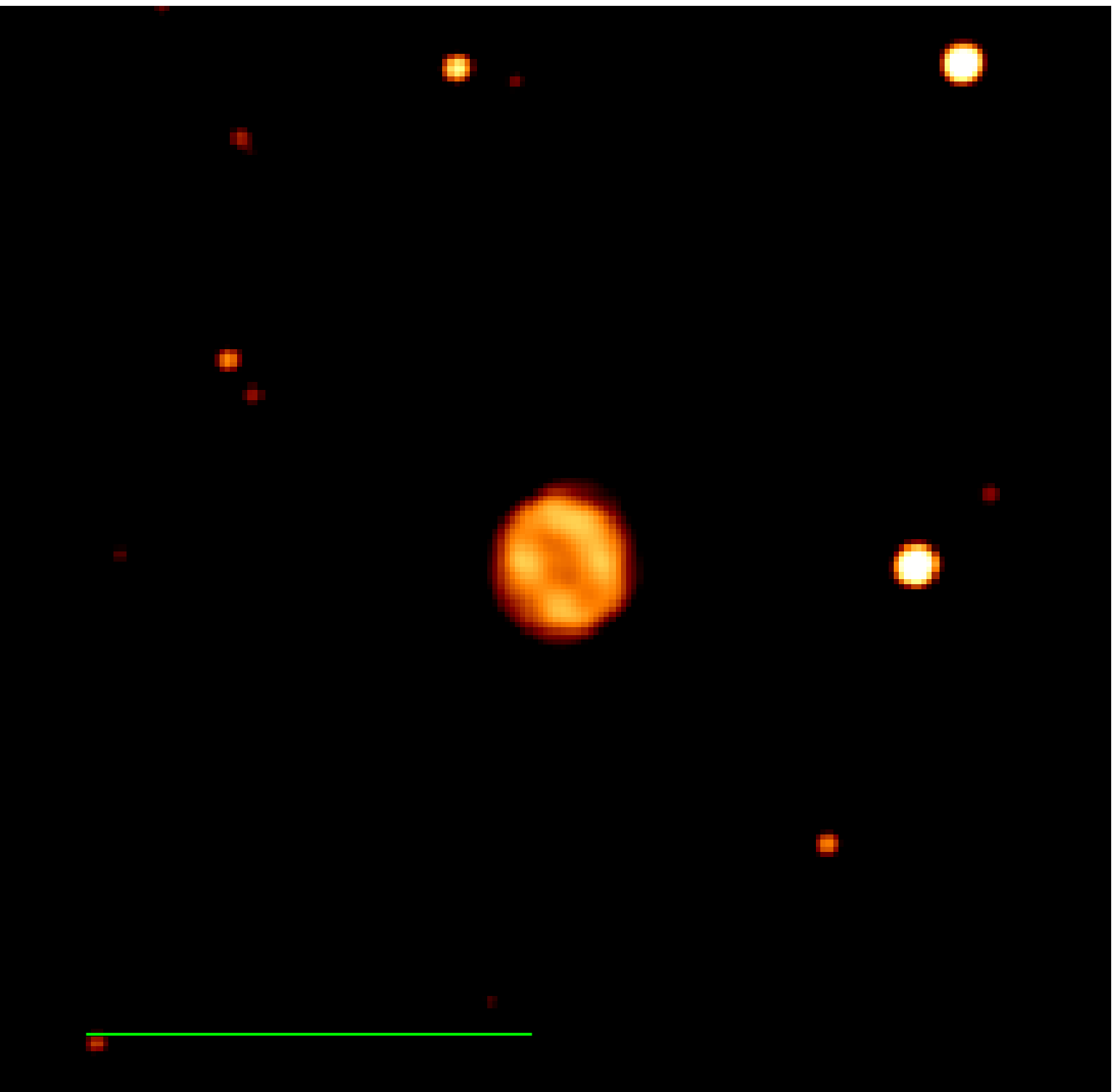}	\hspace{0.8cm}
\includegraphics[width=0.42\textwidth]{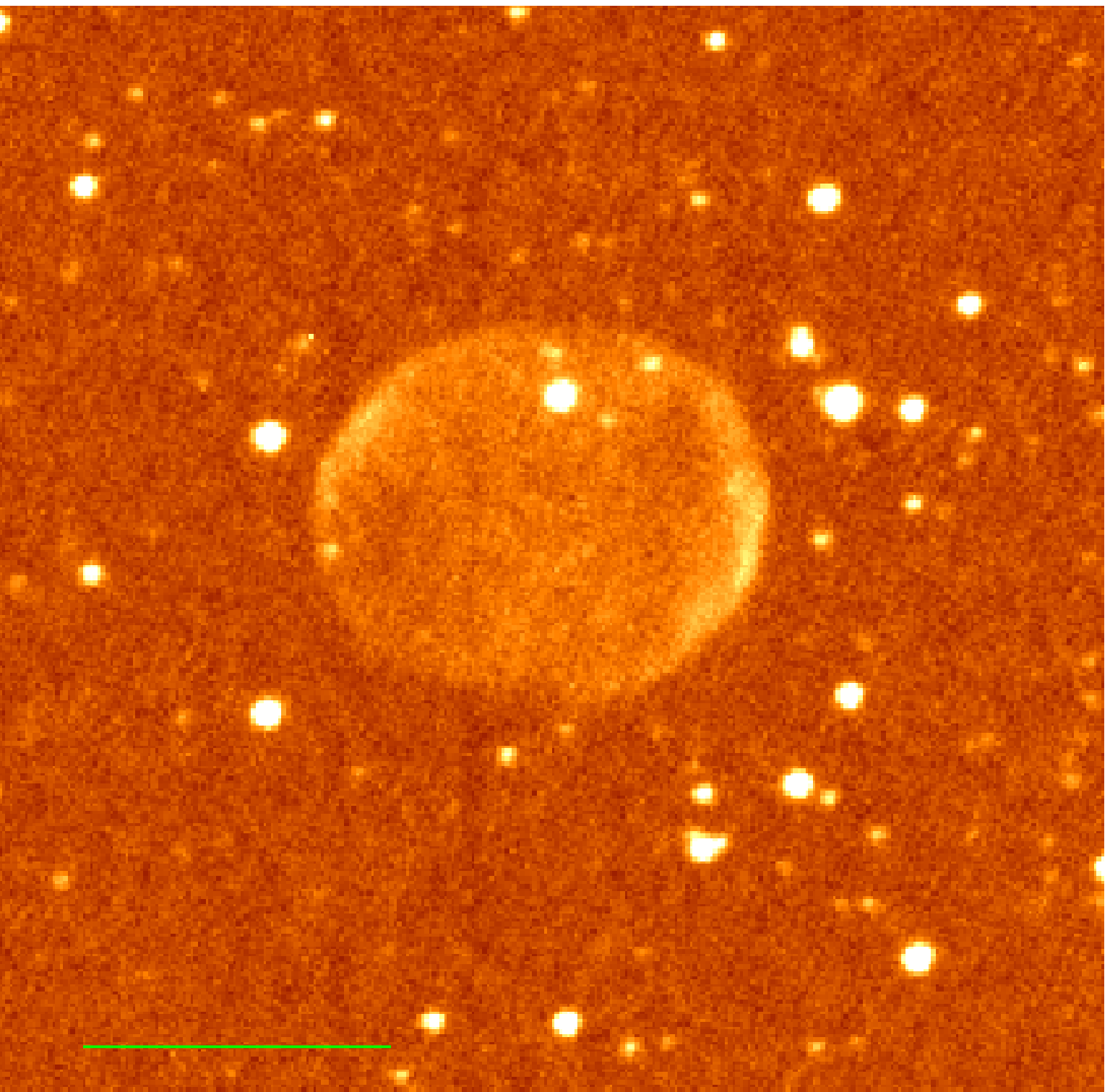}	\vspace{0.25cm}\\
\includegraphics[width=0.42\textwidth]{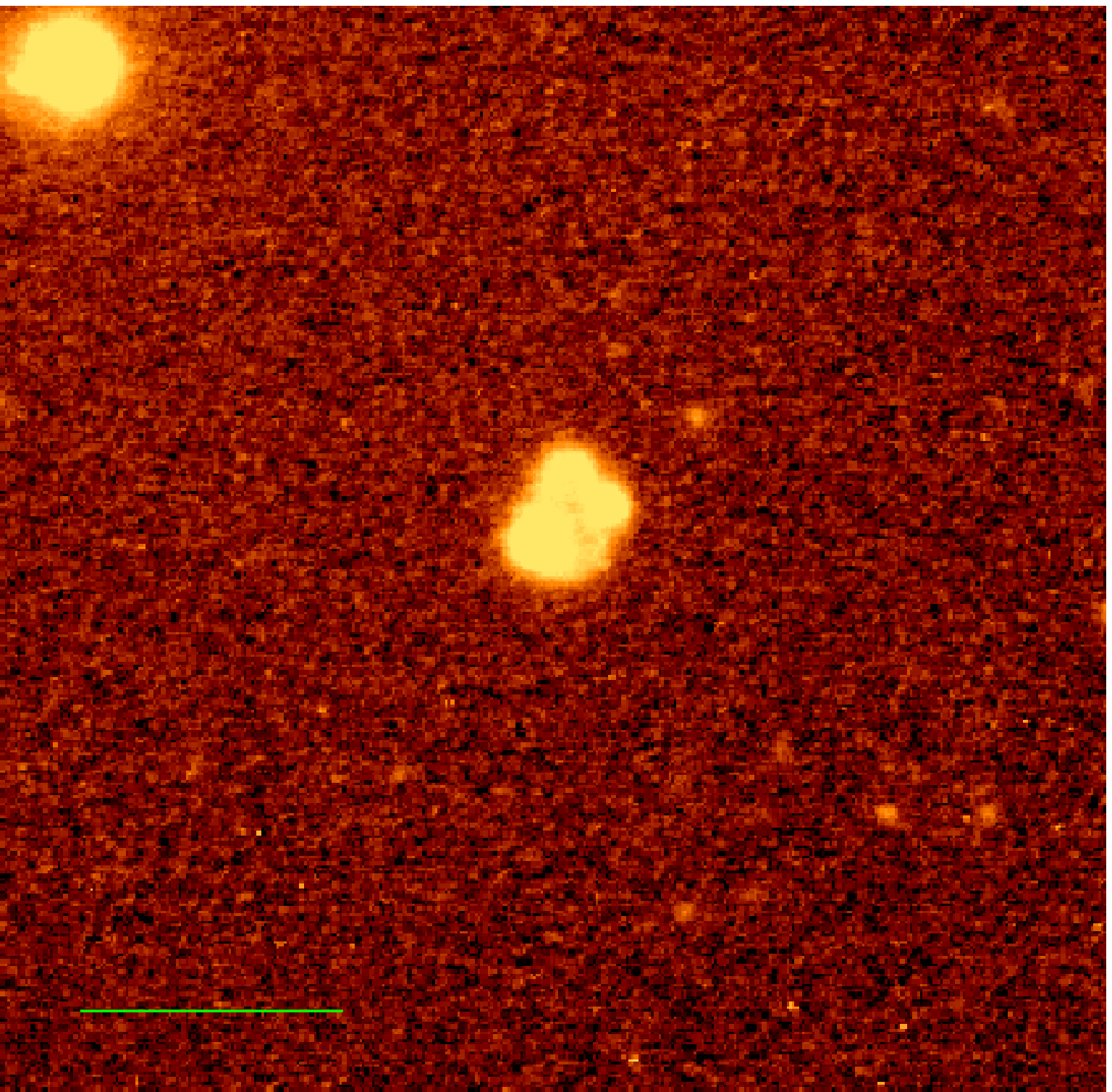}	\hspace{0.8cm}
\includegraphics[width=0.42\textwidth]{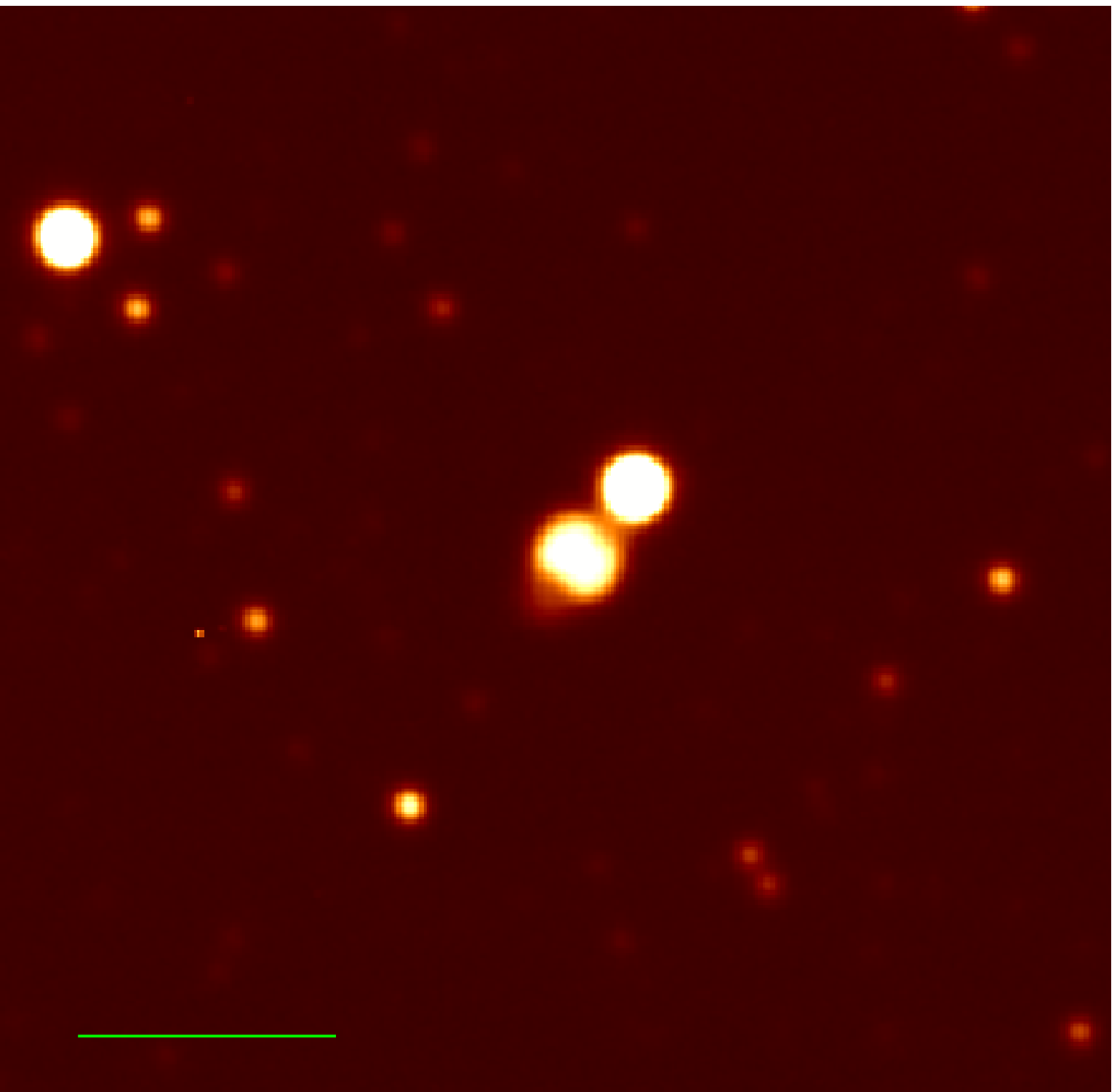}
\caption{Same as Fig.~\ref{Fig_1},
top He~2$-$165 (L) and He~2$-$164, middle IC~4642 (L) and Lo~12, bottom NGC~6153 and H~1$-$3.}
  \label{Fig_17}
\end{figure*}

\begin{figure*}[!ht]
  \centering
\includegraphics[width=0.42\textwidth]{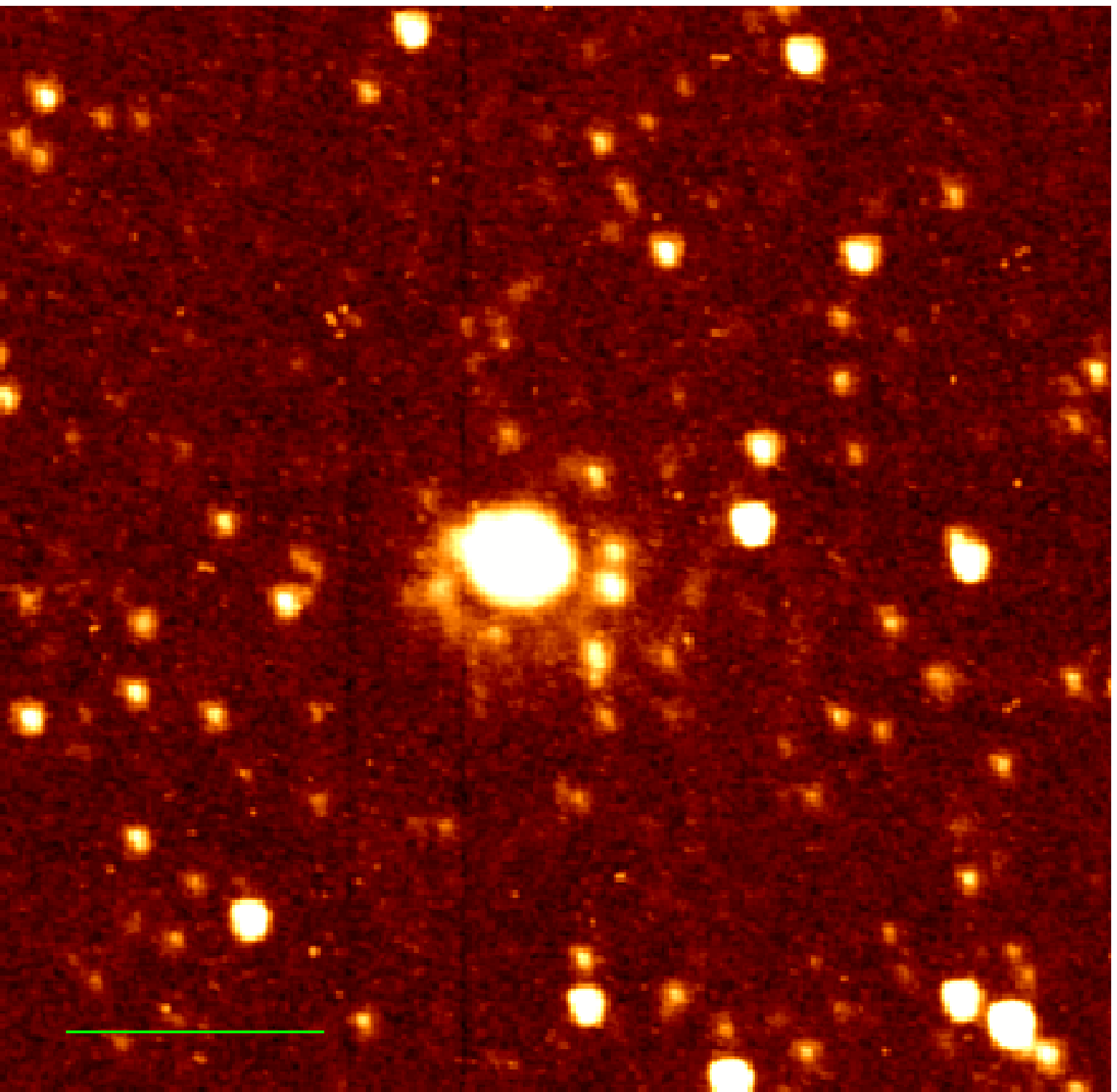}	\hspace{0.8cm}
\includegraphics[width=0.42\textwidth]{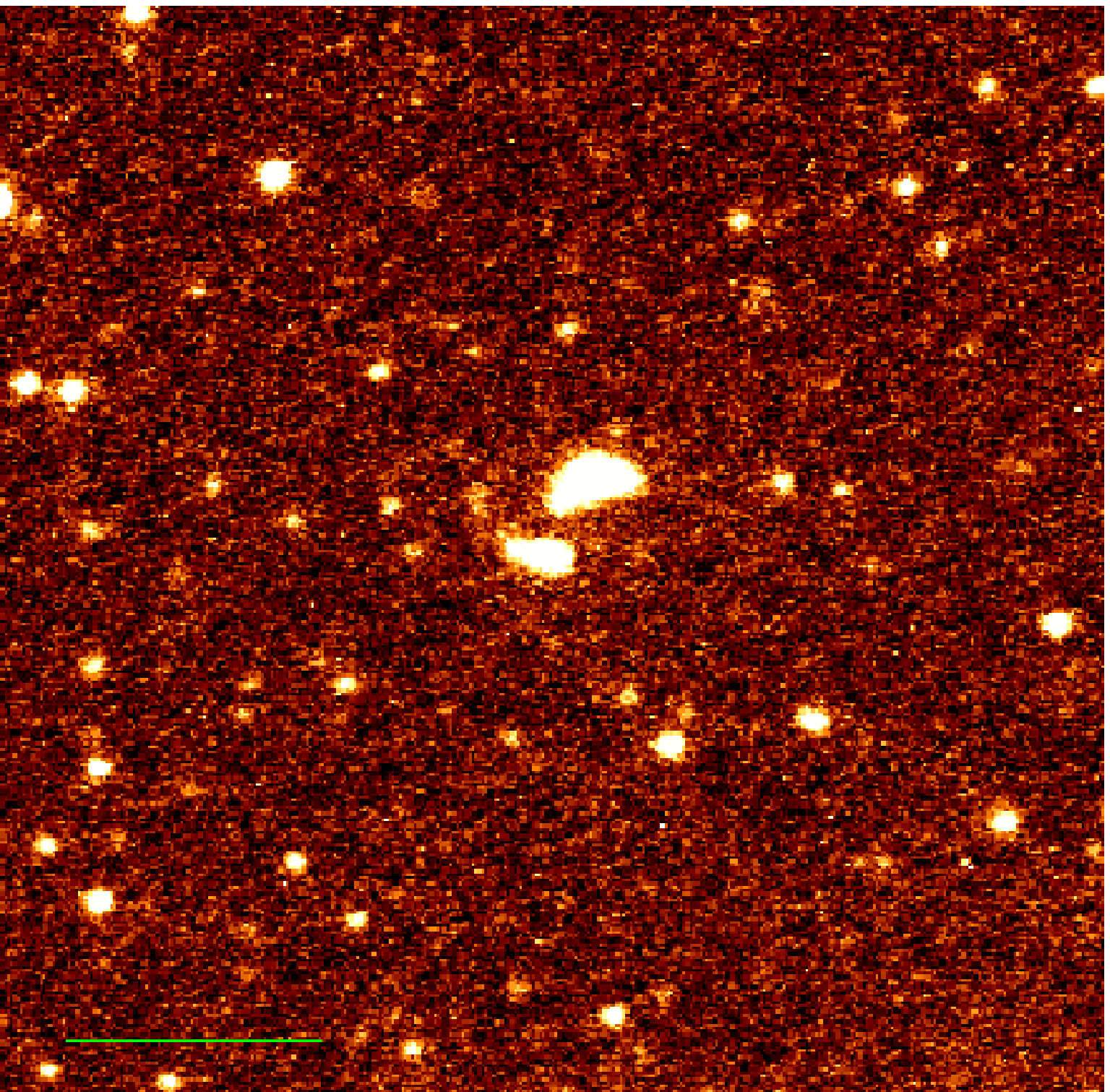}	\vspace{0.25cm}\\
\includegraphics[width=0.42\textwidth]{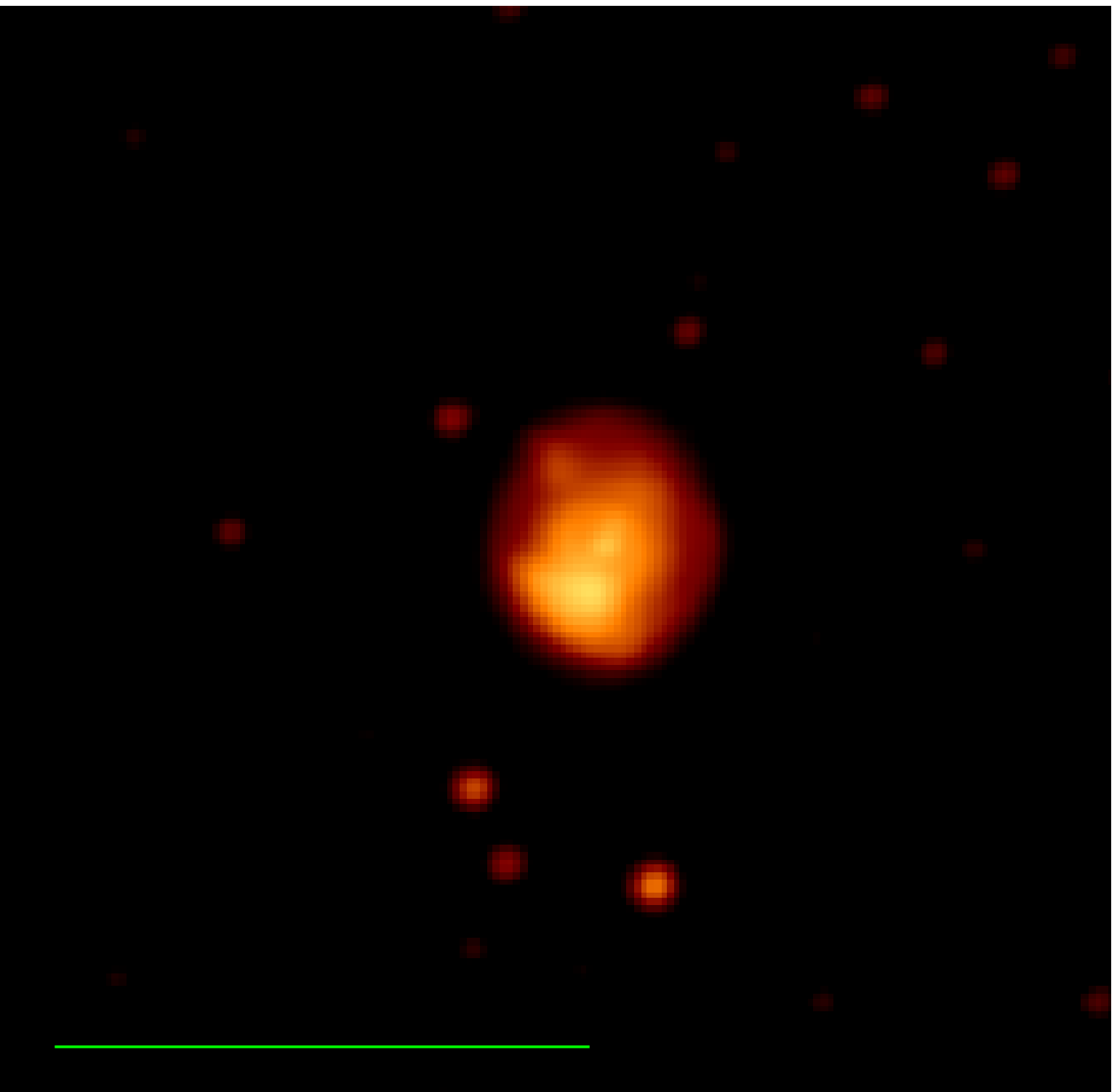}	\hspace{0.8cm}
\includegraphics[width=0.42\textwidth]{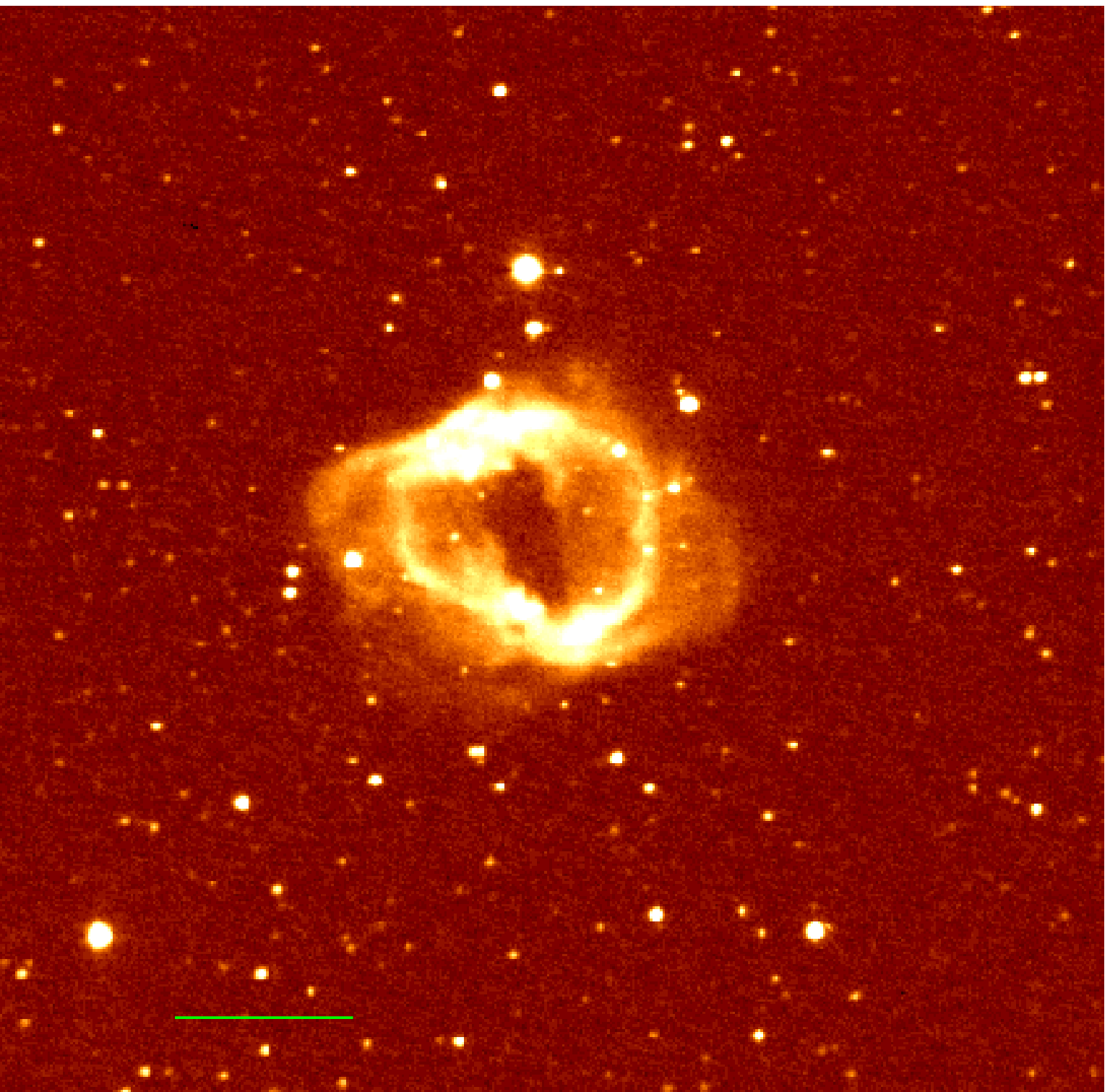}	\vspace{0.25cm}\\
\includegraphics[width=0.42\textwidth]{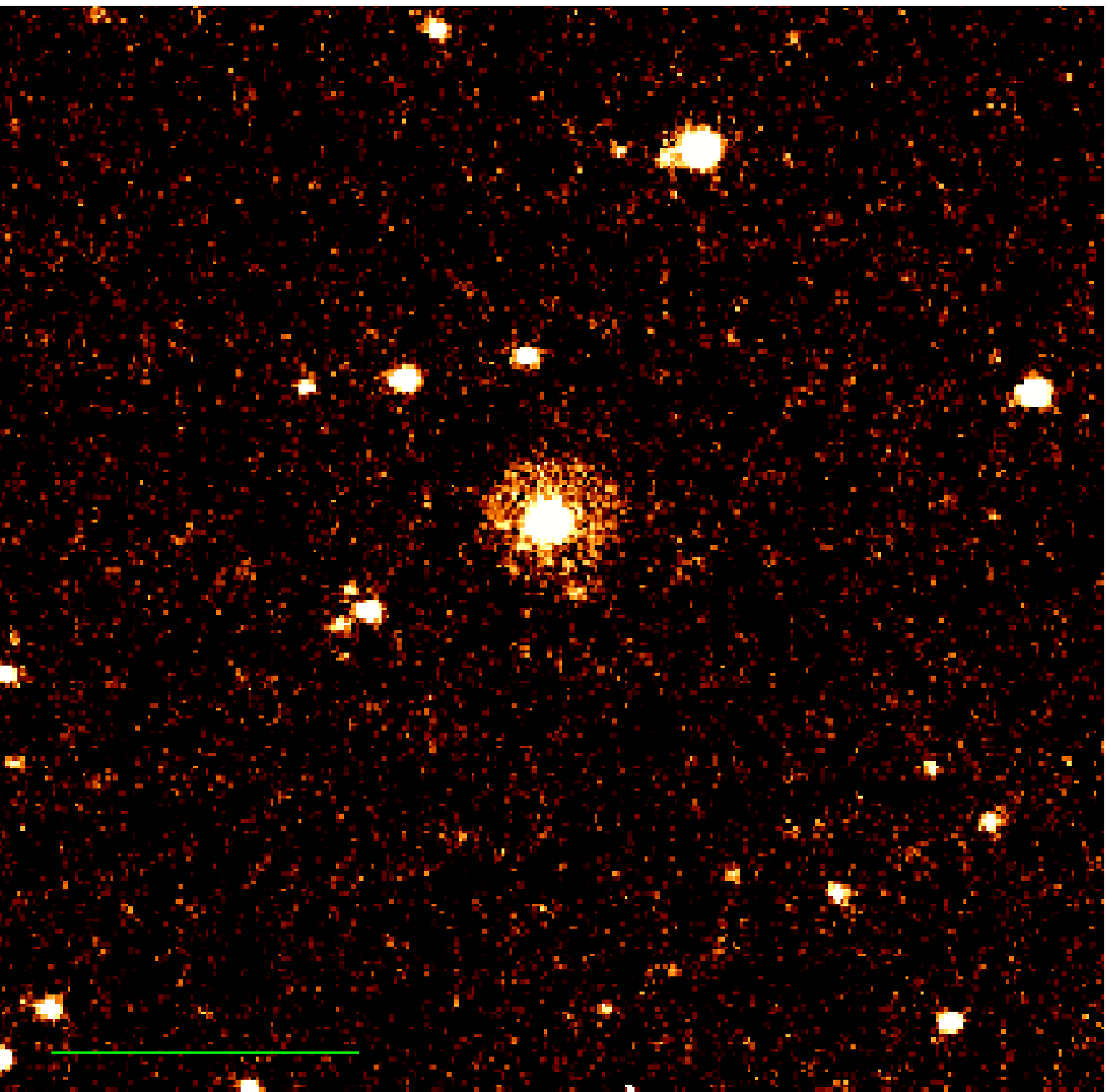}	\hspace{0.8cm}
\includegraphics[width=0.42\textwidth]{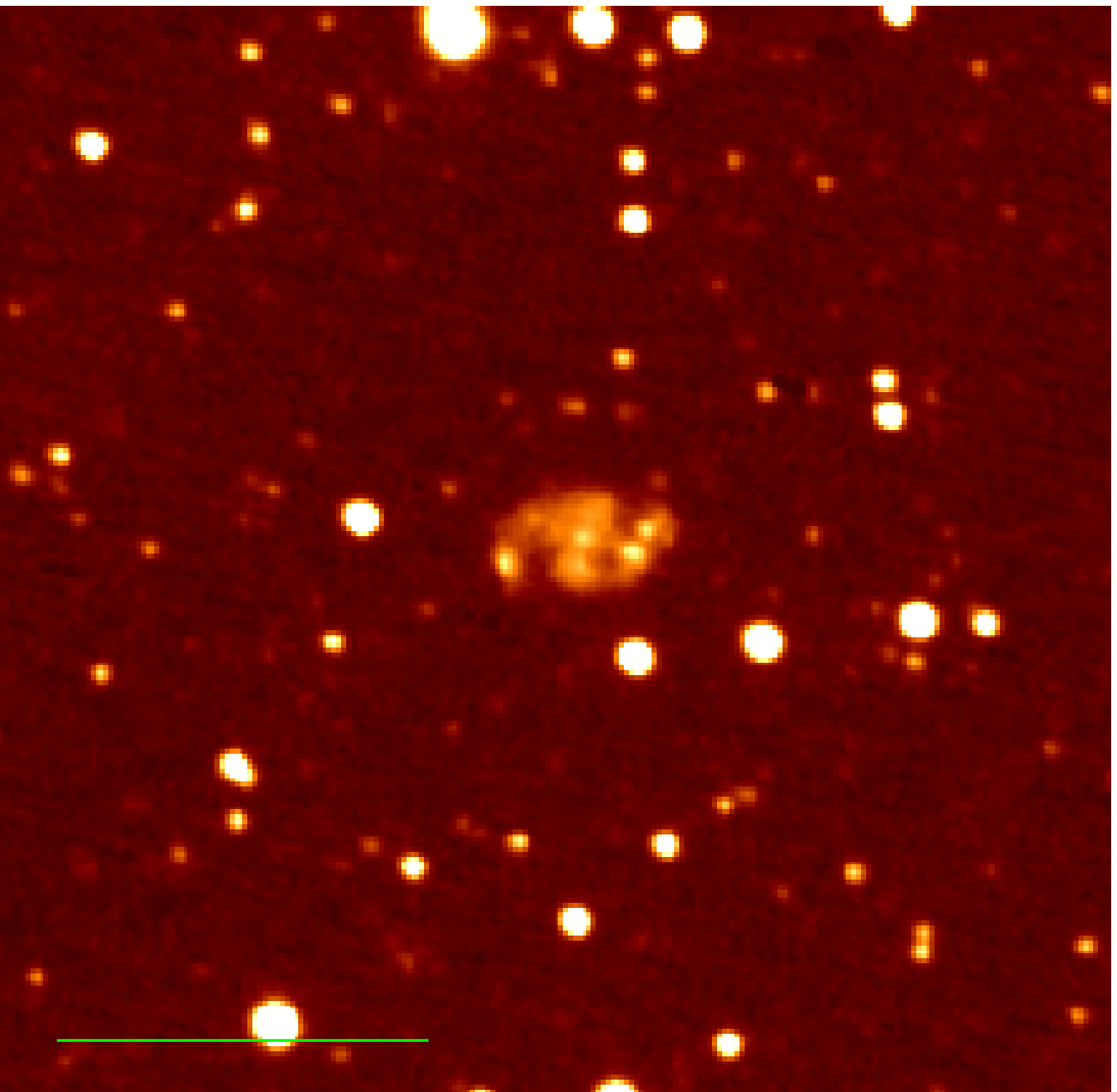}
\caption{Same as Fig.~\ref{Fig_1},
top Pe~1$-$8 (L) and He~2$-$207, middle IC~4637 (L) and K~1$-$3, bottom K~2$-$16 (L) and Wray~16$-$411.}
  \label{Fig_18}
\end{figure*}

The atlas of monochromatic images is presented in Figs.~\ref{Fig_1} to \ref{Fig_18}.
In all the images we tried different brightness and contrast levels to emphasize
the most interesting features of each object.
A logarithmic scale is indicated with an ``(L)'' in the caption
of the figures.

Our atlas includes objects that do not have any morphological classification
(e.g. Wray~16$-$20, ESO~259$-$10, LoTr~7, and VBRC~6) or are
very poorly studied 
(e.g. HaWe~9, SaWe~1, ESO~427$-$19, and ESO~209$-$15);
in particular, it contains images of recently classified PNe \citep{2005MNRAS.362..689P}.
Moreover, this catalogue shows and reveals very interesting structures in several nebulae. 
For example, 
PHR0716$-$1053 displays a non-concentric multishell, 
resembling the mysterious rings of supernova 1987A, or the
outstanding planetary nebulae VBRC~1 and Wray~16$-$121 that shows a
capricious distribution of ionized material.

We compared our images with those available in the literature (even if they were taken in another filter), 
to compare our 
classification morphology with previous ones given by other authors.
As a result, we reclassified some objects and provided one to those nebulae that did not have previous classification.

We hope that these new images and morphological descriptions 
presented here will provide a guide for future research,
thus contributing to a better understanding of the final stages of
stellar evolution.


\begin{longtab}
\begin{longtable}{l l c c c c c c }  
\caption{List of PNe included in our atlas.
The PNe are denoted by their common name and by their PN G designation. 
Sixth column lists the telescope used in each observation, EABA and CASLEO, indicated by ``E'' and ``C'' respectively.
Seventh column lists the number of combined images and 
eighth column shows the figure number for each nebulae.
PNe observed by \cite{1998A&AS..133..361H} and \cite{1987AJ.....94..671B} are marked by ``*'' and
 objects labelled with ``$^{\dag}$'' have probably H$\alpha$ contamination or only the continuum emission is detected.}\\  
\hline\hline
 N 	 &    Name         &  PN G                & RA (J2000)        & Dec (J2000)      & Telescope     & Images     & Figure   \\     
\hline  
\endfirsthead
\caption{continued.}\\
\hline\hline
 N       &   Name          &  PN G                & RA (J2000)        & Dec (J2000)      & Telescope     & Images    & Figure   \\ 
\hline
\endhead
\hline
\endfoot
\hline\hline                 
1   &  IC~5148$-$50$^*$$^{\dag}$ & 002.7$-$52.4 & 21 59 31.7   & $-$39 22 36.2    & E      & 10      &  \ref{Fig_1}      \\
2   &  Hb~4	        & 003.1$+$02.9     & 17 41 52.7        & $-$24 42 08.0    & E	   & 11      &  \ref{Fig_1}    \\
3   &  SaWe~3$^*$$^{\dag}$ & 013.8$-$02.8  & 18 26 03.1        & $-$18 12 05.6    & E      & 15      &  \ref{Fig_1}        \\
4   &  HDW~12           & 014.8$-$25.6     & 19 58 13.2        & $-$26 28 15.4    & C      & 3       &  \ref{Fig_1}     \\
5   &  DeHt~3           & 019.4$-$13.6     & 19 17 04.6        & $-$18 01 34.2    & E      & 17      &  \ref{Fig_1}       \\
6   &  A~66             & 019.8$-$23.7     & 19 57 31.8        & $-$21 36 36.8    & C      & 8       &  \ref{Fig_1}      \\
7   &  A~67             & 043.5$-$13.4     & 19 58 29.3        & $+$03 02 22.6    & C      & 3       &  \ref{Fig_2}         \\
8   &  NGC~246$^*$      & 118.8$-$74.7     & 00 47 03.8        & $-$11 52 21.6    & C      & 3       &  \ref{Fig_2}          \\
9   &  PHR0615$-$0025   & 209.1$-$08.2     & 06 15 20.4        & $-$00 25 49.1    & C      & 3       &  \ref{Fig_2}            \\
10  &  K~1$-$11         & 215.6$+$11.1     & 07 36 07.7        & $+$02 42 17.2    & C      & 8       &  \ref{Fig_2}           \\
11  &  PHR0702$-$0324   & 217.2$+$00.9     & 07 02 34.2        & $-$03 24 34.9    & C      & 3       &  \ref{Fig_2}            \\
12  &  PHR0705$-$0924   & 222.9$-$01.1     & 07 05 51.4        & $-$09 24 11.2    & C      & 4       &  \ref{Fig_2}           \\
13  &  PHR0700$-$1143   & 224.3$-$03.4     & 07 00 05.8        & $-$11 43 50.9    & C      & 5       &  \ref{Fig_3}         \\
14  &  PHR0716$-$1053   & 225.4$+$00.4     & 07 16 08.0        & $-$10 53 06.0    & C      & 6       &  \ref{Fig_3}        \\
15  &  PHR0711$-$1238   & 226.4$-$01.3     & 07 11 43.3        & $-$12 38 03.1    & C      & 6       &  \ref{Fig_3}        \\
16  &  M~1$-$16         & 226.7$+$05.6     & 07 37 18.9        & $-$09 38 48.1    & E      & 10      &  \ref{Fig_3}             \\
17  &  PHR0719$-$1222   & 227.1$+$00.5     & 07 19 46.7        & $-$12 22 46.9    & C      & 3       &  \ref{Fig_3}              \\
18  &  PHR0727$-$1259   & 228.6$+$01.9     & 07 27 49.0        & $-$12 59 30.1    & C      & 3       &  \ref{Fig_3}             \\
19  &  PHR0727$-$1707   & 232.1$-$00.1     & 07 27 08.3        & $-$17 07 14.2    & C      & 5       &  \ref{Fig_4}           \\
20  &  PHR0724$-$1757   & 232.6$-$01.0     & 07 24 43.4        & $-$17 57 51.1    & C      & 3       &  \ref{Fig_4}            \\
21  &  SaWe~1           & 233.0$-$10.1     & 06 50 41.0        & $-$22 26 09.3    & C      & 3       &  \ref{Fig_4}            \\
22  &  HaWe~9           & 236.0$-$10.6     & 06 54 29.0        & $-$25 21 11.6    & C      & 6       &  \ref{Fig_4}             \\
23  &  PHR0730$-$2151   & 236.7$-$01.6     & 07 30 46.9        & $-$21 51 43.9    & C      & 3       &  \ref{Fig_4}              \\
24  &  ESO~427$-$19     & 239.6$-$12.0     & 06 55 12.1        & $-$29 07 28.4    & C      & 3       &  \ref{Fig_4}             \\
25  &  NGC~2610$^{\dag}$ & 239.6$+$13.9    & 08 33 23.5        & $-$16 08 55.9    & E	   & 13      &  \ref{Fig_5}         \\
26  &  PHR0726$-$2858   & 242.5$-$05.9     & 07 26 04.8        & $-$28 58 23.5    & C      & 4       &  \ref{Fig_5}        \\
27  &  PRTM~1           & 243.8$-$37.1     & 05 03 01.7        & $-$39 45 44.0    & C      & 3       &  \ref{Fig_5}           \\
28  &  PHR0742$-$3247   & 247.5$-$04.7     & 07 42 23.6        & $-$32 47 44.9    & C      & 3       &  \ref{Fig_5}        \\
29  &  A~23             & 249.3$-$05.4     & 07 43 18.9        & $-$34 45 13.0    & C      & 3       &  \ref{Fig_5}          \\
30  &  PHR0834$-$2819   & 249.8$+$07.1     & 08 34 18.1        & $-$28 19 03.0    & C      & 3       &  \ref{Fig_5}          \\
31  &  PHR0803$-$3331   & 250.4$-$01.3     & 08 03 12.5        & $-$33 31 01.9    & C      & 4       &  \ref{Fig_6}     \\
32  &  PHR0736$-$3901   & 252.4$-$08.7     & 07 36 25.2        & $-$39 01 31.1    & C      & 3       &  \ref{Fig_6}    \\
33  &  PHR0820$-$3516   & 253.9$+$00.7     & 08 20 52.4        & $-$35 16 32.9    & C      & 3       &  \ref{Fig_6}    \\
34  &  VBRC~1           & 257.5$+$00.6     & 08 30 58.0        & $-$38 19 52.3    & C      & 3       &  \ref{Fig_6}    \\
35  &  Wray~17$-$1      & 258.0$-$15.7     & 07 14 48.0        & $-$46 57 24.4    & C      & 3       &  \ref{Fig_6}   \\
36  &  He~2$-$11        & 259.1$+$00.9     & 08 37 08.2        & $-$39 26 25.5    & C      & 3       &  \ref{Fig_6}    \\
37  &  Wray~16$-$20     & 260.7$-$03.3     & 08 23 40.4        & $-$43 12 44.0    & C      & 3       &  \ref{Fig_7}    \\
38  &  He~2$-$15        & 261.6$+$03.0     & 08 53 30.6        & $-$40 03 34.4    & C      & 3       &  \ref{Fig_7}   \\
39  &  NGC~2818         & 261.9$+$08.5     & 09 16 00.5        & $-$36 37 31.6    & E      & 20      &  \ref{Fig_7}      \\
40  &  Wray~17$-$18     & 262.6$-$04.6	   & 08 23 53.7        & $-$45 31 09.9	  & E 	   & 10      &  \ref{Fig_7}  \\
41  &  ESO~209$-$15     & 263.3$-$08.8	   & 08 05 10.9        & $-$48 23 30.1	  & E	   & 20      &  \ref{Fig_7}   \\
42  &  He~2$-$7$^{\dag}$ & 264.1$-$08.1    & 08 11 31.9        & $-$48 43 14.5    & C      & 3       &  \ref{Fig_7}      \\
43  &  ESO~259$-$10     & 265.1$-$04.2     & 08 34 07.0        & $-$47 16 38.1    & C      & 3       &  \ref{Fig_8}     \\
44  &  PHR0911$-$4205   & 265.4$+$04.2     & 09 11 48.5        & $-$42 05 09.9    & C      & 4       &  \ref{Fig_8}            \\
45  &  PHR0927$-$4347   & 268.6$+$05.0     & 09 27 28.1        & $-$43 47 53.2    & C      & 6       &  \ref{Fig_8}           \\
46  &  PHR0905$-$4753   & 268.9$-$00.4     & 09 05 39.6        & $-$47 53 38.0    & C      & 3       &  \ref{Fig_8}           \\
47  &  He~2$-$37        & 274.6$+$03.5     & 09 47 24.3        & $-$48 58 19.1    & C      & 3       &  \ref{Fig_8}             \\
48  &  PHR0924$-$5506   & 276.1$-$03.3     & 09 24 15.1        & $-$55 06 24.8    & C      & 6       &  \ref{Fig_8}             \\
49  &  NGC~2899         & 277.1$-$03.8     & 09 27 03.5        & $-$56 06 18.3    & E      & 20      &  \ref{Fig_9}          \\
50  &  PHR0958$-$5039   & 277.1$+$03.3     & 09 58 10.2        & $-$50 39 34.9    & C      & 6       &  \ref{Fig_9}               \\
51  &  Wray~17$-$31     & 277.7$-$03.5	   & 09 31 27.1        & $-$56 17 41.1    & E      & 20      &  \ref{Fig_9}	    \\
52  &  He~2$-$32        & 278.5$-$04.5     & 09 30 55.9        & $-$57 36 57.8    & C      & 3       &  \ref{Fig_9}              \\
53  &  PHR0940$-$5658   & 279.1$-$03.1     & 09 40 52.5        & $-$56 58 00.1    & C      & 3       &  \ref{Fig_9}            \\
54  &  PHR1010$-$5146   & 279.2$+$03.5     & 10 10 04.1        & $-$51 46 48.0    & C      & 6       &  \ref{Fig_9}               \\
55  &  He~2$-$36        & 279.6$-$03.1     & 09 43 26.0        & $-$57 16 59.7	  & E	   & 10      &  \ref{Fig_10}   	    \\
56  &  Ste~2$-$1        & 280.0$+$02.9     & 10 11 57.9        & $-$52 38 15.2    & C      & 3       &  \ref{Fig_10}     \\
57  &  PHR1019$-$6059   & 285.5$-$03.3     & 10 19 27.6        & $-$60 59 09.9    & C      & 3       &  \ref{Fig_10}     \\
58  &  PHR1036$-$5909   & 286.3$-$00.7     & 10 36 09.2        & $-$59 09 20.2    & C      & 6       &  \ref{Fig_10}            \\
59  &  He~2$-$55        & 286.3$+$02.8	   & 10 48 43.1        & $-$56 03 23.0	  & E      & 14      &  \ref{Fig_10}     \\
60  &  PHR1046$-$6109   & 288.4$-$01.8     & 10 46 51.1        & $-$61 09 23.0    & C      & 3       &  \ref{Fig_10}       \\
61  &  Hf~38            & 288.4$+$00.3     & 10 54 34.9        & $-$59 09 48.7    & C      & 3       &  \ref{Fig_11}       \\
62  &  PHR1058$-$5853   & 288.7$+$00.8     & 10 58 02.9        & $-$58 53 34.1    & C      & 3       &  \ref{Fig_11}        \\
63  &  Hf~39	        & 288.9$-$00.8	   & 10 53 58.9        & $-$60 26 42.0	  & E	   & 3       &  \ref{Fig_11}      \\
64  &  Hf~48$^*$        & 290.1$-$00.4     & 11 03 55.6        & $-$60 36 05.6    & E      & 20      &  \ref{Fig_11}         \\
65  &  Fg~1 	        & 290.5$+$07.9	   & 11 28 35.9        & $-$52 56 06.4	  & E	   & 10      &  \ref{Fig_11}    \\	
66  &  NGC~3699$^*$     & 292.6$+$01.2     & 11 27 59.2        & $-$59 57 32.0    & C      & 3       &  \ref{Fig_11}       \\
67  &  He~2$-$70$^*$    & 293.6$+$01.2     & 11 35 12.6        & $-$60 16 54.1    & C      & 3       &  \ref{Fig_12}     \\
68  &  BlDz~1	        & 293.6$+$10.9	   & 11 53 03.8        & $-$50 50 41.8	  & E	   & 25      &  \ref{Fig_12} 	    \\
69  &  NGC~3918         & 294.6$+$04.7     & 11 50 18.9        & $-$57 10 51.4    & E      & 20      &  \ref{Fig_12}          \\
70  &  He~2$-$72$^*$    & 294.9$-$00.6     & 11 41 37.9        & $-$62 28 56.9    & C      & 3       &  \ref{Fig_12}         \\
71  &  NGC~3195         & 296.6$-$20.0	   & 10 09 22.0        & $-$80 51 30.6	  & E	   & 10      &  \ref{Fig_12} 	    \\
72  &  He~2$-$76        & 298.2$-$01.7     & 12 08 26.0        & $-$64 12 12.1    & E      & 13      &  \ref{Fig_12}           \\
73  &  PHR1221$-$5907   & 299.0$+$03.5     & 12 21 00.7        & $-$59 07 32.9    & C      & 4       &  \ref{Fig_13}        \\
74  &  K~1$-$23         & 299.0$+$18.4	   & 12 30 52.0        & $-$44 14 22.2    & E	   & 20      &  \ref{Fig_13}     \\
75  &  Wray~16$-$121    & 302.6$-$00.9     & 12 48 31.1        & $-$63 49 56.5    & C      & 3       &  \ref{Fig_13}          \\
76  &  PHR1250$-$6346   & 302.7$-$00.9	   & 12 50 04.4        & $-$63 46 51.9    & E	   & 14      &  \ref{Fig_13}     \\
77  &  Wray~16$-$122    & 304.2$+$05.9     & 13 00 41.4        & $-$56 53 28.3    & C      & 6       &  \ref{Fig_13}       \\
78  &  NGC~5189$^*$     & 307.2$-$03.4     & 13 33 41.9        & $-$65 58 28.9    & E      & 20      &  \ref{Fig_13}       \\
79  &  He~2$-$99        & 309.0$-$04.2     & 13 52 31.0        & $-$66 23 28.0    & E      & 19      &  \ref{Fig_14}      \\
80  &  VBRC~5           & 309.2$+$01.3     & 13 43 59.2        & $-$60 49 40.4    & C      & 3       &  \ref{Fig_14}     \\
81  &  He~2$-$103       & 310.7$-$02.9     & 14 05 36.9        & $-$64 40 57.0    & E      & 20      &  \ref{Fig_14}      \\
82  &  LoTr~7 	        & 310.8$-$05.9	   & 14 15 23.1        & $-$67 31 57.6    & E	   & 14      &  \ref{Fig_14}     \\
83  &  PHR1408$-$6229   & 311.7$-$00.9     & 14 08 47.3        & $-$62 29 57.8    & C      & 3       &  \ref{Fig_14}     \\
84  &  He~2$-$111       & 315.0$-$00.3     & 14 33 18.3        & $-$60 49 44.6    & E      & 14      &  \ref{Fig_14}    \\
85  &  He~2$-$119$^*$   & 317.1$-$05.7     & 15 10 39.9        & $-$64 40 19.1    & E      & 20      &  \ref{Fig_15}       \\
86  &  VBRC~6	        & 317.8$+$03.3	   & 14 41 35.7        & $-$56 15 06.6    & E	   & 25      &  \ref{Fig_15}	   \\
87  &  He~2$-$114       & 318.3$-$02.0     & 15 04 08.8        & $-$60 53 21.2    & E      & 20      &  \ref{Fig_15}    \\
88  &  He~2$-$116       & 318.3$-$02.5     & 15 06 00.8        & $-$61 21 24.4    & E	   & 10      &  \ref{Fig_15}    \\
89  &  ESO~135$-$04     & 318.4$-$03.0     & 15 08 42.8        & $-$61 44 03.9    & E      & 20      &  \ref{Fig_15}    \\
90  &  IC~4406$^*$      & 319.6$+$15.7	   & 14 22 26.5        & $-$44 09 06.0    & E	   & 10      &  \ref{Fig_15}    \\
91  &  He~2$-$120       & 321.8$+$01.9     & 15 11 56.1        & $-$55 39 51.1    & E      & 10      &  \ref{Fig_16}     \\
92  &  Mz~1	        & 322.4$-$02.6	   & 15 34 16.7        & $-$59 08 59.5    & E	   & 10      &  \ref{Fig_16}     \\
93  &  He~2$-$163       & 327.8$-$07.2     & 16 29 30.3        & $-$59 09 22.3    & E      & 12      &  \ref{Fig_16}      \\
94  &  He~2$-$146       & 328.9$-$02.4     & 16 10 40.9        & $-$54 57 31.9    & E      & 20      &  \ref{Fig_16}       \\
95  &  Mz~2             & 329.3$-$02.8     & 16 14 32.1        & $-$54 57 04.0    & E      & 20      &  \ref{Fig_16}    \\
96  &  PHR1557$-$5128 	& 329.7$+$01.4	   & 15 57 07.4        & $-$51 28 00.8    & E	   & 20      &  \ref{Fig_16}   \\
97  &  He~2$-$165	& 331.5$-$03.9	   & 16 29 59.6        & $-$54 09 36.7    & E	   & 12      &  \ref{Fig_17}   \\
98  &  He~2$-$164 	& 332.0$-$03.3	   & 16 29 53.2        & $-$53 23 04.1    & E	   & 14      &  \ref{Fig_17}    \\
99  &  IC~4642 	        & 334.3$-$09.3     & 17 11 45.4        & $-$55 24 02.4	  & E	   & 10      &  \ref{Fig_17}   \\
100 &  Lo~12	 	& 340.8$+$10.8	   & 16 08 26.3        & $-$37 08 48.3	  & E	   & 20      &  \ref{Fig_17}    \\
101 &  NGC~6153         & 341.8$+$05.4     & 16 31 30.9        & $-$40 15 22.5    & E      & 14      &  \ref{Fig_17}     \\
102 &  H~1$-$3		& 342.7$+$00.7	   & 16 53 31.5        & $-$42 39 18.1	  & E	   & 10      &  \ref{Fig_17}     \\
103 &  Pe~1$-$8         & 342.9$-$02.0	   & 17 06 22.8        & $-$44 13 12.1	  & E      & 10      & 	\ref{Fig_18}     \\
104 &  He~2$-$207       & 342.9$-$04.9     & 17 19 32.5        & $-$45 53 10.0    & E      & 18      &  \ref{Fig_18}     \\
105 &  IC~4637          & 345.4$+$00.1	   & 17 05 09.0        & $-$40 52 57.1	  & E	   & 10      &  \ref{Fig_18}     \\
106 &  K~1$-$3$^*$      & 346.9$+$12.4	   & 16 23 17.3        & $-$31 44 56.9	  & E	   & 12      &  \ref{Fig_18}    \\
107 &  K~2$-$16$^{\dag}$ & 352.9$+$11.4    & 16 44 49.1        & $-$28 04 05.4    & E      & 17      &  \ref{Fig_18}     \\
108 &  Wray~16$-$411	& 353.7$-$12.8	   & 18 26 41.5        & $-$40 29 51.1	  & E	   & 14      &  \ref{Fig_18}     \\
 \label{atlas}
\end{longtable}
\end{longtab}


\begin{acknowledgements}
We thank the referee, Romano Corradi, whose very useful remarks helped us to improve this paper.
This work is partially supported by CONICET (PIP 11220120100298).
The CCD and data ac-quisition system at CASLEO has been financed by R. M. Rich through U.S.
NSF grant AST-90-15827.
This work is partially based on observations obtained with the 1.54m
telescope of the Estación Astrofísica de Bosque Alegre,
dependent of the Universidad Nacional de Córdoba, Argentina.
This research has made use of SAO Image DS9, developed by Smithsonian Astrophysical Observatory.
This research made use of the SIMBAD database, operated at the CDS, Strasbourg, France. 
\end{acknowledgements}


\bibliographystyle{bibtex/aa.bst} 
\bibliography{referencias.bib} 


\end{document}